%% file: main.tex
\documentclass[12pt,twoside]{mitthesis}
\usepackage{rotating}
\usepackage{multirow}
\usepackage{amssymb}
\usepackage{amsmath}
\usepackage{lgrind}
\usepackage{subfigure}
\usepackage{lineno}
\usepackage{multirow}
\usepackage{array}
\usepackage{slashed}
\usepackage{cite}
\input{defs.tex}

\usepackage[margin=1.in]{geometry}

\usepackage{url}

\usepackage{cancel}

\usepackage{mathtools}

\usepackage{graphicx}

\usepackage[section] {placeins}

\usepackage{amssymb}
\usepackage{amsmath}

\usepackage{upgreek}

\begin{document}

\include{cover}
\include{signature}
\pagestyle{plain}
\include{contents}

\include{mywords}
\include{Introduction}

\include{CMSExperimentAtLHC}
\include{HggAnaOverview}

\include{Diphoton}

\include{ExclusiveTag}
\include{Class}

\include{StatAna}

\include{Results}

\include{OtherResults}

\include{Conclusion}
\appendix
\include{appa}
\include{appb}
\include{biblio}
\end{document}

%% file: defs.tex
\newcommand{\eV}{\ensuremath{\mathrm{e\kern -0.1em V}}}
\newcommand{\keV}{\ensuremath{\mathrm{ke\kern -0.1em V}}}
\newcommand{\MeV}{\ensuremath{\mathrm{Me\kern -0.1em V}}}
\newcommand{\GeV}{\ensuremath{\mathrm{Ge\kern -0.1em V}}}
\newcommand{\TeV}{\ensuremath{\mathrm{Te\kern -0.1em V}}}
\newcommand{\eVc}{\ensuremath{\mathrm{e\kern -0.1em V/}c}}
\newcommand{\keVc}{\ensuremath{\mathrm{ke\kern -0.1em V/}c}}
\newcommand{\MeVc}{\ensuremath{\mathrm{Me\kern -0.1em V/}c}}
\newcommand{\GeVc}{\ensuremath{\mathrm{Ge\kern -0.1em V/}c}}
\newcommand{\TeVc}{\ensuremath{\mathrm{Te\kern -0.1em V/}c}}
\newcommand{\eVcc}{\ensuremath{\mathrm{e\kern -0.1em V/}c^2}}
\newcommand{\keVcc}{\ensuremath{\mathrm{ke\kern -0.1em V/}c^2}}
\newcommand{\MeVcc}{\ensuremath{\mathrm{Me\kern -0.1em V/}c^2}}
\newcommand{\GeVcc}{\ensuremath{\mathrm{Ge\kern -0.1em V/}c^2}}
\newcommand{\TeVcc}{\ensuremath{\mathrm{Te\kern -0.1em V/}c^2}}

\newcommand{\LB}{\ensuremath{\mathrm{\Lambda_b}}}



%% file: cover.tex
\title{Observation and Measurement of a Standard Model Higgs Boson-like Diphoton Resonance \\
with the CMS Detector}

\author{Mingming Yang}

\department{Department of Physics}

\degree{Doctor of Philosophy}

\degreemonth{September}
\degreeyear{2015}
\thesisdate{June 15, 2015}

\supervisor{Christoph M. E. Paus}{Professor}

\chairman{Professor Nergis Mavalvala}{Associate Department Head for Education}

\maketitle

\cleardoublepage

\setcounter{savepage}{\thepage}
\begin{abstractpage}
\input{abstract}
\end{abstractpage}

\cleardoublepage

\section*{Acknowledgments}
Looking back on the journey of searching for the Higgs boson in its decay into two photons at the CMS experiment at CERN's Large Hadron Collider, I would like to thank first my adviser Christoph Paus, a genuine man with a warm soul, a passionate physicist, and an adviser I would choose again. His unwavering confidence in me has motivated me to pass through this monumental journey and to arrive here. I would also like to thank my other two MIT companions and passionate physicists: Fabian Stoeckli and Josh Bendavid. Starting as their apprentice and growing into their tenured team member has been my great honor.

These three men are not only my teachers and colleagues, but also ``Damon and Pythias''-like friends in my heart. Their guidance, encouragement, support, and collaboration have given me infinite strength and courage, while their devotion has driven me to devote my entire life for this journey as well. The ``24-hour'' discussions with them about physics and technical issues in person, online, or on the phone (Josh only), the many late nights working together with them on the analysis, the secret competitions with them in drinking more coffee and sleeping less, the extreme difficulties and pressures we faced together, and the incomparable excitement we shared when we were getting remarkable results compose an important part of my memory of the past few years. Their unconditional delivery of their knowledge, skills, experiences, and wisdom have nurtured my mind, while their passions, integrities, lively characters, genuine hearts, and warm souls have resonated with my heart. 

I have embedded my deep gratitude and respect to them into all my work, which I want to express now in these limited words.

I would also like to express my sincere gratitude to all my colleagues from the CMS Higgs to Two Photons Working Group. Without them, this journey would not have been as exciting as it was, or simply even not have existed. I am grateful to their constructive competition, great trust, and also strong collaboration,  and I treasure the days working extremely hard together with them to produce and cross-check several rounds of analysis results. It was my deep honor, representing them, to unblind our search result to the entire CMS collaboration, on June 15, 2012, which provided the first convincing evidence for the existence of a new particle. And it is my great pleasure to achieve the end of the journey together with them, with the final result---using the full LHC Run I data with the best calibration---which confirms our results in 2012 and provides the standalone observation of this new particle, with properties consistent with the Higgs boson, from the two photon channel. 

I want to thank the CMS colleagues who have designed, constructed, and calibrated the Electromagnetic Calorimeter. Their huge efforts have provided the great resolution of photon energy measurement, crucial for the Higgs-to-two-photons analysis. And I also want to thank the colleagues for having worked on the tracker which allow us to identify electrons from photons and to use the electrons for validating the signal model. Moreover, I thank all the CMS colleagues for having worked on the different stages and aspects of the experiment during the past 20 years and enabled the final data analysis for the Higgs search. I am grateful for all the advice, help, and encouragements from the colleagues whom I had the opportunity to meet or work with. I would like to thank all the ATLAS colleagues as well for the competition and cross-check. And I also thank all the LHC colleagues for providing the most energetic and intense proton-proton collisions, essential for the creation and observation of the new particle. Furthermore, I thank all the people across the world who have provided support in one way or another to enable the search of the Higgs boson.

I also treasure very much the time that I have spent together with all my MIT colleagues, whose strong support is the irreplaceable source of strength for me during the past few years. The numerous valuable comments and suggestions I having received from them through the group meetings and the MIT analysis email list are integral to this journey. In addition:

--- I am grateful to sit in the office with Guillelmo Gomez Ceballos Retuerto, Marco Zanetti, and Erik Butz. They are not only the experts on analysis, accelerator, and detector to learn from, but also very caring office mates. 

--- I have also learned a lot from the colleagues sitting in the office in front of mine. Si Xie has answered tons of my questions on physics, detector, and computing, with enormous patience and crystal clear explanations. I owe him a big ``Thank You''. I appreciate the advice from Gerry Bauer and Sham Sumorok, who have experienced the progress of high energy physics over a period longer than my life. It has also been my pleasure to work with Jan Veverka, who moved to the office later and became my new Higgs-to-two-photons partner. 

--- I would also like to thank Steve Nahn and Markus Klute, sitting next to my office, for organizing summer BBQs and cheese fondue dinners, which forced the group to stop working and to start talking about topics like favorite novels. 

--- I also want to deliver my many thanks to all my other colleagues for their great help and company: Aram Apyan, Andrew Levin, Duncan Ralph, Katharina Bierwagen, Kristian Hahn, Kevin Sung, Lavinia Darlea, Leonardo Di Matteo, Matthew Chan, Max Goncharov, Olivier Raginel, Phil Harris, Pieter Everaerts, Roger Wolf, Stephanie Brandt, Valentina Dutta, and Xinmei Niu. I miss the afternoon ice cream, the explorations of Geneva restaurants, and the interesting conversations. It is also my pleasure to meet Brandon Allen, Daniel Abercrombie, Dylan Hsu, Sid Narayanan, and Yutaro Iiyama when I am writing this thesis at MIT.

There are more people I need to thank during my PhD period. I want to thank my academic adviser Bolek Wyslouch for all the advice. I would also like to thank all the friends at MIT or CERN, for their help in both academic work and life. I should thank especially Jianbei Liu and Lulu Liu for letting me live on their ``balcony''for free, Hai Chen and Wei Sun for introducing me to the wonderful books in the CERN library, and Lu Feng for her steady friendship and support through the entire period. 

I would also like to thank all the staff working at CERN providing various services. I appreciate the services from the warm staff at the CERN restaurants, and I enjoyed watching Mont Blanc while drinking coffee at Restaurant 1. I also thank the staff at the CERN Fire Brigade for giving me a ride home or helping me open the door of my office. I also want to give a great thanks to the staff at the CERN hostels, for being patient under my many unexpected interruptions near midnight and being able to find me a room. Since I have borrowed so many books from CERN library, I have to thank the staff working there as well.  

\vspace{\baselineskip}
I must thank Gerry again for giving more than one hundred pages of comments on this thesis and flying from Wisconsin to Boston for my thesis defense. I also thank Christoph, Lu, and Eve Sullivan for reading the thesis, and giving valuable comments and corrections.  

\clearpage

I would like to express the deep gratitude from the bottom of my heart to my undergraduate adviser Bing Zhou at the University of Michigan for leading me, a mechanical engineering student, through the gate of high energy physics. She has been not only my adviser in research but also my friend in life. Her help and guidance to me are unmeasurable. I also must give my huge thank to my physics teacher Jean Krisch at the University of Michigan, for her enormous encouragement. I also have to thank Homer Neal at the University of Michigan, together with Jean and Bing, for giving me the opportunity to perform summer research at CERN, which was essential for my decision to come back to CERN for research in graduate school. I also want to thank Alberto Belloni for his guidance during that summer.

I will always remember the year that I spent in the Junior High School Affiliated with Nanjing Normal University when I was 13. I especially thank my beloved teacher Hui Han for her passion and love. I treasure all the wishes from my teachers and classmates when I moved from Nanjing to Shanghai. And I was pleased to explore New York and California together with my old classmate Weisha Zhu many years later, as we used to explore the streets near our school for delicious food.  

I also treasure all the friendships I having received at different times and space in my life. In particular, I should thank Hao Chu for his friendship since elementary school till now. And I thank my dear friend Ziqing Zhai for her understanding and love over the past decade. All her wishes, carried by letters from different places in the world, have been rain drops from the sky dancing cheerfully while deeply into the river of my life, to protect its passion and to follow its adventure.  

\vspace{\baselineskip}
I save this paragraph for my mother Yali Duan and my father Xiaodong Yang. Their infinite love and unconditional support have nurtured my life. I especially thank them for a relaxing and happy childhood with little constraint, and let me grow freely into myself.

I also want to thank all of my family members for their unconditional support during my life. I especially treasure my time at Nanjing saturated with golden color, together with my late Grandparents Lei Zhou and Xingyi Duan, and also my cousin Ran Duan.
 
And I thank the Dingshan Mountain, the Zijinshan Mountain, the Changjiang River, the Xuanwu Lake, and the Xiuqiu Park for nurturing my childhood.

\vspace*{\fill}
\begin{center}
\begin{minipage}{0.4\textwidth}
\textbf{
I also thank\\\\\\\\ 
The grand Jura Mountain\\\\\\\\
For accompanying me\\
In the past few years,\\\\\\
At 1 am, 2 am, 3 am,\\
Under the dark sky\\ 
With infinite unknown,\\\\\\
And also\\\\\\
At 5 am, 6 am, 7 am,\\
In the gentle sunlight\\ 
With infinite hope.\\\\\\\\\\\\\\
I thank deeply\\ 
The Nature.
}
\end{minipage}
\end{center}
\vfill 
\clearpage

This is the end of my story of searching for the Higgs boson. 
\vspace{\baselineskip}

I still remember a conversation with Christoph at CERN Building 32, 4th floor, in 2011. I told him that no matter if we would have a discovery or not, and whether I would get a PhD or not, I would never regret to come and work here. This is my choice of life. And I want to say the same now.

It is thrilling to have been deeply involved with the milestone of the discovery of a Higgs boson-like particle. But having worked with these people in that space at that time itself is thrilling no matter the outcome. To me, connecting to the generations of physicists pursuing a common goal, witnessing and devoting myself to the monumental effort of human beings along with all the colleagues, discovering and interacting with all the beautiful minds and hearts, continuously discovering myself, and feeling the deep harmony with the nature are the most precious parts of the Higgs search Odyssey and far beyond what this thesis could contain. 

All my work is not for a PhD, and even not for the discovery, but for life itself. And it has already been finished long before and has been contained in all the moments. My main motivation to write this thesis is to use this opportunity, to tell these people I have worked with, that I love them. 

The analysis in this thesis could be repeated, but those moments and the stories of these people are not replaceable. Many of the stories are very lively. They probably will never be stated, but they have been detected and stored in my heart. And probably only the people who have experienced and witnessed these stories would feel the deepest resonance.

\vspace{\baselineskip}
The past journey has been wonderful. Especially because I have shared it with some people who, whenever I think of them, bring hot tears to my eyes. I would choose the same way to spend my 20s again and again if I were given the infinite chances to step back and infinite ways to choose. I will continue to discover the future from all the uncertainties, following the sky above me and the road within my heart. I hope this road will lead to the liberation of the soul and enrichment of the spirit of a human being, and of human beings, as what this Higgs search Odyssey should ultimately lead to.

\vspace*{\fill}
\begin{center}
\begin{minipage}{0.6\textwidth}
\textbf{
At this moment,\\\\\\\\\\
My heart\\\\\\
Has already melted\\ 
Into infinite number of protons\\
Colliding at infinite points\\ 
In that space and time,\\ 
Producing infinite number of Higgs bosons\\
Decaying into infinite pairs of photons\\ 
Carrying all my infinite treasuring moments\\\\\\\\ 
Flying\\\\\\\\
Into the future.\\\\\\\\\\\\\\
What is eternity?\\ 
Every moment is eternity.
}
\end{minipage}
\end{center}
\vfill 

%% file: abstract.tex
This thesis concerns the observation of a new particle and the measurements of its properties, from the search of the Higgs boson through its decay into two photons at the CMS experiment at CERN's Large Hadron Collider (LHC), on the full LHC ``Run I'' data collected by the CMS detector during 2011 and 2012, consisting of proton-proton collision events at $\sqrt{s}$ $=$ $7~\mathrm{TeV}$ with $L$ $=$ $5.1~\mathrm{fb^{-1}}$ and at $\sqrt{s}$ $=$ $8~\mathrm{TeV}$ with $L$ $=$ $19.7~\mathrm{fb^{-1}}$, with the final calibration. In particular, an excess of events above the background expectation is observed, with a local significance of 5.7 standard deviations at a mass of 124.7 GeV, which constitutes the observation of a new particle through the two photon decay channel. A further measurement provides the precise mass of this new particle as $124.72_{-0.36}^{+0.35}$ GeV  = 124.72$_{-0.32}^{+0.31}$(stat)$_{-0.16}^{+0.16}$(syst) GeV. Its total production cross section times two photon decay branching ratio relative to that of the Standard Model Higgs boson is determined as $1.12_{-0.23}^{+0.26}$ = 1.12$_{-0.21}^{+0.21}$(stat)$_{-0.09}^{+0.15}$(syst), compatible with the Higgs boson expectation. Further extractions of its properties relative to the Higgs boson, including the production cross section times decay branching ratios for separate Higgs production processes, couplings to bosons and to fermions, and effective couplings to the photon and to the gluon, are all compatible with the expectations for the Standard Model Higgs boson. 

%% file: contents.tex
\tableofcontents

%% file: mywords.tex
\vspace*{\fill}
\begin{center}
\begin{minipage}{1\textwidth}
\textbf{I used to travel at the speed of light till I found the field to slow me down.}
\end{minipage}
\end{center}
\vfill

%% file: Introduction.tex
\chapter{Introduction}
\textbf{The night before Friday, June 15, 2012.} I started writing the ``unblinding'' slides. My colleagues from two different teams continued progressing independently towards the final plots. This was going to be a sleepless night for these searchers in a working group of the Compact Muon Solenoid (CMS) experiment at the Large Hadron Collider (LHC) of the European Organization for Nuclear Research (CERN), searching for the Higgs boson ($H$) through its decay into two photons.

\vspace{\baselineskip}
At this point, the Higgs boson remained as the last undetected elementary particle predicted by the Standard Model (SM) of particle physics\cite{SM1,SM2,SM3,Politzer:1973fx,Gross:1973id}. The Standard Model endeavors to describe the fundamental components of matter and interactions except for gravity---the strong, electromagnetic, and weak interactions---in terms of the corresponding elementary fields and field quanta, the elementary particles, of which spin-1/2 fermions are matter components and spin-1 vector bosons are interaction mediators. The fermions consist of leptons and quarks, while the vector bosons consist of gluons for the strong interaction, photons for the electromagnetic interaction, and $W$ and $Z$ bosons for the weak interaction. The main component of the Standard Model is the electroweak theory, which provides a unified description of the electromagnetic interaction and the weak interaction as the electroweak interaction. The fundamental assumption underlying this theory is the symmetry between these two interactions---the electroweak symmetry. However, the manifest symmetry does not allow the particles associated with the interactions to possess mass. This works for the massless photon, but not for the massive $W$ and $Z$ bosons. To resolve this inconsistency and to formulate the current form of the electroweak theory in the 1960s, a mechanism invented independently by Robert Brout and Fran\c{c}ois Englert, Peter Higgs, and Gerald Guralnik, Carl Hagen, and Tom Kibble\cite{Higgs1,Higgs2,Higgs3,Higgs4,Higgs5,Higgs6} was applied to break the electroweak symmetry. A scalar field permeating over space is introduced, which generates masses of $W$ and $Z$ bosons by interacting with them. This scalar field could also generate masses of quarks and charged leptons through the additional Yukawa interaction. 

The observation of weak neutral currents---mediated by the $Z$ boson---by the Gargamelle experiment at CERN in 1973\cite{Hasert:1973cr,Hasert:1973ff}, and then the direct observations of the $W$ and $Z$ bosons by the UA1 and UA2 experiments at CERN's proton-antiproton collider in 1983\cite{Arnison:1983rp,Banner:1983jy,RevModPhys57699} confirmed the prediction by the electroweak theory of the existences of the $W$ and $Z$ bosons. And these experimental confirmations established this theory as the theoretical cornerstone of particle physics.
Still, the crucial point of the theory lacked experimental evidence---the mechanism for the electroweak symmetry breaking and mass generation. Observation of the quantum of the scalar field, the scalar boson with spin-0, conventionally called the Higgs boson, is the key.

The search for the Higgs boson had been one of the central tasks of experimental particle physicists since the observations of $W$ and $Z$ bosons. The Standard Model predicts the couplings of the Higgs boson to the other elementary particles---proportional to the mass squared of bosons and just to the mass for fermions---so that its production and decay rates at a given mass could be calculated and compared with observations at collider experiments. But the mass of the Higgs boson, $m_{H}$, is not predicted, which adds to the complications of the search. There were indirect constraints on the Higgs mass from the probability conservation of $WW$ scattering, $m_{H}$ $<$ $\sim$$\mathrm{1~TeV}$\cite{TeV1,TeV2,TeV3,TeV4}, and from the precision electroweak measurements, $m_{H}$ $<$ 158 GeV at 95\% confidence level (CL)\cite{PrecisionElectroweakMeasurements}, but still a wide range of Higgs mass hypotheses had to be explored. Before the search at the LHC, direct searches at CERN's Large Electron-Positron Collider (LEP) and Fermilab's proton-antiproton collider Tevatron excluded the mass range $m_{H}$ $<$ 114.4 GeV\cite{LEPHIGGS} and $\mathrm{162~GeV}$ $<$ $m_{H}$ $<$ $\mathrm{166~GeV}$\cite{CombinationofTevatron}, respectively, with no evidence of the Higgs boson. 

The LHC was designed to collide two beams of protons composed of quarks bound by gluons, at center-of-mass energies up to $\sqrt{s}$ = $\mathrm{14~TeV}$ and instantaneous luminosities up to $L_{Inst}$ $=$ $10^{34}~\mathrm{cm}^{-2}\mathrm{s}^{-1}$---about 7 times the collision energy and $\mathcal{O}(10)$ times the intensity of the Tevatron\cite{tevamachine}, the previous most powerful hadron collider---which allows the LHC to create particles with masses up to the TeV scale and to relatively quickly accumulate proton-proton (pp) collision events for physics analyses---with one potential Higgs boson with a mass of 125 GeV produced from about 4 billions of inelastic collisions at 7 TeV\cite{LHCworkinggroup,LHCoutreach}. The design and construction of the LHC\cite{LHC}, along with its two largest experiments CMS\cite{CMS} and ATLAS\cite{ATLAS} (A Toroidal LHC Apparatus) each having a comprehensive particle detector and a collaboration of thousands of physicists and engineers from all over the world, were centered on proving or excluding the existence of the Higgs boson.

Higgs bosons are produced at the LHC through four major processes from pp collisions: gluon fusion (\textit{ggH}), vector boson ($W/Z$ boson) fusion (\textit{VBF}), associated production with a $W$ or $Z$ boson (\textit{VH}), and associated production with a pair of top quarks (\textit{t$\overline{t}$H}). Gluon fusion is the dominant process. The other three processes occur much less frequently than gluon fusion, but with additional particles present along with the Higgs boson, whose features are used to identify the Higgs event. The Higgs boson decays immediately---with a lifetime about $\mathrm{10^{-22}~s}$ for a Higgs mass of $\mathrm{125~GeV}$. The Higgs search is therefore conducted through its decay channels as well as its production processes. There are five main decay channels in terms of the sensitivity to the Higgs search: Higgs decaying into two photons ($H\rightarrow \gamma\gamma$), two $Z$ bosons to four charged leptons ($ZZ \rightarrow 4\ell$) (the charged lepton here refers to electron or muon), two $W$ bosons to two charged leptons and two neutral leptons---two neutrinos ($W^{+}W^{-} \rightarrow 2\ell 2\nu$), and either two tau leptons or two bottom quarks ($\tau^{+}\tau^{-}$ or $ b\overline{b}$). The $H\rightarrow \gamma\gamma$ channel---having a final state of two energetic and isolated photons which are clearly identified and whose energies are measured with excellent resolution---is one of the most promising channels in the search range $\mathrm{110~GeV}$ $<$ $m_{H}$ $<$ $\mathrm{150~GeV}$. The diphoton signature allows the reconstruction of a narrow signal peak in the diphoton mass ($m_{\gamma\gamma}$) spectrum---corresponding to the Higgs resonance with a natural width negligible relative to the detection resolution---on top of a smoothly falling background from known SM physics processes, which yields eloquent evidence if the Higgs boson exists. 

About one year ago before this June night, when the LHC had just ramped up the luminosity of pp collisions at $\sqrt{s}$ $=$ $7~\mathrm{TeV}$ to produce significant amount of data for physics analyses, I came to CERN to work on this thesis---searching for the Higgs boson through the two photon decay channel by analyzing the data collected by the CMS detector, together with my colleagues from MIT and the $H\rightarrow \gamma\gamma$ working group of the CMS collaboration. Despite the appealing feature of a signal peak in the diphoton mass spectrum, the two photon decays are very rare---about one out of five hundred decays from the already rarely produced Higgs boson, assuming a mass of 125 GeV. The great challenge facing this channel is to identify the small peak from a background that is several orders of magnitude larger. 

To fully unfold the power of the diphoton mass spectrum, we made the analysis strategy to classify diphoton events according to the expected signal-to-background ratio ($S/B$) under the peak assuming the existence of the SM Higgs boson. Specifically, we developed the analysis using advanced multivariate analysis (MVA) techniques to fold in all the relevant diphoton information of an event---variables related to photon identification, diphoton kinematics and mass resolution---into a single event classifier, and used it to optimize the classification of the events. The MVA analysis significantly improved the expected Higgs sensitivity, equivalent to adding about more than 40\% of the data, with respect to our initial analysis using the traditional cut-based techniques, which selected and classified diphoton events by applying simple cuts on a subset of MVA input variables. We therefore used the MVA analysis as our main analysis, with the cut-based analysis as a cross-check later. For both analyses, the additional feature of the \textit{VBF} Higgs production process---a pair of jets fragmented from a pair of quarks present in the final state along with the two photons---was utilized to further select events into high $S/B$ classes. Finally, any possible Higgs signal of a mass in $\mathrm{110~GeV}$ $<$ $m_{H}$ $<$ $\mathrm{150~GeV}$ was extracted by a simultaneous likelihood fit to the reconstructed diphoton mass spectra over $\mathrm{100~GeV}$ $<m_{\gamma\gamma}<$ $\mathrm{180~GeV}$ of all the event classes, using parametric signal and background models. For each event class, the signal model for any Higgs mass was derived from simulation, 

and the data/simulation discrepancies are corrected and validated through control samples. The background model was derived directly from the data utilizing the smoothly-falling nature of the background shape. The expected background under the emerging signal peak for any Higgs mass was constrained by the large number of events in the diphoton mass sidebands of the signal region. To cross-check the background modeling, we used an alternative MVA analysis, which extracted the signal by counting events in the signal region---in classes defined by both the diphoton event classifier used in the main MVA analysis and the diphoton mass---and was less sensitive to the exact shape of the background.   

By early 2012, we observed an excess of events above the background expectation with a local significance of about 3 standard deviations at a mass around 125 GeV in $H\rightarrow \gamma\gamma$ channel, from both the cut-based analysis and the newly developed MVA analyses, on the 2011 dataset collected by the CMS detector from pp collision events at $\sqrt{s}$ $=$ $7~\mathrm{TeV}$ with $L$ $=$ $4.8~\mathrm{fb^{-1}}$ (1 barn (b) equals $\mathrm{10^{-24}~cm^{2}}$). Taking into account the probability that the background fluctuated at any mass point within our entire search range, the global significance was below 2 standard deviations\cite{hgg2011cutbased,hgg2011mva}. Among other search channels, the $H\rightarrow ZZ \rightarrow 4\ell$ channel observed an excess near 120 GeV but not as significant\cite{PhysRevLett.108.111804}. And the CMS combined result of the five main decay channels was driven by $H\rightarrow \gamma\gamma$, showing an excess with a local significance just above 3 standard deviations\cite{2011cms}. At the same time, the ATLAS experiment also observed an excess with a local significance above 3 standard deviations near 125 GeV from the combined search, driven by its two photon channel as well, and with a global significance of about 2 standard deviations\cite{2011atlas}. 
   
The excess in $H\rightarrow \gamma\gamma$ channel remained when we rerun our analyses on the full 2011 dataset with the integrated luminosity increased to $L$ $=$ $5.1~\mathrm{fb^{-1}}$ and with improved detector calibration. To determine the source of the excess---an upward fluctuation of background vs. a real signal---the analysis of the 2012 data was critical. We improved and re-optimized our analyses to accommodate the enhanced collision energy to $\sqrt{s}$ $=$ $8~\mathrm{TeV}$ and the increased overlapping pp collisions. To avoid the possibility of biasing the results, we developed the analysis in a strict ``blind'' manner, i.e. we did not look at the diphoton mass spectrum or extract the observed results in the potential signal region $\mathrm{110~GeV}$ $<$ $m_{\gamma\gamma}$ $<$ $\mathrm{150~GeV}$ until our analysis procedure was fixed and fully verified. All the other Higgs search channels also progressed with a ``blind'' procedure as well. 

\vspace{\baselineskip}
Finally, we reached this night before June 15. We had gotten our analyses pre-approved by the collaboration one week ago, by showing the performance of the various components and the expected results of the analyses---both the expected exclusion limit of the signal strength under the background-only hypothesis, and the significance of the excess over background assuming the existence of the SM Higgs boson---on the first 2012 data with $L$ $=$ $1.5~\mathrm{fb^{-1}}$ combined with the 2011 data. And we just obtained the ``green light'' this afternoon after a further review to unblind the cut-based analysis on the 2012 data with $L$ $=$ $3.1~\mathrm{fb^{-1}}$ certified right after the pre-approval. The unblinding of our MVA analyses---the ones into which we were putting the most effort---was postponed until the next week, as we decided to wait for an updated simulation sample important for ``training'' the MVA event classifier earlier this morning around 3:00 am.

I stayed in a room of the CERN hostel this night, within three minutes walk from my office at CERN Building 32. The past couple of weeks were so intense---producing the results for the pre-approval and then working for the ``unblinding''---that I almost lived at CERN, with few hours of sleep at the desk plus $\sim$10 cups of coffee many of the days. What was I feeling when I wrote down the title ``Search For A Standard Model Higgs Boson Decaying Into Two Photons---Unblinding''? Was my heart beating as fast as I am feeling now? Or even faster? Representing the $H\rightarrow \gamma\gamma$ group, I was going to unblind our Higgs search results to the entire CMS collaboration in the coming afternoon---the results still sitting in the dark waiting to be uncovered.

To reach this night, we had gone through many sleepless nights working on the analyses over the past year---from the development of different analysis ingredients, the various corrections and validations for the signal modeling, the large amount of tests and justifications of the background model, to the multiple rounds of producing results on the dataset to keep updated with the increased luminosity and improved calibrations---accompanied by countless meetings, presentations, emails, messages, discussions, and also multiple rounds of documentation preparations. Our different teams, running independent analysis frameworks, had also gone through constructive while sometimes fierce competitions on the analysis methods, but ultimately to striving together for the final results. 

The nights were mixed with mornings for the past few days. To re-optimize and finalize the analysis ingredients for the ``unblinding'' data condition, validate the inputs and outputs, process the datasets and simulation samples to select events and compute variables for the final event classification and diphoton mass spectrum reconstruction, and synchronize the event selection and variable computation among different teams---within a week---equals to a huge effort of the whole group. In the end, two teams synchronized to a satisfactory level. Each team was going to produce a set of ``unblinded'' results to cross-check.

\vspace{\baselineskip}
The night deepened, folding all the sleepless nights into the dark sky, towards the unknown. The slides grew, with analysis descriptions and validation plots, reaching the blank region for the final plots. The last round of cross-checks started between the two teams, with information flowing through an email thread. Finally, the expected results and the event yields agreed. Time to look into the signal region. Around 3:00 am, both teams unblinded the diphoton mass spectra of the 2012 dataset ------

\vspace{\baselineskip}
\vspace{\baselineskip}
\vspace{\baselineskip}
\vspace{\baselineskip}
\vspace{\baselineskip}
\vspace{\baselineskip}
\begin{center}Clear excesses jumping from the falling spectra of multiple event classes\\ around 125 GeV! \\About the same place of the excess that we observed from the 2011 dataset!\end{center}

\vspace*{\fill}
\begin{center}
\begin{minipage}{0.8\textwidth}
\textbf{``A real signal is there!!!'' \\\\\\\\\\
It was not spoken out.\\\\
But I heard\\ 
The yells\\ 
Bursting out,\\ 
From the hearts\\ 
At different ends.\\\\ 
Enormous excitement\\ 
Flowed out,\\ 
Permeating silently\\
In the air.\\\\ 
I would\\
Jump through the window\\
Into the sky,\\ 
With the speed of light;\\\\ 
But in the end,\\ 
Stood together\\ 
With the grand Jura Mountain,\\ 
Quietly upon the ground,\\ 
Watching the pairs of photons\\ 
Passing through layers of nights \\
...}
\end{minipage}
\end{center}
\vfill 

\clearpage

Later in the morning, we had a quick gathering together with more colleagues from the $H\rightarrow \gamma\gamma$ group, in a small meeting room at CERN. Everybody looked extremely excited despite not sleeping much. We tried not to speak loudly, since we had to keep the ``secret'' until the ``unblinding'' event in the afternoon---the event that the unblinded Higgs search results from all the main decay channels were presented to the entire collaboration for the first time. Still more plots to make. We soon went back to the work, with more colleagues from different teams joining to help. At this time, all our hearts were bound together. I kept modifying the slides with suggestions from my colleagues---except for a short break around 11:00 am to meet with the CMS management---while new plots continually came, of various statistical results or diphoton mass spectra, being updated with finer granularity or refined style. As time approached the ``unblinding'' event, I started putting the final version of the plots from my colleagues onto the pages, one after another, each plot a great trust falling silently upon my heart.

Time passed 3:00 pm. I finally finished the slides with the last plot just sent from one of my teammates. Our other colleagues, after this ``super quick and collaborative effort''(quote from said teammate), had left earlier to the ``unblinding'' event, which had already started at the CERN ``Filtration Plant''. The $H\rightarrow W^{+}W^{-} \rightarrow 2\ell 2\nu$ channel was the first to unblind. The $H\rightarrow \gamma\gamma$ channel was the second, starting at 3:30 pm. We saved the slides onto a flash disk and walked quickly to the conference room. Soon, we arrived. My teammate opened the door.

\vspace{\baselineskip}
A hot current flowed out.

The room was packed with CMS colleagues.

\vspace{\baselineskip}
All the seats were taken. 

Many colleagues sat on the floor or stood against the wall. 

There were probably also many colleagues connecting through the video link.

\vspace{\baselineskip} 
``Good Luck!'' said he.

One of the ``Good Luck''s I having received from my colleagues since the morning.

\clearpage
I carried the slides and moved slowly, through the field of colleagues. My body got heavier and heavier, as the lives of more and more people connected to my own

------ My teammates always giving the strongest support; My $H\rightarrow \gamma\gamma$ colleagues striving together for the unblinding over last night and through this day; And the entire group having been working extremely hard together since last year especially during the past couple of weeks; Representing whom I was going to unblind our Higgs search results;

------ The CMS colleagues having worked on different stages and aspects of the experiment during the past 20 years, leading to the final data analysis for the Higgs search---from design, construction, commission, to operation; from hardware, software, to computing; from data taking, calibration, reconstruction, to validation; The colleagues working on the different Higgs decay channels trying to answer the same question; And the colleagues working on the different physics topics from the precision measurements within the Standard Model to the searches beyond the Standard Model, all trying to deepen and enlarge the same drawing of fundamental particle physics; Many of whom were in this room or through the video link, waiting to see and to listen to the results;

------ The ATLAS colleagues working towards the same goal;

------ The LHC colleagues providing the most powerful and intense proton beams; 

------ The generations of experimental physicists searching for the Higgs boson; 

------ The theoretical physicists whose work led to the prediction of this scalar boson about half a century ago; 

------ And all the physicists from the experimental and theoretical sides working together to reach the current understanding of the fundamental components of matter and interactions during the past century;

------ And all the human beings, craving to understand the nature and ourselves, asking and searching, across the vast space and time ...  

Some of us got together, in this space, at this time.

The $H\rightarrow \gamma\gamma$ presentation was starting. 

My heart was beating violently. My mind was calm.

``Please everybody, get ready for the next 15 minutes.'' 

------ These 15 minutes would become a part of our common memories\cite{unblindingtalk}.

This unblinded $H\rightarrow \gamma\gamma$ cut-based results on the combined 2011 ($\sqrt{s}$ $=$ $7~\mathrm{TeV}$, $L$ $=$ $5.1~\mathrm{fb^{-1}}$) and 2012 ($\sqrt{s}$ $=$ $8~\mathrm{TeV}$, $L$ $=$ $3.1~\mathrm{fb^{-1}}$) datasets provided the \textbf{first convincing evidence of the existence of a new particle}---the local significance of the observed excess above the expected SM background was about 4 standard deviations at a mass near $\mathrm{125~GeV}$. 

\vspace{\baselineskip} 
Night of the day, CERN Building 32, fourth floor. 

``We just experienced a historic moment.''  

Yes, we did. Not just in the history of science ...

\vspace{\baselineskip} 
What else from the day I still remember?

The smile from the bottom of everyone's heart.

\vspace{\baselineskip} 
Our final $H\rightarrow \gamma\gamma$ results from the main MVA analysis on the combined 2011 and 2012 datasets, with the 2012 luminosity increased to  $L$ $=$ $5.3~\mathrm{fb^{-1}}$, kept showing an excess of events near 125 GeV, with a local significance of 4.1 standard deviations\cite{longhgg}. This excess was the most significant among all the main decay channels, followed by the excess observed from the $H\rightarrow ZZ \rightarrow 4\ell$ channel with a local significance of 3.2 standard deviations also near 125 GeV\cite{longhgg}. The local significance of the observed excess combining the $H\rightarrow \gamma\gamma$ and $H\rightarrow ZZ \rightarrow 4\ell$ channels reached 5.0 standard deviations. The combined significance of the observed excess of all the five main decay channels was 4.9 standard deviations (updated to 5.0 standard deviations later) near $\mathrm{125~GeV}$\cite{cmsdiscover}. Meanwhile, the ATLAS experiment also observed an excess with a local significance of 5.0 standard deviations (updated to 5.9 standard deviations later) near $\mathrm{125~GeV}$, again with the $H\rightarrow \gamma\gamma$ providing the largest excess with a local significance of 4.5 standard deviations\cite{atlasdiscover}. 

These results led to the announcements of the discovery of a new particle from both experiments at CERN in a joint seminar with the 36th International Conference on High Energy Physics (ICHEP) on July 4, 2012.

This new particle was identified as a boson with integer spin other than 1 because of its decay into two photons.  

Since the discovery of the new particle, we have continued to verify its observation, and to further measure its properties and check its compatibility with the SM Higgs boson, with improved inputs and analyses. In particular, we had about three times more 2012 data collected by the CMS detector before the end of the LHC ``Run I'', with better detector calibration and more accurate simulation. We refined all the major components of the main MVA analysis, from the diphoton event classifier, the event classification procedure, to the modeling of signal and background diphoton mass spectrum. Moreover, we extended the analysis to employ the additional features of all the Higgs production processes to select events in high $S/B$ classes and to separate signal events from different production processes sensitive to Higgs couplings to bosons and to fermions, respectively. These improvements significantly enhance the Higgs search sensitivity---almost doubling the expected significance. They allow precise measurement of the mass of the new particle and extraction of its total production rate relative to that of the Higgs boson (signal strength). They also allow to extract the signal strengths of different Higgs production processes, and to further extract the couplings of the new particle to bosons and to fermions relative to those of the Higgs boson (coupling strengths). 

\vspace{\baselineskip}
This thesis concludes the Odyssey of searching for the Higgs boson through its decay into two photons that I have experienced together with my colleagues since 2011, with a standalone observation of a new particle and the measurements of its mass, signal strengths, and coupling strengths, using the refined and extended main MVA analysis, on the full LHC ``Run I'' data collected by the CMS detector, consisting of pp collision events at $\sqrt{s}$ $=$ $7~\mathrm{TeV}$ with $L$ $=$ $5.1~\mathrm{fb^{-1}}$ in 2011 and at $\sqrt{s}$ $=$ $8~\mathrm{TeV}$ with $L$ $=$ $19.7~\mathrm{fb^{-1}}$ in 2012, with the final calibration.   
\clearpage

More introduction to the Standard Model and the Higgs boson, the Higgs searches at the LHC, and the MVA techniques used in this analysis can be found in the rest of this chapter. The introduction of the CMS detector and the event reconstruction is in Chapter \ref{The CMS Experiment at the LHC}. An overview of this analysis is in Chapter \ref{Higgs Boson to Two Photons Analysis Overview}. The further descriptions of the analysis components are in Chapter \ref{chaper:Diphoton Reconstruction and Selection}-\ref{chap:Statistical Analysis}. The final $H\rightarrow \gamma\gamma$ results are in Chapter \ref{chap:Results}, followed by a review of other Higgs results from the CMS and ATLAS experiments in Chapter \ref{chap:OtherResults}, and the conclusion in Chapter \ref{Conclusion}. The natural units i.e. $\hbar$ = $c$ = 1 are used throughout this thesis.

\vspace{\baselineskip} 
Again, the final results from the main MVA analysis are produced and cross-checked by two highly synchronized analysis frameworks in the CMS $H\rightarrow \gamma\gamma$ group, and cross-checked by alternative cut-based and MVA analyses. More details of the main MVA analysis and the descriptions of the alternative analyses, are in our analysis note\cite{hggan13253} and paper\cite{hggfinalpaper}, where the results presented are randomly chosen from one of the frameworks. Additional results including hypothesis tests between spin-0 and spin-2 models are also in the note/paper, which are all consistent with the SM Higgs boson.  

\vspace*{\fill}
\begin{center}
\begin{minipage}{0.9\textwidth}
Again, there have been many sleepless nights, which are now only in our memories.
\end{minipage}
\end{center}
\vfill 

\clearpage

\section{The Standard Model and the Higgs Boson}
The Standard Model\cite{SM1,SM2,SM3,Politzer:1973fx,Gross:1973id}, based on the relativistic quantum gauge field theory, describes the elementary particles and their interactions except for gravity. Elementary particles are depicted as the quanta of excitation of their corresponding fields, which include spin-1/2 fermions as fundamental components of matter and spin-1 vector bosons as mediators of interactions. The spin-1/2 fermions consisting of leptons and quarks are grouped into three generations with the higher generation a heavier copy of the lower one, as summarized in Table \ref{tab:fermion}. The vector bosons mediating three kinds of interactions, weak, electromagnetic, and strong---listed in increasing strength---are shown in Table \ref{tab:boson}. All quarks and leptons participate in weak interactions. The electrically charged particles including charged leptons, quarks, and $W^{\pm}$ participate in electromagnetic interactions. Quarks and gluons, which carry color charge, participate in strong interactions. 

\begin{table}[htbp]
  \renewcommand{\arraystretch}{1.2}
  \newcolumntype{L}[1]{>{\raggedright\arraybackslash}p{#1}}
  \newcolumntype{C}[1]{>{\centering\arraybackslash}p{#1}}
  \newcolumntype{R}[1]{>{\raggedleft\arraybackslash}p{#1}}
  \caption{Spin-1/2 fermions: leptons and quarks (and corresponding anti-particles) in three generations.}
  \begin{center}
    \begin{tabular}{|l|C{1.8cm}C{1.8cm}|C{1.8cm}C{1.8cm}|C{1.8cm}C{1.8cm}|}
      \hline
      Generation &  \multicolumn{2}{ |c| }{I}  &   \multicolumn{2}{ |c| }{II}  &   \multicolumn{2}{ |c| }{III} \\
      \hline
     
      \multirow{2}{*}{Leptons} & Electron Neutrino & $\nu_{e}$ ($\bar{\nu}_{e}$) &  Muon Neutrino & $\nu_{\mu}$ ($\bar{\nu}_{\mu}$) & Tau Neutrino & $\nu_{\tau}$ ($\bar{\nu}_{\tau}$)  \\ 
      \cline{2-7}
     
      & Electron & $e^{-}$ ($e^{+}$) & Muon & $\mu^{-}$ ($\mu^{+}$) & Tau & $\tau^{-}$ ($\tau^{+}$) \\ 
      \hline
      \multirow{2}{*}{Quarks}& Up & $u$ ($\bar{u}$) & Charm & $c$ ($\bar{c}$) & Top & $t$ ($\bar{t}$)\\ 
      \cline{2-7}
      & Down & $d$ ($\bar{d}$) & Strange & $s$ ($\bar{s}$) & Bottom & $b$ ($\bar{b}$) \\ 
      \hline
    \end{tabular}
  \end{center} \label{tab:fermion}
\end{table}   
\begin{table}[htbp]
  \renewcommand{\arraystretch}{1.1}
  \newcolumntype{L}[1]{>{\raggedright\arraybackslash}p{#1}}
  \newcolumntype{C}[1]{>{\centering\arraybackslash}p{#1}}
  \newcolumntype{R}[1]{>{\raggedleft\arraybackslash}p{#1}}
  \caption{Spin-1 vector bosons and their corresponding interactions.}
  \begin{center}
    \begin{tabular}{|cc|c|}
      \hline
      \multicolumn{2}{ |c| }{ Vector Boson} &  Interaction  \\
      \hline
      $W$ boson & $W^{\pm}$ & Weak\\ 
      \hline
      $Z$ boson & $Z$ & Weak\\ 
      \hline
      Photon & $\gamma$ & Electromagnetic\\
      \hline
      Gluon & $g$ & Strong\\
      \hline
    \end{tabular}
  \end{center} \label{tab:boson}
\end{table}   

The fundamental mechanism underlying the Standard Model is to generate interactions by requiring local gauge symmetries. In particular, its symmetry group is $SU(3)_{c}$ $\otimes$ $SU(2)_{L}$ $\otimes$ $U(1)_{Y}$, in which $SU(3)_{c}$ determines the strong interaction while $SU(2)_{L}$ $\otimes$ $U(1)_{Y}$ determines the electroweak interaction. Though the insight of symmetry enables the derivation of the interactions in a systematic way, it forces the bosons mediating the interactions to be massless, which is consistent with the massless photon and gluon but apparently not with the massive $W$ and $Z$ bosons. A direct breaking of the symmetry would allow for massive $W$ and $Z$ bosons but make the theory no longer renormalizable, i.e. the infinities in the calculation of observables are not removed. To solve this inconsistency, the Higgs mechanism\cite{Higgs1,Higgs2,Higgs3,Higgs4,Higgs5,Higgs6} is employed instead to preserve the renormalizability and at the same time allow for massive $W$ and $Z$ bosons. It introduces a doublet of complex scalar fields, which has a symmetric potential under $SU(2)_{L}$ $\otimes$ $U(1)_{Y}$ and degenerate vacuum states with non-zero expectation values. The $SU(2)_{L}$ $\otimes$ $U(1)_{Y}$ (electroweak) symmetry is spontaneously broken by choosing a particular vacuum state, while the renormalizability of the theory is kept\cite{Hooft1971167}. Only one of the four real scalar fields in the doublet remains, which is the Higgs field. $W$ and $Z$ bosons then get mass through the interaction with the Higgs field, and the degrees of freedom of the three disappearing scalar fields turn into the longitudinal polarizations of $W$ and $Z$. This spontaneous symmetry breaking would also provide mass for fermions, except for neutrinos whose mass generation mechanism is unknown, by adding Yukawa interaction between fermions and the Higgs field. The particle corresponding to the excitation of the Higgs field is the Higgs boson ($H$). It is neutral, colorless, and has spin ($J$), parity ($P$) and charge conjugation ($C$) $J^{PC}=0^{++}$. The Standard Model does not predict the mass of the Higgs boson but its couplings to bosons and fermions, which are proportional to the boson mass squared and to the fermion mass, respectively. With the couplings provided, the Higgs cross section for any production process, and its width and corresponding branching ratio for any decay mode are predicted for any Higgs mass hypothesis, $m_{H}$. For more detailed introduction on the Standard Model and Higgs boson see References\cite{HM,bettini,nonacc}. 

In case a signal is observed, the Standard Model Higgs production cross sections and decay branching ratios at a given mass hypothesis, and its couplings are compared to the experimental observations to quantify the compatibility between the signal and the Higgs boson. For example, in the search of the Higgs boson through one of its decay modes, the compatibility between the observed signal and the Higgs boson is first quantified by extracting its relative total cross section for all production processes times branching ratio with respect to the Standard Model Higgs prediction, namely the signal strength, $\mu_{H}$. Given sufficient data, the signal strength for each production process is extracted to make a more detailed comparison. Depending on the available production processes and the decay mode, the compatibility is further quantified by measuring the relative coupling (coupling strength) to bosons, $\kappa_{V}$, the relative coupling to fermions, $\kappa_{f}$, or both\cite{LHCHiggsCrossSectionWorkingGroup3}.

The last elementary particle of the Standard Model missing experimental confirmation has been the Higgs boson. The search of the Higgs boson is one of the central tasks for experimental particle physics, as its experimental observation is crucial to verify the current understanding of the electroweak symmetry breaking. Prior to the observation of the Higgs boson-like excess in 2012 at the LHC, searches at the Large Electron-Positron Collider (LEP) excluded the Standard Model Higgs boson below a mass of 114.4 GeV (95$\%$ confidence level)\cite{LEPHIGGS}. These exclusions were extended in 2012 by searches at the Tevatron, which excluded 100 GeV $<$ $m_{H}$ $<$ 103 GeV and 147 GeV $<$ $m_{H}$ $<$ 180 GeV (95\% confidence level); but also reported a small (3.0 standard deviations) excess at $m_{H}$ $=$ 120 GeV shortly before the LHC observation\cite{tevahiggs}.    

\section{Search for the Higgs Boson at LHC}
\label{sec:Search for the Higgs Boson at LHC}
The Large Hadron Collider (LHC)\cite{LHC}, constructed by the European Organization for Nuclear Research (CERN), is the highest energy collider of protons (or heavy ions) and allows the study of the physics at the TeV scale. Four major experiments are conducted at LHC, ALICE\cite{ALICE}, ATLAS\cite{ATLAS}, CMS\cite{CMS} and LHCb\cite{LHCb}. ATLAS and CMS use multi-purpose detectors and explore a broad range of particle physics topics, with the search for Higgs boson as one of the main goals.    
  
\begin{figure}[hbpt] 
  \begin{center}
    \includegraphics[width=1\textwidth]{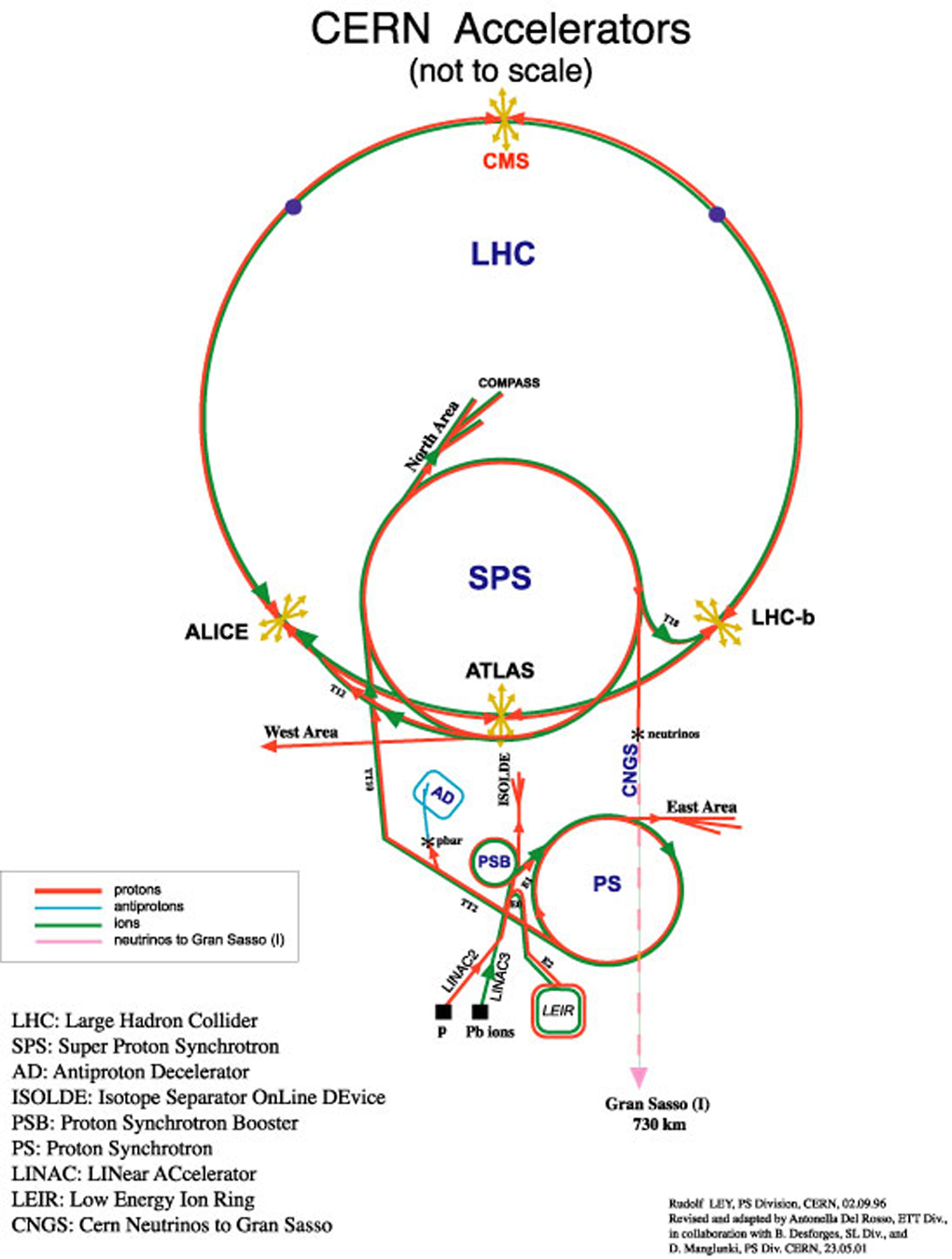}
  \end{center}
  \caption{The accelerator complex of CERN.}
  \label{fig:lhccomplex}
\end{figure}

The LHC is the last element of the CERN accelerator complex as shown in Figure \ref{fig:lhccomplex}\cite{lhccomplex}. It is installed in a circular tunnel with 27 km in circumference, which ranges from 45 m to 170 m in depth beneath the surface at the outskirts of Geneva. It mainly consists of 8 radio frequency cavities for acceleration of each particle beam, 1232 superconducting dipole magnets for beam bending, and 392 superconducting quadruple magnets for beam focusing. The magnets, cooled by superfluid helium to 1.9 K, are designed to provide a magnetic field of 8.33 T. Protons, extracted from hydrogen gas, are first accelerated by a successive set of accelerators and then injected separately into the two beam pipes of the LHC. The two proton beams are designed to run oppositely with 2808 proton bunches per beam and about 10$^{11}$ protons per bunch, which collide (bunch crossing) every 25 ns at center-of-mass energy of up to $\sqrt{s}$ = 14 TeV and with a peak instantaneous luminosity $L_{Inst}$ $=$ $10^{34}~\mathrm{cm}^{-2}\mathrm{s}^{-1}$. The actual bunch crossing rate is every 50 ns, and the collision energy and peak instantaneous luminosities are 7 TeV and about $4~\times~10^{33}~\mathrm{cm}^{-2}\mathrm{s}^{-1}$ in 2011, and $\mathrm{8~TeV}$ and about $8~\times~10^{33}~\mathrm{cm}^{-2}\mathrm{s}^{-1}$ in 2012. The high instantaneous luminosity leads to the presence of inelastic pp interactions with low momentum transfer (pileup interactions) in the same bunch crossing with the interesting inelastic pp interaction with large momentum transfer (hard interaction). The interactions are distributed in space approximately following three dimensional Gaussian distribution. The corresponding standard deviation in the beam direction and in its perpendicular directions, is about 6 cm (5 cm) for 7 TeV (8 TeV), and $\mathcal{O}(10\:\mu m)$, respectively.

\begin{figure}[hbpt] 
  \begin{center}
    \includegraphics[width=0.75\textwidth]{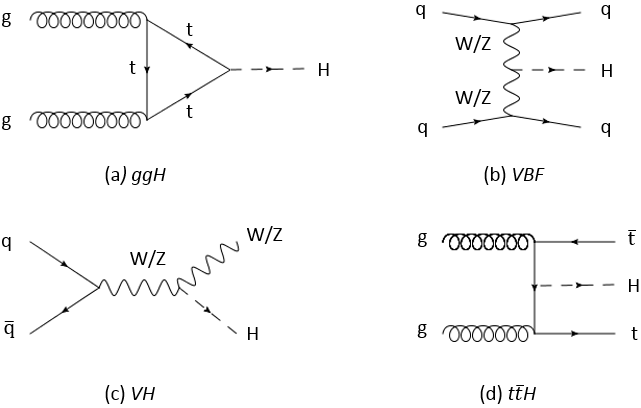}
  \end{center}
  \caption{Leading order Feynman diagrams for Higgs production processes: (a) \textit{ggH} (gluon fusion) (b) \textit{VBF} (vector boson fusion) (c) \textit{VH} (associated production with a $W$ or $Z$) (d) $t\bar{t}H$ (associated production with $t\bar{t}$).}
  \label{fig:higgsproduction}. 
\end{figure}

\begin{figure}[hbpt] 
  \begin{center}
    \includegraphics[width=0.49\textwidth]{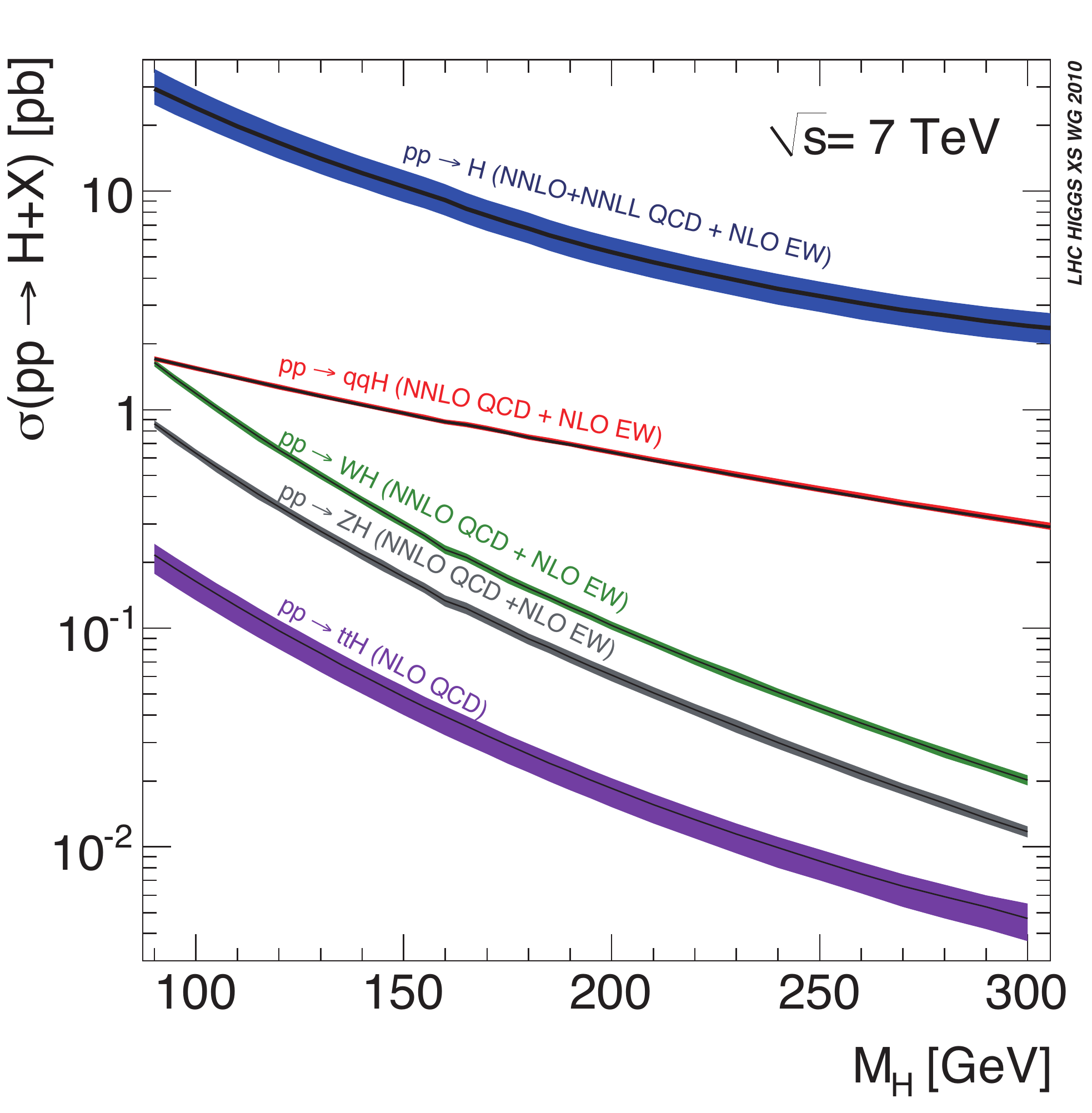}
    \includegraphics[width=0.49\textwidth]{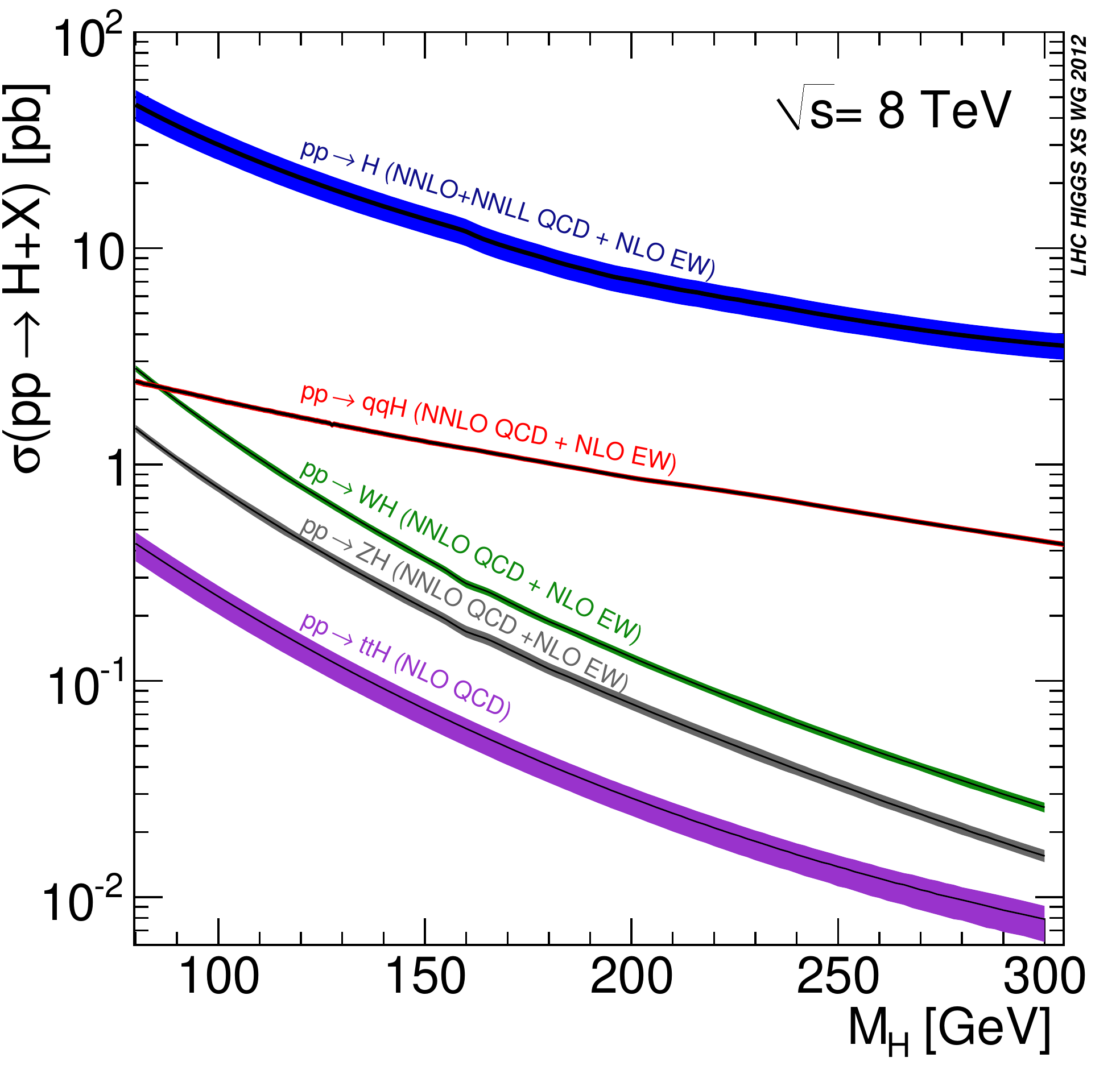}
  \end{center}
  \caption{The Higgs production cross section for each process as a function of Higgs mass $m_{H}$ at 7 TeV (left) and 8 TeV (right), along with the theoretical uncertainty bands. From top to bottom: \textit{ggH}, \textit{VBF}, \textit{WH}, \textit{ZH}, \textit{t$\overline{t}$H}.}
  \label{fig:higgscrosssection}. 
\end{figure}

Higgs bosons are produced at the LHC through the interactions of the partons from the incoming protons. The main production processes, in decreasing order of cross sections, are gluon fusion (\textit{ggH}), vector boson fusion (\textit{VBF}), associated production with a $W$ or $Z$ (\textit{WH} or \textit{ZH}, \textit{VH} for the combined \textit{WH} and \textit{ZH}) and associated production with $t\overline{t}$ (\textit{t$\overline{t}$H}). The corresponding leading order Feynman diagrams are shown in Figure \ref{fig:higgsproduction}, and the cross section for each process as a function of Higgs mass $m_{H}$ at 7 TeV (8 TeV) is shown on the left (right) in Figure \ref{fig:higgscrosssection}\cite{LHCHiggsCrossSectionWorkingGroup3,LHCworkinggroup}. 

Gluon fusion is the dominant process, whose cross section for $m_{H}$ = 125 GeV at 8 TeV is 19.27 pb, about 7 times the sum of the cross sections of all the other processes. In this process, two gluons produces a Higgs boson through a loop of quarks, mainly the heavy top quark. This indirect production is due to the fact that the gluon is massless and the Higgs boson couples to a boson proportional to its mass squared. The production rate of \textit{ggH} process relative to the Standard Model expectation is proportional to $\kappa_{f}^{2}$. It is also sensitive to the existence of any new colored particles in the loop too heavy to be produced directly, which manifests as a deviation of the effective Higgs coupling strength to gluon, $\kappa_{g}$\cite{LHCHiggsCrossSectionWorkingGroup3}, $\mathrm{from~1}$. 

The cross sections for \textit{VBF}, \textit{VH} and \textit{t$\overline{t}$H} processes are much smaller than that for \textit{ggH} process, which for $m_{H}$ = 125 GeV at 8 TeV are 1.578 pb, 1.1199 pb and 0.1293 pb, respectively. Despite the low production rates of \textit{VBF}, \textit{VH} and \textit{t$\overline{t}$H} processes, they are interesting processes to be deployed for two reasons. First, in these processes, the Higgs boson is produced along with other particles whose signature is used to identify the events, and thus improves the signal to background ratio. Furthermore, these processes provide additional information in Higgs coupling to bosons and fermions. In the \textit{VBF} process, two quarks radiate $W$ bosons or $Z$ bosons, which annihilate to produce the Higgs boson. A pair of quarks are present in the final state moving oppositely close to the beam direction, which fragment into two jets with large opening angle. In the \textit{VH} process, a quark and an anti-quark produces a $W$ or $Z$ boson which in turn radiates a Higgs boson. The $W$ or $Z$ further decays leptonically or hadronically. For the leptonic decay, a lepton (muon or electron) plus a neutrino are produced from the $W$ while a pair of leptons are produced from the $Z$. For the hadronic decay, a pair of quarks are produced which fragment to a pair of jets. Both the production rates of \textit{VBF} and \textit{VH} processes relative to the Standard Model expectation are proportional to $\kappa_{V}^{2}$. In the \textit{t$\overline{t}$H} process, two gluons produce a pair of top and anti-top quarks, and a Higgs boson in association. Each top quark decays to a bottom quark plus a $W$ boson. The bottom quark fragments to a so called b-jet, while the $W$ boson decays in the way as mentioned above. The production rate of \textit{t$\overline{t}$H}  relative to the Standard Model expectation is proportional $\kappa_{f}^{2}$, as that of \textit{ggH}. 

\begin{figure}[hbpt] 
  \begin{center}
    \includegraphics[width=0.49\textwidth]{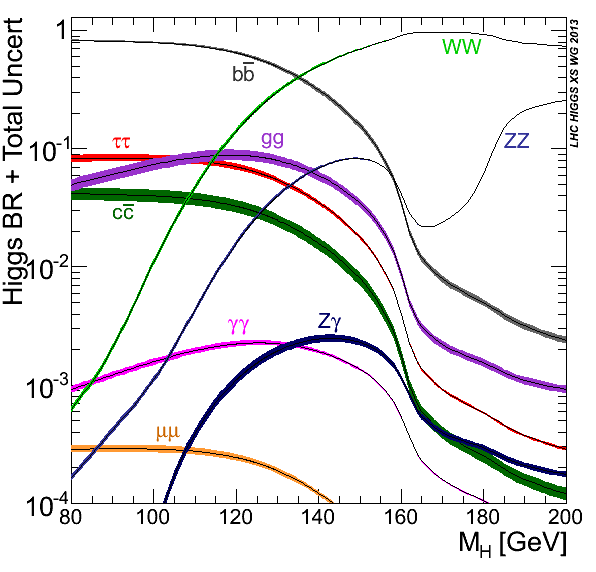}
    \includegraphics[width=0.49\textwidth]{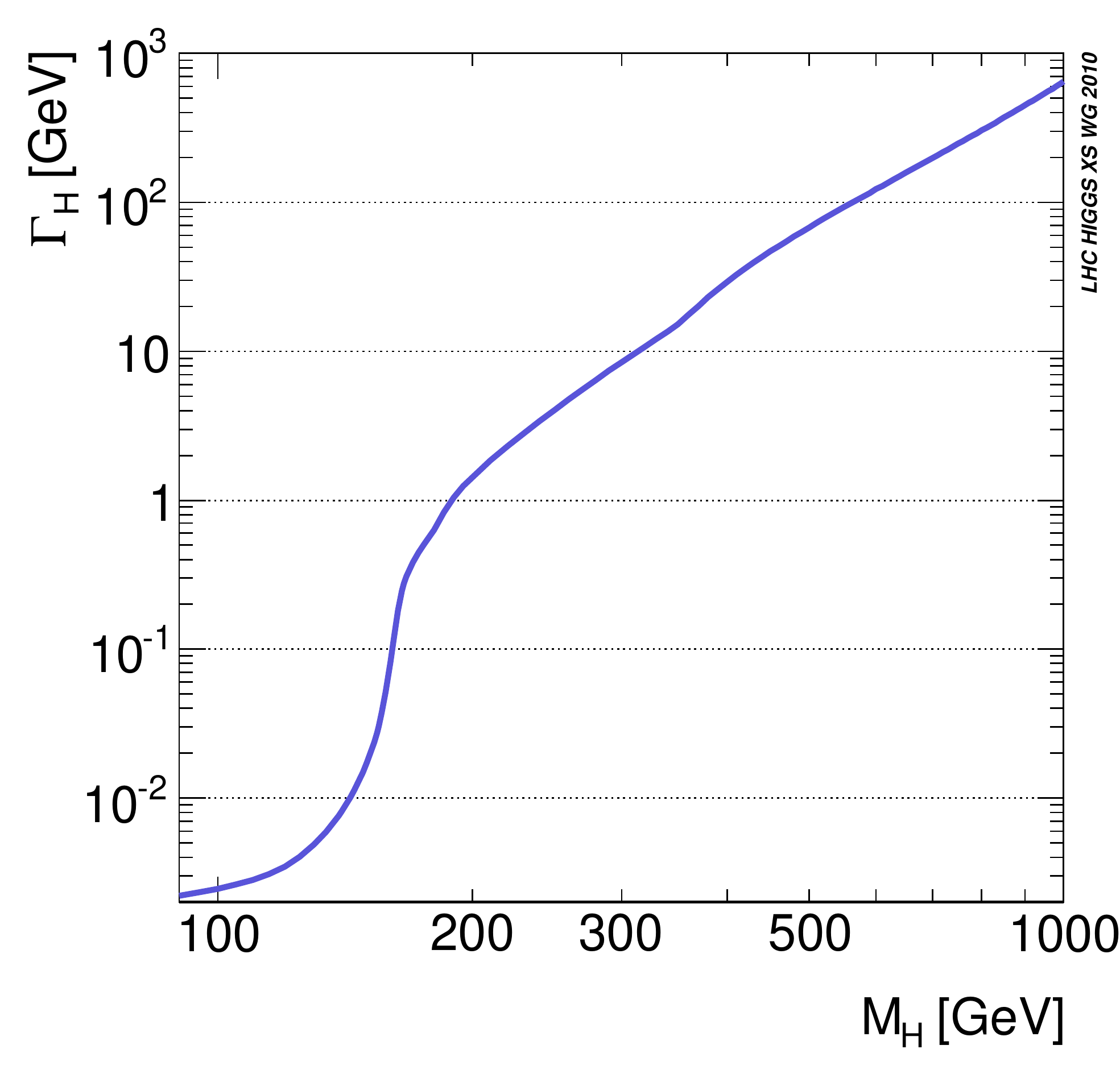}
  \end{center}
  \caption{Higgs decay branching ratios for the various channels (left), and total decay width (right).}
  \label{fig:higgsbr}. 
\end{figure}

The Higgs boson---whose lifetime is about $\mathrm{10^{-22}~s}$ at $m_{H}$ = 125 GeV---decays immediately after its production. The Higgs search is therefore conducted through its decay channels as well as production processes as explained. The main channels in terms of the sensitivity include $H\rightarrow W^{+}W^{-}$, $H \rightarrow ZZ$, $H\rightarrow \gamma\gamma$ where the Higgs boson decays to a pair of bosons, and $H \rightarrow b\overline{b}$ and $H\rightarrow \tau^{+}\tau^{-}$ where the Higgs boson decays to a pair of fermions. The Higgs decay branching ratios in the $m_{H}$ range between 80 GeV and 200 GeV are shown on the left in Figure \ref{fig:higgsbr}\cite{LHCHiggsCrossSectionWorkingGroup3,LHCworkinggroup}. The $H \rightarrow b\overline{b}$ channel dominates in the $m_{H}$ range well below the $W$$W$ production threshold. The $H\rightarrow W^{+}W^{-}$ channel and $H \rightarrow ZZ$ channel are dominant in the $m_{H}$ range just below and beyond this threshold, because $W$ and $Z$ have much larger mass than the other decay particles and so larger couplings to the Higgs boson. Comparing to the other four channels, the $H\rightarrow \gamma\gamma$ channel has a much smaller branching ratio across the mass range, which reaches the maximum between $\mathrm{120~GeV}$ and 130 GeV and has the value 0.228$\%$ at $m_{H}$ = 125 GeV. Despite its small branching ratio, the $H\rightarrow \gamma\gamma$ channel has a clear signature with two energetic and isolated photons, and allows the reconstruction of the narrow Higgs resonance in the diphoton mass spectrum. This makes it one of the most sensitive channels for the Higgs discovery in the low mass range, and also one of the only two channels---the other is $H \rightarrow ZZ$ with four leptons in the final state ($H\rightarrow ZZ \rightarrow 4\ell$)---for the precision measurement of the Higgs mass. In addition, its production rate is sensitive to both the Higgs coupling to bosons and fermions as well as the existence of new charged heavy particles. We therefore choose to search for the Higgs boson through the $H\rightarrow \gamma\gamma$ channel. More details of this channel and our search strategy are given below.

\subsection{Higgs Boson to Two Photons Decay Channel}
\label{sec:Higgs Boson to Two Photons Decay Channel}

\subsubsection{Signal and Background}
\label{sec:Signal and Background}
\begin{figure}[hbpt] 
  \begin{center}
    \includegraphics[width=0.96\textwidth]{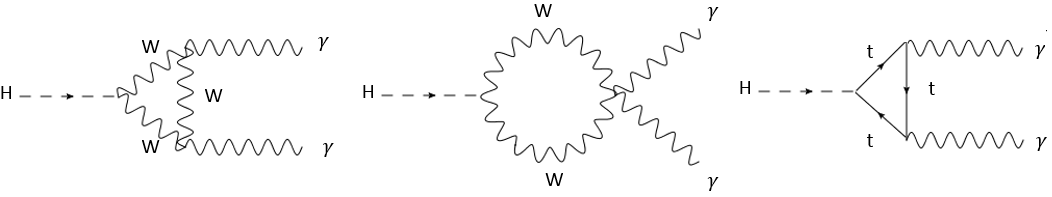}
  \end{center}
  \caption{Leading order Feynman diagrams for Higgs boson decaying to two photons.}
  \label{fig:hggfeynman}. 
\end{figure}

The Higgs boson decays to two photons through a loop of massive charged particles, mainly $W$ boson and top quark, since the photon is massless while the Higgs boson only couples to massive particles. The leading order Feynman diagrams are shown in Figure \ref{fig:hggfeynman}, where the $W$ loop and top quark loop interfere destructively. The loop makes the decay rate of the two photon channel smaller than those of the other four main channels, for which the Higgs boson couples directly to the vector boson or the fermion at the leading order.   

Even though the Higgs boson decays to two photons at a very small rate, this channel is one of the most sensitive channels for the Higgs search in the low $m_{H}$ region thanks to the two isolated high energy photons in the final state. The photons---each carrying an energy of 62.5 GeV for a Higgs boson with $m_{H}$ = 125 GeV decaying at the rest---are clearly detected and identified. Their energies and momenta are well measured, from which the diphoton mass, $m_{\gamma\gamma}$, is reconstructed using the kinematic formula: 
\begin{equation}
  m_{\gamma\gamma} = \sqrt{2E^{\gamma1}E^{\gamma2}(1-cos(\theta_{\gamma\gamma}))},
  \label{eqn:diphotonmass}
\end{equation}    
where $E^{\gamma1}$ and $E^{\gamma2}$ are the measured single photon energies and $\theta_{\gamma\gamma}$ is the measured angle between the momenta of the two photons. 

The total decay width of the Higgs boson in the interested $m_{H}$ range is very narrow---about 4 MeV at $m_{H}$ = 125 GeV---as shown on the right in Figure \ref{fig:higgsbr}\cite{LHCHiggsCrossSectionWorkingGroup3,LHCworkinggroup}. A good resolution of both the measured photon energy and the measured open angle therefore leads to a narrow peak of diphoton mass spectrum associated with the Higgs resonance. Because the distribution of the background events is expected to be continuously falling, this narrow peak provides an eloquent evidence to the existence of the Higgs boson. 

From the amplitude and the location of the peak, the relative total Higgs production cross section times the branching ratio to two photons with respect to the Standard Model Higgs expectation---the signal strength, and the Higgs mass are measured precisely. The rate of the decay, mediated through a loop of particles involving $W$ boson and top quark, is sensitive to the magnitudes of both $\kappa_{V}$ and $\kappa_{f}$ as well as their relative sign---same sign for destructive interference between $W$ loop and top quark loop as expected by the Standard Model while opposite sign for constructive interference. It is also sensitive to any possible new heavy charged particles in the loop, whose existence is quantified through measuring the effective Higgs coupling strength to photon, $\kappa_{\gamma}$\cite{LHCHiggsCrossSectionWorkingGroup3}.

The dominant background consists of ``irreducible'' and ``reducible'' components. The ``irreducible'' component is real (prompt) diphoton events. The ``reducible'' component includes dijet and $\gamma$ + jet events, in which jets are misidentified as photons. A jet typically fakes a photon when it results in a narrow concentration of photonic energy in the detector due to the decays of high energy neutral mesons, especially $\pi^{0}$'s. The $\pi^{0}$ decays into two photons with small opening angle, which may appear as a single photon. 

\subsubsection{Factors for Sensitivity}
\label{sec:Factors for Sensitivity}
The main challenge for the Higgs search through the two photon channel is that the signal is much smaller than the background. The expected inclusive signal ($S$) to background ($B$) ratio $S/B$ under the signal peak, at a Higgs mass of $\mathrm{125~GeV}$, is about 2\% for events at $\mathrm{8~TeV}$ preselected for the final analysis, as evaluated from the numbers in $\mathrm{Table~\ref{tab:sigbkg}}$. In order to achieve optimal sensitivity of the Higgs search and properties measurements, we need to separate the signal and background as much as possible, and further we need to understand the background under the signal peak, well.   

The good separation between signal and background depends on the following factors related to photons: 
\begin{itemize}  
\item Good diphoton mass resolution for a narrow diphoton mass peak---requiring good resolution of both single photon energy and diphoton opening angle.
\item Effective separation between prompt photon and a jet faking a photon.
\item Utilization of differences in diphoton kinematics between signal and background.
\end{itemize}
In addition, the selections of \textit{VBF}, \textit{VH} and \textit{t$\overline{t}$H} events according to the features of other physics objects produced along with the diphoton, the so called Higgs production tags, are another important factors for signal/background separation. Furthermore, these production process tags also separate the different signal production processes sensitive to different Higgs couplings, which allows the measurement of the signal strengths for individual processes and improves the sensitivity of the Higgs coupling strengths. 

\subsubsection{Search Strategy}
\label{sec:Search Strategy}
We use the Compact Muon Solenoid (CMS) detector to detect diphoton events from the pp collisions. The CMS detector, with a homogeneous and fine-grained electromagnetic calorimeter (ECAL), allows us to identify the photons and to measure their energies with high resolution. The photon momentum is obtained using the direction from the reconstructed vertex of the associated diphoton production to the photon location in the ECAL, since the photon trajectory is not directly measured. The diphoton vertex is selected from all the  vertices in a bunch crossing, and the efficiency of selecting the correct vertex drives the resolution of the diphoton opening angle. The multiple sub-detectors of CMS further allow the reconstruction of other physics objects used for the Higgs production tags,  including electrons, muons, jets and the signature of neutrinos---the imbalance to the total momentum projection in the transverse plane with respect to the beam direction (the transverse missing energy).

We design our analysis to maximally separate the signal and background by optimizing the diphoton mass resolution for a given phase space and classifying diphoton events according to expected $S/B$ under the signal diphoton mass peak. We use Multivariate Analysis (MVA) techniques, especially Boosted Decision Trees (BDT)\cite{friedman2000,friedman2001,tmva} as introduced in Section \ref{sec:Boosted Decision Trees}, to address the key photon factors as follows:
\begin{itemize}  
\item Correct the single photon energy and select the diphoton vertex with high efficiency to narrow the expected diphoton mass peak for a given phase space. 
\item Estimate the energy resolution of each photon and the probability of selecting the right diphoton vertex to build a diphoton mass resolution estimator.
\item Combine all the single photon level information into a photon identification classifier between prompt photon and fake photon.
\item Combine all the diphoton event level information, including the diphoton mass resolution estimator, the photon identification classifier for each photon, and diphoton kinematics, into a diphoton event classifier which provides a measure of $S/B$.
\end{itemize}
We then use the features of other physics objects produced along with the diphoton to select the events into high $S/B$ Higgs production tagged classes, and use the diphoton event classifier to select the untagged events into classes with boundaries optimized for the Higgs sensitivity.          

We finally extract the Higgs signal by simultaneous likelihood fit to the diphoton mass spectra of all event classes. The expected background under the emerging signal mass peak for each event class is constrained directly by the large number of events from data in the sidebands of signal region, utilizing the smoothly falling nature of the background shape.

\section{Boosted Decision Trees}
\label{sec:Boosted Decision Trees}
Boosted Decision Trees\cite{friedman2000,friedman2001,tmva} is one of the popular MVA techniques, which are employed in experimental particle physics to estimate a function mapping a set of input variables of an event to its identity as signal or background (classification), or to the value of its certain property (regression). We choose to use BDT in this analysis for its ability to handle large number of input variables and their correlations, as well as its simple mechanism. We use BDT to combine all the relevant information in an event into a single variable, which maximally separates signal from background for classification, or precisely and accurately estimates the target property for regression.

\begin{figure}[h] 
  \begin{center}
    \includegraphics[width=0.8\textwidth]{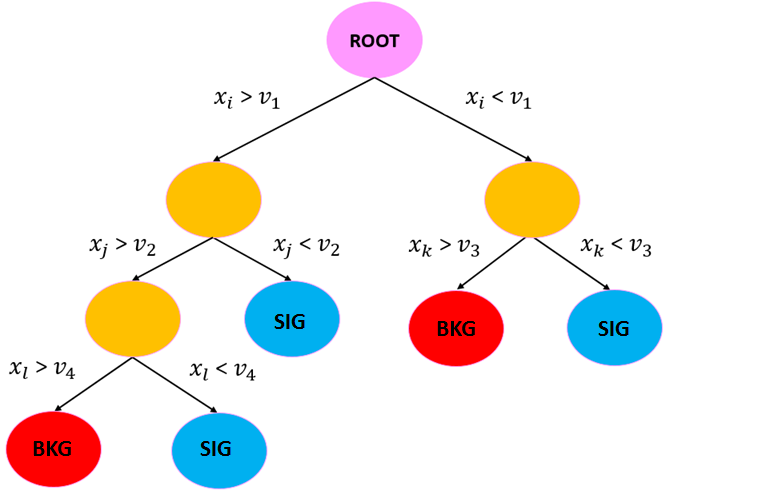}
  \end{center}
  \caption{A demonstration plot of a single decision tree.}
  \label{fig:singledt}
\end{figure}

To construct a classification BDT\cite{tmva}, or to train a BDT, we provide a signal sample and a background sample from Monte Carlo simulated events with known identity, and a selected set of input variables with distinguishing power $\overrightarrow{x}$ = $\{x_{1}, x_{2}, ..., x_{n}\}$. A single decision tree is first trained, which is to cut the variable phase space into several signal dominated or background dominated hypercube regions, following a certain rule to optimize the separation between signal and background, and to label the events in the regions accordingly as ``signal'' or ``background''. A demonstration plot of a single decision tree is shown in Figure \ref{fig:singledt}. It has a tree structure, with a root node in magenta representing the entire variable phase space, intermediate nodes in yellow representing the split phase spaces, and terminal nodes in blue for ``signal'' regions (SIG) while in red for ``background'' regions (BKG). The nodes are connected by arrows labeled with a variable $x_{i}$ under consideration and the corresponding cut value, which specifies how a parent node is split into two daughter nodes. The tree building starts from the root node, with the number of signal and background events reweighted such that both the total weights for signal and for background equal to the number of the signal events. The node is then split by selecting a single variable and a cut value on it. There are several possible splitting criteria. We use the Gini Index defined as:
\begin{equation}
  \mathrm{Gini\:Index} = p_{s}\cdot(1-p_{s}),  
  \label{eqn:gini}
\end{equation}  
where $p_{s}$ represents the fraction of the signal weights of the total signal plus background weights in a node. The Gini Index is maximal at the root node with $p_{s}$ equal to 0.5. The splitting variable and the cut value are chosen to maximize the decrease of the Gini Index from the parent node to the two daughter nodes, for which the relative fraction weighted sum of the Gini Indices of the two daughter nodes is used. The splitting continues iteratively until the predetermined limit is reached, such as the maximum depth of the tree or the minimum number of events in a node. The limit is set to decrease the bias due to statistical fluctuation of the training samples, the overtraining. The terminal nodes with $p_{s}$ greater (less) than 0.5 are labeled as ``signal'' (``background''), and the events in the nodes are assigned a score +1 ($-1$).     

For a single decision tree, some of the events in the terminal nodes are easily misclassified, and the classification result is susceptible to overtraining. To decrease the misclassification rate and the effect of overtraining as much as possible, a procedure called ``boosting'' is used, which is basically to train a set of trees and assign a score to an event as the weighted average of the scores of all the trees. In our analysis, we employ two boosting procedures, Gradient Boost and Adaptive Boost (AdaBoost). The expression for the Gradient Boost is as following:       
\begin{equation}
  F(\overrightarrow{x};P) = \sum_{m=1}^{M}\beta_{m}f(\overrightarrow{x};\alpha_{m}); \: P \in \{\beta_{m};\alpha_{m}\}_{m=1}^{M},
  \label{eqn:boost}
\end{equation}  
where $F(\overrightarrow{x};P)$ represents the function with the set of parameters $P$ corresponding to the BDT made up of $M$ trees, $f(\overrightarrow{x};\alpha_{m})$ represents the function corresponding to the $m_{th}$ tree, $\alpha_{m}$ represents the parameters of the $m_{th}$ tree including the splitting variables and cut values at each node, and $\beta_{m}$ is the weight on the $m_{th}$ tree. The parameter set $P$ is determined by minimizing the deviation between the estimates provided by $F(\overrightarrow{x};P)$ and the true identities of the training events, measured by the loss function:  
\begin{equation}
  L(F(\overrightarrow{x};P),y) = \sum_{n=1}^{N} \mathrm{ln}(1+e^{-2F_{n}(\overrightarrow{x};P)y_{n}}); \: F(\overrightarrow{x};P) \in \{F_{n}(\overrightarrow{x};P)\}_{n=1}^{N}, \: y \in \{y_{n}\}_{n=1}^{N},
  \label{eqn:boost}
\end{equation}  
where $F_{n}(\overrightarrow{x};P)$ represents the estimated value for the $n_{th}$ event, and $y_{n}$ is the true value $+1$ or $-1$ of the $n_{th}$ event, and $N$ is the total number of events. The AdaBoost is obtained by minimizing a different type of loss function: 
\begin{equation}
  L(F(\overrightarrow{x};P),y) = \sum_{n=1}^{N} e^{-F_{n}(\overrightarrow{x};P)y_{n}}. 
  \label{eqn:boost}
\end{equation} 
The trained BDT function is then used to assign a score to any event, given its values of input variables. The score is a quasi-continuous variable, varying from $-1$ to $+1$. The more signal-like an event is, the higher value it gets. 

To train a regression BDT\cite{tmva}, we typically provide a sample of Monte Carlo simulated events, a target variable corresponding to the desired event property---whose value is known for a training event, and a set of other input variables related to the property. The trained BDT function provides an evaluation of the property for any event based on its input variables, which is the weighted average of values estimated by individual decision trees. In the case of this analysis, we use the regression in a more generalized way, which regress a probability density function of the reconstructed energy over the true energy for a photon. We provide a known functional form, and set the parameters of the function as target variables.          

For our analysis, we use Toolkit for Multivariate Data Analysis (TMVA)\cite{tmva} within CERN's ROOT framework\cite{root} to train the classification BDTs, while the approach described in Reference\cite{hggan13253} to train the regression BDT.

%% file: CMSExperimentAtLHC.tex
\chapter{The CMS Experiment at the LHC}
\label{The CMS Experiment at the LHC}

\section{The Compact Muon Solenoid Detector}
The Compact Muon Solenoid (CMS) detector\cite{CMS,cmstdr1} was built to shed light on the mechanism of electroweak symmetry breaking by searching for the Higgs boson, to look for deviations from the Standard Model by making precise measurements of the Standard Model processes, and to search for direct evidence of new physics such as supersymmetry, dark matter and extra dimensions.

\begin{figure}[h] 
  \begin{center}
    \includegraphics[width=1\textwidth]{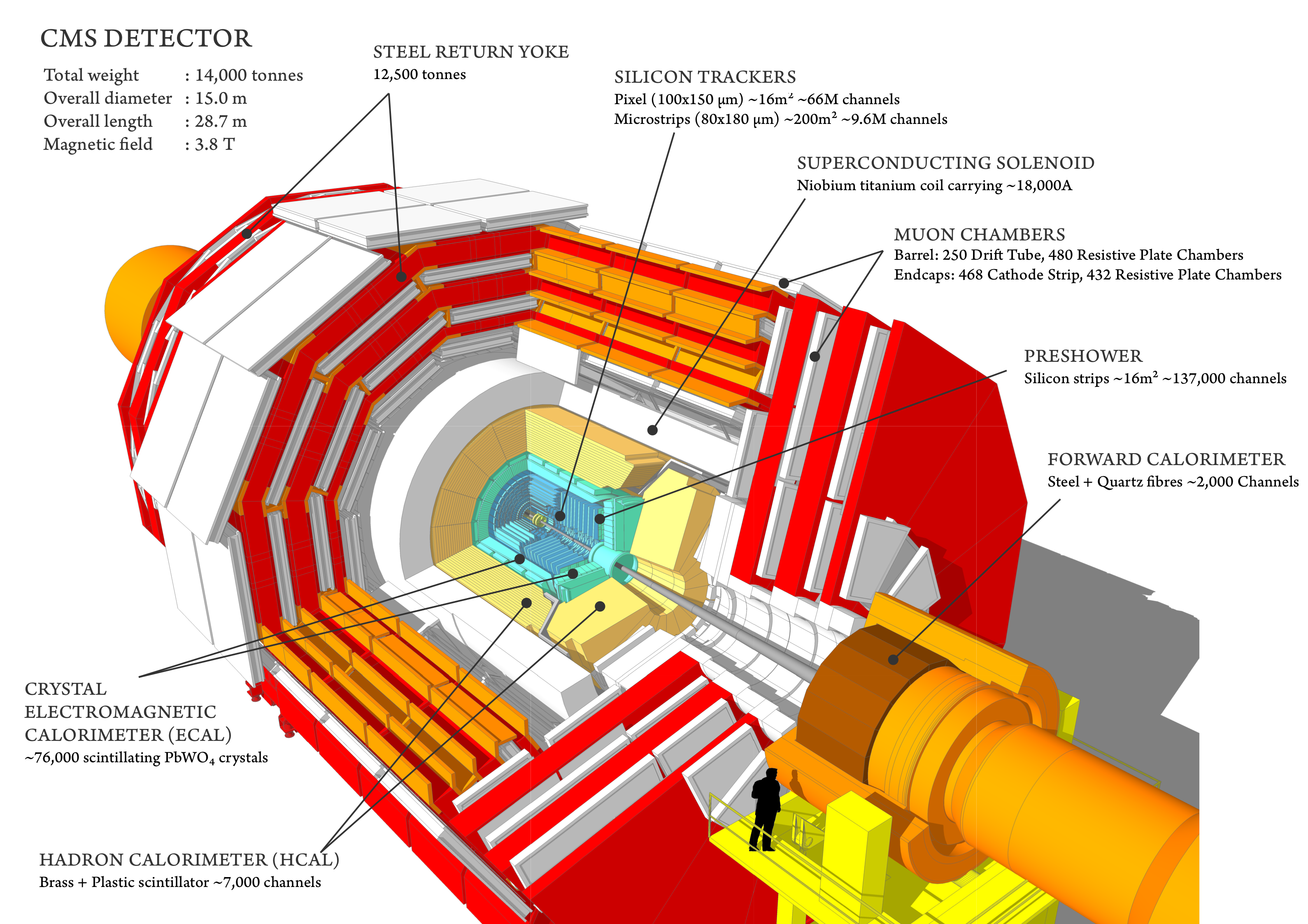}
  \end{center}
  \caption{An overview of Compact Muon Solenoid detector.}
  \label{fig:cms}
\end{figure}

An overview of the CMS detector is shown in Figure \ref{fig:cms}\cite{cmssimulation}. It is a cylindrical detector 28.7 m long and 15 m in diameter, which is centered at the collision point (LHC point 5) with its longitudinal axis along the beam pipe. It is composed of a superconducting solenoid magnet and multiple sub-detectors inside and surrounding the magnet. The solenoid provides a 3.8 T magnetic field along the longitudinal direction of the detector to bend charged particles in the transverse direction. Going from the beam pipe to the solenoid, there is a tracker measuring the momenta of charged particles, an electromagnetic calorimeter to primarily measure the energies of photons and electrons, and a hadronic calorimeter for measuring the energies of charged and neutral hadrons. Outside the solenoid, there are muon chambers measuring momenta of muons, which are interleaved with the steel return yoke of magnetic flux return.

To describe the CMS detector we use both right-handed Cartesian coordinates and polar coordinates, with the nominal collision point as the origin in both cases. For the Cartesian coordinates, the $x$-axis and $y$-axis are in the transverse plane pointing along the inward radial direction of the LHC ring and along the upward vertical direction, respectively, while the $z$-axis is parallel to the beam. For the polar coordinates, $\phi$, $r$ and $\theta$ represent the azimuthal angle from the $x$-axis in the transverse plane, the radial distance in the plane, and the polar angle from the $z$-axis in the $y$-$z$ plane, respectively. 

The fractions of proton momenta carried by the two colliding partons are generally unequal in pp collisions, which leads to the non-zero total collision momentum along the $z$-axis. Under the boost in the $z$ direction, to approximately make an Lorentz-invariant description of the hard collision events, with highly relativistic incoming and outgoing particles, the pseudorapidity $\eta$, defined as $-\mathrm{ln[tan(\theta/2)]}$, and the transverse momentum $\overrightarrow{p}_{T}$, defined as the projection of the momentum $\overrightarrow{p}$ in the transverse plane, are used. The magnitude of the transverse momentum is denoted as $p_{T}$. The presence of high $p_{T}$ particles is a signature of hard collision events, which is used later for the event selection. 

\subsection{Tracker}
\begin{figure}[h] 
  \begin{center}
    \includegraphics[width=1\textwidth]{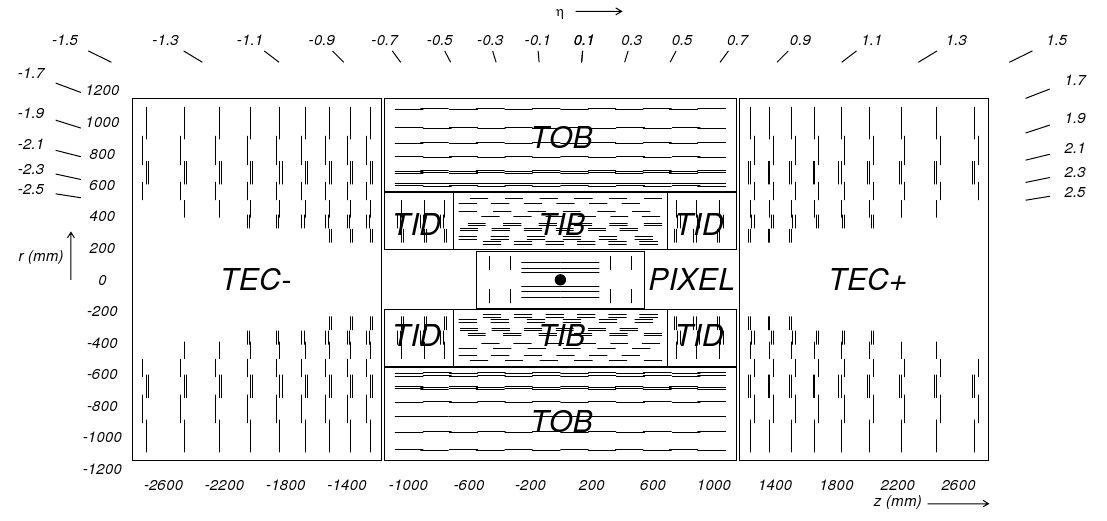}

  \end{center}\
  \caption{The cross section of the tracker in $r$-$z$ plane.}
  \label{fig:cmstrk}
\end{figure}

The tracker\cite{CMS,trktdr1,trktdr2} measures the hit positions of charged particles along their trajectories passing through it, which are used to reconstruct the trajectories (tracks), momenta and production vertices of the particles. The $r$-$z$ plane cross section of the tracker is shown in Figure \ref{fig:cmstrk}\cite{CMS}. It is a full-silicon based detector, consisting of an inner silicon pixel detector and a silicon strip detector with acceptance $|\eta|$ $<$ 2.5. Silicon is used due to its fast response, desired for making measurements from the high luminosity LHC pp collisions, and its good spatial resolution. The granularity of the detector decreases with an increase of distance from the collision point, which corresponds to a decrease of particle flux.  

The pixel detector has three cylindrical layers in the barrel region at effective radii of $r=$ 4.4 cm, 7.3 cm and 10.2 cm within $|z|$ $\leq$ 26.5 cm, and two disks at $|z|$ = 34.5 cm, $46.5~\mathrm{cm}$ in the endcap on each side within about 6 cm $\leq$ $|r|$ $\leq$ 15 cm. It has 66 million pixels each with dimensions of 100 ${\mu m}$ $\times$ 150 ${\mu m}$, which results in an occupancy of about 0.1 permill per pixel per bunch crossing. It measures the positions of charge particles hitting its silicon wafers with a single point resolution from 15 $\mu m$ to 20 $\mu m$.

The silicon strip detector consists of inner and outer parts within $|z|$ $\leq$ 282 cm and $20~\mathrm{cm}$ $<$ $|r|$ $<$ 116 cm. The inner part has 4 layers in the Tracker Inner Barrel (TIB), and 3 disks in the Tracker Inner Endcap (TIE) on each side. The outer part has 6 layers in the Tracker Outer Barrel (TOB), and 9 disks in the Tracker EndCap (TEC) on each side. The whole silicon strip detector has 9.3 million strips with thickness of 320 ${\mu m}$ or 500 ${\mu m}$ and pitches from 80 ${\mu m}$ to 184 ${\mu m}$. It measures the $r$-$\phi$ or $z$-$\phi$ positions of charged particles hitting the strip detector, with resolutions of 23 $\mu m$ to 53 $\mu m$ in the $\phi$ direction.  

\begin{figure}[h] 
  \begin{center}
    \includegraphics[width=0.5\textwidth]{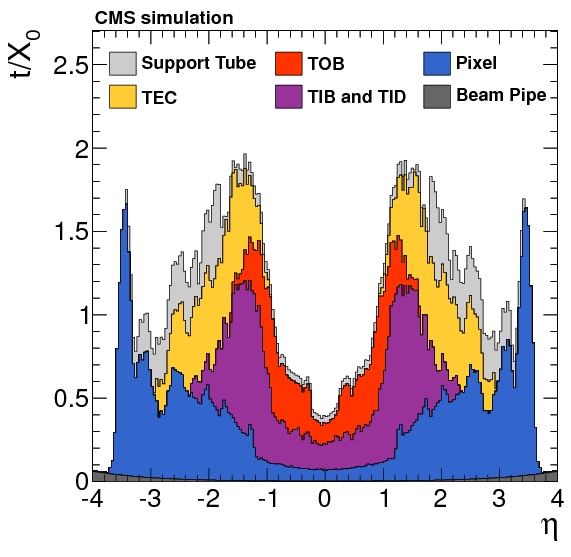}
  \end{center}
  \caption{The thickness of the tracker along with the beam pipe and the support tube, $t$, measured in number of radiation length, $X_{0}$, as a function of $\eta$ from simulation.}
  \label{fig:trkx0}
\end{figure}

The thickness of the tracker material $t$ measured in number of radiation length $X_{0}$ as a function of $\eta$ from simulation is shown in Figure \ref{fig:trkx0}\cite{trkperformance}, and has a maximum of about 2. The amount of material of the CMS full-silicon based tracker, is much larger than that of a tracker utilizing gas detector, e.g. the tracking system of CDF detector at Tevatron has a thickness of $\mathcal{O}(1\%~\mathrm{X_{0}})$\cite{cdf}. As a result, the measurement of electron momentum from the tracker, and the measurement of electron or photon energy from the electromagnetic calorimeter, suffer more from the effects degrading the measurement resolution, including multiple scattering, electron bremsstrahlung or photon conversion. The silicon is chosen as the tracker material despite of this disadvantage because its fast response and good spatial resolution are must for the high luminosity LHC environment.       

For tracks with $p_{T}$ = 100 GeV, the momentum resolution is about 1-2$\%$ in the barrel. The resolutions for transverse impact parameter $d_{xy}$ and for longitudinal impact parameter $d_{z}$ are $\mathcal{O}(10~\mathrm{\mu m})$.  

\subsection{The Electromagnetic Calorimeter}
\label{sec:The Electromagnetic Calorimeter}
The Electromagnetic Calorimeter (ECAL)\cite{CMS,ecaltdr1,ecaltdr2} measures the energies of photons and electrons through the electromagnetic (EM) shower they produce traversing the calorimeter. An electromagnetic shower for a photon or an electron starts as an electron-positron pair production by the impacting photon or the Bremsstrahlung by the impacting electron, and develops to a cascade of electrons, positrons and photons through repeating processes of pair productions and Bremsstrahlung. The CMS ECAL is designed to measure the photon and electron energies with high resolution, which is essential for $H\rightarrow \gamma\gamma$ sensitivity. It is homogeneous, fine-grained, and almost hermetic. It is also compact enough to be put inside the solenoid to reduce the number of radiation lengths in front, and thus reduce the probability of photon conversion and electron Bremsstrahlung before a photon or electron entering the ECAL, which improves the resolution of the photon and electron energy measurements.   

\begin{figure}[h] 
  \begin{center}
    \includegraphics[width=1\textwidth]{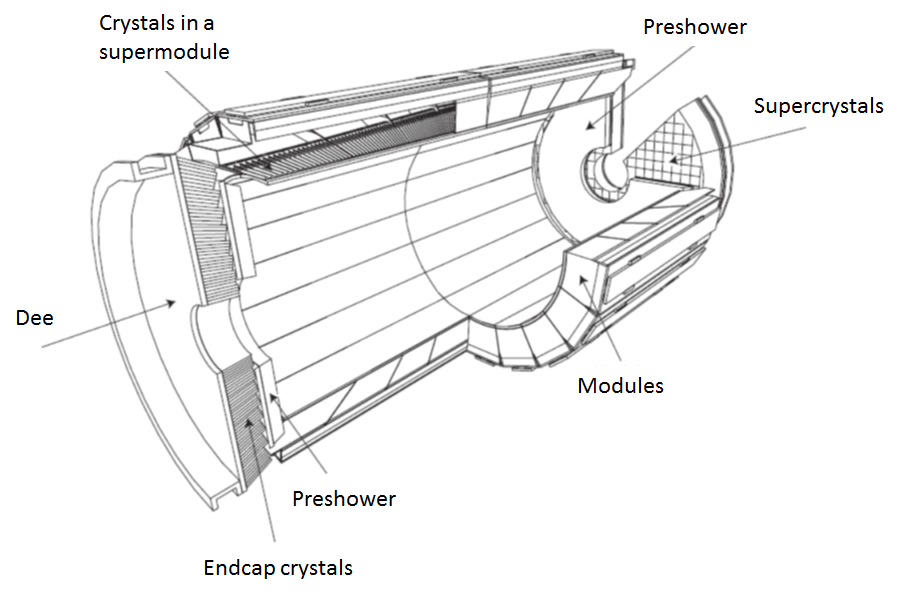}
  \end{center}
  \caption{An overview of the electromagnetic calorimeter.}
  \label{fig:ecal}
\end{figure}

An overview of the ECAL is shown in Figure \ref{fig:ecal}\cite{CMS}. It includes a barrel component covering $|\eta|$ $<$ 1.479 and an endcap component on each side covering 1.479 $<$ $|\eta|$ $<$ 3. Both barrel and each endcap consist of one layer of lead tungstate (PbWO$_{4}$) crystals, which have short radiation length, small Moli$\grave{e}$re radius, good transparency and fast response as desired. Each crystal is coupled to a photodetector: an avalanche photodiode (APD) in the barrel and a vacuum phototriode (VPT) in the endcap---which is subject to higher radiation. An impacting photon or electron generates an electromagnetic shower through the interaction with the crystal and transfers its energy into the shower. The energy of the developed shower is then deposited into the crystals, and the crystals emit scintillation light in proportion to the deposited energy. The scintillation light in each crystal is converted into photoelectrons and amplified by its coupled photodetector, which are further converted into voltages and ADC (Analogue-to-Digital Converter) counts. The ADC counts are finally converted to energy as the measurement of the energy deposit in the crystal, which is later used to reconstruct the total energy of the photon or electron. 
        
The ECAL barrel component starts from $r$ = 129 cm. It has 61200 crystals grouped into 36 supermodules, and each supermodule covers a half barrel in $z$ and 20$^{\circ}$ in $\phi$. Four modules are inside each supermodule, and each module has 400 or 500 crystals. Each crystal is in a truncated-pyramid shape covering 0.0174 ($\Delta\eta$) $\times$ 0.0174 ($\Delta\phi$), with a cross section of 22 mm $\times$ 22 mm in the front end and 26 mm $\times$ 26 mm in the back end comparable to the square of Moli$\grave{e}$re radius. The crystal is 23 cm deep, corresponding to 25.8 radiation lengths, which well contains the total energy of the electromagnetic shower. The crystal is oriented with its axis 3$^{\circ}$ off from pointing to the nominal collision vertex to avoid particles from passing through inter-crystal gaps. 

The ECAL endcaps start from $|z|$ $=$ 315.4 cm. Each endcap has 7324 crystals grouped into two semi-circular parts (``Dees''). Each part consists of 138 units of $5\times5~\mathrm{crystals}$ (supercrystals) and 18 partial supercrystals. Each crystal has a cross section of 28.62 mm $\times$ 28.62 mm in the front end and 30 mm $\times$ 30 mm in the back end. The crystals are 22 cm deep, corresponding to 24.7 radiation lengths. The crystals are oriented with their axes 2$^{\circ}$ to 8$^{\circ}$ off from pointing to the nominal collision vertex. To help resolving the two photons from a neutral meson, a preshower detector is added in front of each ECAL endcap covering 1.653 $<$ $|\eta|$ $<$ 2.6. It has two lead disks with thickness of 2 $X_{0}$ and 1 $X_{0}$ to generate electromagnetic showers. A silicon strip detector with a pitch of 1.9 mm is behind each lead disk to measure the shower shape. 

The variation of a crystal response with time and the variation of responses among crystals are calibrated and corrected as described References\cite{CMS,ecalcali}. The time variation of the response of each crystal is mainly due to changes in crystal transparency from irradiation, and subsequent recovery. These variations are monitored by a laser system consisting of lasers with wavelength $\lambda$ = 440 nm  (near the wavelength of scintillation peak) and $\lambda$ = $796~\mathrm{nm}$. For each crystal, the laser pulse is injected during the beam gap, and a time dependent correction factor is computed from the change of the crystal response. The variation of the relative responses among crystals are calibrated by a series of methods (intercalibration), which use the energy deposition symmetry in $\phi$, the mass of diphotons from $\pi^{0}$ ($\eta^{0}$) decays, and the ratio between ECAL energy and tracker momentum of electrons from $W$ and $Z$ decays, respectively. A correction factor (intercalibration constant) is obtained for each crystal, which is the weighted average of the correction factors from all methods.

The relative energy resolution $\sigma_{E}/{E}$ measured in 2006 from the test beam of electrons with energy $E$ reconstructed by summing energy deposits in $3\times3~\mathrm{crystals}$ is\cite{CMS}
\begin{equation}
  (\sigma_{E}/{E})^{2} = (2.8\%/\sqrt{E})^{2} + (0.12/E)^{2} + (0.30\%)^{2},
  \label{eqn:ecalreso}
\end{equation}
where the first term is the stochastic term mainly associated with the shower fluctuation, the second term is due to the noise from the readout electronics, the third term is determined by the accuracy of calibration. 

The impact positions of photons and electrons in the ECAL are also measured. The position resolution in the barrel is 3 mrad in $\phi$ and 0.001 in $\eta$, and the resolution in the endcap is 5 mrad in $\phi$ and 0.002 in $\eta$\cite{ecalcali}. 

\subsection{The Hadronic Calorimeter}
\begin{figure}[h] 
  \begin{center}
    \includegraphics[width=1\textwidth]{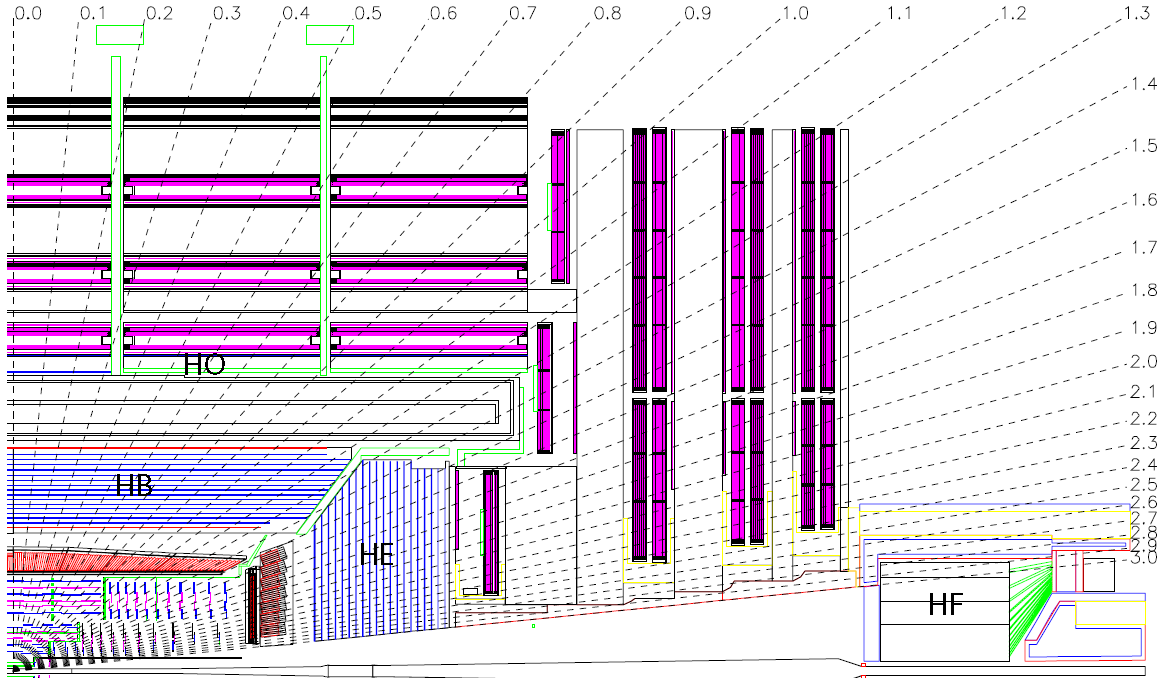}
  \end{center}
  \caption{A quarter of CMS $r$-$z$ cross section. The components of the hadronic calorimeter are labeled, including HCAL Barrel (HB), HCAL Endcap (HE), HCAL Outer (HO) and HCAL Forward (HF).}
  \label{fig:hcal}
\end{figure}

The Hadronic Calorimeter (HCAL)\cite{CMS,hcaltdr} measures energies of hadrons through the hadronic showers they produce passing through the calorimeter. The HCAL is a sampling calorimeter which includes four parts: HCAL Barrel (HB), HCAL Endcaps (HE), HCAL Outer (HO) and HCAL Forward (HF). The various sub-components of the HCAL are shown in Figure \ref{fig:hcal}\cite{CMS}, where a quarter of CMS is portrayed in the $r$-$z$ cross section. The HB and the HE cover respectively $|\eta|$ $<$ 1.3 and 1.3 $<$ $|\eta|$ $<$ 3, which use brass plates as absorbers to generate showers and plastic scintillators as the active material to measure the shower energies. The HO is added in the central barrel outside the solenoid to increase the HCAL thickness, which consists of scintillators and uses the solenoid as absorber. The HF, which uses steel as absorber and quartz fibers as active material, covers 3 $<$ $|\eta|$ $<$ 5.2. The total thickness measured in number of nuclear interaction lengths, including the ECAL, ranges from 10 to 15, depending on $\eta$.       

\subsection{Muon Detector}
\begin{figure}[h] 
  \begin{center}
    \includegraphics[width=1\textwidth]{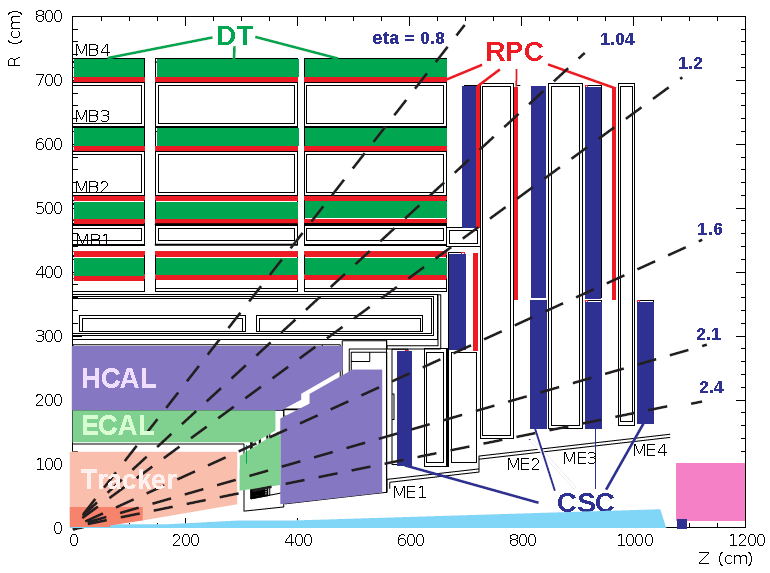}
  \end{center}
  \caption{A quarter view of CMS in the $r$-$z$ cross section. The components of the muon detector are labeled, including Drift Tube (DT) Chambers, Cathode Strip Chambers (CSC) and Resistive Plate Chambers (RPC).}
  \label{fig:muon}
\end{figure}

The muon detector\cite{CMS,muontdr} measures the hit positions of muons along their trajectories passing through the detector, which are used to reconstruct their trajectories and momenta. A quarter view of CMS in the $r$-$z$ cross section is shown in Figure \ref{fig:muon}\cite{CMS}, with the various sub-components of the muon system labeled. There is a barrel system covering $|\eta|$ $<$ 1.2 and endcaps covering 0.9 $<$ $|\eta|$ $<$ 2.4. The barrel is composed of Drift Tube (DT) Chambers and Resistive Plate Chambers (RPC). The endcap includes Cathode Strip Chambers (CSC) and RPCs as well. The DTs and CSCs are used for precision position measurements, and the RPCs are used for fast triggering.

\subsection{Trigger}
The trigger system\cite{CMS,triggertdr1,triggertdr2} is used to select potentially interesting physics events to be read out and recorded for offline use out of the design rate of 40 MHz pp bunch crossings (the actual bunch crossing rate is 20 MHz). The accept rate is subjected to the constraints of the detector readout speed, event processing power, and storage space. The trigger consists of two levels: the Level-1 (L1) trigger based on hardware and the High-Level Trigger (HLT) based on software. The L1 uses coarse information from ECAL, HCAL and muon detectors to determine approximate candidates of physics objects such as electrons, photons, muons, jets and transverse missing energy, and it selects events at a rate of up to 100~kHz. The HLT uses detailed information from the entire detector and reconstructs the physics objects in a similar way to the offline reconstruction used in the final analysis. It further selects events with good quality and high $p_{T}$ objects at a rate of $\mathcal{O}(100~\mathrm{Hz})$.    

\section{Event Reconstruction} 
\label{sec:Reconstruction} 

\subsection{Tracks and Vertices}
Tracks, the trajectories of the charged particles in the tracker, propagate as helices between tracker layers. They are obtained by fitting the tracker hits using the Kalman Filter method\cite{Fruhwirth:1987fm,kalman1,kalman2}, which takes into account multiple scattering, energy loss and the uncertainty of hit positions. A track's initial momentum, its impact parameter with respect to the nominal collision vertex, and its charge result from the fit. Primary vertices are reconstructed by grouping tracks compatible with the region of primary interactions according to their distance in $z$, following a deterministic annealing (DA) algorithm\cite{vertexcluster}. The position for each vertex is fitted from its corresponding tracks using adaptive vertex fitting\cite{vertexfit}. A detailed description of track and primary vertex reconstruction is given in Reference\cite{trkvertex}.   

\subsection{Photons}
\label{sec:Photon}
A photon, produced at the interaction point, first passes through the tracker, and then enters ECAL and loses all its energy through electromagnetic shower. There are two cases. In the first case, the photon traverses the tracker without interaction and deposits about 94$\%$ (97$\%$) of its energy into $3\times3$ ($5\times5$) crystals in the ECAL. Such photon is called unconverted photon. In the second case, the photon converts to electron and positron pair before entering the ECAL, the electron and positron pair bend under the magnetic field and deposit their energies in a larger range in $\phi$. Such photon is called converted photon. To include all the photon energy deposits, photons are reconstructed by clustering the energy deposits in the ECAL crystals into the so-called superclusters\cite{cmstdr1,supercluster}. 

Superclusters in ECAL barrel and those in ECAL endcap are constructed following different algorithms:
\begin{itemize} 
 \item For ECAL barrel, a seed crystal is first located, which is the crystal with highest E$_{T}$ above a certain threshold among the crystals not included in any other supercluster yet. Then $5\times1$ matrices of crystals (bars) each centered at the same $\eta$ with the seed crystal are built, within the range $\pm$17 crystals in $\phi$ from the seed crystal. The bars with total energy above certain threshold connected in $\phi$ are further grouped into clusters called basic clusters. The basic clusters with the highest bar energy above certain threshold are finally grouped to form a supercluster. 
 \item For ECAL endcap, a basic cluster, a $5\times5$ matrix of crystals centered at the seed crystal, is first built. The crystals at the boundary of the matrix are allowed to seed new basic clusters from the crystals not included in any cluster yet. A supercluster is then formed from the connected basic clusters. 
\end{itemize}   

The raw photon energy $E_{Raw}$ is obtained by summing the energy deposits in the crystals of the supercluster calibrated as described in Section \ref{sec:The Electromagnetic Calorimeter}, and the energy deposited in the preshower detector is added to it for photons in the endcap. The photon position in $\eta$-$\phi$ is obtained from the mean position of basic clusters weighted by energy, and the position of basic clusters is calculated from mean positions of crystals corresponding to the shower depth weighted by the logarithm of the crystal energy\cite{photonpos2}.     

For a converted photon, if the conversion happens early enough in the tracker such that the tracks of the electron and positron pair are well reconstructed, the conversion is reconstructed by fitting the conversion vertex from the pair of tracks\cite{conversion}. For the vertex fitting, the two tracks under consideration are required to have opposite charges, and to be parallel at the conversion vertex because the photon is massless, which removes the pollution from the random combination of two tracks from the primary interactions (prompt tracks). The reconstructed conversion tracks are then each matched to an ECAL supercluster, which completes the information about a converted photon.  
 
\subsection{Electrons}
Electrons are reconstructed by matching an ECAL supercluster, the same as used for photon reconstruction, to a track\cite{electron}. The candidate track is obtained by fitting the tracker hits using the Gaussian-sum filter (GSF) algorithm\cite{GSFElectron}, which models the bremsstrahlung energy loss distribution by a weighted sum of Gaussians. 

\subsection{Muons}
Muons used for this analysis are reconstructed following the so-called global muon reconstruction method\cite{cmstdr1,muon}, which uses both information from the muon detector and the tracker. A muon track only using the muon detector information (standalone muon track) is constructed first. It starts from building short track traces (segments) from aligned hits in individual DT chambers and CSC chambers. These segments are then used for the fit of the standalone muon track, following the Kalman Filter method. The obtained standalone muon track is matched to a tracker track, and a global muon track is finally fitted from hits of both tracks again using Kalman Filter method.

\subsection{Jets and Transverse Missing Energy}
Jets and transverse missing energy $\overrightarrow{\text{MET}}$ are reconstructed from electron, muon, photon, charged hadron and neutral hadron candidates built from the particle-flow algorithm\cite{pf1,pf2}. The particle-flow algorithm is designed to reconstruct and distinguish all the stable particles by effectively grouping the information from the entire detector and associating the grouped information to each particle candidate, with no information double counted in two different candidates. This algorithm provides reconstructed particle candidates (particle-flow candidates) as ideal input to reconstruct higher level objects like jets and event level quantity like $\overrightarrow{\text{MET}}$, but not the optimal reconstructed photon. And so we still use the photons from more specialized reconstruction algorithm as described in Section \ref{sec:Photon} for the diphotons candidates in the analysis.

Jets are built through clustering particle-flow candidates. The anti-$k_{T}$ algorithm\cite{antikt} is used and the size parameter is $\Delta$R = 0.5. Jets from bottom quarks (b-jets) are identified using Combined Secondary Vertex algorithm \cite{bjet}, which identifies the decay vertex displaced away from the primary vertex. $\overrightarrow{\text{MET}}$ is computed as the opposite $\overrightarrow{p}_{T}$ sum of all the particle-flow candidates. 

%% file: HggAnaOverview.tex
\chapter{Higgs Boson to Two Photons Analysis Overview}
\label{Higgs Boson to Two Photons Analysis Overview}
We perform the analysis to observe the production and decay of the Higgs boson into two photons, in the Higgs mass hypotheses range 115 GeV $\leq$ $m_{H}$ $\leq$ 135 GeV. The basic flow of our analysis is to reconstruct diphotons from the events with at least two reconstructed photons, preselect a potentially signal-rich sample of diphoton events, obtain their masses, and fit the mass spectrum to search for an excess of signal over background. In the case an excess is observed, we further measure its corresponding Higgs mass, and the signal and coupling strengths to quantify its compatibility with the Standard Model Higgs boson. To optimize the Higgs search sensitivity and measurement precision, we classify the events into several categories according to the expected $S/B$ under the mass peak. We use several BDTs, trained on the Monte Carlo simulated events, to both improve the diphoton mass reconstruction, and to combine all the rest of the diphoton information into a powerful diphoton event classifier, which provides a measure of $S/B$. We also use signatures of Higgs production processes to select events into high $S/B$ classes. For the diphoton mass fit, we model the signal from Monte Carlo simulated Higgs events, and the background directly from the data. 

\begin{figure}[h] 
  \begin{center}
    \includegraphics[width=0.99\textwidth]{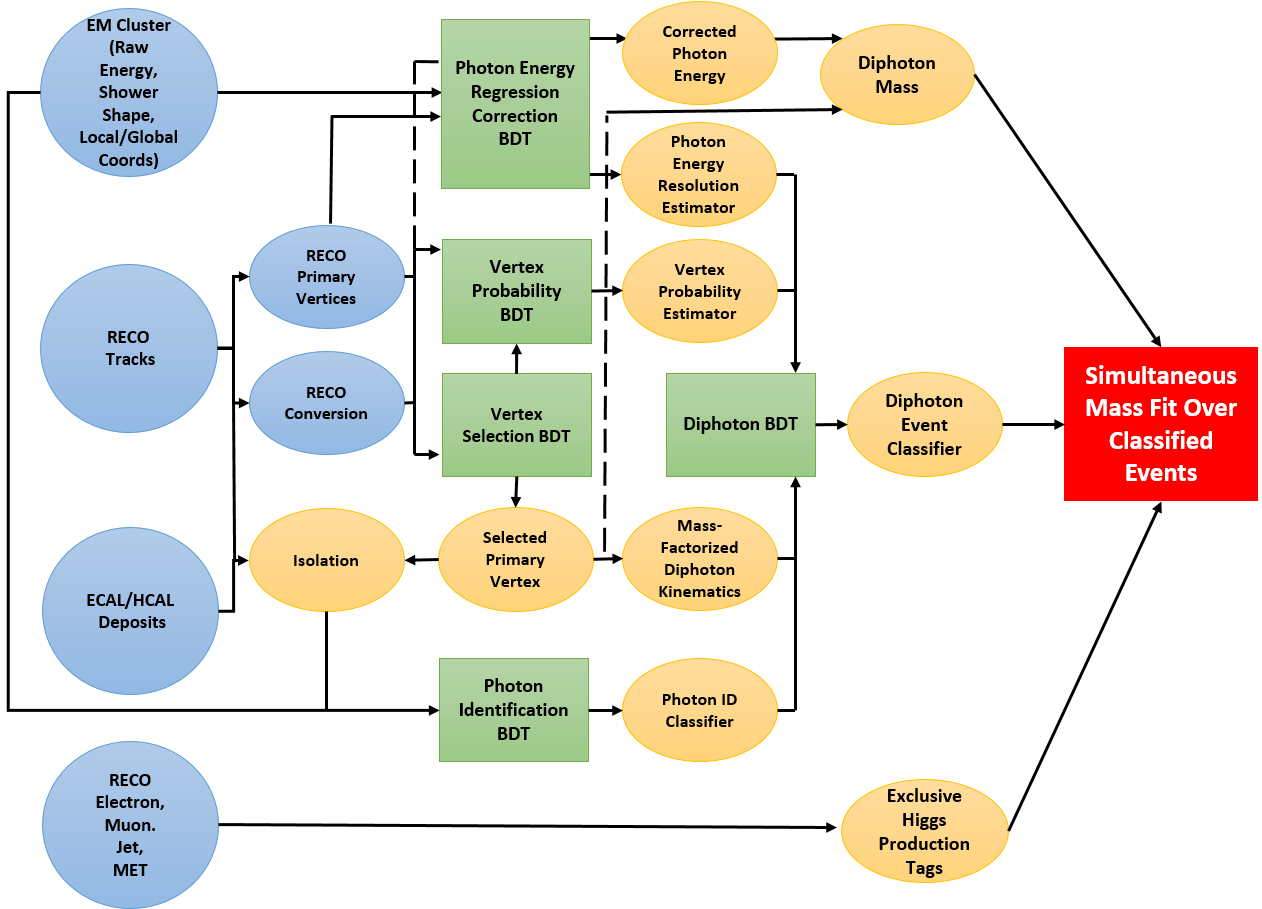}
  \end{center}
  \caption{Higgs boson to two photons analysis workflow. The blue circles represent the input elements to the analysis from the event reconstruction. The green boxes represent BDTs used for information processing. The yellow circles represent the quantities built from input information. The red box represents the process for Higgs signal extraction.}
  \label{fig:higgsworkflow}
\end{figure}

The main components for this analysis are summarized in Figure \ref{fig:higgsworkflow}. Conceptual descriptions of these components are provided in the following of this chapter. The data and Monte Carlo simulation samples used for the analysis are introduced afterwards. 

\section{Analysis Components} 

\subsection{Diphoton Reconstruction}
\label{sec:Diphoton Reconstruction}
To reconstruct diphotons and their masses, we correct the single photon energies and select the diphoton production vertex for each diphoton from all the vertices in the same bunch crossing.

\subsubsection{Photon Energy Correction}
\label{sec:Photon Energy Correction}
The raw energy of a reconstructed photon $E_{Raw}$ needs correction, as it is deviated from the true photon energy $E_{True}$, mainly due to the combined effect of photon shower loss and the pileup contamination. The shower lost consists of the part outside of the supercluster window, especially for converted photons, and the part passing through the inter-crystal gaps or inter-module cracks within the window. The fraction of photon energy lost therefore depends on whether it is converted, and the location and detailed pattern of its shower in the ECAL. The fraction of energy contaminated depends on the energy density due to the pileup interactions in the event. We train a BDT (``photon energy correction regression BDT'') to regress the photon energy correction factor, taking the above factors into consideration. The target is the probability density of the ratio between the true photon energy and the reconstructed raw photon energy $E_{True}/E_{Raw}$, and the input variables are chosen such that all the relevant information is included: the supercluster energy, the global detector coordinates and local ECAL coordinates of the ECAL clusters, the shower shape variables as measures for photon conversion and shower pattern, and pileup information. The trained BDT provides an estimation of the probability density of $E_{True}/E_{Raw}$ for any given photon, and the most probable $E_{True}/E_{Raw}$ is used as the correction factor.

\subsubsection{Vertex Selection}
\label{sec:Vertex Selection}
The diphoton production vertex needs to be selected from an average of 9 (21) pp collision vertices for 7 TeV (8 TeV) distributed in $z$ with an RMS of about 6 cm (5 cm). To keep the effect of the vertex selection on the diphoton mass resolution negligible with respect to the single photon energy resolution, the selected diphoton vertex is required to be within 1 cm in $z$ from the true diphoton vertex. For the discrimination between the diphoton production vertex and the pileup vertices, we use the knowledge that the total transverse momentum of the recoiling tracks, mainly from the underlying events associated with the diphoton production vertex, roughly balances the diphoton transverse momentum. The balance is not exact as we do not have the association between neutral particles and vertices, so the total transverse momentum of neutral particles recoiling against the diphoton for a given vertex is unknown. Nevertheless, comparing between the recoiling tracks of the diphoton production vertex and those for the pileup vertices, for the former, on average, the sum of their transverse momentum square is larger, the relative difference in the magnitude between their total transverse momentum and the diphoton transverse momentum is smaller, and the projection of their total transverse momentum onto the direction of the diphoton transverse momentum is larger. Besides the correlation between the kinematics of the recoiling tracks and that of the diphoton, in the case that at least one photon is converted, the position of the conversion vertex, together with either the direction of the conversion momentum or the position of the ECAL supercluster, provides an extrapolation of the position of the diphoton vertex, which is used for the vertex selection. We train a BDT (``vertex selection BDT''), using the above information, to distinguish between the prompt vertex and the pileup vertices. The BDT assigns scores to the vertices according to how likely it is a prompt vertex. The vertex with the highest score is selected.  

\subsection{Signal to Background Separation}
\label{sec:Signal to Background Separation}
The reconstructed diphoton events include potential Higgs signal events and a mixture of background events. In the background events, there are mainly ``irreducible'' prompt diphoton events, and ``reducible'' $\gamma$ + jet and dijet events with jets faking photons. The fake photons are majorly due to energetic neutral mesons, from jet fragmentation, decaying into two photons, which end up in the same supercluster and are reconstructed as a single photon. The task of the rest of the analysis is to maximize the separation between the signal and background.

\subsubsection{Diphoton Event Preselection}
\label{sec:Diphoton Preselection}
We first preselect a sample of diphoton events. We design the preselection mainly to select the maximum common phase space between the data and Monte Carlo simulated events, such that the BDTs trained on the Monte Carlo events are optimal for the data as well, the signal model derived from the Higgs Monte Carlo simulation is for the correct phase space in data, and the acceptance and efficiency for the Higgs signal is maintained as large as possible. We also apply an electron veto to distinguish electrons from photons.

To select the common phase space, we apply geometric and kinematic acceptance cuts, and very loose photon identification cuts on the reconstructed photons to remove fake photons. The photon identification depends on two different features between the fake photon and the prompt photon. First, the ECAL shower of the fake photon is expected to be wider than that of the prompt photon since it is supposed to be the combined shower of the two photons. Second, the fake photon is not isolated as other jet fragments leave traces in the detector around the photon supercluster. These fragments are reconstructed in the form of tracks, energy deposits in the ECAL and HCAL (detector isolation), or the particle-flow candidates (particle-flow isolation). We use a set of ECAL shower shape and isolation variables for the discrimination between the prompt photon and fake photon, and choose the corresponding cut values to simulate the effects of the trigger cuts on data, and the generator level cuts on Monte Carlo simulated dijet and $\gamma$ + jet events. This removes most of dijet events and a significant amount of $\gamma$ + jet events, while remaining almost fully efficient for events with two prompt photons. 

\subsubsection{Event Classification}
\label{sec:Event Classification}
We then classify the preselected events into classes in the order of roughly $S/B$ under the signal mass peak:
\begin{itemize} 
\item We first select the events into the exclusive tagged classes, based on the signatures of the Higgs production processes including vector boson fusion (\textit{VBF} tag), associated production with a $W$ or $Z$ boson (\textit{VH} tag), and associated production with $t\overline{t}$ (\textit{t$\overline{t}$H} tag):  
  \begin{itemize} 
  \item \textit{VBF} tag: it tags the \textit{VBF} like events by identifying the additional pair of energetic jets with large separation in $\eta$.
  \item \textit{VH} tag: it tags the \textit{VH} like events by identifying the additional $W$ or $Z$ boson in its decays to lepton (electron or muon), dijet, or neutrino manifesting as transverse missing energy. 
  \item \textit{t$\overline{t}$H} tag: it tags the \textit{t$\overline{t}$H} like events by identifying the additional pair of top quarks in their decays to lepton (electron or muon) or multijet.   
  \end{itemize}   
\item We further classify the untagged events according to their diphoton quality, measured by the following elements:
  \begin{itemize} 
  \item Single photon energy resolution: it depends on the same factors as for the photon energy correction.
  \item Diphoton opening angle resolution: it improves as the probability of selecting the right vertex (vertex probability) increases. The vertex probability depends on the transverse momentum of the diphoton, the total number of vertices, the number of converted photons, and how close the scores of the top ranked vertices and their distances to each other. 
  \item Photon identification: the further discrimination, between prompt photons and the more photon like fake photons passing the preselection, depends on finer photon shower and isolation information, which vary with respect to the energy density due to pileup interactions, the photon energy and the photon location.  
  \item Diphoton kinematics: the two photons from Higgs events have different kinematic distributions than those from the background events, because the former are the decays from scalar particles, and the initial states for Higgs production are different from those of the background events. This provides a way to distinguish the Higgs events from the ``irreducible'' diphoton background. We construct a set of variables, which contains full kinematic information of the two photons but with diphoton mass factorized as explained below.       
  \end{itemize}
  We train BDTs, optimally using the information for individual elements, to build a single photon energy resolution estimator from the width of the probability density of $E_{True}/E_{Raw}$ (``photon energy correction regression BDT''), a vertex probability estimator (``vertex probability BDT''), and a classifier between the prompt and fake photons (``photon identification BDT''). We finally use a BDT (``diphoton BDT'') to construct an optimal diphoton event classifier, combining the outputs of all the BDTs for individual elements and the diphoton kinematic information. To maximize the expected Higgs sensitivity, we classify the events according to the diphoton event classifier. 

The training variables are built such that the diphoton BDT cannot reconstruct the diphoton mass to use it distinguishing the signal from the background. This is to achieve the same BDT performance for different Higgs mass hypotheses since the true Higgs mass is unknown. This is also to avoid the preference in selecting background events, with diphoton mass close to the Higgs mass of the signal training sample, into the high $S/B$ event classes to produce an unwanted peak in the background diphoton mass spectrum. There is no loss of sensitivity for this ``diphoton mass factorization'' in the diphoton BDT because the diphoton mass information is used later in the diphoton mass fit for the signal extraction.
\end{itemize}   

\subsection{Higgs Signal Extraction from Diphoton Mass Fit}
\label{sec:Higgs Signal Extraction from Diphoton Mass Fit}
After the event classes are determined, we construct the diphoton mass spectrum for each event class, and the corresponding Higgs signal model and background model:
\begin{itemize} 
\item The expected diphoton mass spectrum of Higgs signal events is modeled by parametric functions, fitted from Monte Carlo simulated events with four Higgs production processes mixed according to their cross sections. The discrepancies between data and Monte Carlo simulation on photons are evaluated mainly using $Z\rightarrow e^{+}e^{-}$ events with the electron reconstructed as the photon. The photons from $Z\rightarrow \mu^{+}\mu^{-}\gamma$ events are used for the validation of Monte Carlo simulation as well, which, though the transverse momentum is on average lower and the statistical uncertainty is larger, provides a valuable cross check to the validation using electrons. The Monte Carlo simulation related to the vertex selection, which mainly depends on the number of interaction vertices and recoiling tracks from the underlying events for a given vertex, is validated using $Z\rightarrow \mu^{+}\mu^{-}$ events. The differences between data and Monte Carlo simulation are either corrected for or treated as systematic uncertainties for the signal model. 
\item The expected diphoton mass spectrum of background events is modeled by parametric functions with a smoothly falling feature, fitted directly from data. The background model in the signal region for any Higgs mass hypothesis under consideration is constrained by the background events in the sidebands. The fitting range is set as $100~\mathrm{GeV}$ $<$ $m_{\gamma\gamma}$ $<$ 180 GeV, to get the signal region well contained, and to get sufficient number of background events in the sidebands. The uncertainty on the Higgs signal extraction and measurements due to the limited knowledge of the exact background shape is evaluated by profiling over a set of functions well describing the data and general enough to cover the true background function.      
\end{itemize} 

The Higgs signal is finally extracted by statistical procedures based on simultaneous likelihood fit to the diphoton mass spectra over all event classes.

\section{Data and Monte Carlo Simulation Samples}
\label{sec:Data and Monte Carlo Samples}
We analyze the full datasets collected by the CMS detector in 2011 and 2012 LHC run periods. The 2011 and 2012 datasets consist of pp collision events respectively at center of mass energy $\sqrt{s}$ $=$ $7~\mathrm{TeV}$ with a integrated luminosity $L$ $=$ $5.1~\mathrm{fb^{-1}}$, and at $\sqrt{s}$ $=$ $8~\mathrm{TeV}$ with $L$ $=$ $19.7~\mathrm{fb^{-1}}$. An event only gets selected if it passes either of the following two classes of diphoton High-Level Triggers designed for $H \rightarrow \gamma\gamma$:
\begin{itemize}  
\item Trigger 1: the photon energy projected in the transverse plane $E_{T}^{\gamma1}$ $>$ $26~\mathrm{GeV}$ for the photon with the highest $E_{T}$ (leading photon), $E_{T}^{\gamma2}$ $>$ $18~\mathrm{GeV}$ for the photon with the second highest $E_{T}$ (sub-leading photon), and both photons passing Level-1 trigger.
\item Trigger 2: $E_{T}^{\gamma1}$ $>$ $36~\mathrm{GeV}$ for the leading photon, $E_{T}^{\gamma2}$ $>$ $22~\mathrm{GeV}$ for the sub-leading photon, and at least one photon passing Level-1 trigger.
\end{itemize} 
 For both types of trigger, the leading and sub-leading photons are required to pass loose photon identification requirements based on shower shape and isolation. The trigger efficiency is 99.4$\%$ for events selected for the final statistical analysis, evaluated using the ``Tag and Probe'' method \cite{tagandprobe} on $Z\rightarrow e^{+}e^{-}$ events.   

We use Monte Carlo simulation samples of $H \rightarrow \gamma\gamma$ to train the BDTs, optimize the event classification, and build the signal model of the diphoton mass distribution. The $H \rightarrow \gamma\gamma$ samples are produced for all the four production processes \textit{ggH}, \textit{VBF}, \textit{VH} and \textit{t$\overline{t}$H} at Higgs mass hypotheses ranging from $115~\mathrm{GeV}$ to $135~\mathrm{GeV}$, at both $\sqrt{s}$ $=$ $7~\mathrm{TeV}$ and $\sqrt{s}$ $=$ $8~\mathrm{TeV}$. To decrease the effect of the statistical fluctuation of any particular sample, samples at different Higgs masses are used in general for the BDT trainings, event class optimization and signal modeling, respectively. For \textit{ggH} and \textit{VBF} processes, POWHEG\cite{powheg1,powheg2,powheg3,powheg-ggH,powheg-VBF} is used for matrix element generation at next-to-leading order (NLO), and PYTHIA\cite{pythia} is used for parton showering and hadronization. For \textit{VH} and \textit{t$\overline{t}$H} processes, PYTHIA is used for both matrix element generation at leading order (LO), and parton showering and hadronization. The production cross sections for the Standard Model Higgs boson, and the branching ratio for its decay to two photons that are used are from the LHC Higgs boson Cross Section Working Group \cite{LHCHiggsCrossSectionWorkingGroup3}. To describe the Higgs kinematics, we match the distribution of the transverse momentum of the Higgs boson from \textit{ggH} process to the next-to-next-to-leading logarithmic resummation (NNLL) plus NLO calculations from HqT\cite{HqT1,HqT2,deFlorian:2011xf}, by reweighting the produced events at 7 TeV, and tuning POWHEG for event generation at $8~\mathrm{TeV}$ according to Reference\cite{LHCHiggsCrossSectionWorkingGroup2} respectively. To account for the effect of interference between the \textit{ggH} process and the continuum $gg \rightarrow \gamma\gamma$ process, we reduce the the cross section for \textit{ggH} process by $2.5\%$\cite{Dixon:2003yb}.   

We use Monte Carlo simulation samples of background processes to train the BDTs and to optimize the event classification. For the ``irreducible'' diphoton background at 7 TeV, the sample of diphoton Born process is generated using MADGRAPH\cite{Alwall:2011uj} interfaced with PYTHIA, and the sample of Box process is generated using PYTHIA. The ``irreducible'' diphoton background at 8 TeV both Born and Box processes are generated using SHERPA\cite{Gleisberg:2008ta}, which provides a better description of the events in the phase space with additional jets from Initial State Radiation(ISR). For the ``reducible'' background, the samples of $\gamma~+~\mathrm{jet}$ process and dijet process are generated using PYTHIA. A ``double EM-enriched filter'' including loose isolation cuts is applied to select the events which are likely to pass the later diphoton selection of the analysis, in order to save computing power for the further simulation of interactions between the particles and the detector. The background cross sections are calculated at LO and corrected by a scale factor from 1.0 to 1.3 obtained from CMS measurements\cite{kfact1,kfact2}.   

\begin{figure}[h] 
  \begin{center}
    \includegraphics[width=0.45\textwidth]{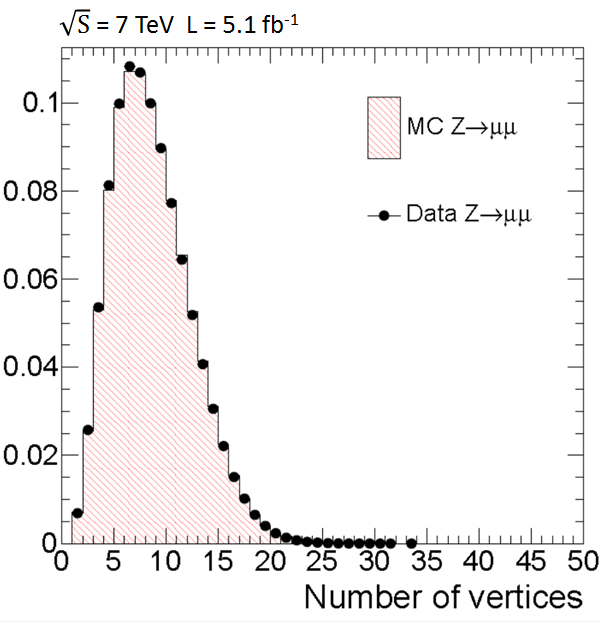}
    \includegraphics[width=0.45\textwidth]{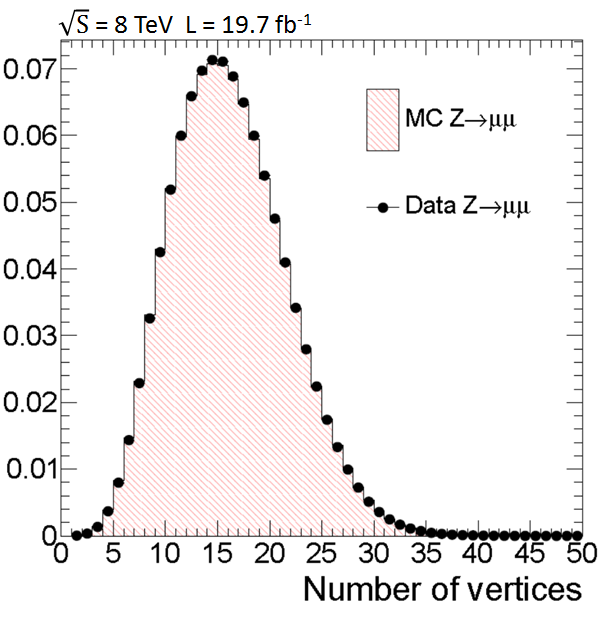}\\
  \end{center}
  \caption{The comparison between data and Monte Carlo distributions of the number of reconstructed vertices in $Z\rightarrow \mu^{+}\mu^{-}$ events after pileup reweighting for 7 TeV (left) and $8~\mathrm{TeV}$ (right) are shown.}
  \label{fig:zmumu}
\end{figure}

For both signal and background Monte Carlo samples pileup interactions are simulated using PYTHIA. For event class optimization and signal modeling, the Monte Carlo events are reweighted to match the pileup distribution in data.  The detector response is simulated using GEANT4\cite{Agostinelli:2002hh}. The discrepancy between data and Monte Carlo simulation is evaluated using events from $Z\rightarrow e^{+}e^{-}$, $Z\rightarrow \mu^{+}\mu^{-}\gamma$, and $Z\rightarrow \mu^{+}\mu^{-}$ data and Monte Carlo simulation generated using POWHEG. Comparison between data and Monte Carlo distributions of the number of reconstructed vertices in $Z\rightarrow \mu^{+}\mu^{-}$ events after pileup reweighting for 7 TeV and 8 TeV are shown on the left and right of Figure \ref{fig:zmumu} \cite{hggan13253}. Good agreement is observed. The data and Monte Carlo samples including the additional background samples used for the \textit{VH} and \textit{t$\overline{t}$H} tags are listed in detail in the analysis note of Reference\cite{hggan13253}.              

%% file: Diphoton.tex
\chapter{Diphoton Reconstruction and Selection}
\label{chaper:Diphoton Reconstruction and Selection}
We preselect a potentially signal-rich sample of diphoton events as described in Section \ref{sec:Diphoton Event Preselection}. The diphotons are reconstructed with corrected photon energy and selected diphoton vertex. The photon energy correction and the diphoton vertex selection are described in Section \ref{sec:Photon Energy Correction} and Section \ref{sec:Vertex Selection BDT}, respectively. To further classify the diphoton events according to $S/B$ under the diphoton mass peak, individual BDTs are first trained to provide a single photon energy resolution estimator as described in $\mathrm{Section}~\mathrm{\ref{sec:Photon Energy Correction Regression}}$, a diphoton vertex probability estimator as described in $\mathrm{Section}~\mathrm{\ref{sec:Vertex Probability BDT}}$, and a single photon identification classifier as described in Section \ref{sec:Photon Identification BDT}. A diphoton BDT is then trained to combine the outputs of the above BDTs into a single diphoton event classifier as described in Section \ref{sec:Diphoton BDT}, which provides a measure of expected $S/B$ for each diphoton event and is used for event classification later.

\section{Diphoton Event Preselection}
\label{sec:Diphoton Event Preselection}
For each event with at least two reconstructed photons, the diphoton pairs are first reconstructed by grouping the reconstructed photons into all possible two photon combinations. For each diphoton pair, a primary vertex is selected as described in Section \ref{sec:Vertex Selection BDT}. The momentum of each photon is constructed with its magnitude obtained from the corrected photon energy as described in Section \ref{sec:Photon Energy Correction} and its direction pointing from the selected vertex to the supercluster. A preselection is then applied to the diphotons, which consists of a so called single photon preselection on each photon, and a set of cuts on the diphoton kinematic acceptance. If more than one diphoton pair pass the preselection, the diphoton with the maximum scalar sum of photon transverse momentum is used for the analysis. A detailed description of the preselection is given below.

\subsection{Single Photon Preselection}
\label{sec:single photon preselection}
The single photon preselection includes a cut on the acceptance of supercluster pseudorapidity measured with respect to the origin of the detector coordinate $\eta_{SC}$, a set of loose photon identification cuts against jets faking photons and an electron veto.

\subsubsection{Acceptance of Supercluster Pseudorapidity}
\label{sec:Acceptance of supercluster pseudorapidity}
The acceptance on the supercluster pseudorapidity is determined to exclude the transition region between the ECAL barrel and endcap, and the region outside the tracker acceptance, which is $|\eta_{SC}|$ $<$ 1.4442 in the barrel or 1.566 $<$ $|\eta_{SC}|$ $<$ 2.5 in the endcap.

\subsubsection{Loose Photon Identification Cuts}
\label{sec:Loose Photon Identification Cuts}
The loose photon identification cuts are applied to a set of ECAL shower shape and isolation variables defined as following:
\begin{itemize}    
\item Shower shape variables
  \begin{itemize} 
  \item $R_{9}$: the ratio between the energy in the $3\times3$ crystals centered at the seed crystal and supercluster energy. 
  \item $\sigma_{i\eta i\eta}$: the log-energy weighted standard deviation of single crystal $\eta$ in crystal index within the $5\times5$ crystals centered at the seed crystal. The weight per-crystal is 4.7 plus the logarithm of the ratio between the energy in the crystal to the energy in the $5\times5$ crystals. If the weight is negative then 0 is used instead. 
  \end{itemize}   
\item $\Delta R$: the separation in the $\eta$-$\phi$ plane $\sqrt{{\Delta\eta}^2+{\Delta\phi}^2}$.  
\item Detector isolation variables
  \begin{itemize}
  \item H/E: the ratio between the sum of energies of deposits in HCAL within $\Delta R\:<~\mathrm{0.15}$ from the ECAL supercluster, and the ECAL supercluster energy.
  \item ISO$_{Trk}$: the sum of $p_{T}$ of tracks within 0.04 $<$ $\Delta R$ $<$ 0.3 from the photon momentum direction. The photon momentum direction used in this case is obtained with respect to the vertex with the maximum sum of track $p_{T}^2$, and only tracks matching this vertex are included in the isolation computation. 
  \item ISO$_{TrkPtCorr}$: ISO$_{Trk}$ ${-}$ 0.002$p_{T}$
  \item ISO$_{HCAL}$: the scalar sum of transverse energies of deposits in HCAL within 0.15 $<$ $\Delta R$ $<$ 0.3 from the ECAL supercluster. 
  \item ISO$_{HCALPtCorr}$: ISO$_{HCAL}$ ${-}$ 0.005$p_{T}$
  \end{itemize}   
\item Particle-flow isolation variable\cite{pf1,pf2}
  \begin{itemize}
  \item ISO$_{PFChargedSelVtx02}$: the sum of $p_{T}$ of particle-flow charged hadron within 0.02 $\leq$ $\Delta R$ $<$ 0.2 from the photon momentum direction. Only the particle-flow charged hadrons with impact parameter along $z$ direction $|d_{z}|\leq$ 0.2 cm and transverse impact parameter $|d_{xy}|\leq$ 0.1 cm with respect to the selected photon vertex are included for the isolation computation.
  \end{itemize}   
\end{itemize}
To apply the loose photon identification cuts, the photons are classified into four categories according to the photon supercluster location in the ECAL (barrel or endcap) and the value of $R_{9}$ ($>$ 0.9 or $\le$ 0.9). The photons in the barrel and endcap are treated separately because the geometry of the crystals and the amount of tracker materials in front are different for ECAL barrel and endcap. The value of $R_{9}$ is used as a measure of the shower width, and the photons with higher $R_{9}$ are more likely to be prompt photons. 
The cut values are in Table \ref{tab:preselCuts}.
\begin{table}[hbtp]
  \noindent
  \small\addtolength{\tabcolsep}{-6pt}
  \caption{The loose photon identification cuts for single photon preselection. The photons are divided into four categories according to the photon supercluster location in the ECAL (barrel or endcap) and the value of $R_{9}$ ($>$ 0.9 or $\le$ 0.9). The cut values vary with the photon categories.}
  \begin{center}
    \setlength{\tabcolsep}{20pt}
    \begin{tabular}{|l|c|c|} 
      \hline
       $R_{9}$ $>$ 0.9  & Barrel & Endcap\\
      \hline
      H/E & $<$ 0.082 & $<$ 0.075\\
      \hline
      $\sigma_{i\eta i\eta}$ & $<$ 0.014 & $<$ 0.034\\
      \hline
      ISO$_{HCALPtCorr}$ & $<$ 50 GeV & $<$ 50 GeV \\
      \hline
      ISO$_{TrkPtCorr}$ & $<$ 50 GeV & $<$ 50 GeV \\
      \hline
      ISO$_{PFChargedSelVtx02}$ & $<$ 4 GeV & $<$ 4 GeV \\
      \hline
      \hline
      $R_{9}$ $\le$ 0.9  & \:Barrel & Endcap\\
      \hline
      H/E & $<$ 0.075 & $<$ 0.075\\
      \hline
      $\sigma_{i\eta i\eta}$ & $<$ 0.014 & $<$ 0.034\\
      \hline
      ISO$_{HCALPtCorr}$ & $<$ 4 GeV & $<$ 4 GeV \\
      \hline
      ISO$_{TrkPtCorr}$ & $<$ 4 GeV & $<$ 4 GeV \\
      \hline
      ISO$_{PFChargedSelVtx02}$ & $<$ 4 GeV & $<$ 4 GeV \\
      \hline
    \end{tabular}
    \label{tab:preselCuts}
  \end{center}
\end{table}

\subsubsection{Electron Veto}
\label{sec:Electron Veto}
The electron veto is used to distinguish electrons from photons. The photon candidates having the same supercluster with a GSF electron candidate are removed. To avoid rejecting the converted photons, the electron track is required to have no missing hits in the tracker before its first hit, and not to match an identified conversion.

\subsection{Diphoton Kinematic Acceptance}
\label{sec:Diphoton Kinematic Acceptance}
The cuts of diphoton kinematic acceptance are determined to select the phase space right above the trigger threshold and to define a region for the diphoton mass fit. The cuts include $p_{T}^{\gamma1}/m_{\gamma\gamma}$ $>$ 1/3 and $p_{T}^{\gamma2}/m_{\gamma\gamma}$ $>$ 1/4, for leading photon $\gamma1$ and sub-leading photon $\gamma2$ respectively, and 100 GeV $<m_{\gamma\gamma}<$ 180 GeV. The threshold of the transverse momentum for photons entering the analysis is thus 100 GeV/4 = 25 GeV.  

\subsection{Selection Efficiencies and Scale Factors Between Data and Monte Carlo Simulation}
\label{sec:Diphoton Kinematic Acceptance}
The efficiencies of the loose photon identification cuts for the prompt photons in the four photon categories are evaluated using electrons from $Z\rightarrow e^{+}e^{-}$ events, for which the electron $R_{9}$ is rescaled to match the photon $R_{9}$ distribution. The ``Tag and Probe'' method \cite{tagandprobe} is used to evaluate the efficiencies on data and Monte Carlo simulation at 7 TeV and 8 TeV respectively. The efficiencies as well as the corresponding efficiency scale factors, ratios between efficiencies on data and Monte Carlo simulation, are in Table \ref{tab:preseleff}.  
\begin{table}[htbp]
\caption{The loose photon identification efficiencies for prompt photons from data and Monte Carlo simulation at 7 TeV and at $8~\mathrm{TeV}$ as well as the corresponding efficiency scale factors between data and Monte Carlo simulation. The photons are classified into four categories according to the photon supercluster location in the ECAL (barrel or endcap) and the value of $R_{9}$ ($>$ 0.9 or $\le$ 0.9). The efficiencies are evaluated using electrons from $Z\rightarrow e^{+}e^{-}$ events.}
\begin{center}
\begin{tabular}{|l|c|c|c|}
\hline
$\sqrt{s}$ $=$ $7~\mathrm{TeV}$ & Data & Monte Carlo & Data/Monte Carlo Scale Factor \\
\hline 
$R_{9}$ $>$ 0.9 Barrel & 0.9872 $\pm$ 0.0025 & 0.9908 $\pm$ 0.0002 & 0.996 $\pm$ 0.003 \\
$R_{9}$ $\le$ 0.9 Barrel & 0.9619 $\pm$ 0.0050 & 0.9670 $\pm$ 0.0005 & 0.995 $\pm$ 0.006 \\
$R_{9}$ $>$ 0.9 Endcap & 0.9906 $\pm$ 0.0085 & 0.9824 $\pm$ 0.0004 & 1.008 $\pm$ 0.009 \\
$R_{9}$ $\le$ 0.9 Endcap & 0.9606 $\pm$ 0.0150 & 0.9560 $\pm$ 0.0011 & 1.005 $\pm$ 0.018 \\
\hline
\hline
$\sqrt{s}$ $=$ $8~\mathrm{TeV}$ & Data & Monte Carlo & Data/Monte Carlo Scale Factor \\
\hline 
$R_{9}$ $>$ 0.9 Barrel & 0.9879 $\pm$ 0.0030 & 0.9864 $\pm$ 0.0001 & 0.999 $\pm$ 0.003 \\
$R_{9}$ $\le$ 0.9 Barrel & 0.9566 $\pm$ 0.0055 & 0.9610 $\pm$ 0.0002 & 0.995 $\pm$ 0.006 \\
$R_{9}$ $>$ 0.9 Endcap & 0.9838 $\pm$ 0.0090 & 0.9789 $\pm$ 0.0002 & 1.005 $\pm$ 0.009 \\
$R_{9}$ $\le$ 0.9 Endcap & 0.9545 $\pm$ 0.0170 & 0.9445 $\pm$ 0.0005 & 1.011 $\pm$ 0.018 \\
\hline
\end{tabular}
\end{center}
\label{tab:preseleff}
\end{table}

The electron veto efficiencies for the prompt photons are evaluated using photons from $Z\rightarrow \mu^{+}\mu^{-}\gamma$ events. The photons are classified into four categories according to the photon supercluster location in the ECAL (barrel or endcap) and the value of $R_{9}$ ($>$ 0.94 or $\le$ 0.94). The value of $R_{9}$ is used as a measure of the likelihood of photon conversion. The photons with $R_{9}$ $>$ 0.94 are dominated by unconverted photons while photons with $R_{9}$ $\le$ 0.94 are dominated by converted photons. The efficiencies on data and Monte Carlo simulation at 8 TeV as well as the corresponding efficiency scale factors between data and Monte Carlo simulation are shown in Table \ref{tab:elevetoeff}. The efficiencies on data and Monte Carlo simulation at 7 TeV and the corresponding scale factors come out to be 1.
\begin{table}[htbp]
\caption{The electron veto efficiencies for prompt photons from data and Monte Carlo at $8~\mathrm{TeV}$ as well as the corresponding efficiency scale factors between data and Monte Carlo. The photons are classified into four categories according to the photon supercluster location in the ECAL (barrel or endcap) and the value of $R_{9}$ ($>$ 0.94 or $\le$ 0.94). The efficiencies are evaluated using photons from $Z\rightarrow \mu^{+}\mu^{-}\gamma$ events.}
\begin{center}
\begin{tabular}{|l|c|c|c|}
\hline
$\sqrt{s}$ $=$ $8~\mathrm{TeV}$ & Data & Monte Carlo & Data/Monte Carlo Scale Factor \\
\hline 
$R_{9}$ $>$ 0.94 Barrel &  0.9984 $\pm$ 0.0003 &  0.9991 $\pm$ 0.0003 & 0.9994 $\pm$ 0.0004\\
$R_{9}$ $\le$ 0.94 Barrel &  0.9867 $\pm$ 0.0012 &  0.9930 $\pm$ 0.0009 & 0.9937 $\pm$ 0.0014\\
$R_{9}$ $>$ 0.94 Endcap &  0.9893 $\pm$ 0.0016 &  0.9938 $\pm$ 0.0012 & 0.9955 $\pm$ 0.0020\\
$R_{9}$ $\le$ 0.94 Endcap &  0.9639 $\pm$ 0.0033 &  0.9738 $\pm$ 0.0030 & 0.9899 $\pm$ 0.0045\\
\hline
\end{tabular}
\end{center} \label{tab:elevetoeff}
\end{table}    

\section{Photon Energy Correction}
\label{sec:Photon Energy Correction}
\subsection{Photon Energy Correction Regression BDT}
\label{sec:Photon Energy Correction Regression}
The photon energy correction regression BDT is trained to provide each photon a correction factor to its raw energy, and a per-photon energy resolution estimator, which is used for the diphoton BDT as described in Section \ref{sec:Diphoton BDT}.

\subsubsection{Training Samples}
\label{sec:Training Samples}    
The training sample for the BDT is composed of reconstructed photons from Monte Carlo $\gamma$ + jet events. Each photon is required to match a prompt photon at the generator level, and the generated energy of the prompt photon is used as the true photon energy $E_{True}$. In addition, the photon is required to pass the single photon preselection with $p_{T}$ $>$ 15 GeV, looser than the analysis threshold of 25 GeV to increase the size of training sample. The trainings are performed separately for photons from pp collisions with different center-of-mass energies (7 TeV or 8 TeV) and in different ECAL locations (barrel or endcap). 

\subsubsection{Input Variables}
\label{sec:Input Variables}
The input variables are summarized as following:
\begin{itemize}    
\item Supercluster variables:
  \begin{itemize} 
  \item $E_{SC}$: energy deposit in the ECAL supercluster.
  \item $\eta_{SC}$: pseudorapidity of ECAL supercluster measured with respect to the origin of the detector coordinate.
  \item $R_{9}$: the ratio between the energy in the $3\times3$ crystals centered at the seed crystal and supercluster energy.
  \item $H/E$: the ratio between the sum of energies of deposits in HCAL within $\Delta R\:<~\mathrm{0.15}$ from the ECAL supercluster, and the ECAL supercluster energy.
  \item SC $\eta$-Width: the energy-weighted standard deviation of single crystal eta in detector coordinate within supercluster. The weight per-crystal is the ratio of the single crystal energy to the supercluster energy.
  \item SC $\phi$-Width: the energy-weighted standard deviation of single crystal phi in detector coordinate within supercluster. The weight per-crystal is the ratio of the single crystal energy to the supercluster energy. 
  \item The number of basic clusters.
  \item The supercluster azimuthal angle $\phi_{SC}$. (This is only used for the barrel since its inclusion does not improve the resolution for electrons in the endcap from $Z\rightarrow e^{+}e^{-}$ events in data.)    
  \item Ratio between preshower energy and supercluster energy (endcap only). 
  \end{itemize}   
\item Seed basic cluster variables:
  \begin{itemize}
  \item Ratio between seed basic cluster energy and supercluster energy. 
  \item Seed basic cluster $\eta$ and $\phi$ relative to the supercluster
  \item $\sigma_{i\eta i\eta}$: the log-energy weighted standard deviation of single crystal $\eta$ in crystal index within the $5\times5$ crystals centered at the seed crystal. The weight per-crystal is 4.7 plus the logarithm of the ratio between the energy in the crystal to the energy in the $5\times5$ crystals. If the weight is negative then 0 is used instead. 
  \item $\sigma_{i\phi i\phi}$: the log-energy weighted standard deviation of single crystal $\phi$ in crystal index within the $5\times5$ crystals centered at the seed crystal.
  \item cov$_{i\eta i\phi}$: the log-energy weighted covariance of single crystal $\eta$-$\phi$ in crystal index within the $5\times5$ crystals centered at the seed crystal.
  \item Ratios between energies of various combinations of crystals within the seed basic cluster and seed basic cluster energy.    
  \end{itemize}   
\item Seed crystal variables:
  \begin{itemize}
  \item Seed crystal $\eta$ and $\phi$ relative to the seed basic cluster.
  \end{itemize}   
\item Pileup variables:
  \begin{itemize}
  \item $\rho_{Event}$: the estimate of transverse energy per unit area in the $\eta$-$\phi$ plane contributed by the pileup interactions and underlying-event effects in the event. It is the median $\frac{Jet\:p_{T}}{Jet\:Area}$ of jets constructed using the $k_{T}$ algorithm\cite{Cacciari:2007fd}.   
  \item $N_{Vtx}$: the number of reconstructed vertices.
  \end{itemize}   
\end{itemize}

\subsubsection{Output and Performance}
\label{sec:Output and Performance}
The target is the probability density of $E_{True}/E_{Raw}$ for any photon with input variable $\overrightarrow{x}$. It is parametrized empirically using a modified Crystal Ball function ($\mathrm{CB_{double-sided}}$)\cite{CB} consisting of a Gaussian core and power law tails on both sides:
\begin{equation} 
  \text{Target} = \mathrm{CB_{double-sided}}(E_{True}/E_{Raw}\mid\mu(\overrightarrow{x}),\sigma(\overrightarrow{x}),\alpha_{L}(\overrightarrow{x}),n_{L}(\overrightarrow{x}),\alpha_{R}(\overrightarrow{x}),n_{R}(\overrightarrow{x})),
  \label{eqn:double CB}
\end{equation}
where $\mu(\overrightarrow{x})$ and $\sigma(\overrightarrow{x})$ are the mean and standard deviation of Gaussian core, and $\alpha_{L(R)}(\overrightarrow{x})$ and $n_{L(R)}(\overrightarrow{x})$ are the cut off and power of left (right) tail. The parameters are functions of the input variables $\overrightarrow{x}$ estimated by BDT and are determined by maximum likelihood fit.   

\begin{figure}[hbpt] 
  \begin{center}
    \includegraphics[width=0.4\textwidth]{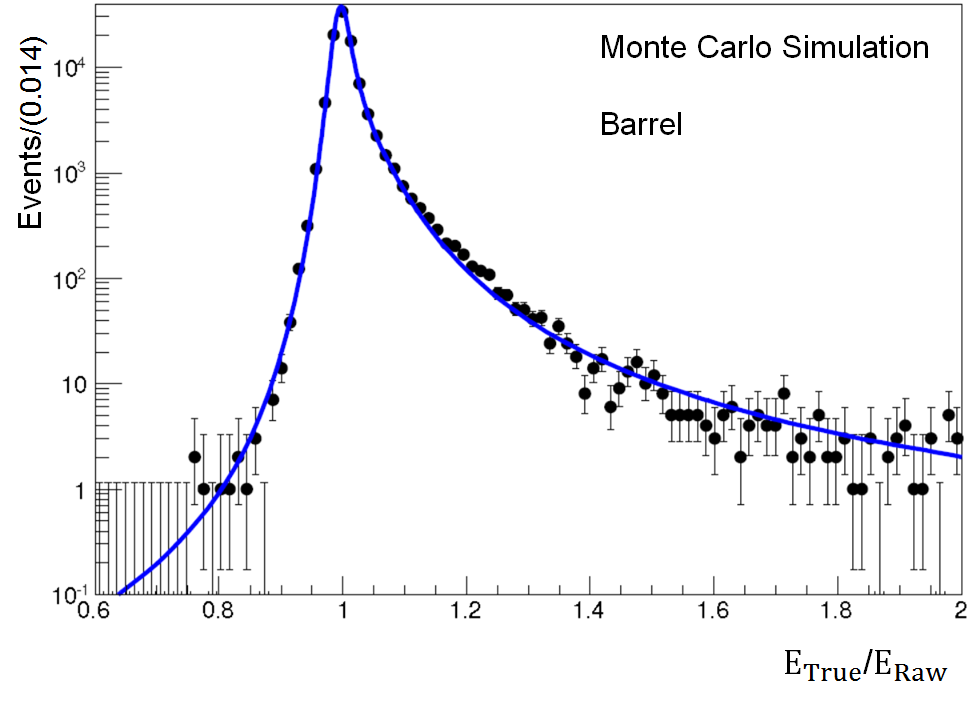}
    \includegraphics[width=0.4\textwidth]{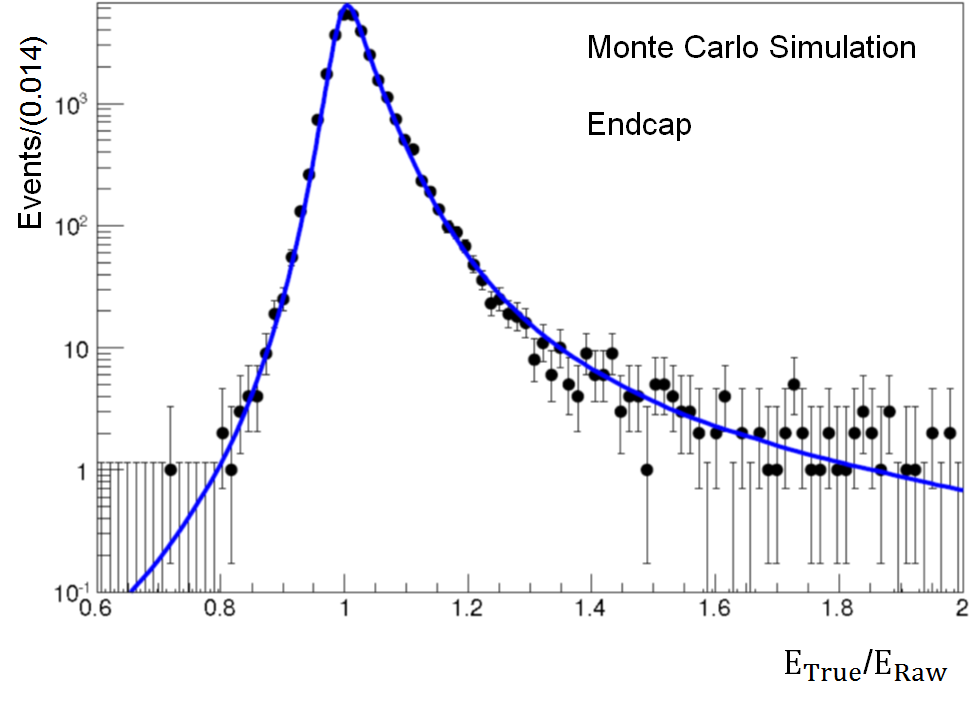}
  \end{center}
  \caption{The normalized sum of the individual photon $E_{True}/E_{Raw}$ distributions estimated by the regression BDT (blue line), compared to the true $E_{True}/E_{Raw}$ distribution (points), for photons in the barrel (left) and in the endcap (right) of a Monte Carlo sample independent from the training sample.}
  \label{fig:regression output}. 
\end{figure}

The trained BDT estimates the probability density of $E_{True}/E_{Raw}$ for each photon according to its input variable $\overrightarrow{x}$. The performance of the estimation is evaluated on a testing Monte Carlo sample of photons, independent from the training sample. As shown on the left (right) in Figure \ref{fig:regression output}\cite{hggan13253}, for photons in the barrel (endcap), the normalized sum of the estimated $E_{True}/E_{Raw}$ distribution for each photon (blue line) agrees well with the true $E_{True}/E_{Raw}$ distribution of the sample (points).      

For each photon, its energy is corrected to the most probable value of the true energy $E(E_{Raw},\overrightarrow{x})$ by multiplying the correction factor $\mu(\overrightarrow{x})$ as:
\begin{equation}
  E(E_{Raw},\overrightarrow{x}) =  \mu(\overrightarrow{x})E_{Raw}.
  \label{eqn:regression energy}
\end{equation}
The per-photon energy resolution estimator $(\sigma_{E}/{E})(\overrightarrow{x})$ is assigned as:
\begin{equation}
  (\sigma_{E}/{E})(\overrightarrow{x}) = \sigma(\overrightarrow{x})/\mu(\overrightarrow{x}).
  \label{eqn:regression resolution}
\end{equation}

\subsection{Energy Correction Between Data and Monte Carlo Simulation}
\label{sec:Energy Correction Between Data and Monte Carlo}
The imperfect simulation of detector effects causes discrepancies in the scale and resolution of regression photon energy between data and Monte Carlo simulation. The discrepancies are corrected for building the model of diphoton mass spectrum for the Higgs boson from Monte Carlo simulation, which is used in the signal extraction and is crucial for the Higgs mass measurement. The corrections are derived from $Z\rightarrow e^{+}e^{-}$ events from data and Monte Carlo simulation with the electron ECAL supercluster reconstructed the same way as the photon supercluster, and performed in a three-step procedure. The first step corrects the energy scale difference mainly due to the imperfect correction for the crystal transparency loss in data, which varies with time and photon location. The second step mainly corrects the underestimation of the energy resolution in Monte Carlo, which varies with the photon location and whether a photon is converted. The third step corrects the residual energy scale difference as a function of photon energy for photons from 8 TeV data, for which enough statistics is available for the derivation of this fine-grained correction. A detailed description of these three steps are as follows.

First, the energies of photons from data are scaled to match the energy scale of Monte Carlo simulated photons. The scale factors are derived separately for photons from different LHC run ranges and located in different pseudorapidities, which are classified into 59 run ranges $\times$ 4 $|\eta_{SC}|$ ranges, 2 $|\eta_{SC}|$ ranges for barrel and 2 $|\eta_{SC}|$ ranges for endcap, single photon categories. To derive the scale factor $ R_{step 1}$ for each category, the mass spectra are built for $Z\rightarrow e^{+}e^{-}$ events from both data and Monte Carlo simulation, with both electrons from the same single photon category. Each spectrum is fitted by an expected mass distribution p($m_{ee}$), parametrized as a Breit-Wigner (BW) function convoluted with a Crystal Ball (CB) function:
\begin{equation}
  \text{p}(m_{ee}) = \text{BW}(m_{ee}\mid m_{Z},\Gamma_{Z})\:\star\:\text{CB}(m_{ee}\mid\Delta M,\Delta \sigma,\alpha,n),
  \label{eqn:BW convoluted CB}
\end{equation}    
where the Breit-Wigner function models the intrinsic distribution of $Z\rightarrow e^{+}e^{-}$, with the peak mass $m_{Z}$ and width $\Gamma_{Z}$ parameters fixed to the Particle Data Group values\cite{pdg}. The Crystal Ball function models effects from the detector measurement with the parameters mean and standard deviation of Gaussian core $\Delta M$ and $\Delta \sigma$, and  cut off and power of the power law tail $\alpha$ and $n$, floating during the fit. The scale factor $R_{step 1}$ is then obtained as the ratio between the measured mass $m_{peak\:MC}$ of Monte Carlo (MC) simulation and $m_{peak\:Data}$ of data as: 
\begin{equation}
  R_{step 1} = \frac{m_{peak\:MC}}{m_{peak\:Data}} = \frac{ m_{Z} + \Delta M_{MC} }{ m_{Z} + \Delta M_{Data} },
  \label{eqn:scale factor}
\end{equation}
where $\Delta M_{MC(Data)}$ is the fitted mean of the Gaussian core of the Crystal Ball function for Monte Carlo simulation (data). 
 
Second, the energies of photons from Monte Carlo simulation are smeared to match the energy resolution of photons from data, while the energies of photons from data are further scaled to correct the residual scale difference with Monte Carlo simulation. The corrections are derived for 2 $R_{9}$ $\times$ 4 $|\eta_{SC}|$ single photon categories, with a smearing parameter $\sigma_{smear}$ and a scale factor $R_{step 2}$ for each category. To derive the corrections, the double electrons from data and Monte Carlo simulation are classified into 36 categories according to the single photon categories of the two electrons. For each double electron category, the energy of each Monte Carlo electron is scaled by a random factor from a Gaussian distribution with mean at 1 and standard deviation $\sigma_{smear}$ corresponding to its single photon category. The histogram of the smeared double electron mass $m_{ee}$ for Monte Carlo events is constructed correspondingly, which is a function of the smearing values of both electrons ($\sigma_{smear}^{i}$, $\sigma_{smear}^{j}$), where $i$ and $j$ represent the related single photon category numbers. The energy of each electron from data is scaled by $R_{step 2}$ corresponding to its single photon category, and the scaled double electron mass spectrum is built as a function of ($R_{step 2}^{i}$, $R_{step 2}^{j}$). The smeared Monte Carlo histogram of $m_{ee}$ is then fitted to the scaled data. The 8 pairs of ($\sigma_{smear}$, $R_{step 2}$) for each single photon category are determined by maximizing the total likelihood of the 36 double electron categories. 

The smearing parameter $\sigma_{smear}$ is parametrized as a constant for electrons in the endcap at 7 TeV and 8 TeV. An additional energy dependent term is added in quadrature for electrons in the barrel at 8 TeV, where more statistics are available to fit the improved parametrization:   
\begin{equation}
\sigma_{smear} = \left\{
  \begin{array}{lr}
    C_{1}                                   & \text{(7 TeV and 8 TeV Endcap)}, \\
    \sqrt{C_{1}^2+(C_{2}/\sqrt{E_{T}})^2}&             \text{(8 TeV Barrel)}, 
  \end{array}
\right.
\label{eqn:energy dependent smearing}
\end{equation}
where $C_{1}$ and $C_{2}$ are constants for each single photon category, and $E_{T}$ is the photon transverse energy.

Third, a residual energy dependent scale factor is applied to the energies of photons in the barrel from data at 8 TeV. The scale factors are derived for $\:$20$\:$$R_{9}$ $\times$ $|\eta_{SC}|$ $\times~E_{T}$ categories following the same method as in the second step.

\begin{figure}[h] 
  \begin{center}
    \includegraphics[width=0.9\textwidth]{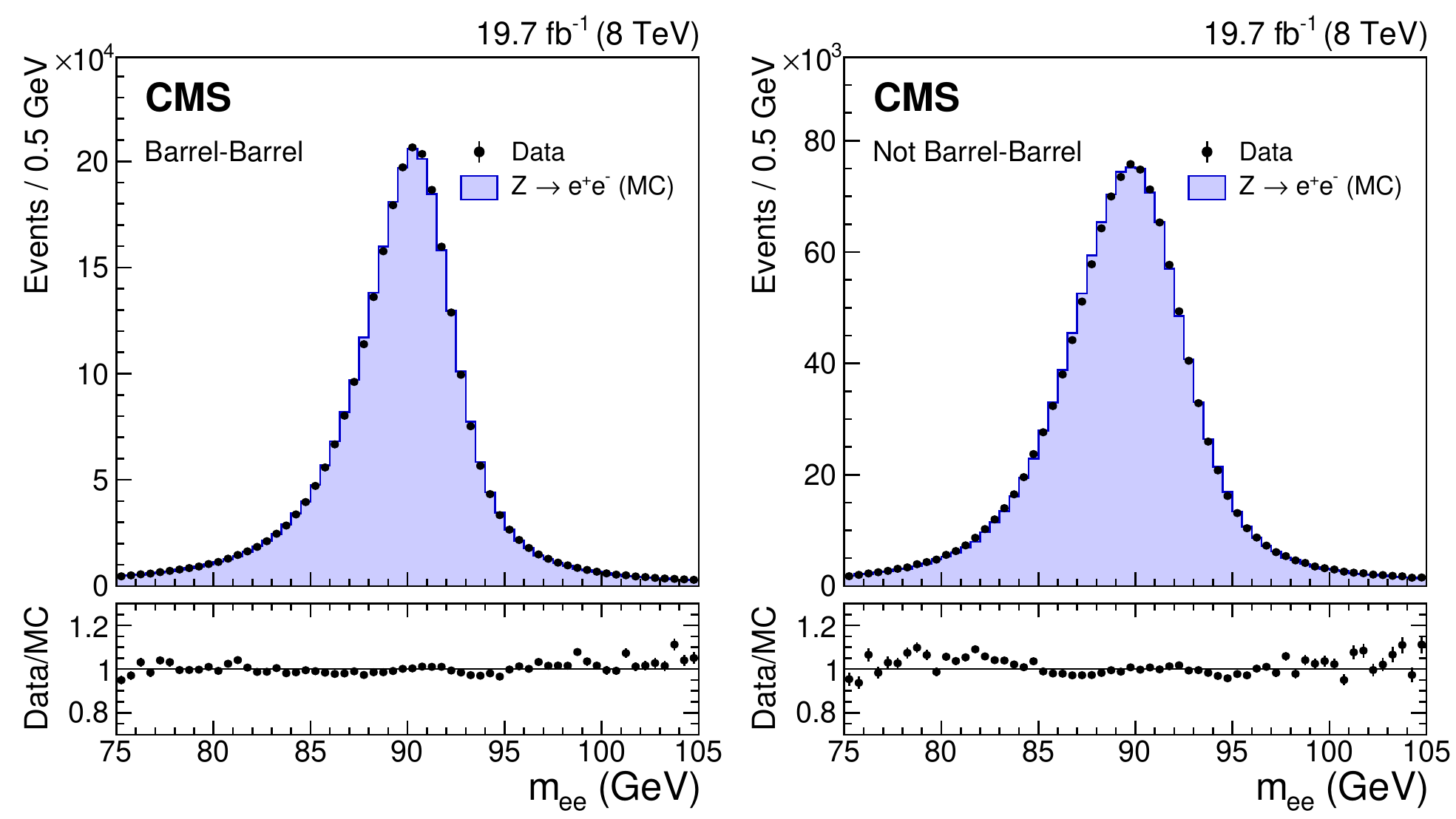}
  \end{center}
  \caption{The $Z\rightarrow e^{+}e^{-}$ mass distributions of data (points) and Monte Carlo simulated events (histogram) at 8 TeV after energy corrections with both electrons in the barrel (left) and at least one electron in the endcap (right). The electron ECAL superclusters are reconstructed in the same way as the photon superclusters, and the events are requested to pass the preselection with inverted electron veto.}
  \label{fig:zeemassenergy}
\end{figure}

The comparison between 8 TeV data and Monte Carlo simulated $Z\rightarrow e^{+}e^{-}$ mass distributions after energy corrections are shown in Figure \ref{fig:zeemassenergy}\cite{hggfinalpaper}. The events are requested to pass the preselection with inverted electron veto. The distributions for the events with both electrons in the barrel are shown on the left, and the distributions for the events with at least one electron in the endcap are shown on the right. Good agreement is observed and the residual difference is taken into account as the systematic uncertainties due to correction method in the signal modeling for the final statistical analysis as described in Section \ref{sec:Systematic Uncertainties}.  

The relative resolution estimator $\sigma_{E}/{E}$ for each single photon is smeared as well for both data and Monte Carlo simulation, by adding in quadrature the smearing parameter $\sigma_{smear}$ for the corresponding photon category. The discrepancy between the $\sigma_{E}/{E}$ distributions of Monte Carlo simulation and data due to the imperfect simulation of detector response are evaluated using electrons from $Z\rightarrow e^{+}e^{-}$ events and photons from $Z\rightarrow \mu^{+}\mu^{-}\gamma$ events. A scaling of $\pm$10$\%$ of the Monte Carlo $\sigma_{E}/{E}$ is shown to cover the discrepancy. 

\section{Vertex Selection}
\label{sec:Vertex Selection}

\subsection{Vertex Selection BDT}
\label{sec:Vertex Selection BDT}
The vertex selection BDT is trained to select the diphoton production vertex in an event, which has an average of 21 (9) pp collision vertices for 8 TeV (7 TeV) as a result of pileup interactions.  

\subsubsection{Training Samples}
\label{sec:Training Samples}
The training is performed on a Monte Carlo simulation of $H\rightarrow \gamma\gamma$ events. The signal sample consists of the reconstructed vertices of diphotons from Higgs decays, while the background sample consists of the pileup vertices. 

\subsubsection{Input Variables}
\label{sec:Input Variables}
The input variables are the following:     
\begin{itemize}
\item $\sum_{i}^{}({p_{T i}^{Trk}})^2$: the sum of the square of transverse momentum of each track associated with the vertex, ${p_{T i}^{Trk}}$. This quantity is expected to be larger for the diphoton vertex than for pileup vertices. 
\item $(\sum_{i}\overrightarrow{p}_{T i}^{Trk})\cdot(-\overrightarrow{p}_{T}^{\gamma\gamma}/p_{T}^{\gamma\gamma})$: the projection of the sum of transverse momenta of tracks associated with the vertex $\sum_{i}\overrightarrow{p}_{T i}^{Trk}$ onto the opposite direction of the diphoton transverse momentum $\overrightarrow{p}_{T}^{\gamma\gamma}$. This quantity is expected to be near 0 for the pileup vertices while near $p_{T}^{\gamma\gamma}$ for the diphoton vertex.
\item $(|\sum_{i}\overrightarrow{p}_{T i}^{Trk}|-p_{T}^{\gamma\gamma})/(|\sum_{i}\overrightarrow{p}_{T i}^{Trk}|+p_{T}^{\gamma\gamma})$: the asymmetry between the magnitude of the vector sum of transverse momenta of tracks associated with the vertex, $|\sum_{i}\overrightarrow{p}_{T i}^{Trk}|$, and the magnitude of diphoton transverse momentum, $p_{T}^{\gamma\gamma}$. This quantity is expected to be near $-$1 for the pileup vertices while near 0 for the diphoton vertex.
\item $|z_{Vtx}-z_{Conv}|/\sigma_{z_{Conv}}$ only for cases with at least one converted photon: the distance between the $z$ position of the vertex, $z_{Vtx}$, and the estimated position of diphoton vertex, $z_{Conv}$, from the conversion and normalized by the uncertainty of the estimation, $\sigma_{z_{Conv}}$. This quantity is expected to be near 0 for the diphoton vertex while larger for pileup vertices. 
\end{itemize}

\subsubsection{Output}
\label{Output}
The BDT output is a score assigned to each vertex which ranges from $-$1 to 1. The higher the score assigned to a vertex, the more likely the vertex is the diphoton production vertex. The vertex with the highest BDT score is selected as the diphoton vertex.
   
\subsection{Vertex Probability BDT}
\label{sec:Vertex Probability BDT}
The vertex probability BDT is trained to estimate the probability that the selected vertex is the correct diphoton vertex for each event. The criteria for being correct is that the distance between the selected vertex and the true diphoton vertex is within 1 cm in the $z$ direction, in which case the diphoton mass resolution is insensitive to the exact position of the vertex. The vertex probability is a measure of the diphoton opening angle resolution. It is used for the diphoton BDT as described in Section \ref{sec:Diphoton BDT}.

\subsubsection{Training Samples}
\label{sec:Training Samples}
The training is performed on a Monte Carlo simulation of $H\rightarrow \gamma\gamma$ events. The signal sample consists of events with correct vertex selected, and the background sample consists of events with wrong vertex selected. 

\subsubsection{Input Variables}
\label{sec:Input Variables}
The input variables are the following:       
\begin{itemize}
\item $N_{Vtx}$: the number of reconstructed vertices.
\item $p_{T}^{\gamma\gamma}$: the magnitude of diphoton transverse momentum.
\item The top three vertex selection BDT scores for the vertices in the event.
\item The distances in $z$ between the selected vertex and the vertices with the second and the third highest BDT scores.
\item The number of conversions in the diphoton (0, 1 or 2).
\end{itemize}

\subsubsection{Output}
\label{sec:Output}     
The BDT output is a score assigned to each event, which ranges from $-$1 to 1. Events are binned according to BDT score, and the diphoton vertex selection efficiency in each bin, defined as the fraction of events with diphoton vertex selected correctly in the bin, is measured. A linear relation between the vertex selection efficiency and the BDT score is derived, which is used to transform a BDT score to a per-event vertex probability between 0 and 1.  

\subsection{Performance}
\label{sec:Performance}     
\begin{figure}[h] 
  \begin{center}
    \includegraphics[width=0.6\textwidth]{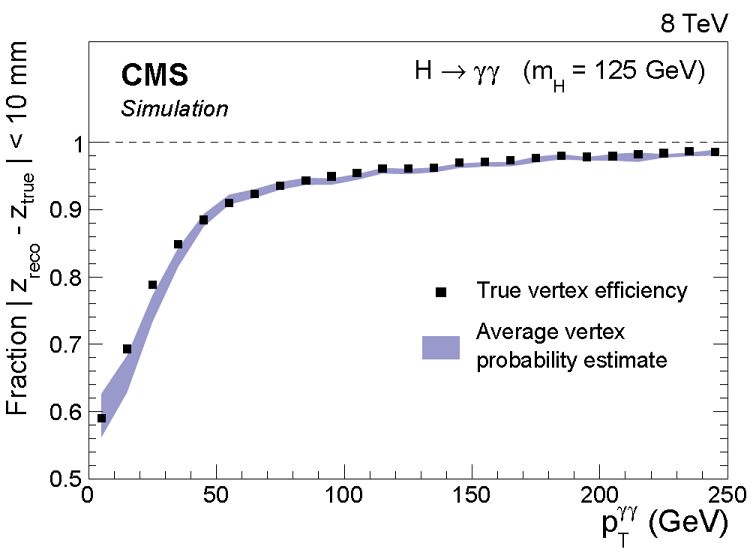}
  \end{center}
  \caption{The measured vertex selection efficiency (points) and the average vertex probability along with uncertainty (blue band) evaluated from the BDT on Monte Carlo simulated $H\rightarrow \gamma\gamma$ events at a Higgs mass of 125 GeV at 8 TeV, in bins of $p_{T}^{\gamma\gamma}$.}
  \label{fig:ptvertex}
\end{figure}

To measure the performance of both the vertex selection BDT and the vertex probability BDT, the diphoton vertex selection efficiency and the average vertex probability are evaluated on Monte Carlo simulated $H\rightarrow \gamma\gamma$ events at a Higgs mass of 125 GeV, in bins of $p_{T}^{\gamma\gamma}$. As shown in Figure \ref{fig:ptvertex}\cite{hggfinalpaper}, the average vertex probability along with uncertainty (blue band) predicts well the measured vertex selection efficiency (data points), and both increase with the increasing $p_{T}^{\gamma\gamma}$. The total vertex selection efficiency is 79.6$\%$ (85.4$\%$) for the $H\rightarrow \gamma\gamma$ events at a Higgs mass of 125 GeV at 8 TeV (7 TeV). The efficiency at 8 TeV is lower than that at 7 TeV because of higher number of pilup interactions.           

\section{Photon Identification BDT}
\label{sec:Photon Identification BDT}
The photon identification BDT is trained to provide each photon with a score, measuring how likely it is a prompt photon rather than a jet faking a photon (fake photon), which is used as an input to the diphoton BDT as described in Section \ref{sec:Diphoton BDT}. 

\subsection{Training Samples}
\label{sec:Training Samples}
The training is performed on a Monte Carlo simulation of $\gamma$ + jet events passing the preselection with $p_{T}$ $>$ 15 GeV, looser than the analysis threshold of 25 GeV to increase the size of the training sample. The training is tested with another training on a sample passing the preselection with $p_{T}$ $>$ 25 GeV. The performances of BDTs from both trainings agree well. The signal sample consists of the reconstructed photons which match prompt photons at the generator level, while the background sample consists of the ones that do not match. The trainings are performed separately for photons from pp collisions with different center-of-mass energies (7 TeV or 8 TeV) and in different ECAL locations (barrel or endcap). 

\subsection{Input Variables}
\label{sec:Input Variables}
The input variables for the photon identification BDT are listed as following:
\begin{itemize}    
\item Shower shape variables:
  \begin{itemize} 
  \item $R_{9}$: the ratio between the energy in the $3\times3$ crystals centered at the seed crystal and the supercluster energy.
  \item SC $\eta$-Width: the energy-weighted standard deviation of single crystal $\eta$ in detector coordinate within supercluster. The weight per-crystal is the ratio of the single crystal energy to the supercluster energy.
  \item SC $\phi$-Width: the energy-weighted standard deviation of single crystal $\phi$ in detector coordinate within supercluster. The weight per-crystal is the ratio of the single crystal energy to the supercluster energy.
  \item $\sigma_{i\eta i\eta}$: the log-energy weighted standard deviation of single crystal $\eta$ in crystal index within the $5\times5$ crystals centered at the seed crystal. The weight per-crystal is 4.7 plus the logarithm of the ratio between the energy in the crystal to the energy in the $5\times5$ crystals. If the weight is negative then 0 is used instead. 
  \item cov$_{i\eta i\phi}$: the log-energy weighted covariance of single crystal $\eta$-$\phi$ in crystal index within the $5\times5$ crystals centered at the seed crystal. 
  \item $E_{2\times2}/E_{5\times5}$: the ratio of the energy in the $2\times2$ crystal array containing the seed crystal (the $2\times2$ crystal array with the highest energy in all the possible combinations) to the energy in the $5\times5$ crystals centered at the crystal.
  \item Preshower $\sigma_{RR}$ (endcap only): the sum in quadrature of the energy-weighted standard deviation of the strip index in the $x$ and $y$ planes of the preshower detector. 
  \end{itemize}   
\item Particle-flow based isolation variables\cite{pf1,pf2}:
  \begin{itemize}
  \item ISO$_{PFChargedSelVtx03}$: defined in the same way as with ISO$_{PFChargedSelVtx02}$, which is defined in Section \ref{sec:Diphoton Event Preselection} but using here a different annulus of 0.02 $\leq$ $\Delta R$ $<$ 0.3.   
  \item ISO$_{PFChargedWorstVtx03}$: defined in the same way as with ISO$_{PFChargedSelVtx03}$ but using the vertex with the maximum isolation for photon momentum direction and isolation computation.
  \item ISO$_{PFPhoton}$: the $p_{T}$ sum of particle-flow photon within annulus 0.07 $\leq$ $\Delta R$ $<$ 0.3 ($\Delta R$ $<$ 0.3 and $|\Delta\eta|$ $\geq$ 0.015) from the photon momentum direction for photon in the endcap (barrel). The photon momentum direction used in this case is obtained with respect to the vertex associated with each particle-flow photon.  
  \end{itemize}   
\item Auxiliary variables:
  \begin{itemize}
  \item $\rho_{Event}$: the estimate of transverse energy per unit area in the $\eta$-$\phi$ plane contributed by the pileup interactions and underlying-event effects in the event. It is the median $\frac{Jet\:p_{T}}{Jet\:Area}$ of jets constructed using the kt algorithm\cite{Cacciari:2007fd}. 
  \item $\eta_{SC}$: pseudorapidity of the ECAL supercluster measured with respect to the origin of the detector coordinate. 
  \item $E_{SC}$: energy deposit in the ECAL supercluster.
  \end{itemize}    
\end{itemize}
The shower shape variables and the isolation variables are used as they are related to the two intrinsic differences between a prompt photon and a fake photon, respectively. One is that the shower of a fake photon is wider on average since it is the combined shower of the two photons from a neutral meson decay. The other is that the isolation for a fake photon is larger due to the traces of other fragments of the associated jet leaving in the detector around the photon supercluster. The auxiliary variables are included such that the distributions of shower shape and isolation variables are used differentially as functions of pileup contamination measured by $\rho_{Event}$ and photon kinematics measured by $\eta_{SC}$ and $E_{SC}$. In order to reduce the photon kinematic dependence of the photon identification BDT and the associated mass dependence in the diphoton BDT, explicit use of kinematic differences between prompt photons and fake photons in the training sample is avoided, by reweighting the 2D $p_{T}$-$\eta_{SC}$ distribution of the signal to that of the background. 

\begin{figure}[hbpt] 
  \begin{center}
    \includegraphics[width=0.31\textwidth]{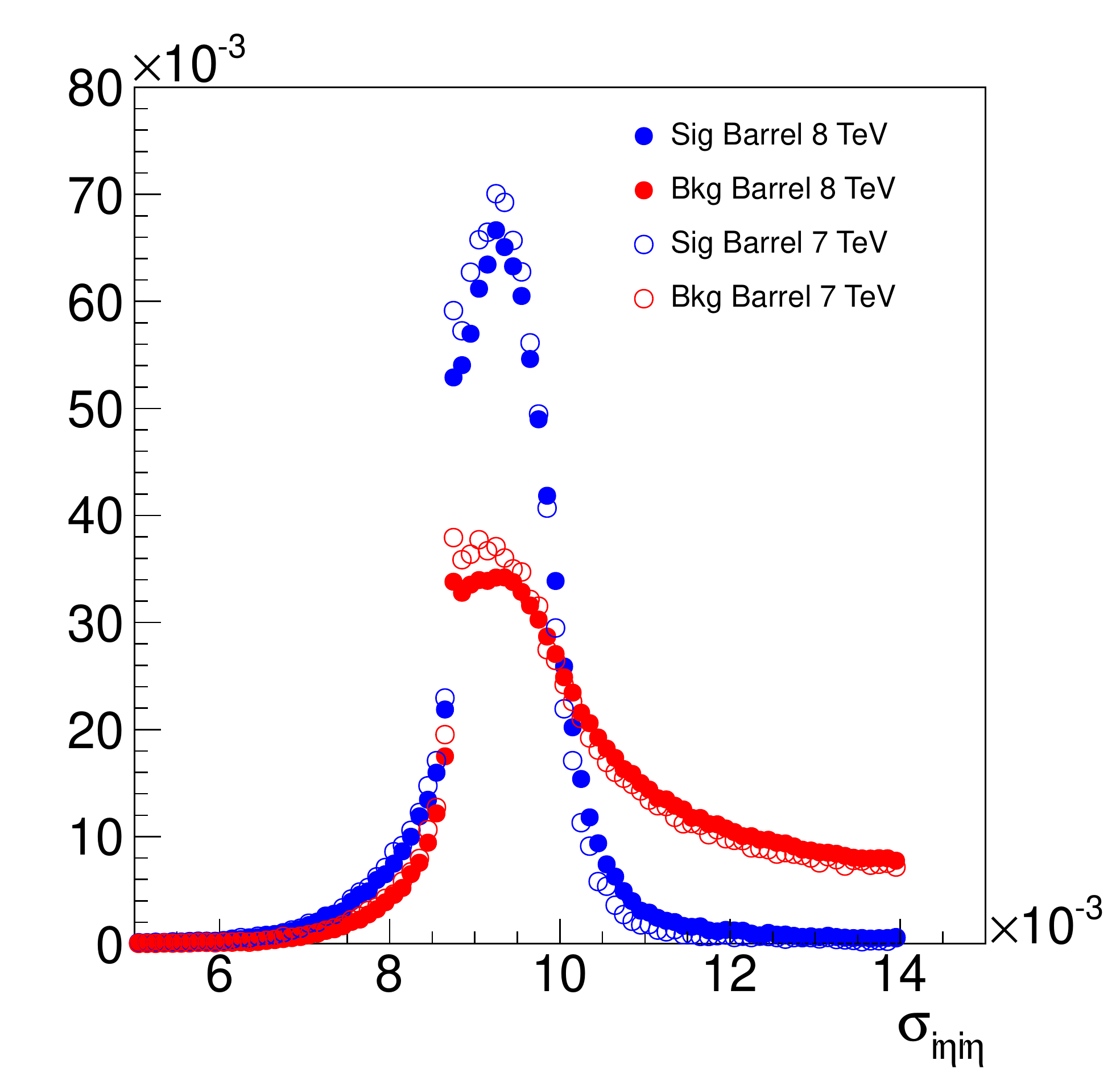}
    \includegraphics[width=0.31\textwidth]{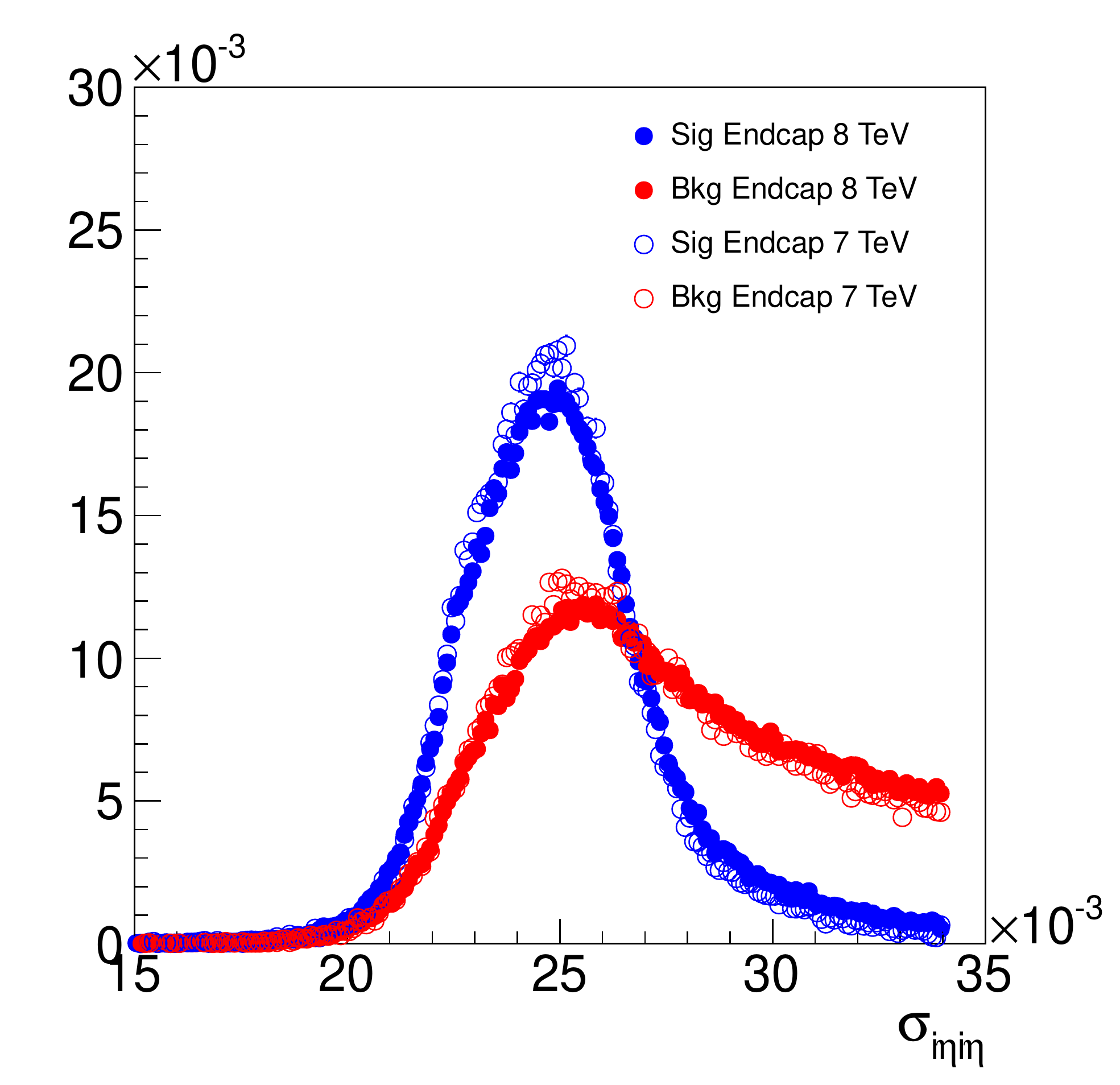}\\
    \includegraphics[width=0.31\textwidth]{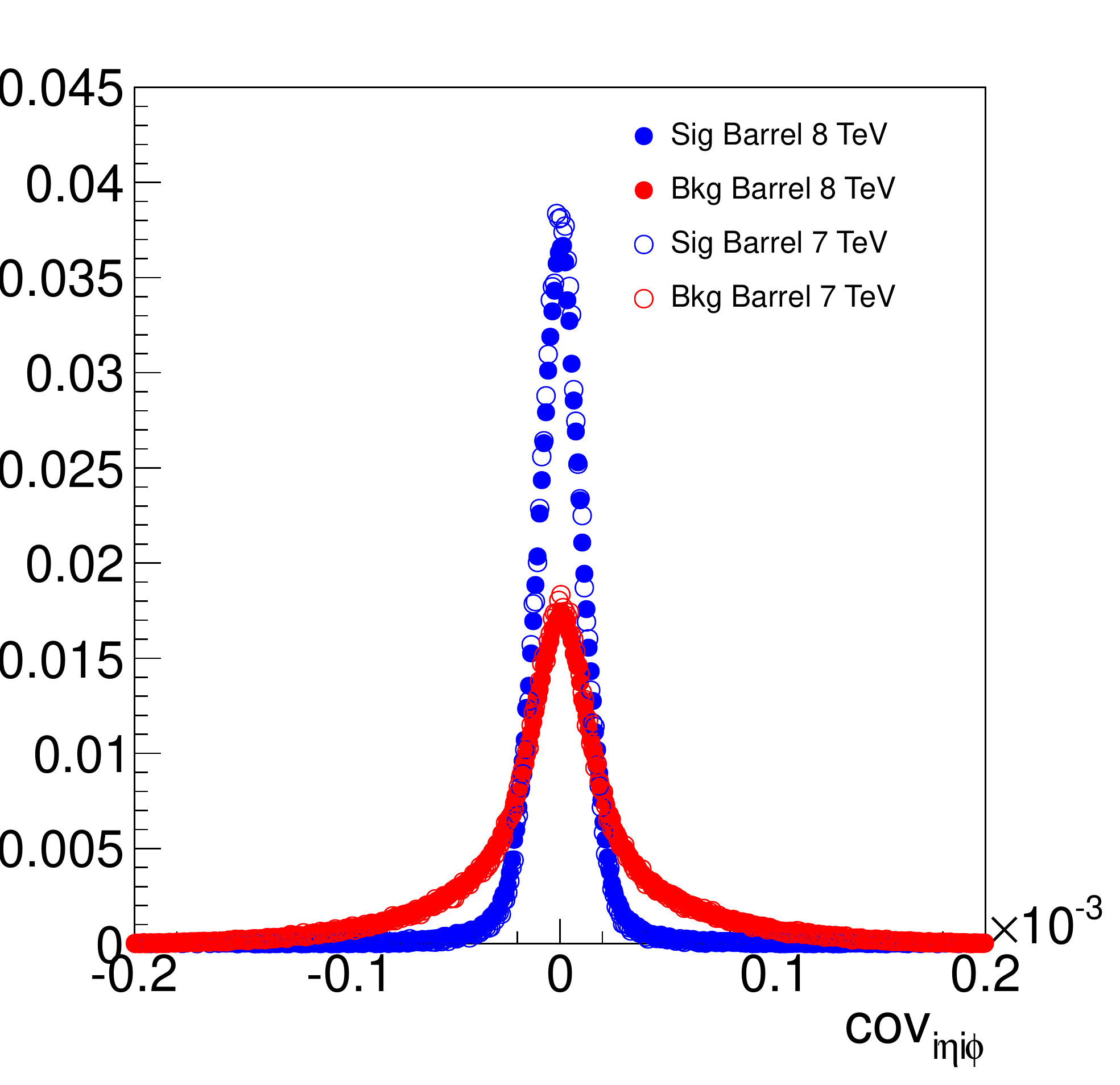}
    \includegraphics[width=0.31\textwidth]{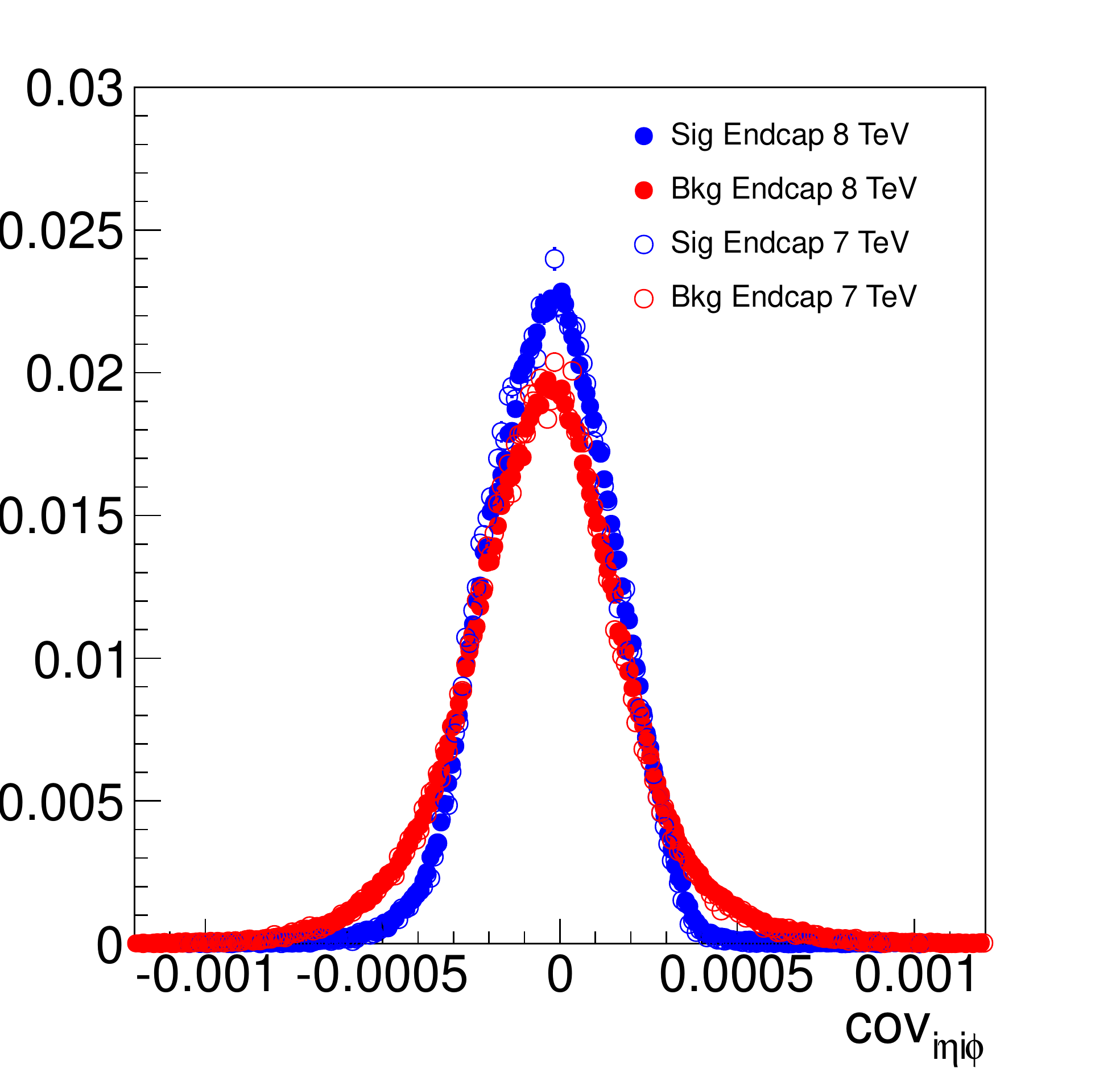}\\
    \includegraphics[width=0.31\textwidth]{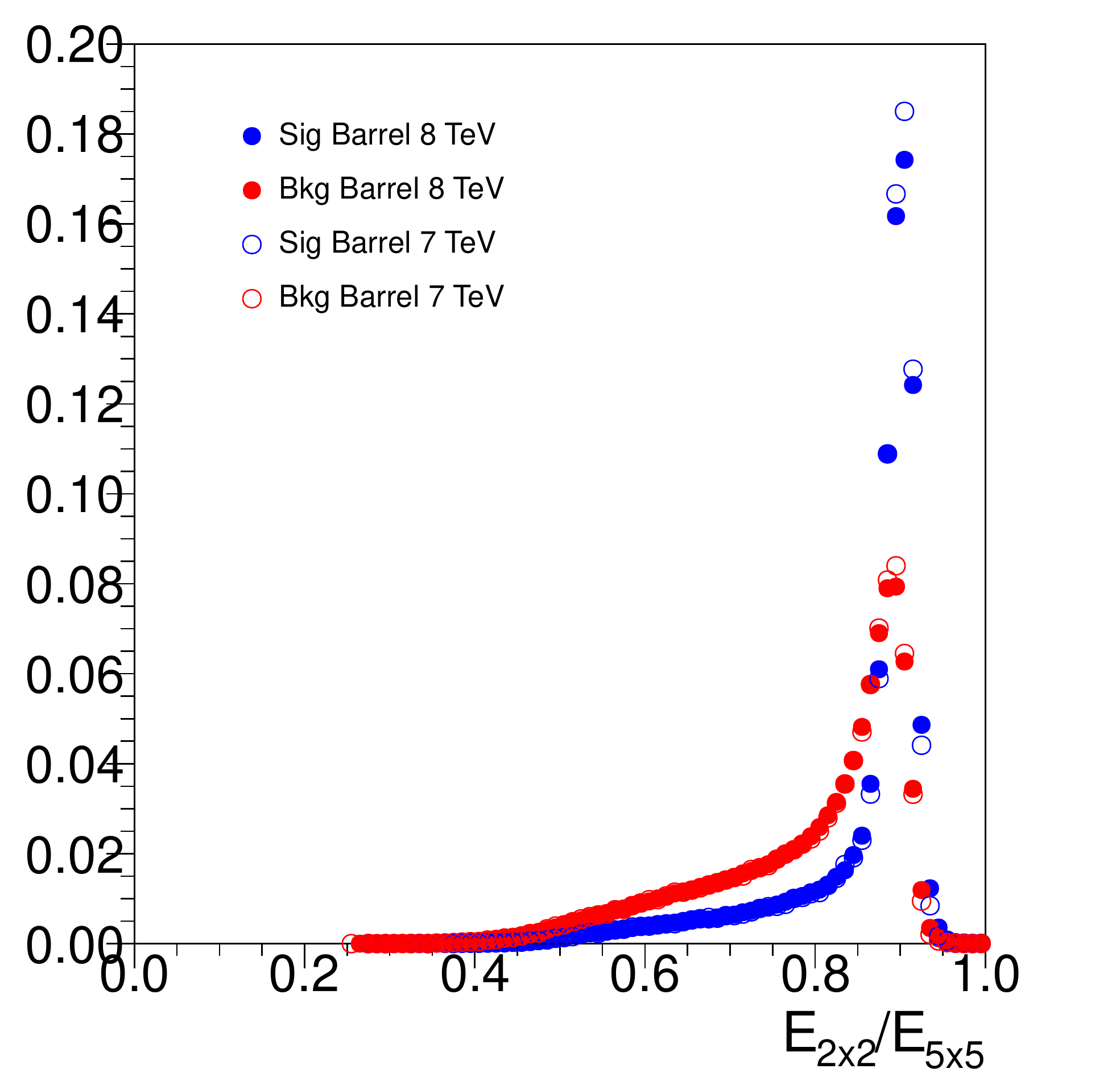}
    \includegraphics[width=0.31\textwidth]{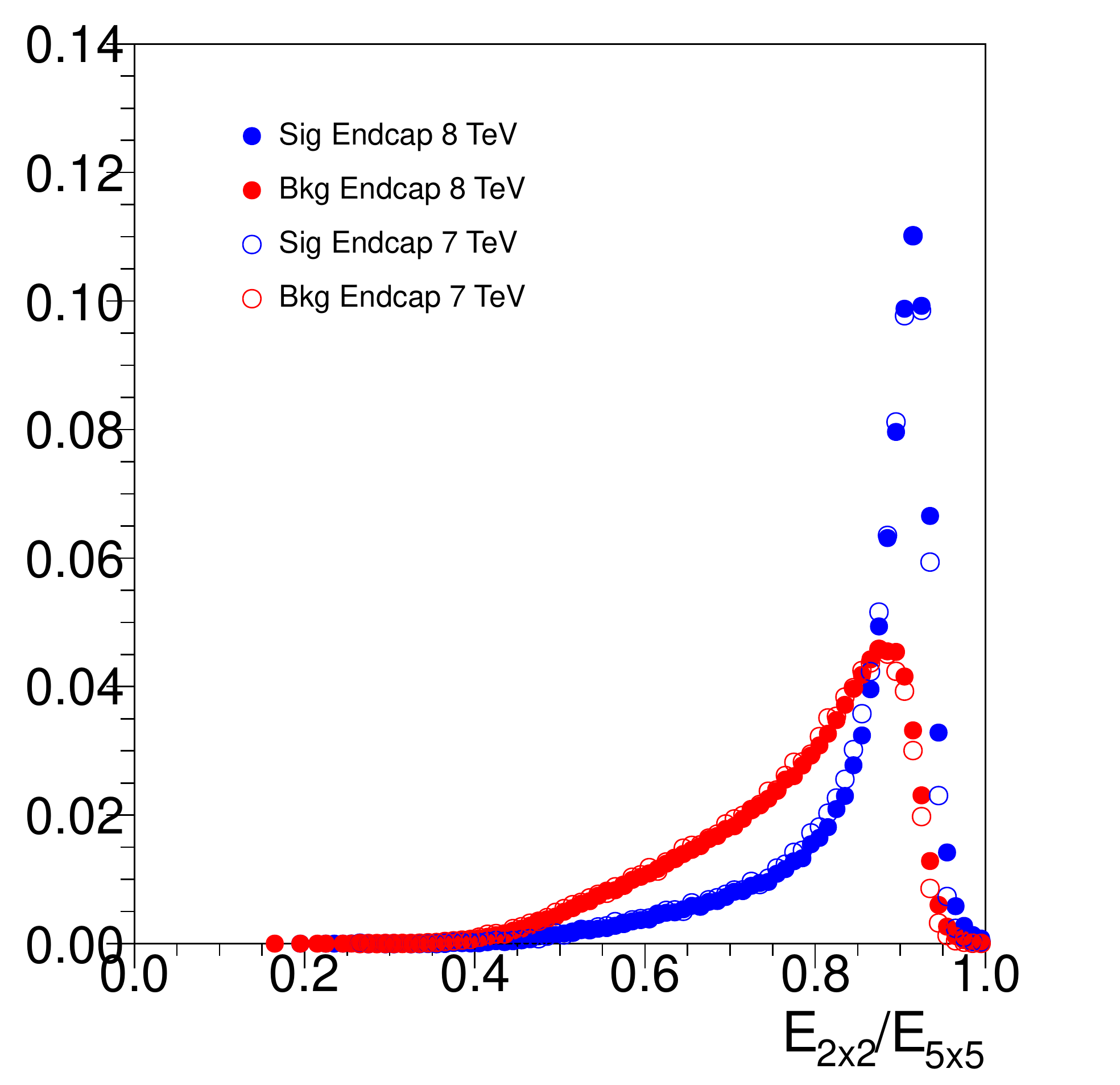}\\
  \end{center}
  \caption{The distributions of photon identification BDT input variables $\sigma_{i\eta i\eta}$ (first row), cov$_{i\eta i\phi}$ (second row), and $E_{2\times2}/E_{5\times5}$ (third row) for signal prompt photons (blue) and background fake photons (red) in the barrel (left) and in the endcap (right) from pp collisions at 7 TeV (hollow) and $\mathrm{8~TeV}$ (solid). The photons are from the training samples passing the preselection with $p_{T}$ $>$ $15~\mathrm{GeV}$ and after $p_{T}$-$\eta_{SC}$ reweighting.}
  \label{fig:idmva input basic cluster}
\end{figure}

\begin{figure}[hbpt] 
  \begin{center}
    \includegraphics[width=0.31\textwidth]{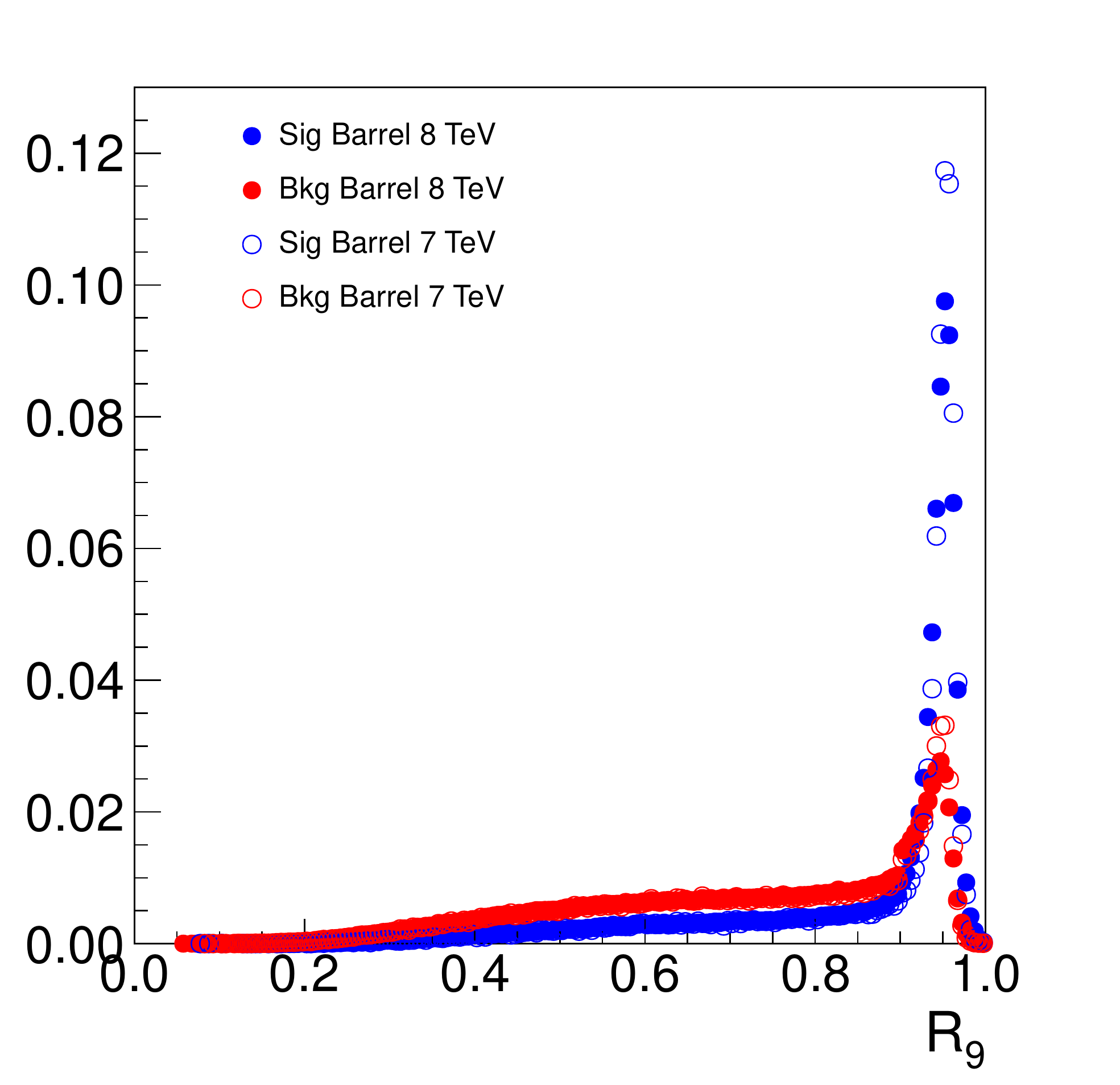}
    \includegraphics[width=0.31\textwidth]{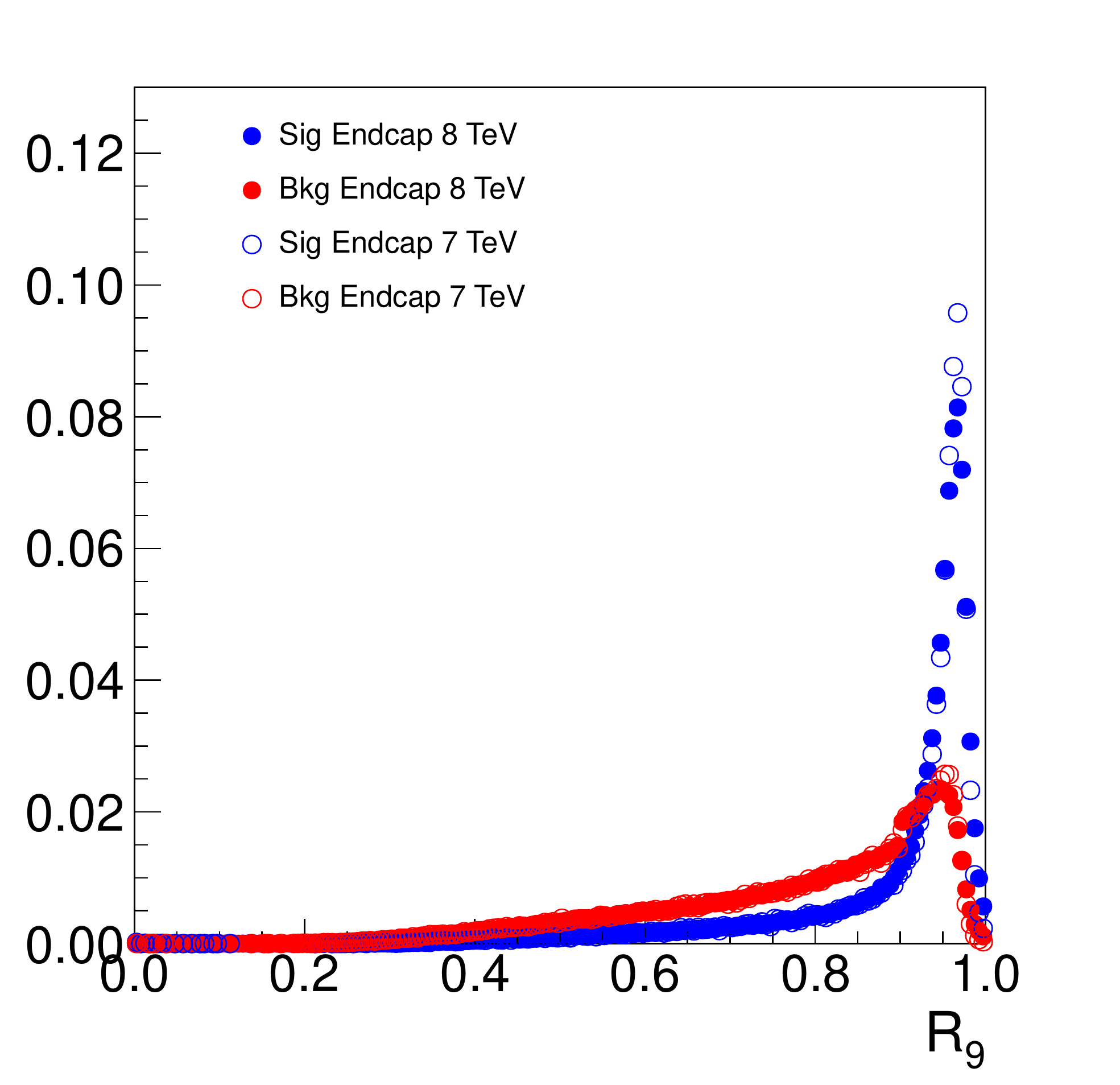}\\
    \includegraphics[width=0.31\textwidth]{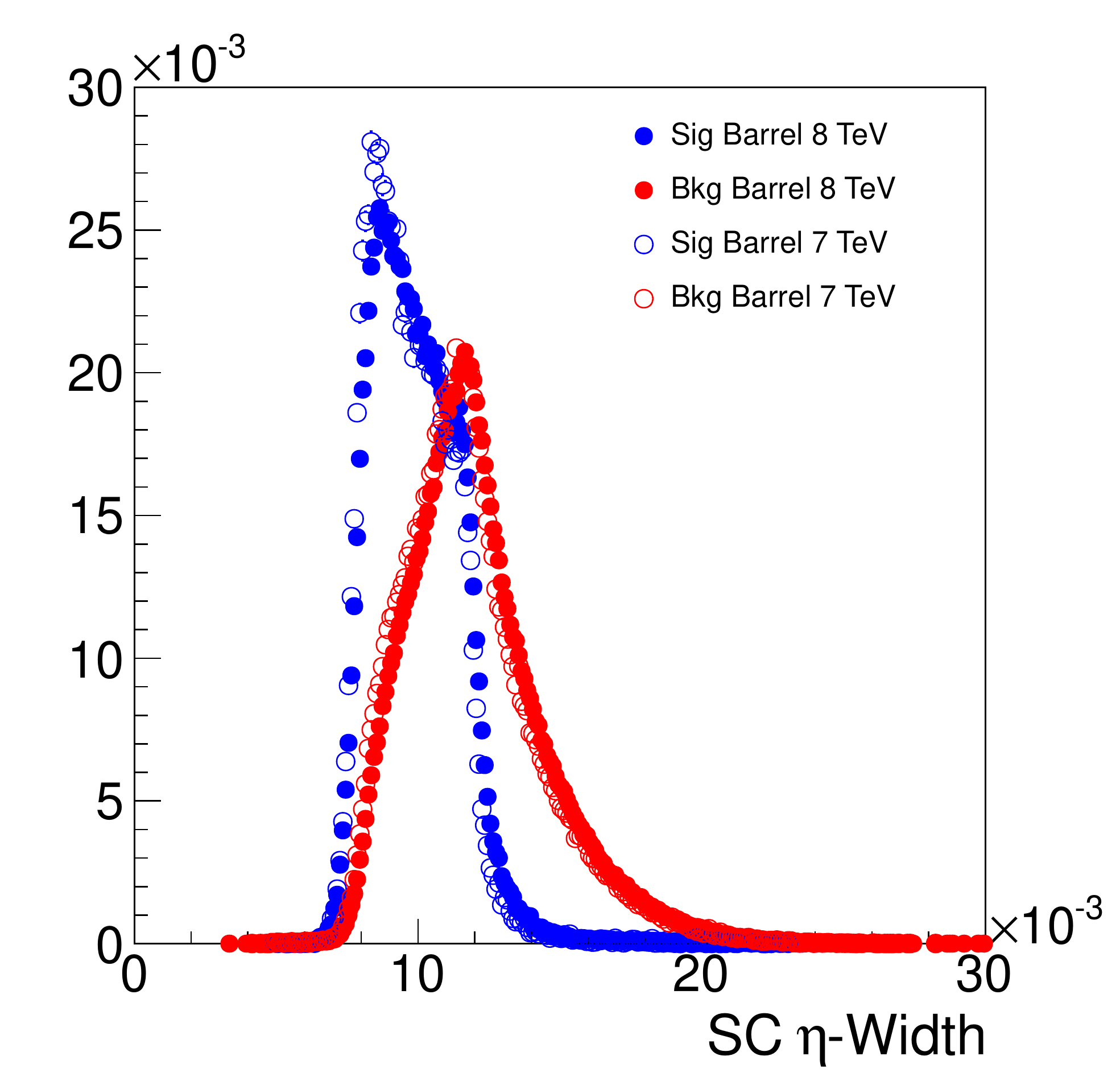}
    \includegraphics[width=0.31\textwidth]{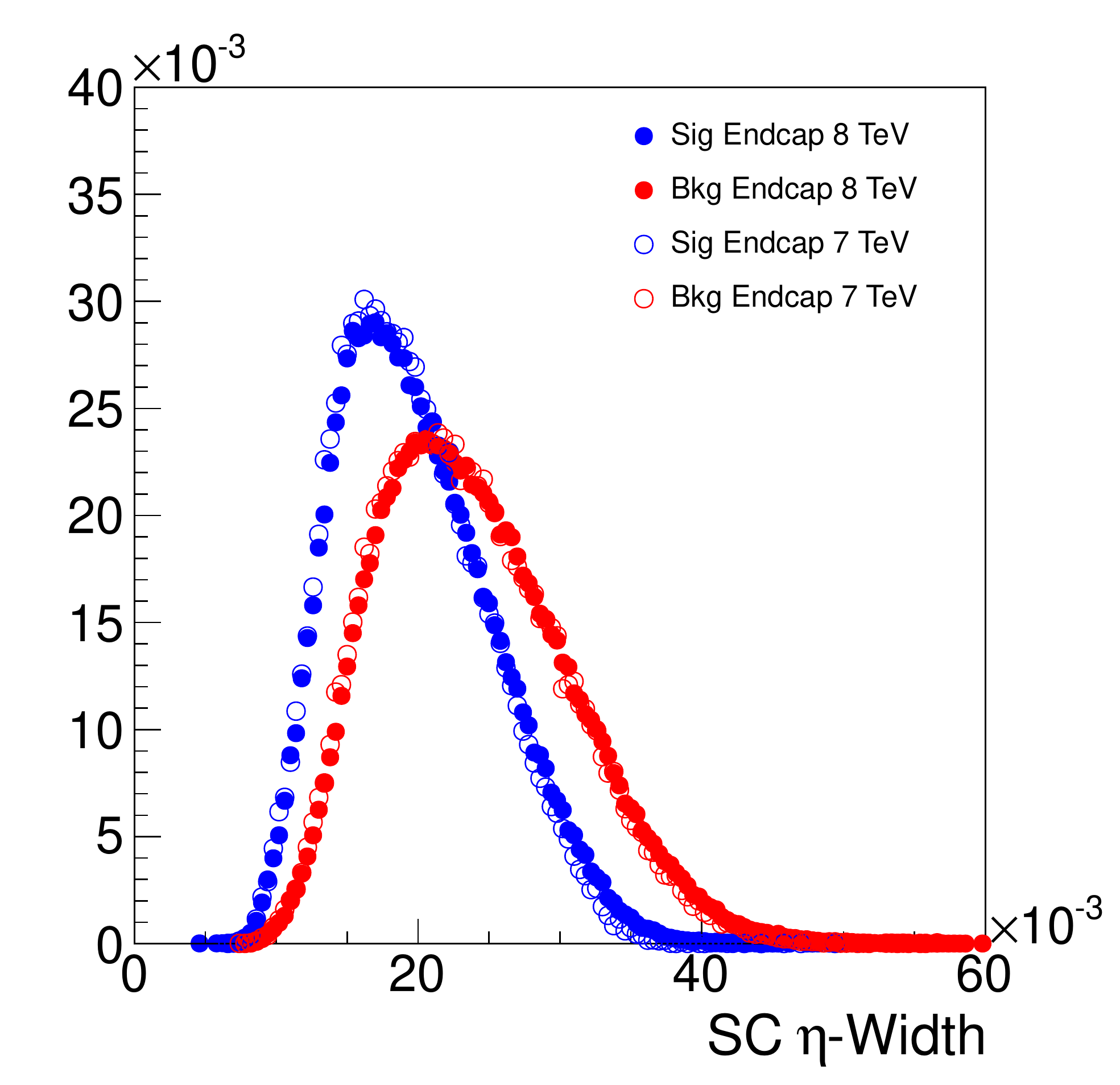}\\
    \includegraphics[width=0.31\textwidth]{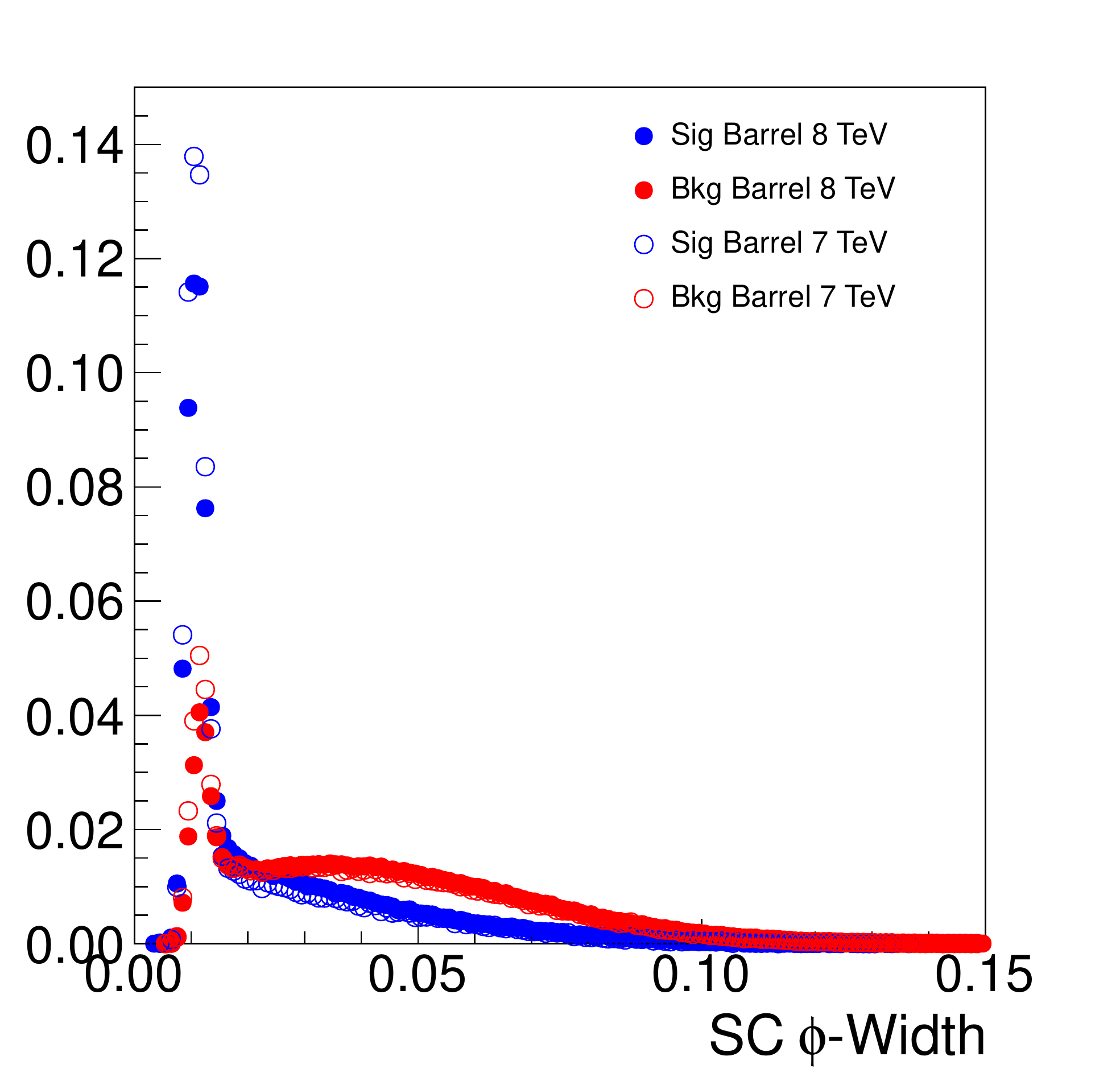}
    \includegraphics[width=0.31\textwidth]{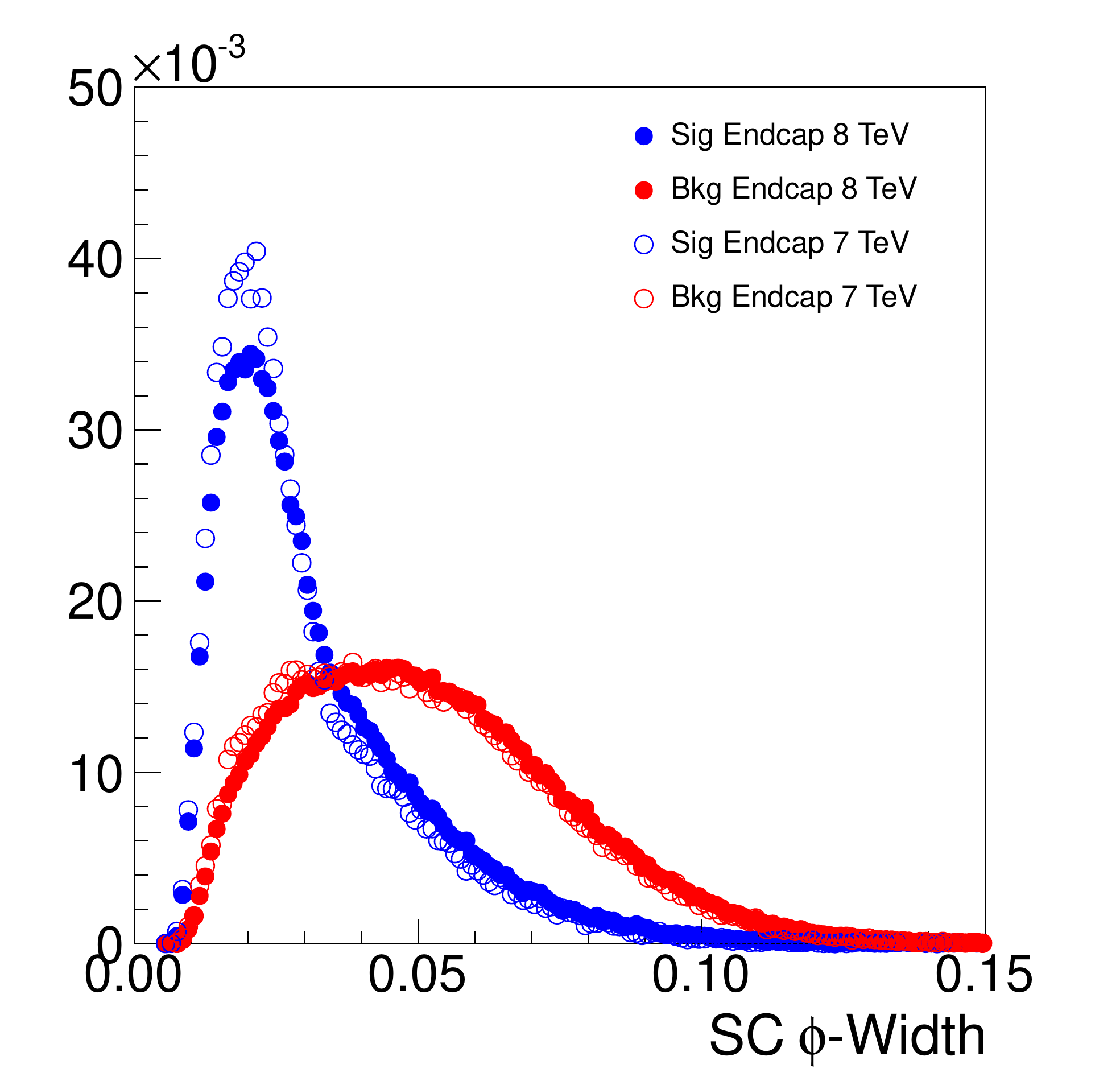}\\
    \includegraphics[width=0.31\textwidth]{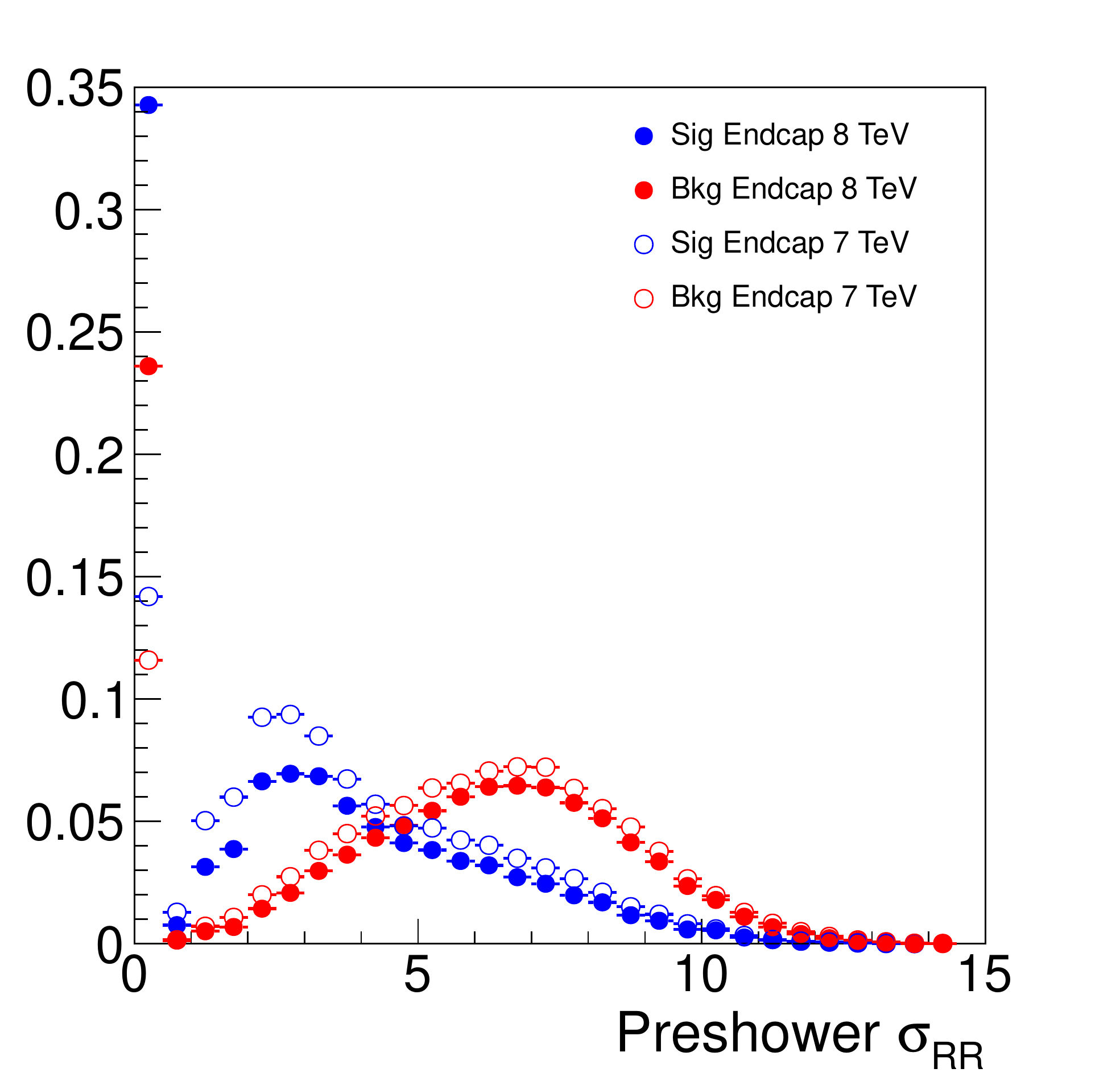}\\
  \end{center}
  \caption{The distributions of photon identification BDT input variables $R_{9}$ (first row), SC $\eta$-Width (second row), and SC $\phi$-Width (third row) for signal prompt photons (blue) and background fake photons (red) in the barrel (left) and in the endcap (right), along with the distribution of  Preshower $\sigma_{RR}$ (fourth row) for photons in the endcap only, from pp collisions at 7 TeV (hollow) and $\mathrm{8~TeV}$ (solid). The photons are from the training samples passing the preselection with $p_{T}$ $>$ $15~\mathrm{GeV}$ and after $p_{T}$-$\eta_{SC}$ reweighting.}
  \label{fig:idmva input supercluster}
\end{figure}

\begin{figure}[hbpt] 
  \begin{center}
    \includegraphics[width=0.31\textwidth]{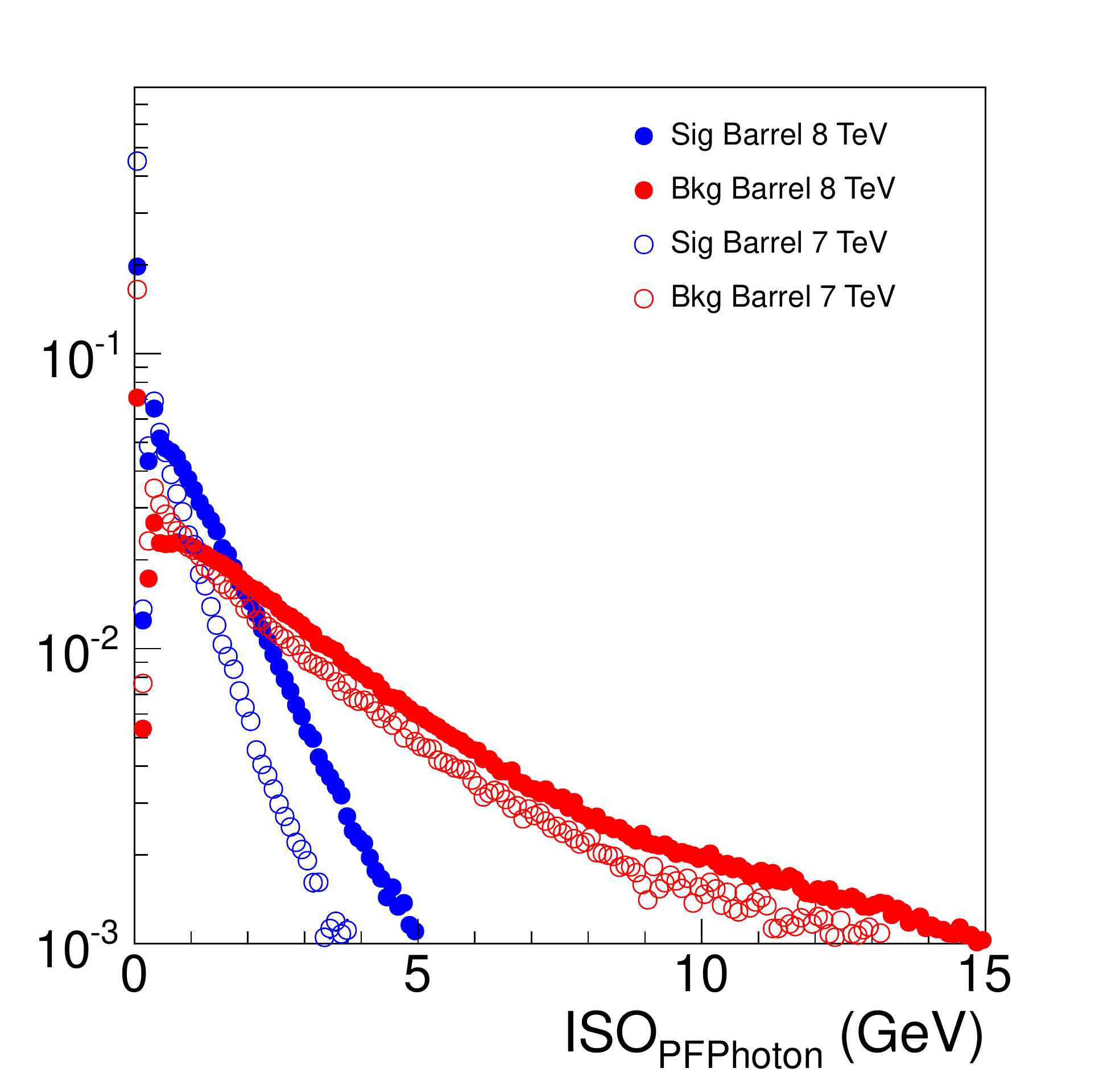}
    \includegraphics[width=0.31\textwidth]{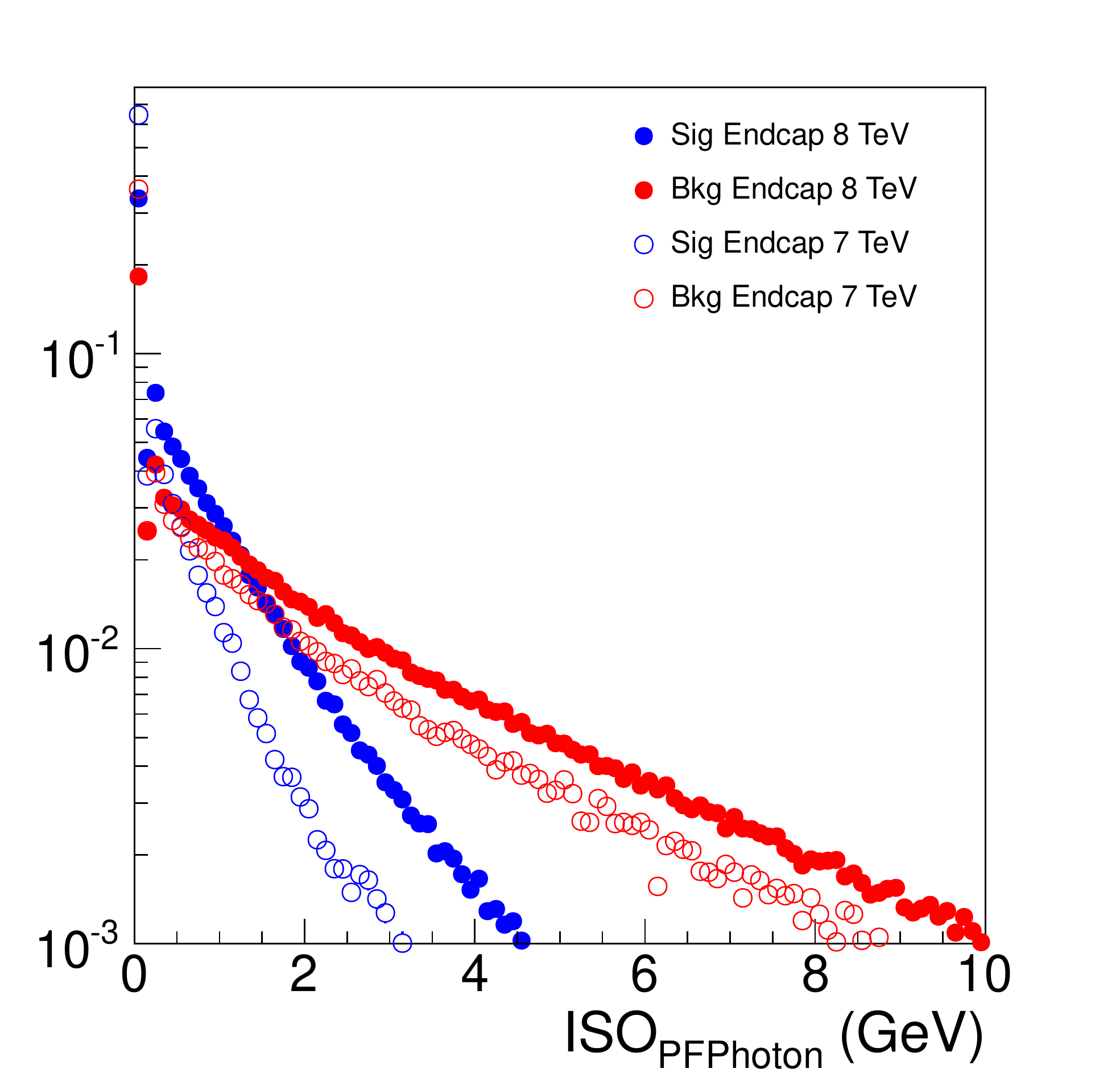}\\
    \includegraphics[width=0.31\textwidth]{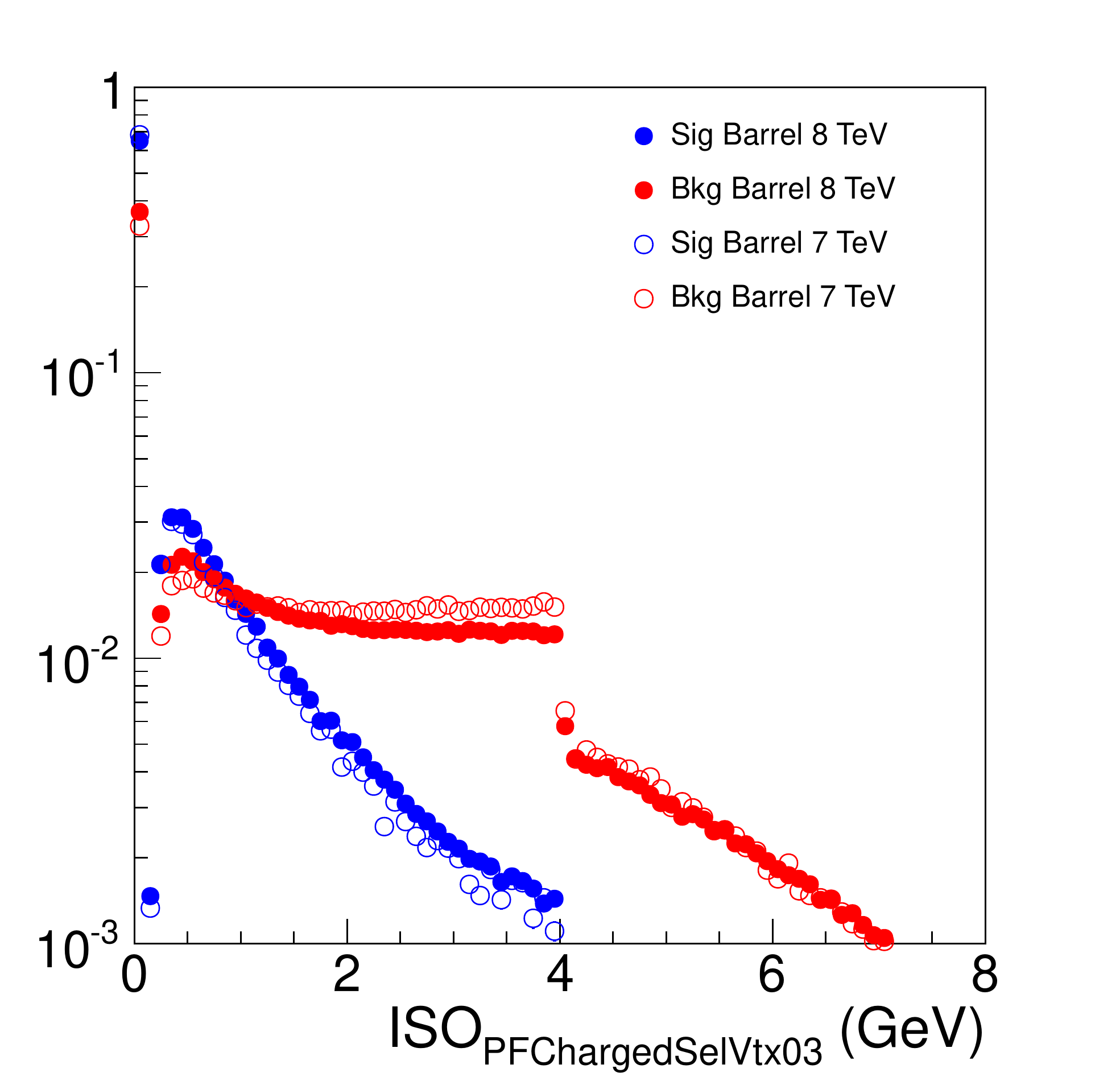}
    \includegraphics[width=0.31\textwidth]{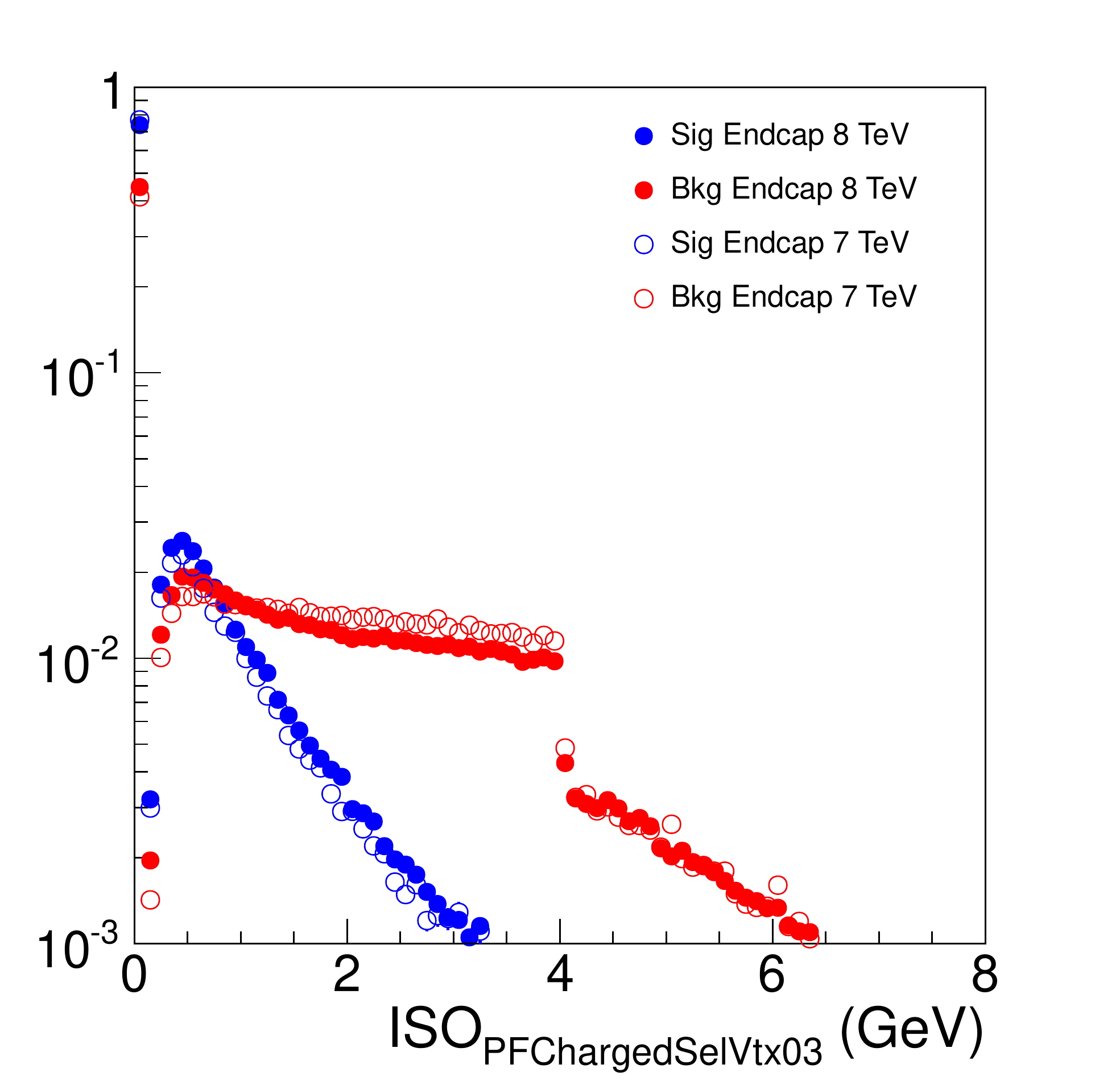}\\
    \includegraphics[width=0.31\textwidth]{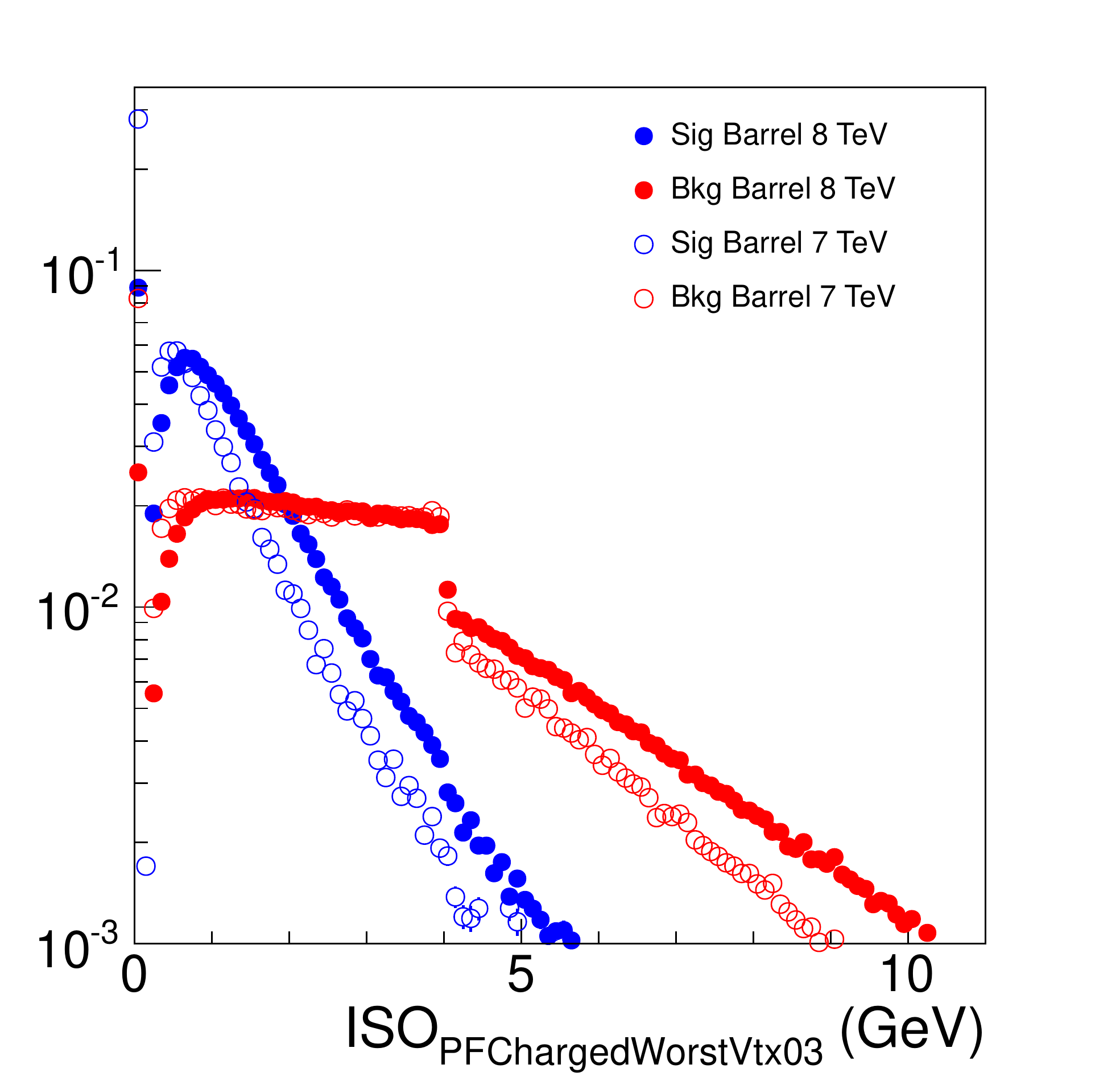}
    \includegraphics[width=0.31\textwidth]{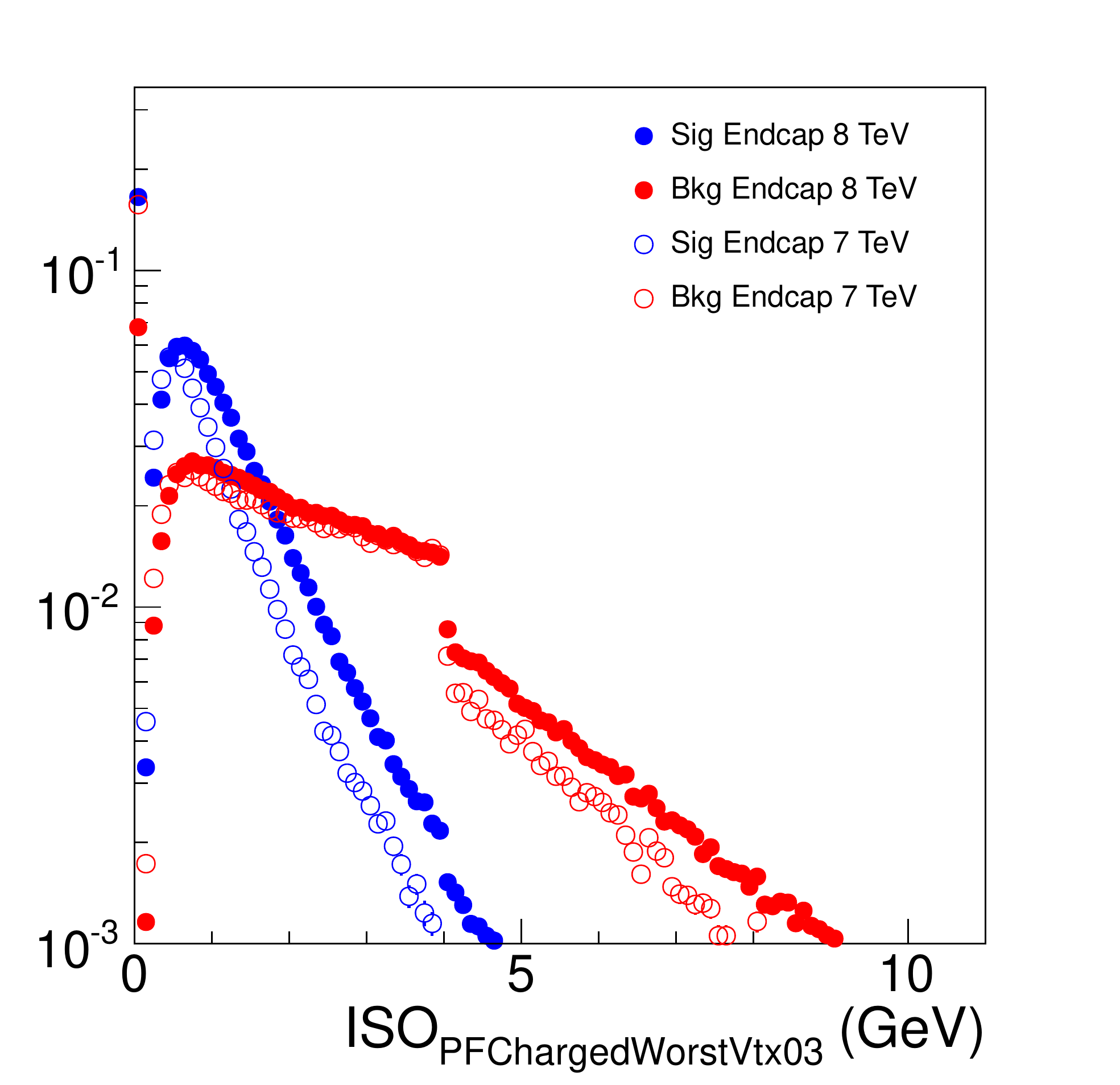}\\
  \end{center}
  \caption{The distributions of photon identification BDT input variables ISO$_{PFPhoton}$ (first row), ISO$_{PFChargedSelVtx03}$ (second row), and ISO$_{PFChargedWorstVtx03}$ (third row) for signal prompt photons (blue) and background fake photons (red) in the barrel (left) and in the endcap (right) from pp collisions at 7 TeV (hollow) and $\mathrm{8~TeV}$ (solid). The photons are from the training samples passing the preselection with $p_{T}$ $>$ $15~\mathrm{GeV}$ and after $p_{T}$-$\eta_{SC}$ reweighting.}
  \label{fig:idmva input isolation}
\end{figure}

\begin{figure}[hbpt] 
  \begin{center}
    \includegraphics[width=0.31\textwidth]{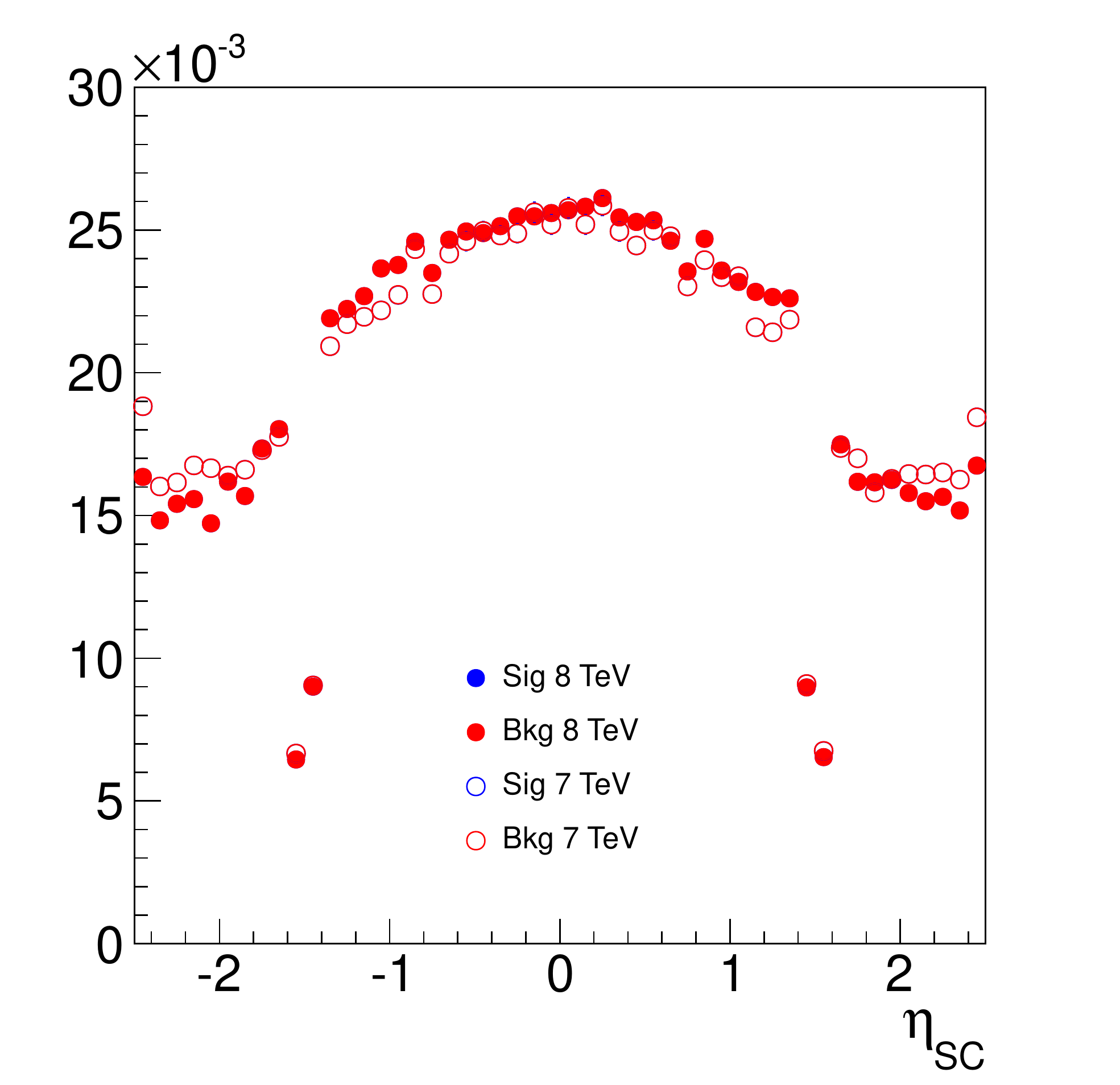}
  \end{center}
  \begin{center}
    \includegraphics[width=0.31\textwidth]{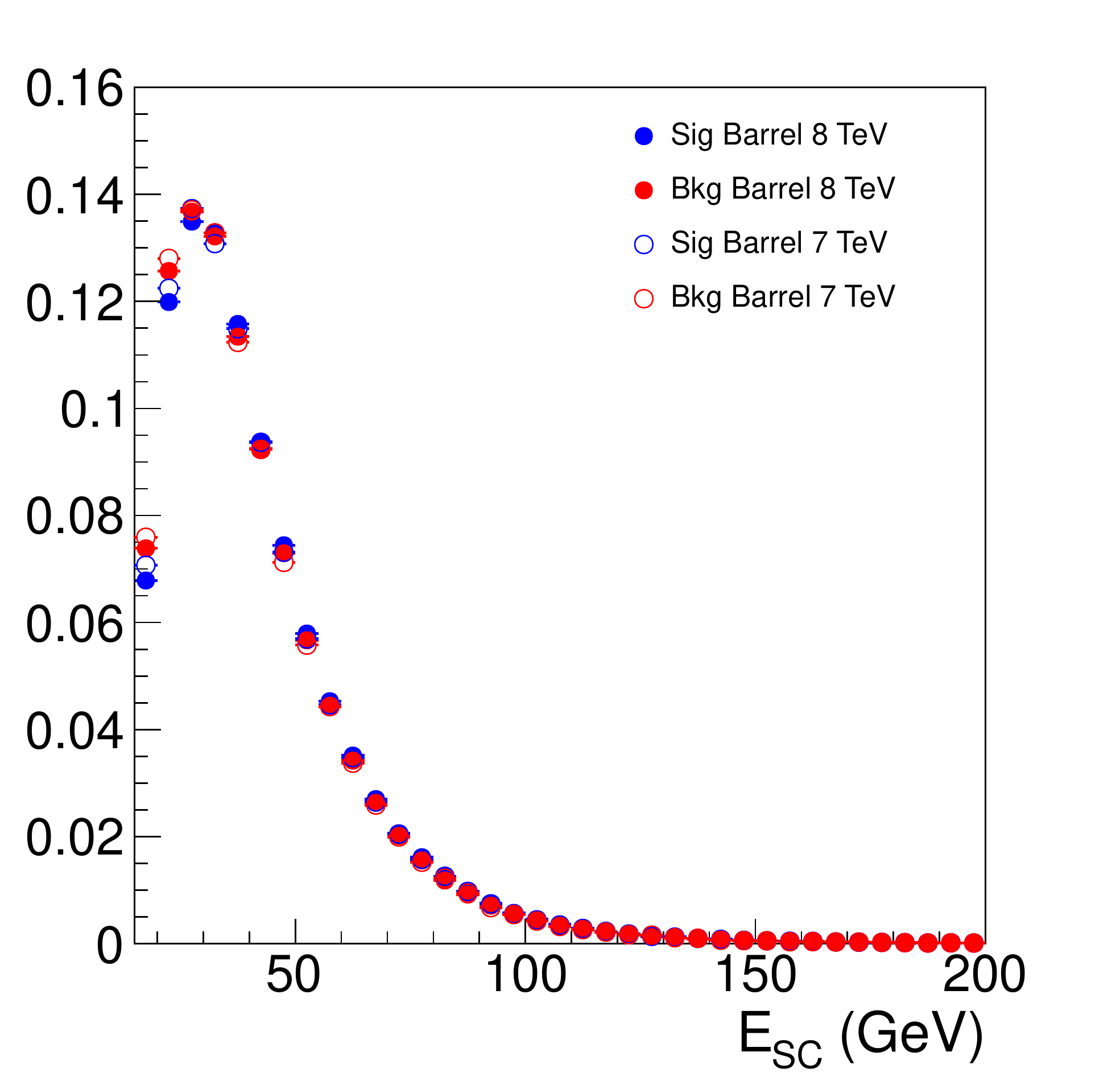}
    \includegraphics[width=0.31\textwidth]{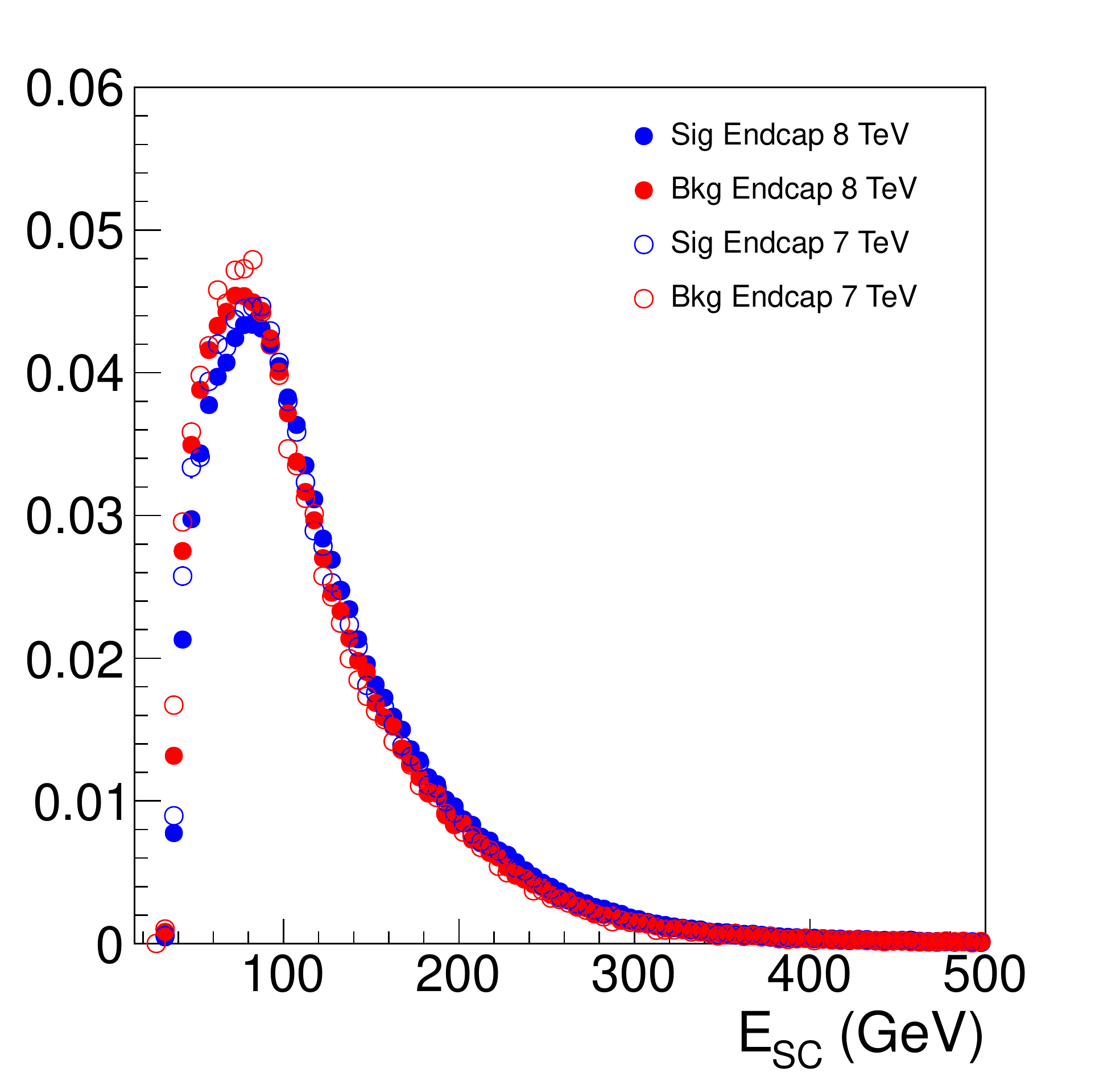}\\
    \includegraphics[width=0.31\textwidth]{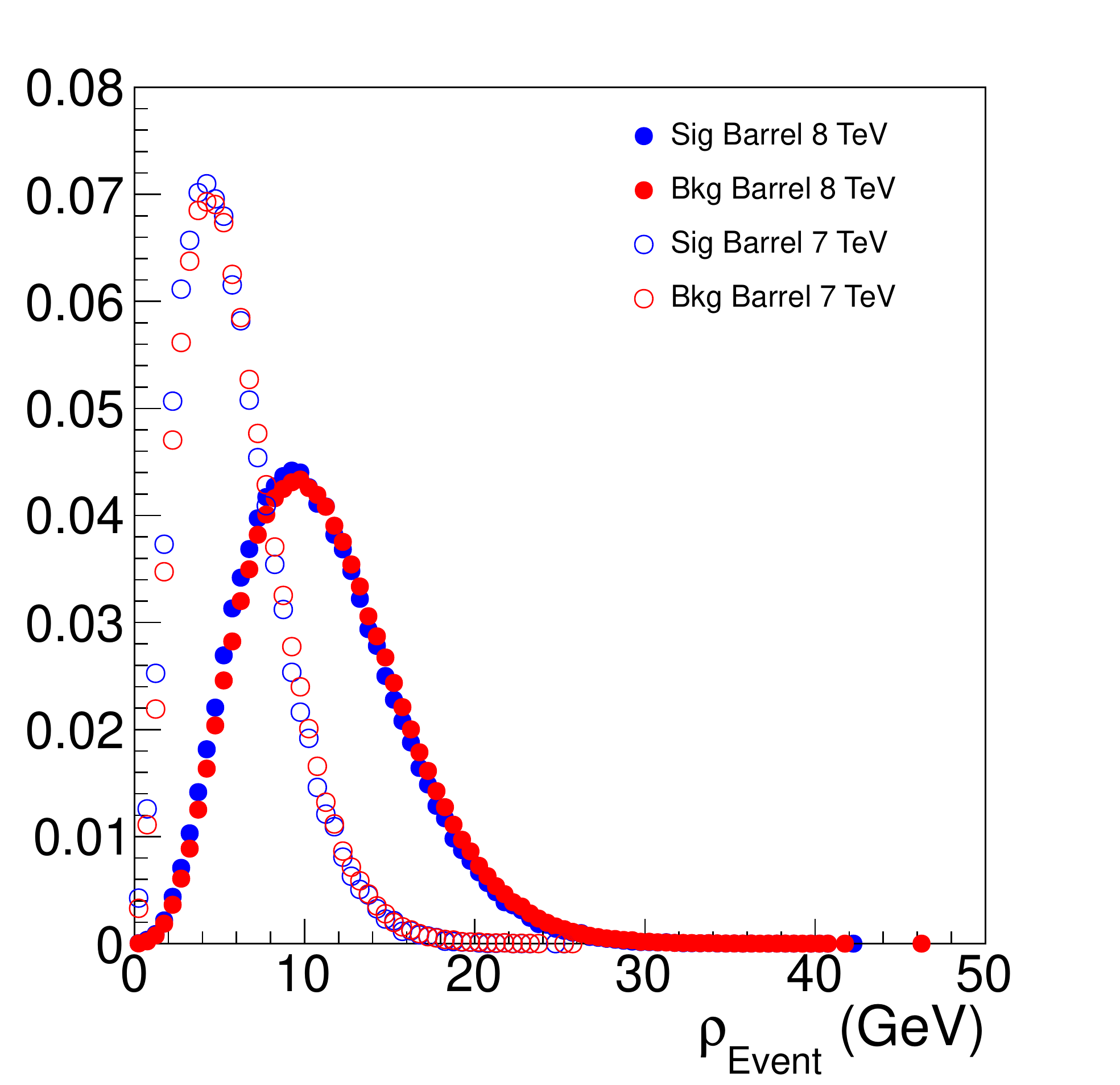}
    \includegraphics[width=0.31\textwidth]{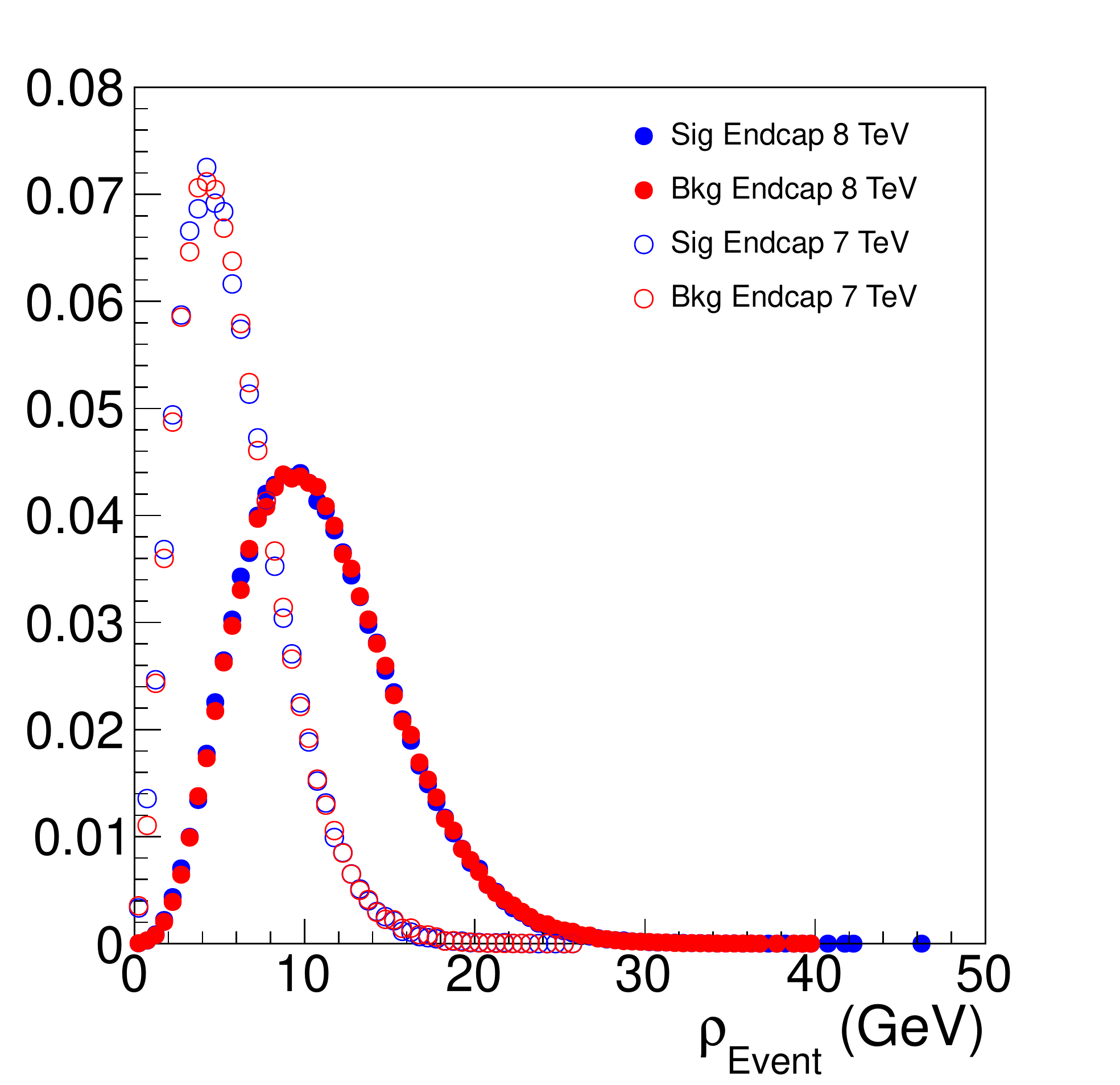}\\
  \end{center}  
  \caption{The distributions of photon identification BDT input variables $\eta_{SC}$ (first row), $E_{SC}$ (second row), and $\rho_{Event}$ (third row) for signal prompt photons (blue) and background fake photons (red) in the barrel (left for $E_{SC}$ and $\rho_{Event}$) and in the endcap (right for $E_{SC}$ and $\rho_{Event}$) from pp collisions at 7 TeV (hollow) and $\mathrm{8~TeV}$ (solid). The photons are from the training samples passing the preselection with $p_{T}$ $>$ $15~\mathrm{GeV}$ and after $p_{T}$-$\eta_{SC}$ reweighting.}
  \label{fig:idmva input auxiliary}
\end{figure}

The distributions of the input variables for the signal and background training samples after the reweighting are shown in Figure \ref{fig:idmva input basic cluster}, Figure \ref{fig:idmva input supercluster}, Figure \ref{fig:idmva input isolation} and Figure \ref{fig:idmva input auxiliary}. The discontinuities in the ISO$_{PFChargedSelVtx03}$ distribution and in the ISO$_{PFChargedWorstVtx03}$ distribution for the background, shown in Figure \ref{fig:idmva input isolation}, are due to the cut ISO$_{PFChargedSelVtx02}$ $<$ $4~\mathrm{GeV}$ in the preselection. The significant drops in the $\eta_{SC}$ distribution around the transition regions between ECAL barrel and endcap for both signal and background, shown in Figure \ref{fig:idmva input auxiliary}, are due to the acceptance cut which removes the photons in the region 1.4442 $<$ $|\eta_{SC}|$ $<$ 1.566.          

The reweighting is only done for the training process, but not for the evaluation of BDT output and performance as introduced below.

\subsection{Output and Performance}
\label{sec:Output and Performance}

\begin{figure}[hbpt] 
  \begin{center}
    \includegraphics[width=0.45\textwidth]{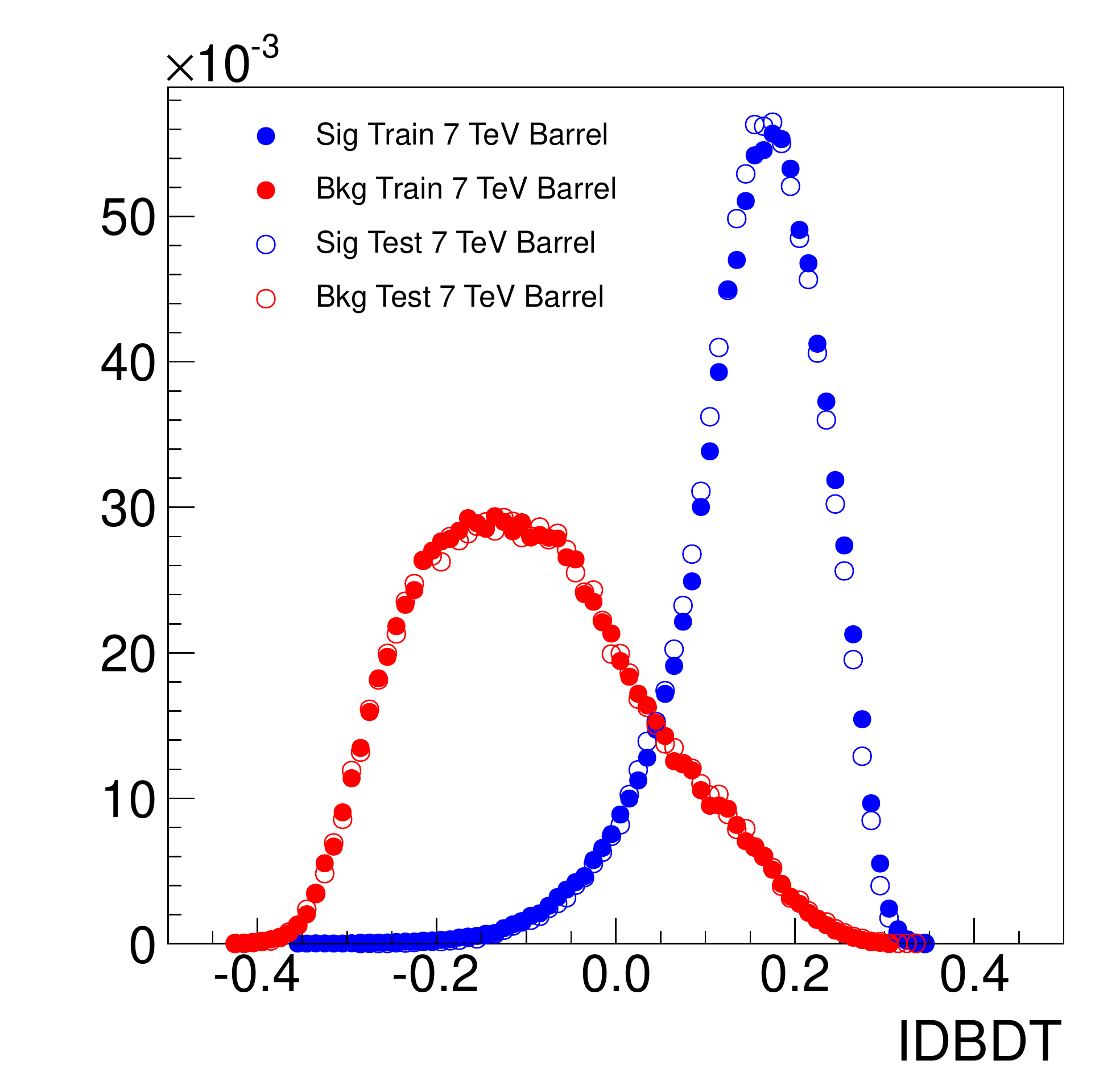}
    \includegraphics[width=0.45\textwidth]{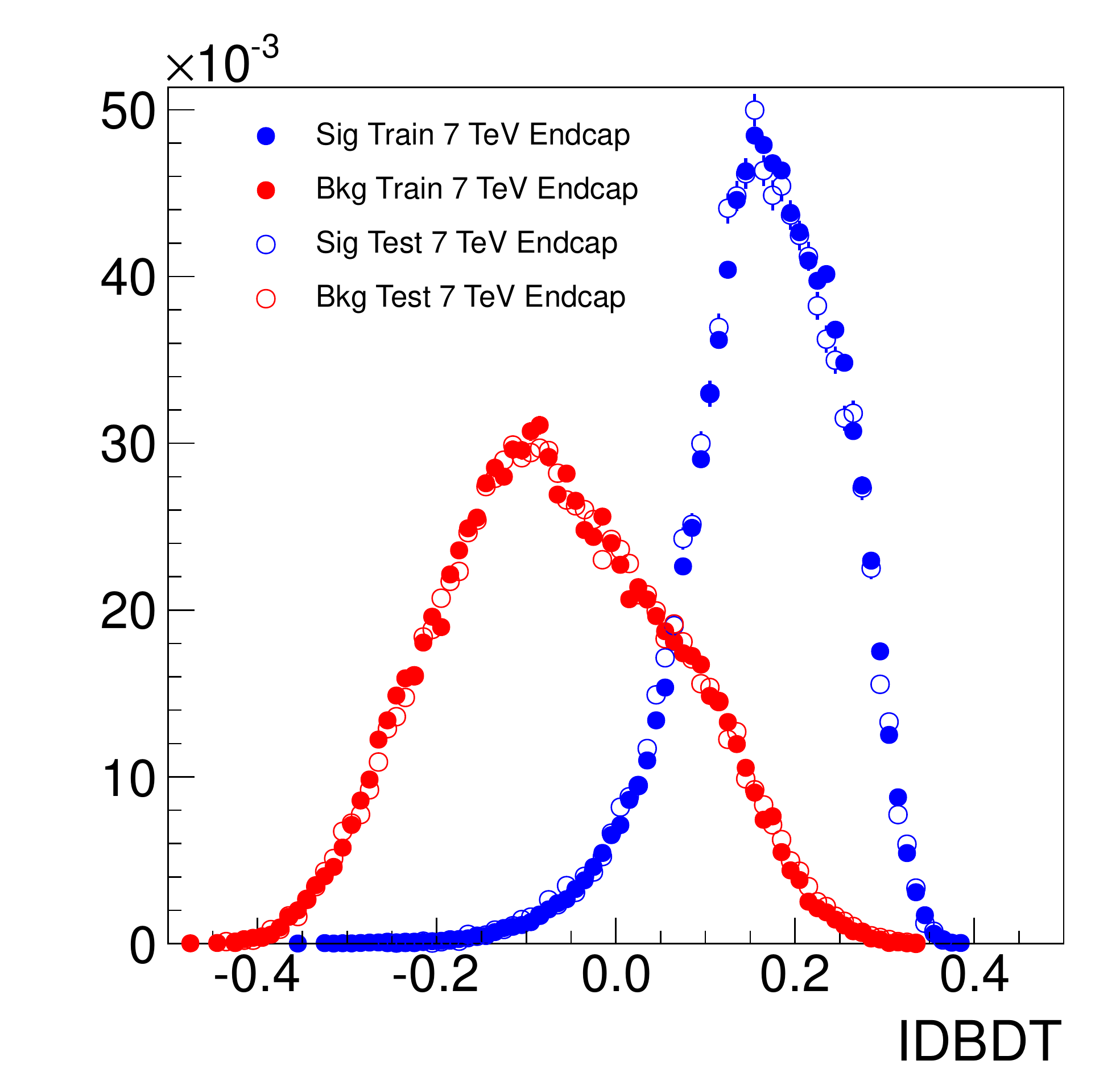}\\
    \includegraphics[width=0.45\textwidth]{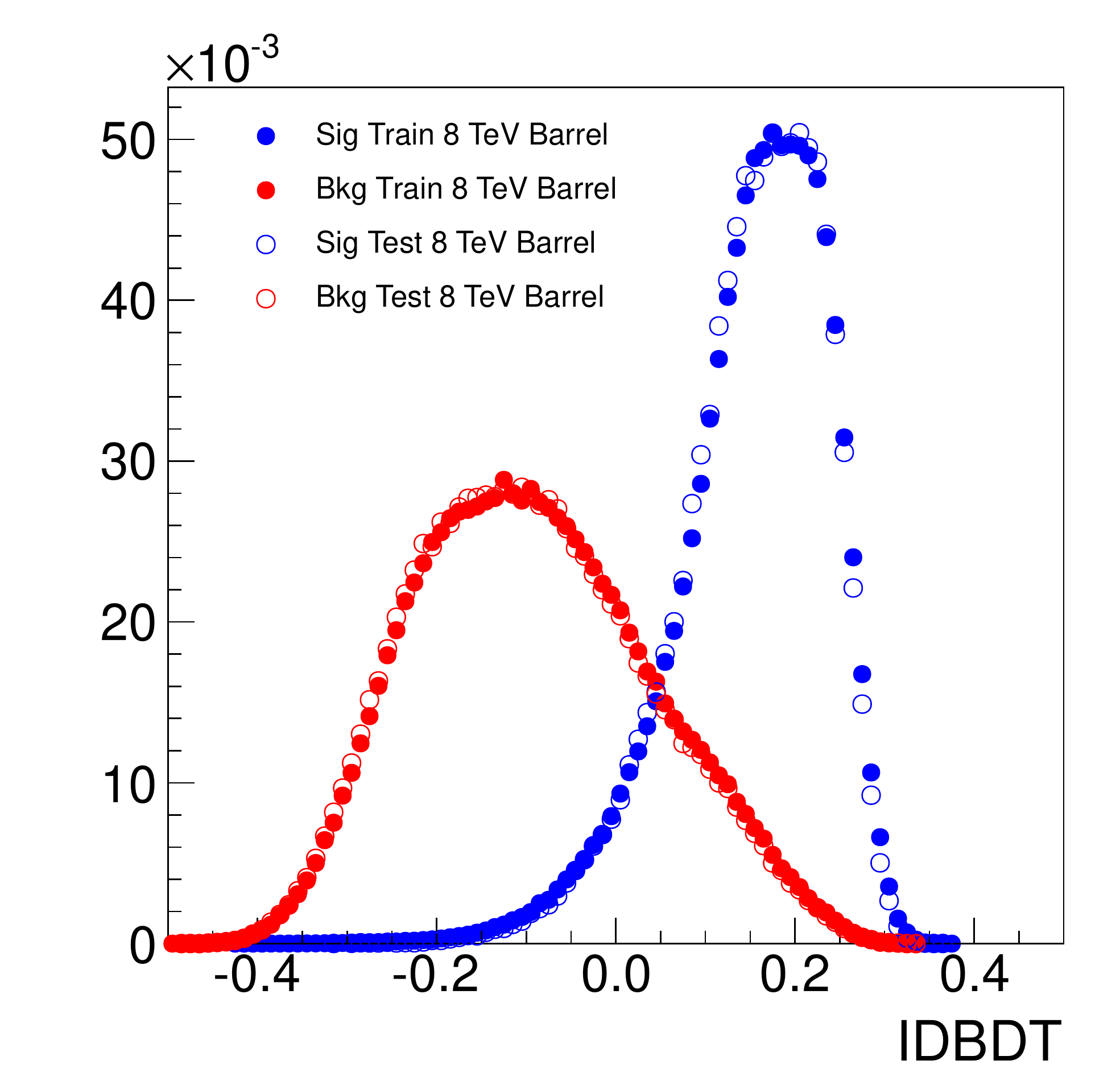}
    \includegraphics[width=0.45\textwidth]{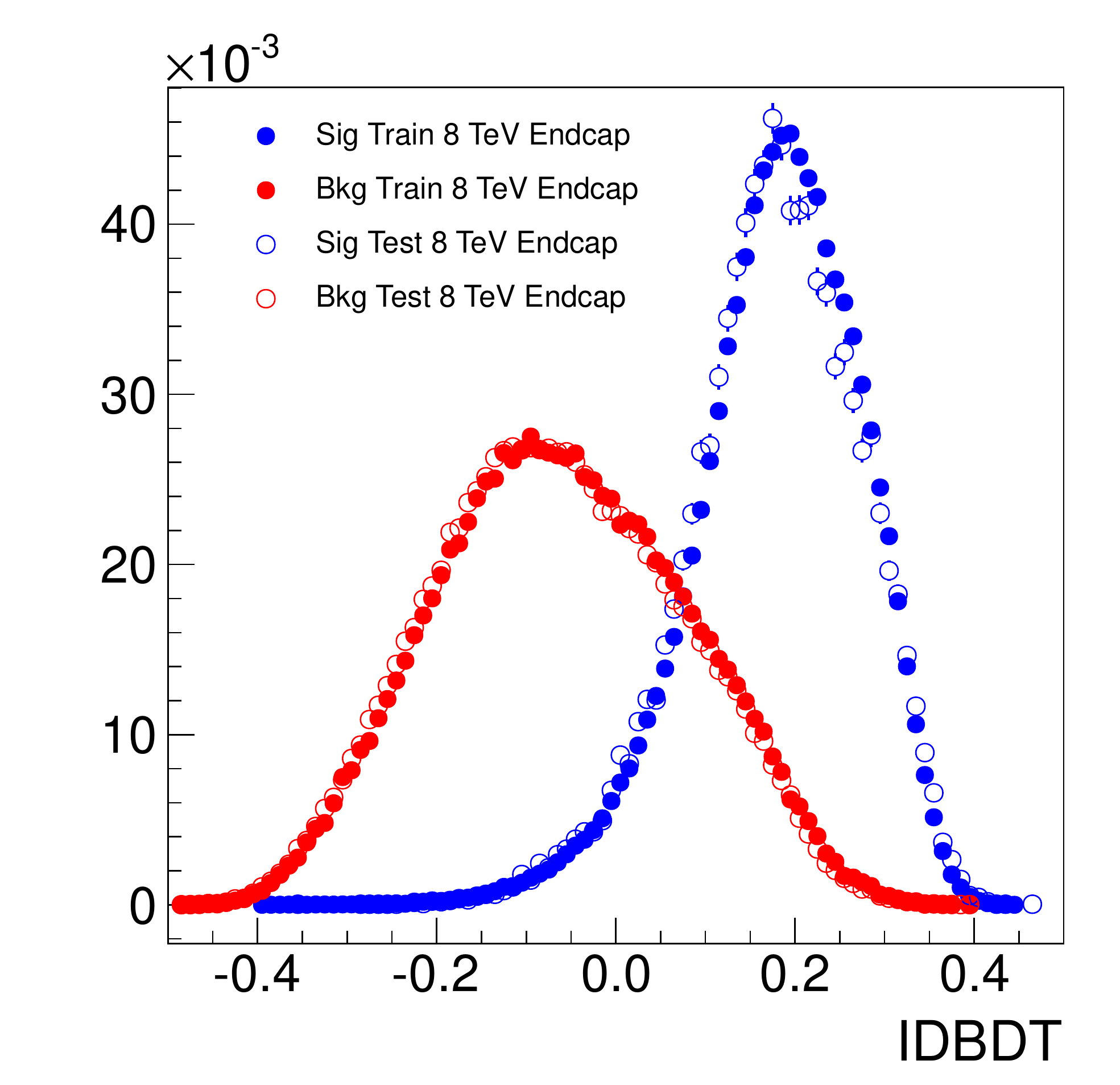}\\
  \end{center}
  \caption{The distributions of IDBDT for signal prompt photons (blue) and background fake photons (red) in the barrel (left) and endcap (right) from 7 TeV (top) and 8 TeV (bottom) pp collisions. The photons are from the training samples (solid) and the testing samples (hollow) passing the preselection with $p_{T}$ $>$ $25~\mathrm{GeV}$.}
  \label{fig:idmva output}
\end{figure}

The photon identification BDT output is a score named IDBDT assigned to each photon which ranges from $-1$ to 1. The higher the score assigned to a photon, the more likely the photon is a prompt photon rather than a fake photon. Figure \ref{fig:idmva output} shows the IDBDT distributions of the signal (blue) and background (red) training samples (solid circles), and of the corresponding testing samples (hollow circles), separately for photons in the barrel (left) and for photons in the endcap (right), at 7 TeV (up) and 8 TeV (down). The testing signal sample consists of prompt photons from a Monte Carlo simulation of $H\rightarrow \gamma\gamma$ events at a Higgs mass of $121~\mathrm{GeV}$ (124 GeV) at 7 TeV (8 TeV). The testing background sample consists of fake photons from Monte Carlo $\gamma$ + jet events not used for training. Both training and testing samples of photons for the plots pass the preselection with $p_{T}$ $>$ 25 GeV. Good agreement between the distributions of the testing samples and those of the training samples is shown, which verifies the statistical stability of the IDBDT.

\begin{figure}[h] 
  \begin{center}
    \includegraphics[width=0.49\textwidth]{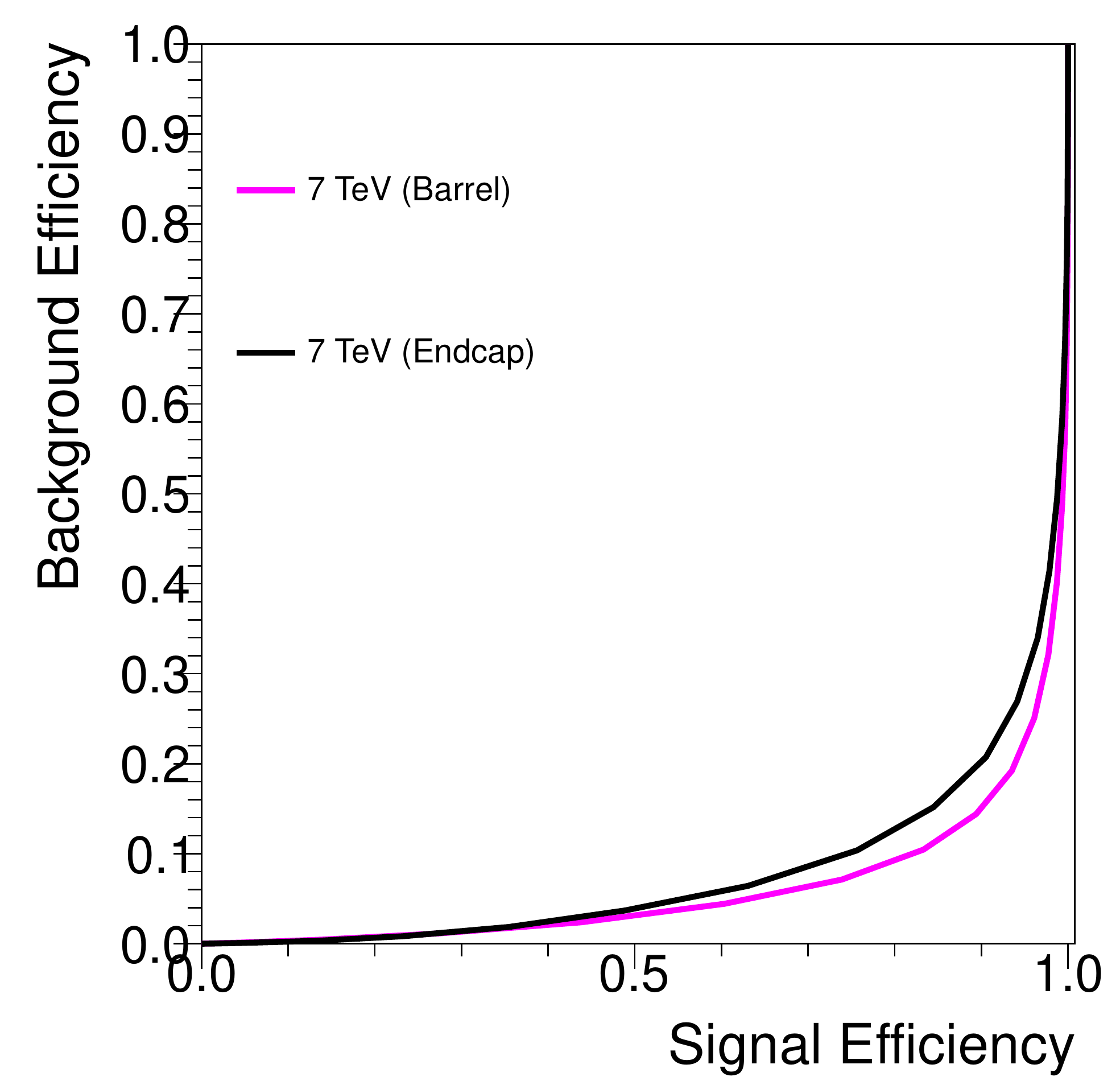}
    \includegraphics[width=0.49\textwidth]{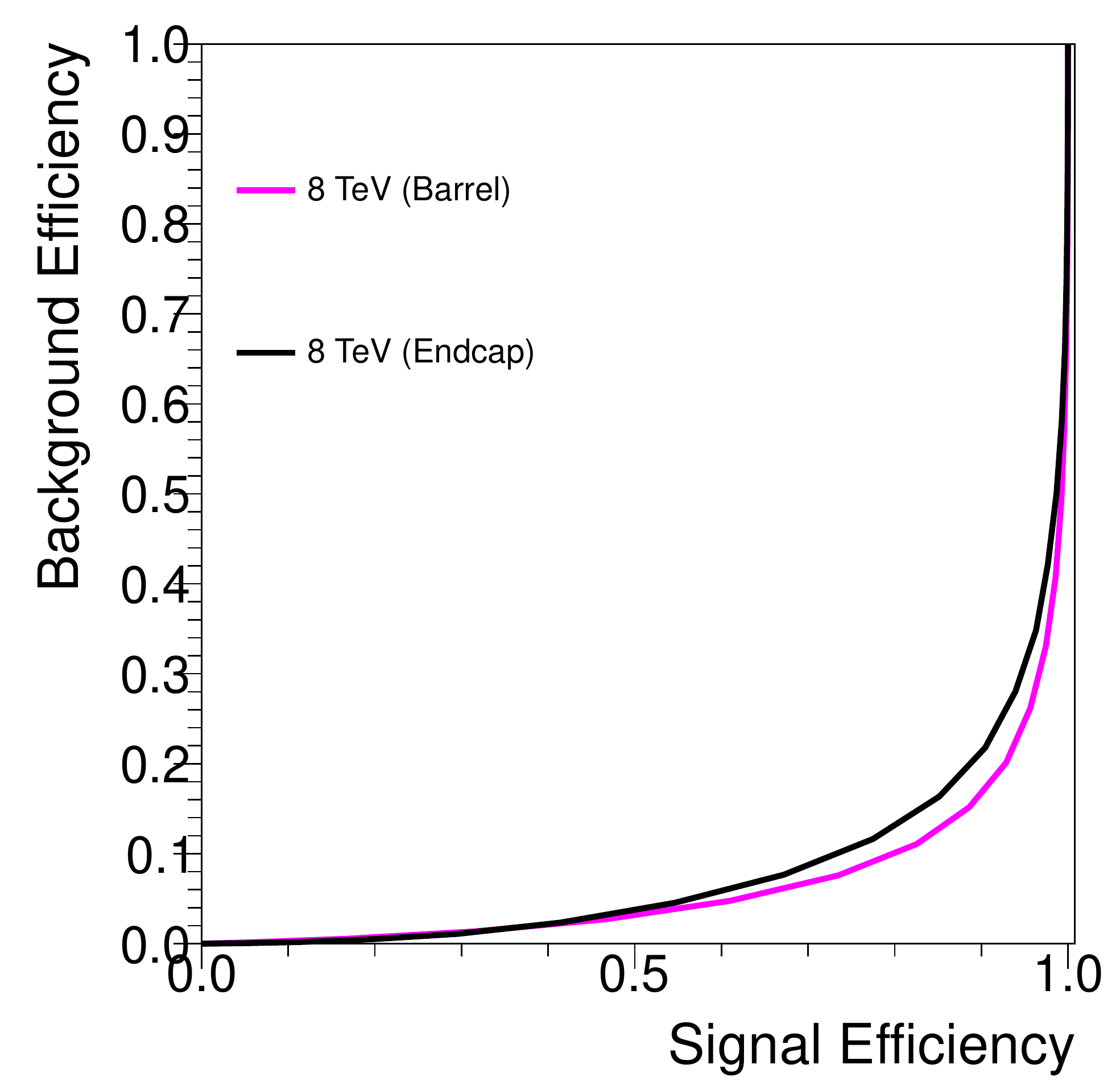}\\
  \end{center}
  \caption{The efficiency for background fake photons versus the efficiency for signal prompt photons in the barrel (magenta) and in the endcap (black) from 7 TeV (left) and 8 TeV (right) pp collisions. The photons are from testing samples passing the preselection with $p_{T}~>25~\mathrm{GeV}$.}
  \label{fig:idmva bkg vs sig eff}
\end{figure}

\begin{figure}[hbpt] 
  \begin{center}
    \includegraphics[width=0.4\textwidth]{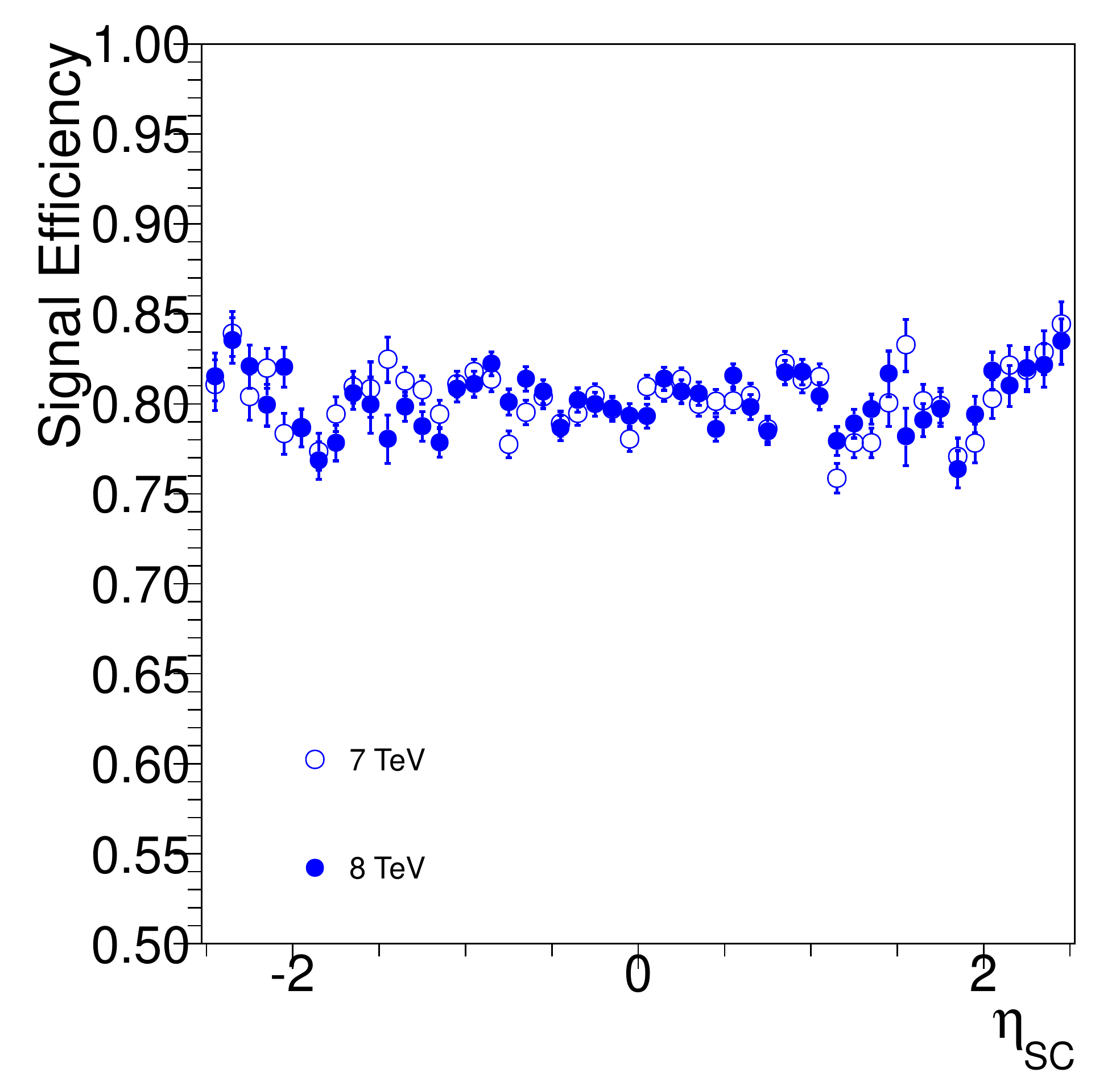}
    \includegraphics[width=0.4\textwidth]{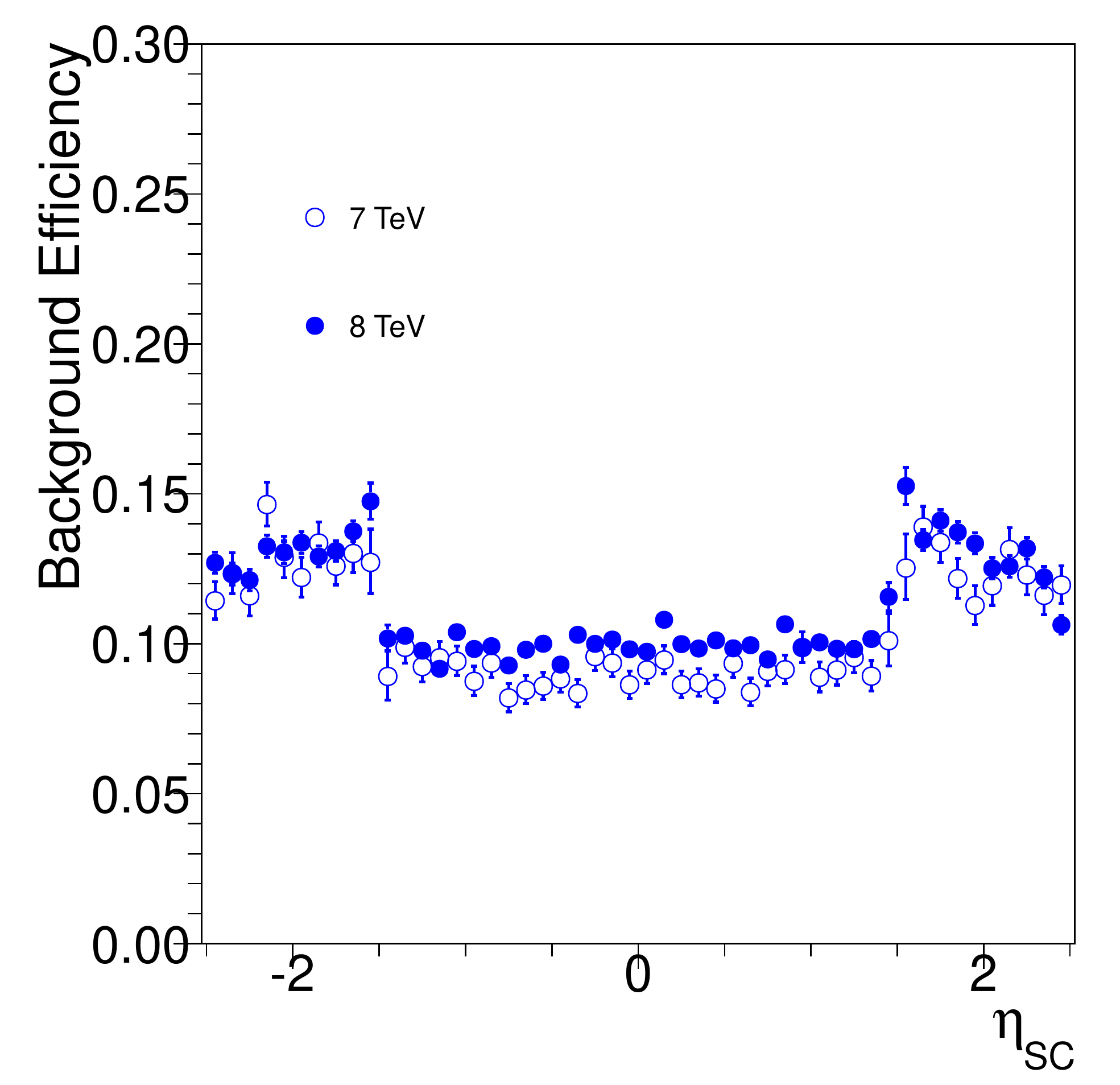}\\
    \includegraphics[width=0.4\textwidth]{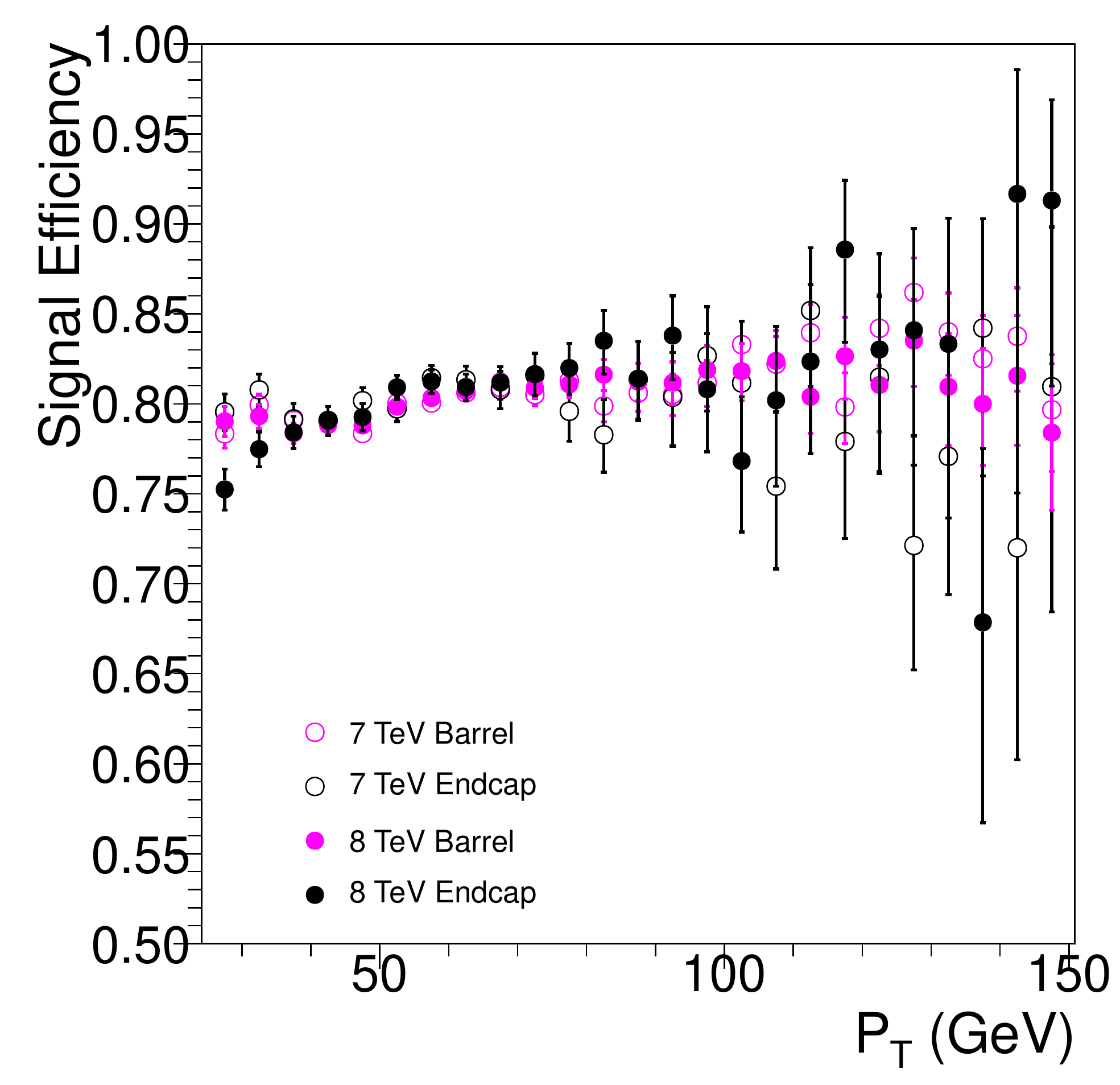}
    \includegraphics[width=0.4\textwidth]{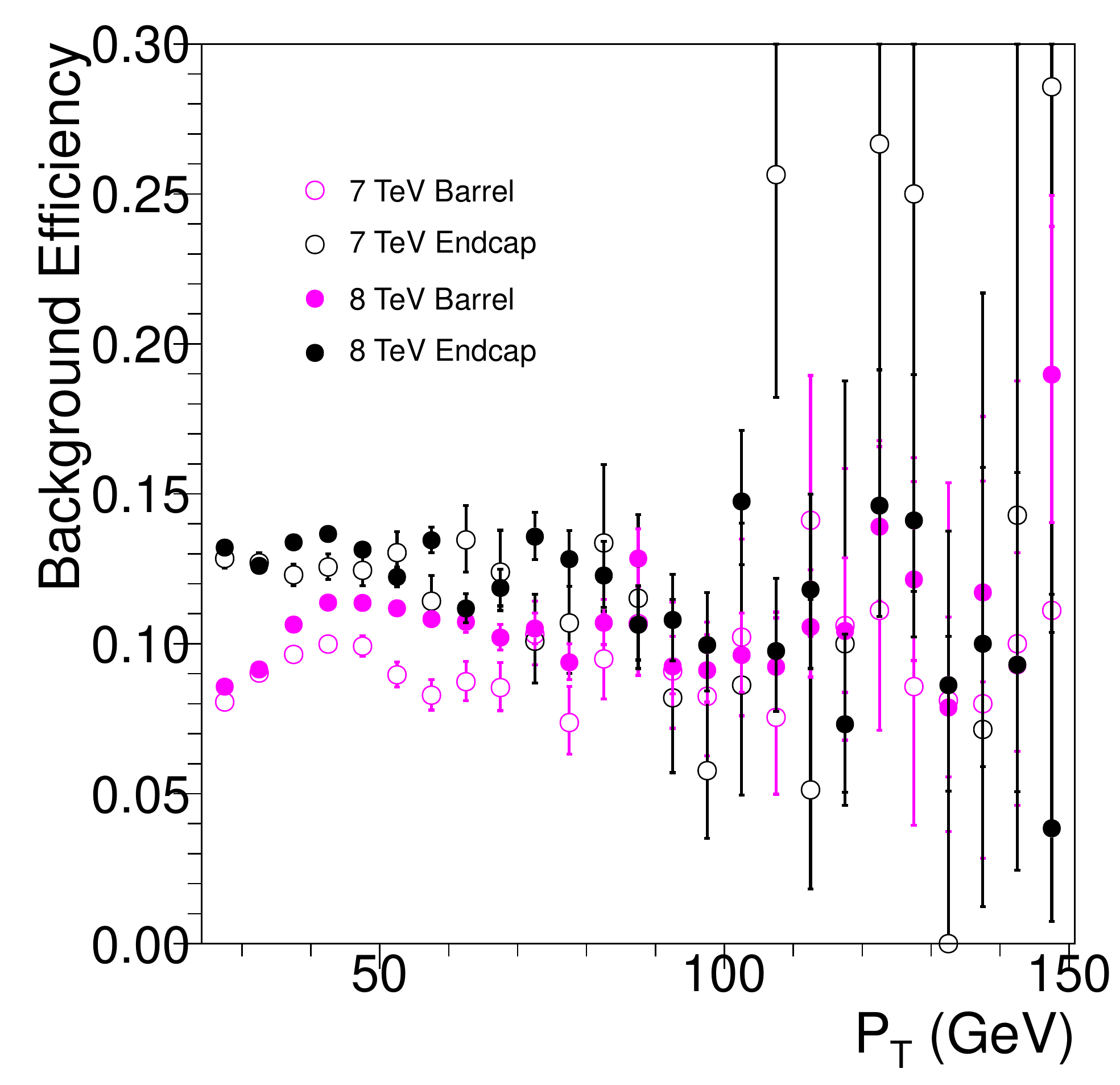}\\
    \includegraphics[width=0.4\textwidth]{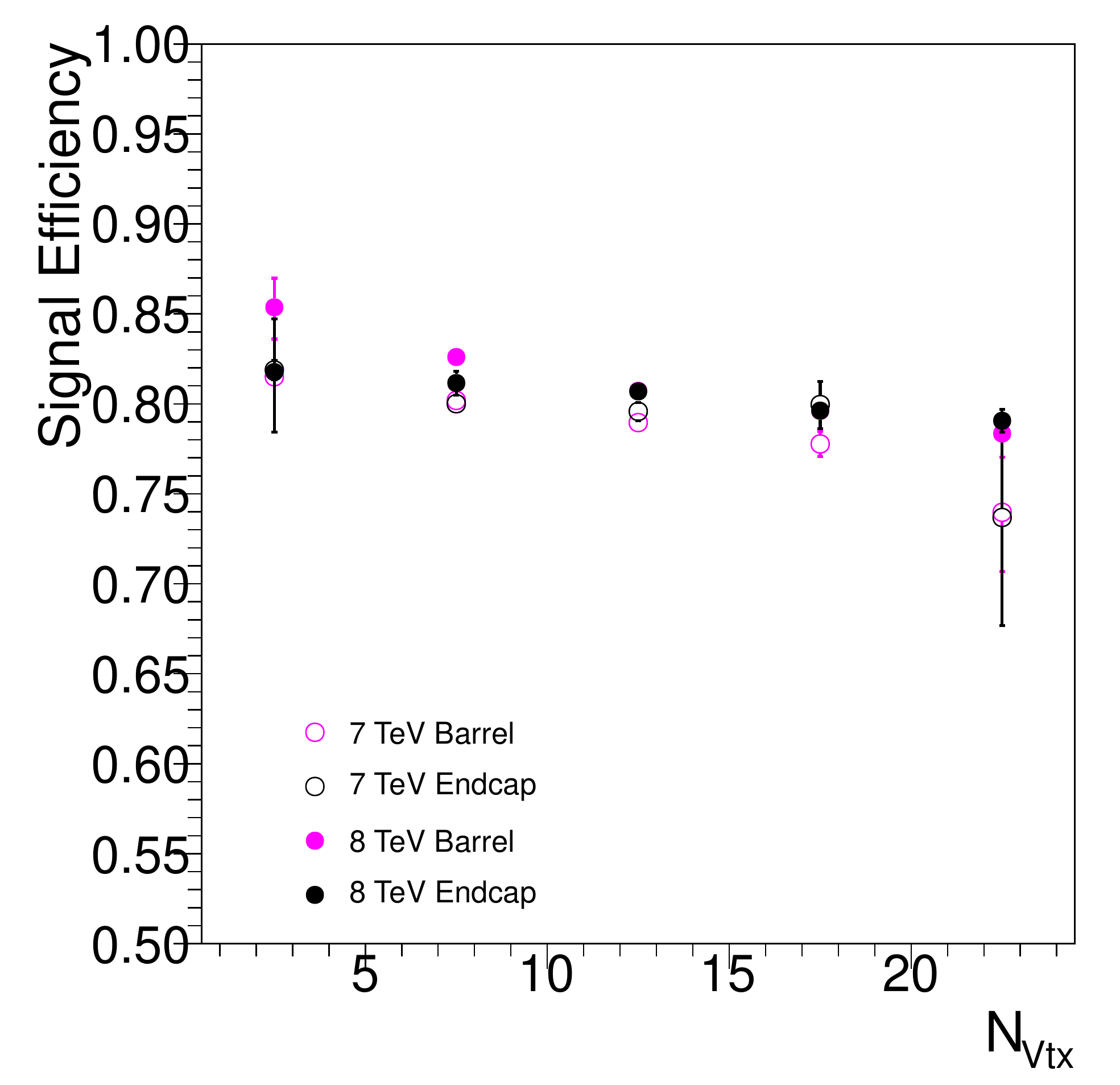}
    \includegraphics[width=0.4\textwidth]{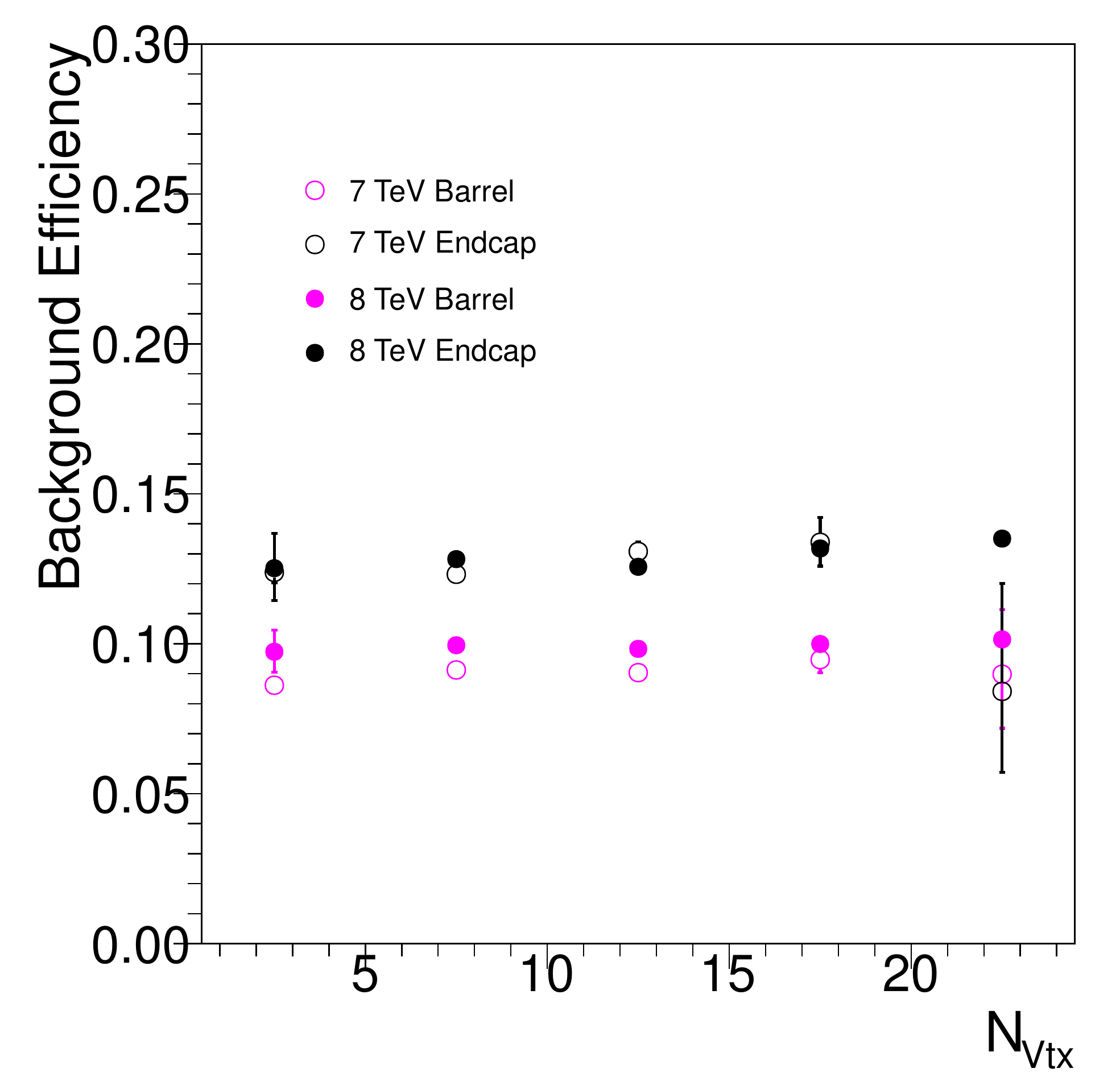}\\
  \end{center}
  \caption{The differential efficiencies versus $\eta_{SC}$ (top), $p_{T}$ (middle) and $N_{Vtx}$ (bottom) for signal prompt (left) and background fake (right) photons at 7 TeV (hollow) and 8 TeV (solid) pp collisions. The efficiencies are evaluated at overall signal efficiency 80\%. The efficiencies versus $\eta_{SC}$ are in blue for both photons in the barrel and endcap. The efficiencies versus $p_{T}$ and $N_{Vtx}$ are in magenta for the photons in the barrel and in black for photons in the endcap. The photons are from testing samples passing the preselection with $p_{T}$ $>$ $25~\mathrm{GeV}$.}
  \label{fig:idmva eff vs pt eta vtx}
\end{figure}

The photon identification BDT performance is evaluated using the testing samples. The curves of overall background efficiency versus signal efficiency, corresponding to IDBDT cuts, for photons in the barrel and endcap at 7 TeV (8 TeV) are shown on the left (right) in Figure \ref{fig:idmva bkg vs sig eff}. As a reference, the background efficiency for photons in the barrel (endcap) at $7~\mathrm{TeV}$ ($8~\mathrm{TeV}$), at 80$\%$ signal efficiency, is listed in Table \ref{tab:idmva eff}. The corresponding differential signal and background efficiencies versus $\eta_{SC}$, $p_{T}$ and $N_{Vtx}$ are shown in Figure \ref{fig:idmva eff vs pt eta vtx}. The efficiencies versus  $\eta_{SC}$ and $p_{T}$ are reasonably flat for photons in the barrel (endcap). This is a desirable feature due to the inclusion of $\eta_{SC}$ and $E_{SC}$ into the input variables, and the 2D $p_{T}$-$\eta_{SC}$ reweighting in the training. The efficiencies are also reasonably flat as a function of $N_{Vtx}$, which is expected as a result of using $\rho_{Event}$ as an input variable.     
\begin{table}[h]
 \caption{The efficiency of background fake photons at the signal prompt photon efficiency 80\% . The photons are from testing samples passing the preselection with $p_{T}~>~25~\mathrm{GeV}$.}
 \begin{center}
   \begin{tabular}{|l|c|c|} 
    \hline
           &  7 TeV (\%) & 8 TeV (\%)\\
    \hline
    Barrel &  9.0  $\pm$ 0.2 & 10.0 $\pm$ 0.2\\
    \hline
    Endcap & 12.4 $\pm$ 0.2 & 13.0 $\pm$ 0.2\\
    \hline
    \end{tabular}
       \label{tab:idmva eff}
  \end{center}
\end{table}

\begin{figure}[h] 
  \begin{center}
    \includegraphics[width=1\textwidth]{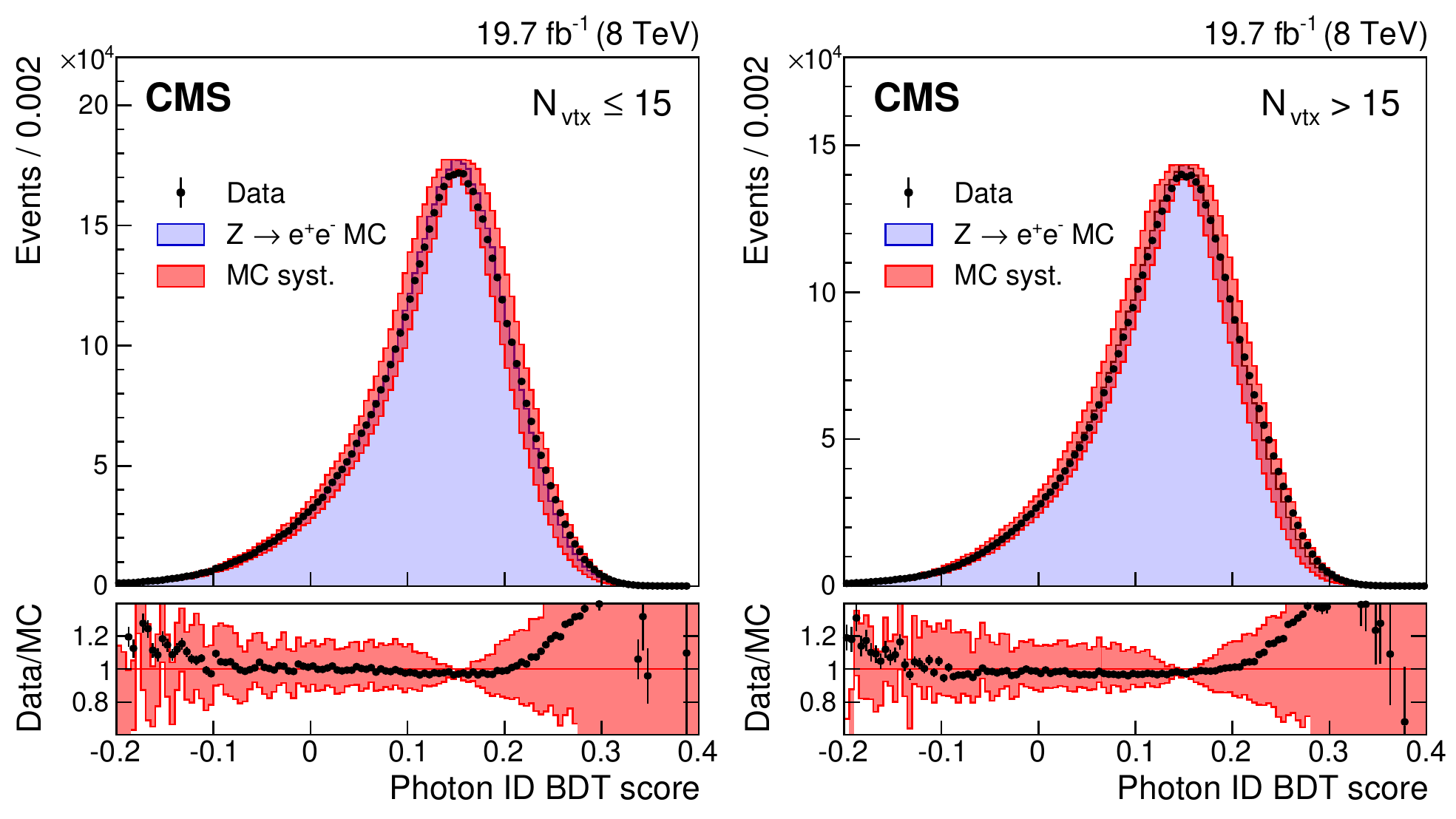}
  \end{center}
  \caption{IDBDT (Photon ID BDT score) distributions for electrons in the barrel, from $Z\rightarrow e^{+}e^{-}$ events from data (points) and Monte Carlo simulation (histogram) at 8 TeV with $N_{Vtx}$ $\leq$ 15 (left) and $N_{Vtx}$ $>$ 15 (right) are shown. Electrons are required to pass the preselection with inverted electron veto and IDBDT $>$ $-$0.2. The $\pm$0.01 shift of the  Monte Carlo distribution is shown as the red band.}
  \label{fig:zeemassid}
\end{figure}

The discrepancy between the IDBDT distributions of Monte Carlo simulation and data due to the imperfect simulation of detector response is evaluated using electrons from $Z\rightarrow e^{+}e^{-}$ events and photons from $Z\rightarrow \mu^{+}\mu^{-}\gamma$ events, and is treated as a systematic uncertainty in the Higgs signal extraction. Photons are required to pass the cut IDBDT $>$ $-0.2$, since in IDBDT region below $-$0.2 the agreement between data and Monte Carlo simulation is relatively poor, and signal to background ratio is very small. The efficiency for the cut on the prompt photons is rounded to 1, as well as the efficiency scale factor from Monte Carlo simulation to data. A shift of $\pm$0.01 of the Monte Carlo IDBDT is shown to cover the discrepancy in the region IDBDT $>$ $-$0.2. Figure \ref{fig:zeemassid}\cite{hggfinalpaper} shows the IDBDT distributions for electrons in the barrel, from $Z\rightarrow e^{+}e^{-}$ events from data (points) and Monte Carlo simulation (histogram) at 8 TeV with $N_{Vtx}$ $\leq$ 15 (left) and $N_{Vtx}$ $>$ 15 (right), passing the preselection with inverted electron veto and IDBDT $>$ $-$0.2. Good agreement between data and Monte Carlo simulation is observed within the $\pm$0.01 variation band of the Monte Carlo simulation.   

\section{Diphoton BDT}
\label{sec:Diphoton BDT}
The diphoton BDT is trained to provide each preselected diphoton pair a score, as a measure of its expected $S/B$ under the signal diphoton mass peak in the existence of the Higgs boson, which is used for event classification as described in Chapter \ref{chap:Event Classification}.

\subsection{Training Samples}
\label{sec:Training Samples}
The training is performed on Monte Carlo simulation of reconstructed diphoton events which pass the preselection and IDBDT $>$ $-$0.2 for both photons. The signal sample consists of $H\rightarrow \gamma\gamma$ events at a Higgs mass of 123 GeV, with all four production processes weighted by cross section. The background sample consists of a proper mixture of prompt diphoton, $\gamma$ + jet and dijet events. The trainings are performed separately for events at 7 TeV and $8~\mathrm{TeV}$.

\subsection{Input Variables}
\label{sec:Input Variables}
The input variables are described as following:
\begin{itemize}
\item Diphoton mass resolution variables:
  \begin{itemize}
  \item $(\sigma_{m}/m)_{R}$: the diphoton mass resolution estimator assuming the correct vertex is selected. It is the sum in quadrature of the per-photon energy resolution estimators of the leading and sub-leading photon $(\sigma_{E}/E)^{\gamma1}$ and $(\sigma_{E}/E)^{\gamma2}$ as:
    \begin{equation}
      (\sigma_{m}/m)_{R} = \frac{1}{2} \sqrt{\{(\sigma_{E}/E)^{\gamma1}\}^2+\{(\sigma_{E}/E)^{\gamma2}\}^2}.
      \label{eqn:mass resolution right}
    \end{equation} 
  \item $(\sigma_{m}/m)_{W}$: the diphoton mass resolution estimator assuming the wrong vertex is selected. It is the sum in quadrature of $(\sigma_{m}/m)_{R}$ and the mass resolution contributed by vertex selection $(\sigma_{m}^{Vtx}/m)$ as:
    \begin{equation}
      (\sigma_{m}/m)_{W} = \sqrt{(\sigma_{m}/m)_{R}^2+(\sigma_{m}^{Vtx}/m)^2},
      \label{eqn:mass resolution wrong}
    \end{equation}
    where $\sigma_{m}^{Vtx}/m$ is computed by propagating the uncertainty of the distance between the selected vertex and true vertex, approximated by $\sqrt{2}$ times the average standard deviation of the pp interaction region in $z$.   
  \item $p_{Vtx}$: the probability of the selected vertex being the right vertex estimated from vertex probability BDT.
  \end{itemize}  
\item Photon identification variables:
  \begin{itemize}
  \item IDBDT$^{\gamma1}$: the score assigned to the leading photon from the photon identification BDT.
  \item IDBDT$^{\gamma2}$: the score assigned to the sub-leading photon from the photon identification BDT.
  \end{itemize}
\item Diphoton kinematics variables:
  \begin{itemize}
  \item $p_{T}^{\gamma1}/m_{\gamma\gamma}$: the transverse momentum of the leading photon divided by the diphoton mass.  
  \item $p_{T}^{\gamma2}/m_{\gamma\gamma}$: the transverse momentum of the sub-leading photon divided by the diphoton mass.  
  \item $\eta^{\gamma1}$: the pseudorapidity of the leading photon momentum.  
  \item $\eta^{\gamma2}$: the pseudorapidity of the sub-leading photon momentum.  
  \item $cos(\Delta\phi_{\gamma\gamma})$: the cosine of the separation in the azimuthal angle between the leading and sub-leading photon.
  \end{itemize}
\end{itemize}     
The diphoton mass resolution variables are not used directly in the training but combined into a weight $1/\sigma_{Eff}$ as in Equation \ref{eqn:effective resolution weight}, where $\sigma_{Eff}$ is the effective diphoton mass resolution estimator. The $H\rightarrow \gamma\gamma$ events in the training sample are weighted by $1/\sigma_{Eff}$, such that the events with better resolution get higher weights and appear more signal like.     
\begin{equation}
  1/\sigma_{Eff} =  \frac{p_{Vtx}}{(\sigma_{m}/m)_{R}} + \frac{1-p_{Vtx}}{(\sigma_{m}/m)_{W} }
  \label{eqn:effective resolution weight}
\end{equation}

\begin{figure}[hbpt] 
  \begin{center}
    \includegraphics[width=0.37\textwidth]{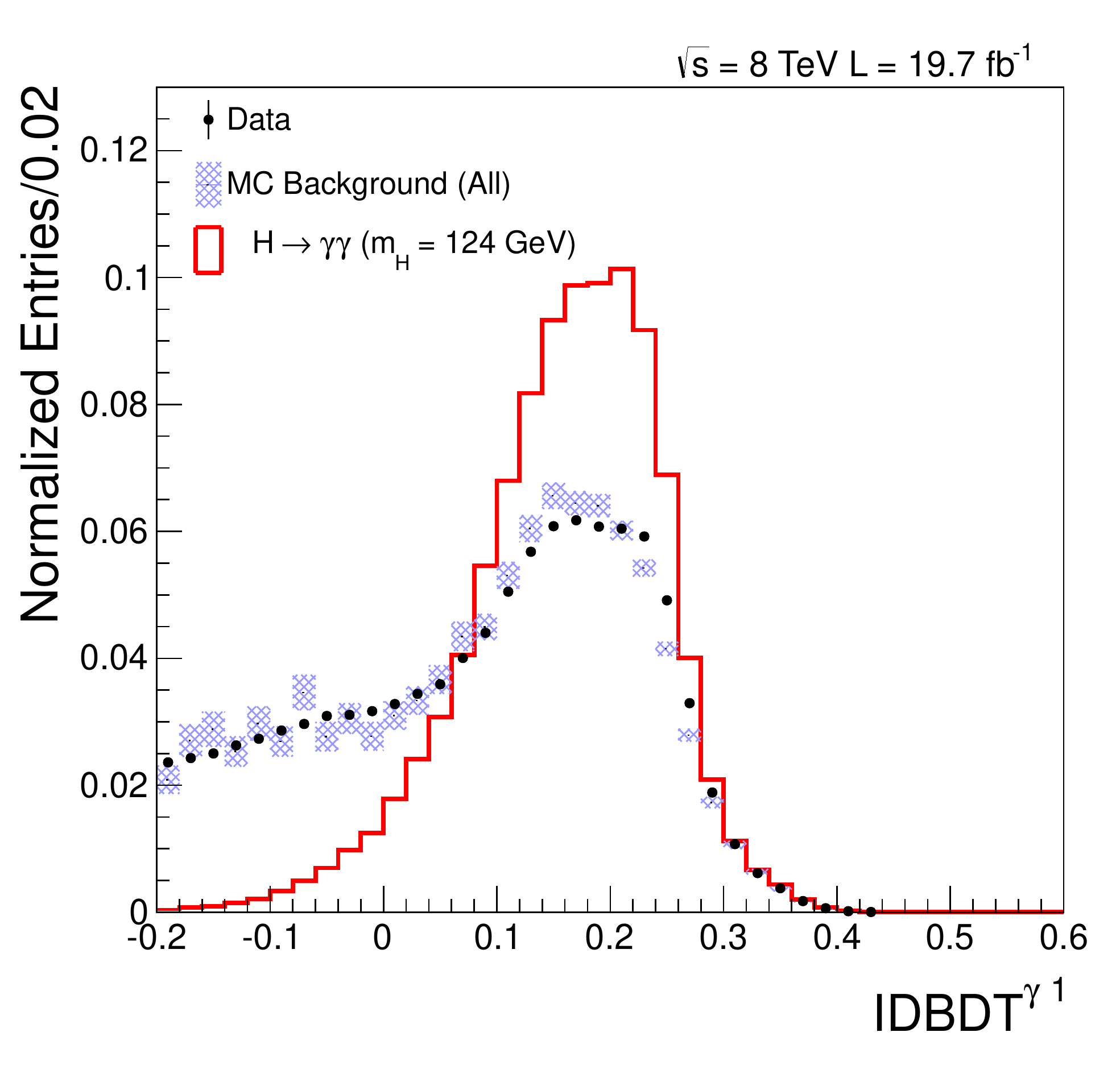}
    \includegraphics[width=0.37\textwidth]{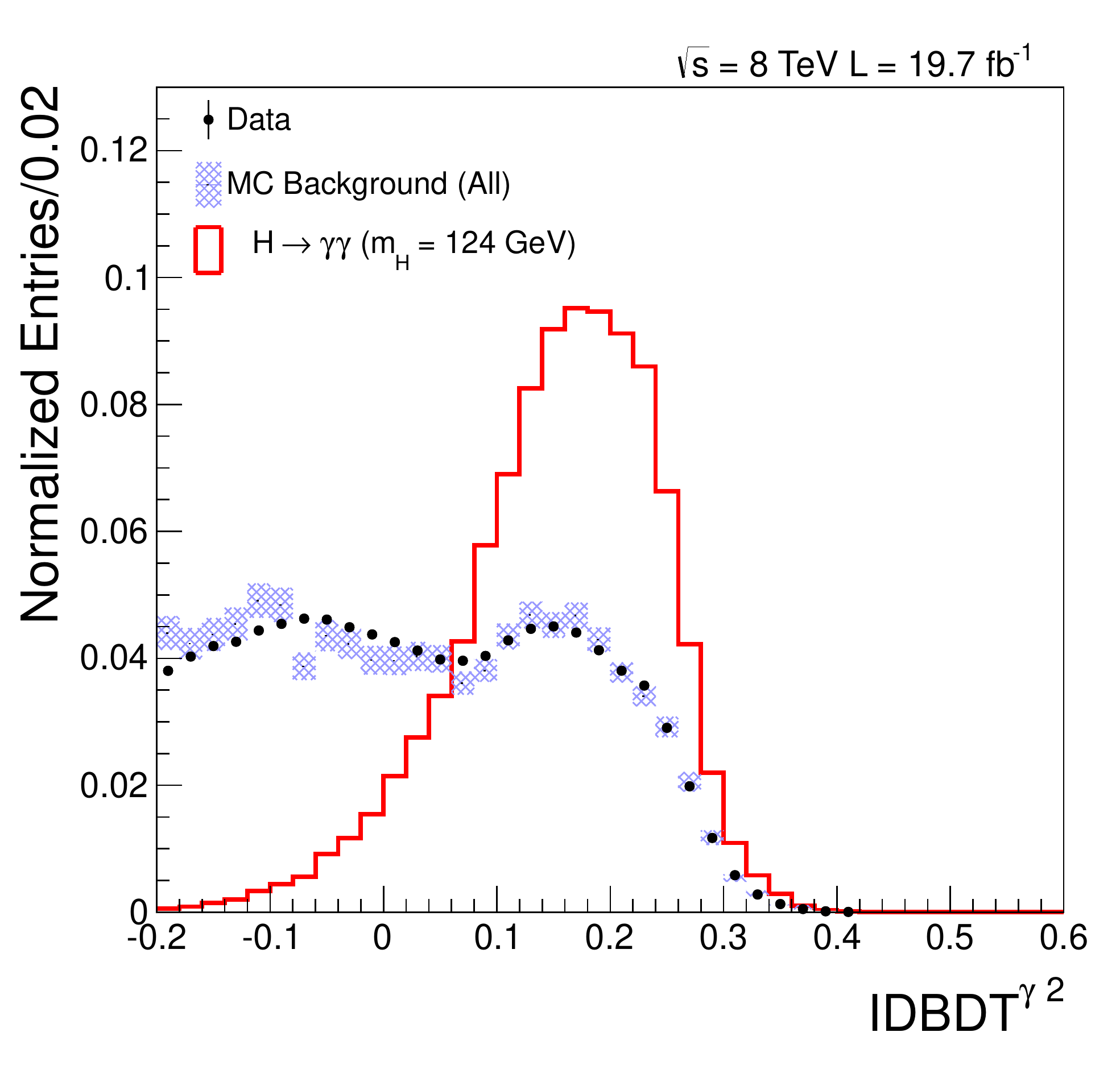}
    \includegraphics[width=0.37\textwidth]{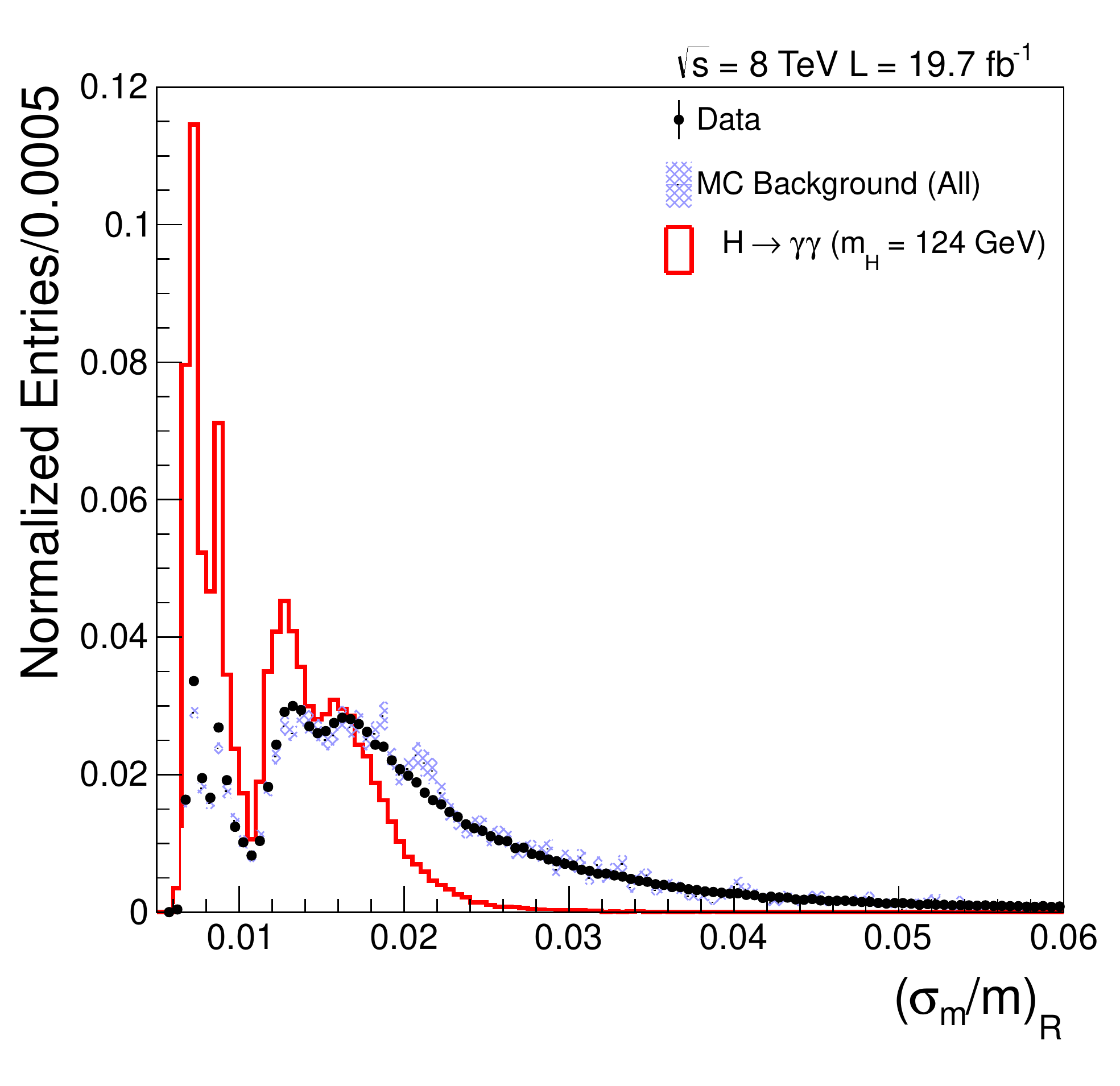}
    \includegraphics[width=0.37\textwidth]{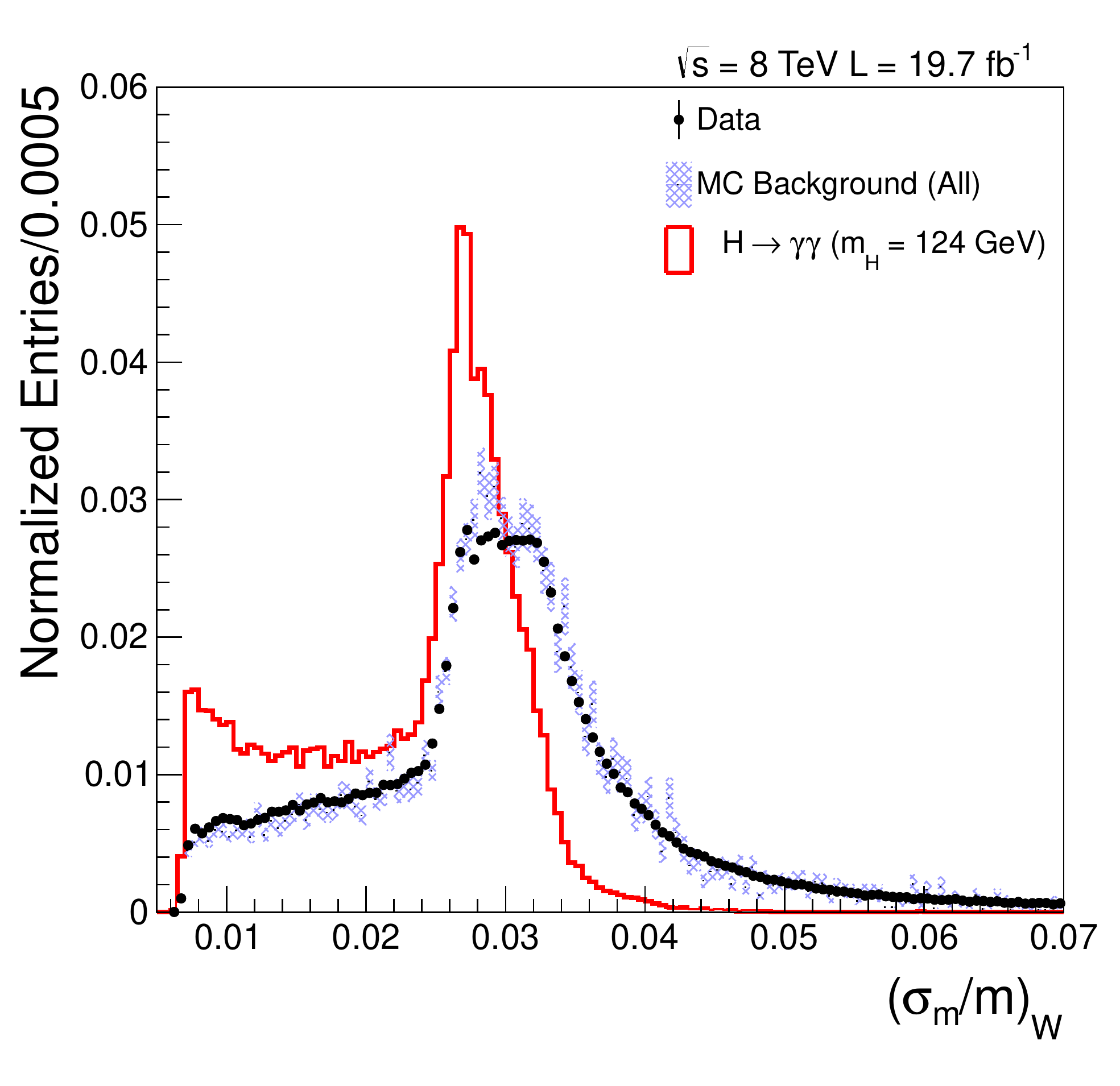}
    \includegraphics[width=0.37\textwidth]{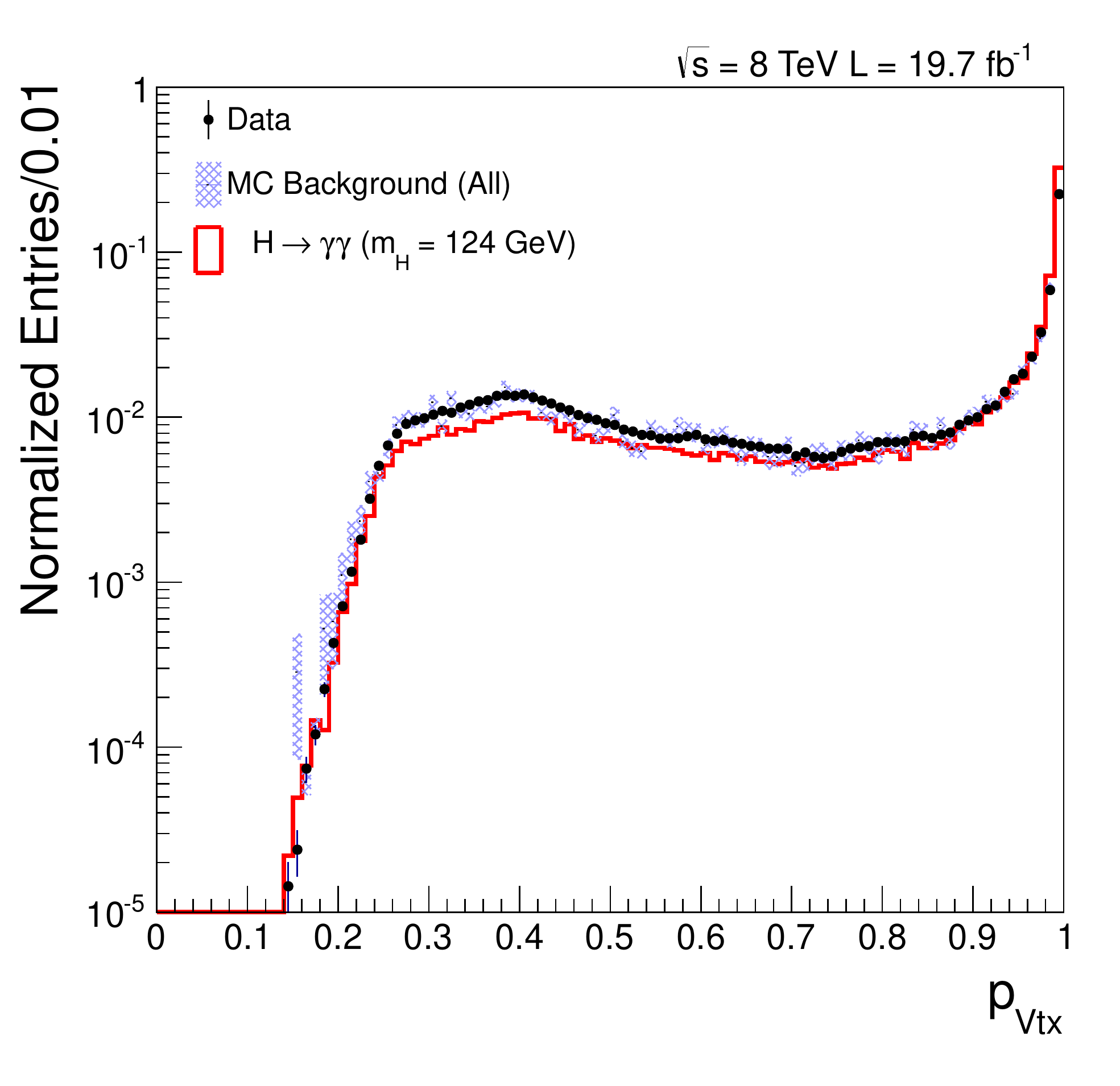}
  \end{center}
  \caption{The distributions of the diphoton BDT input variables IDBDT$^{\gamma1}$ (top left), IDBDT$^{\gamma2}$ (top right), $(\sigma_{m}/m)_{R}$ (middle left), $(\sigma_{m}/m)_{W}$ (middle right), and $p_{Vtx}$ (bottom) for data (points), Monte Carlo background (histogram with blue band for statistical uncertainty) consisting of prompt diphoton, $\gamma$ + jet and dijet events weighted by cross section, and Monte Carlo signal (red line) consisting of $H\rightarrow \gamma\gamma$ events at a Higgs mass of 124 GeV with all four production processes weighted by cross section, at $8~\mathrm{TeV}$. The data and Monte Carlo background events in the signal region 120 GeV $<$ $m_{\gamma\gamma}$ $<$ 130 GeV are removed. All events pass the preselection with IDBDT $>$ $-$0.2 for both photons. Photon energy corrections are applied to both data and Monte Carlo events, and additional corrections are applied to Monte Carlo events including efficiency scaling and pileup reweighting.}
  \label{fig:diphotonmva input 1}
\end{figure}

\begin{figure}[hbpt] 
  \begin{center}
    \includegraphics[width=0.37\textwidth]{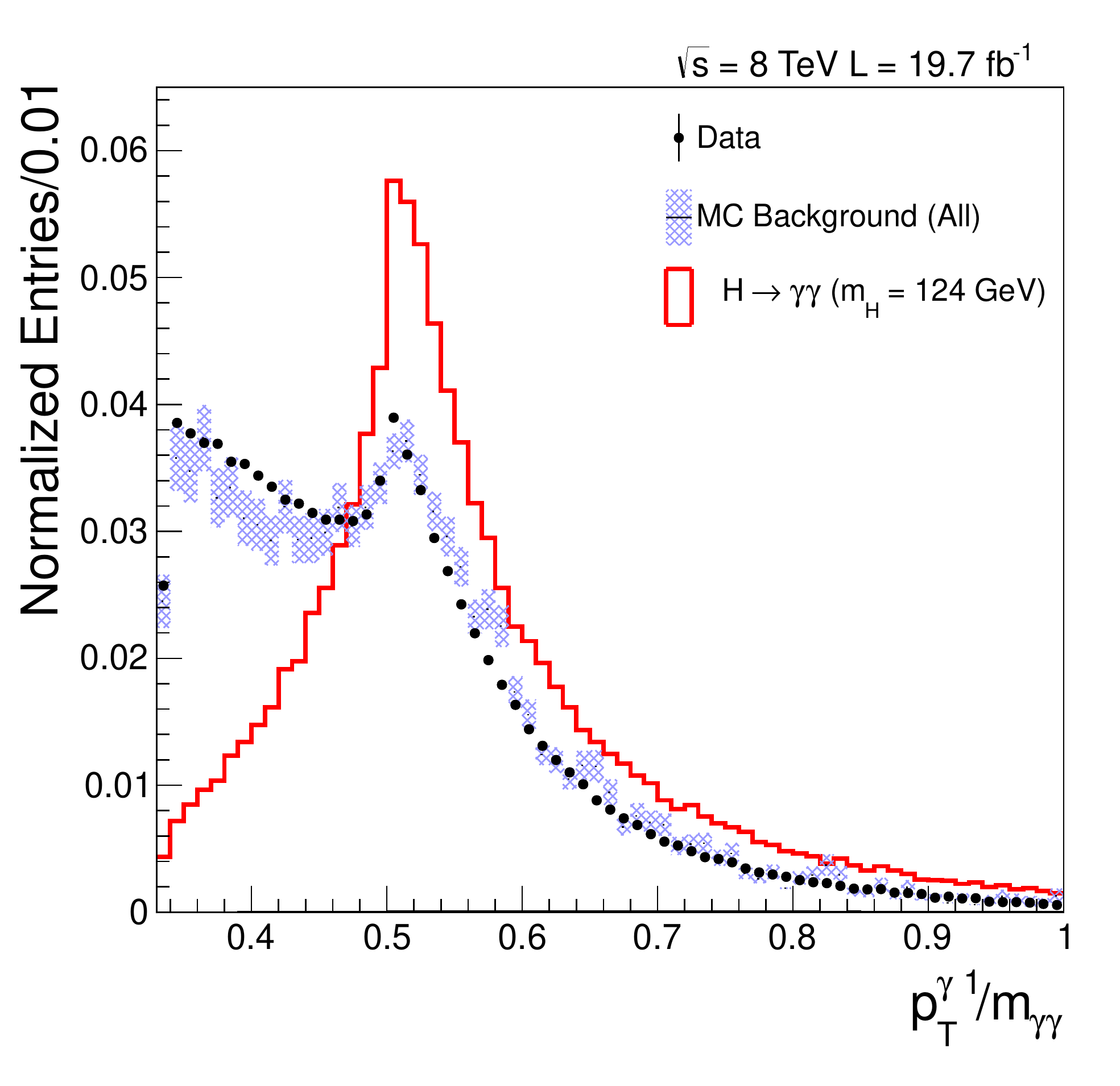}
    \includegraphics[width=0.37\textwidth]{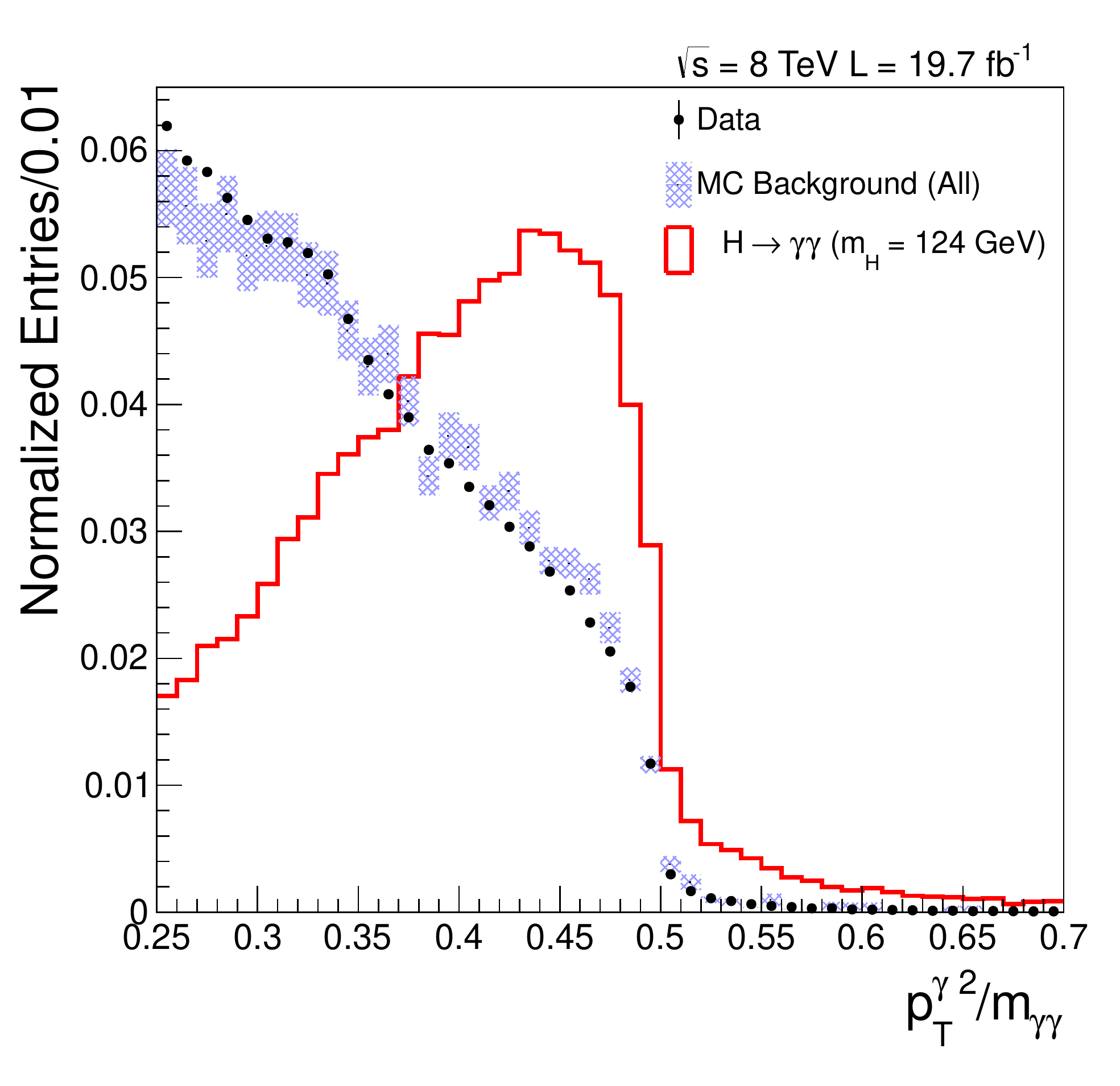}
    \includegraphics[width=0.37\textwidth]{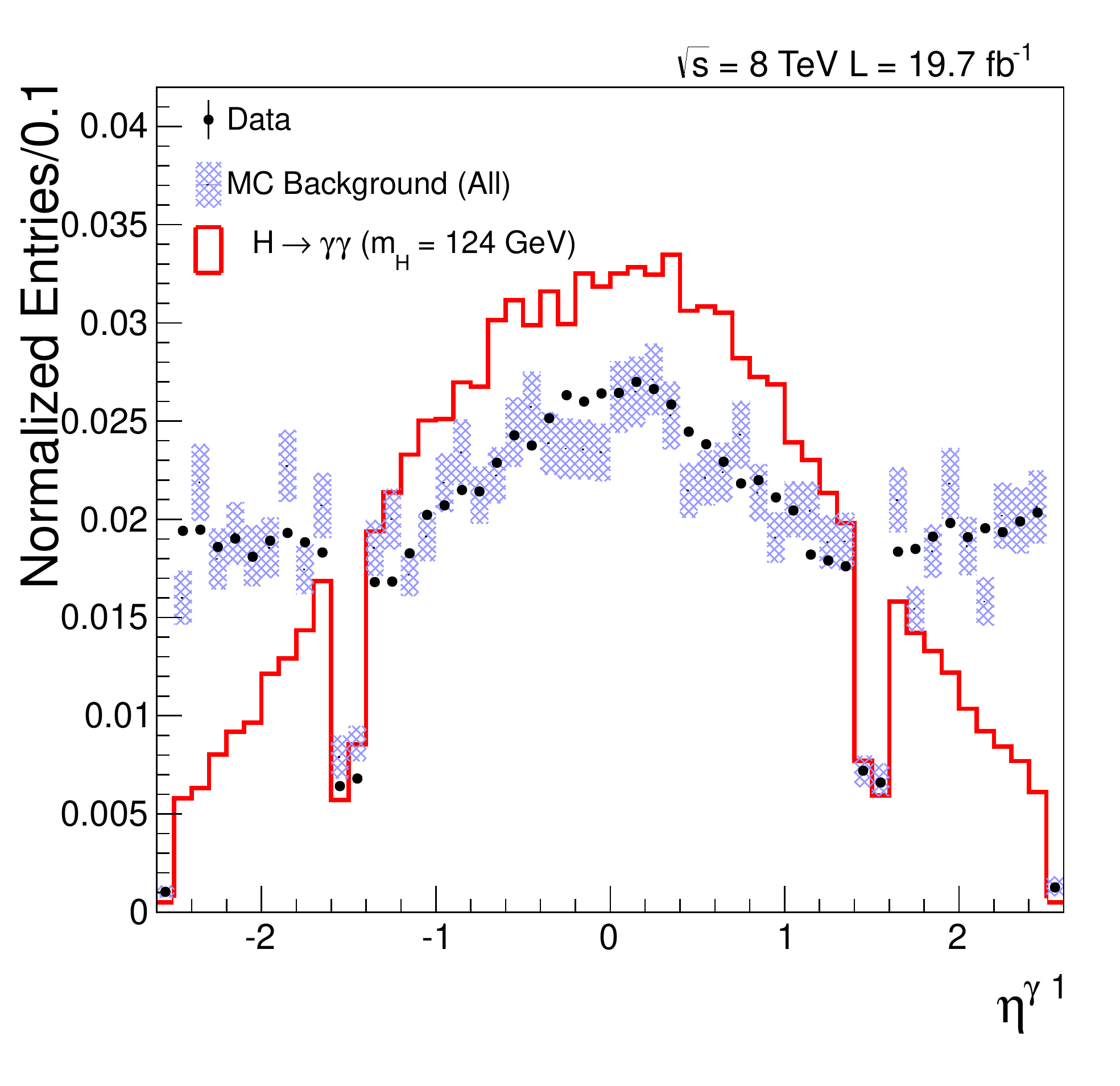}
    \includegraphics[width=0.37\textwidth]{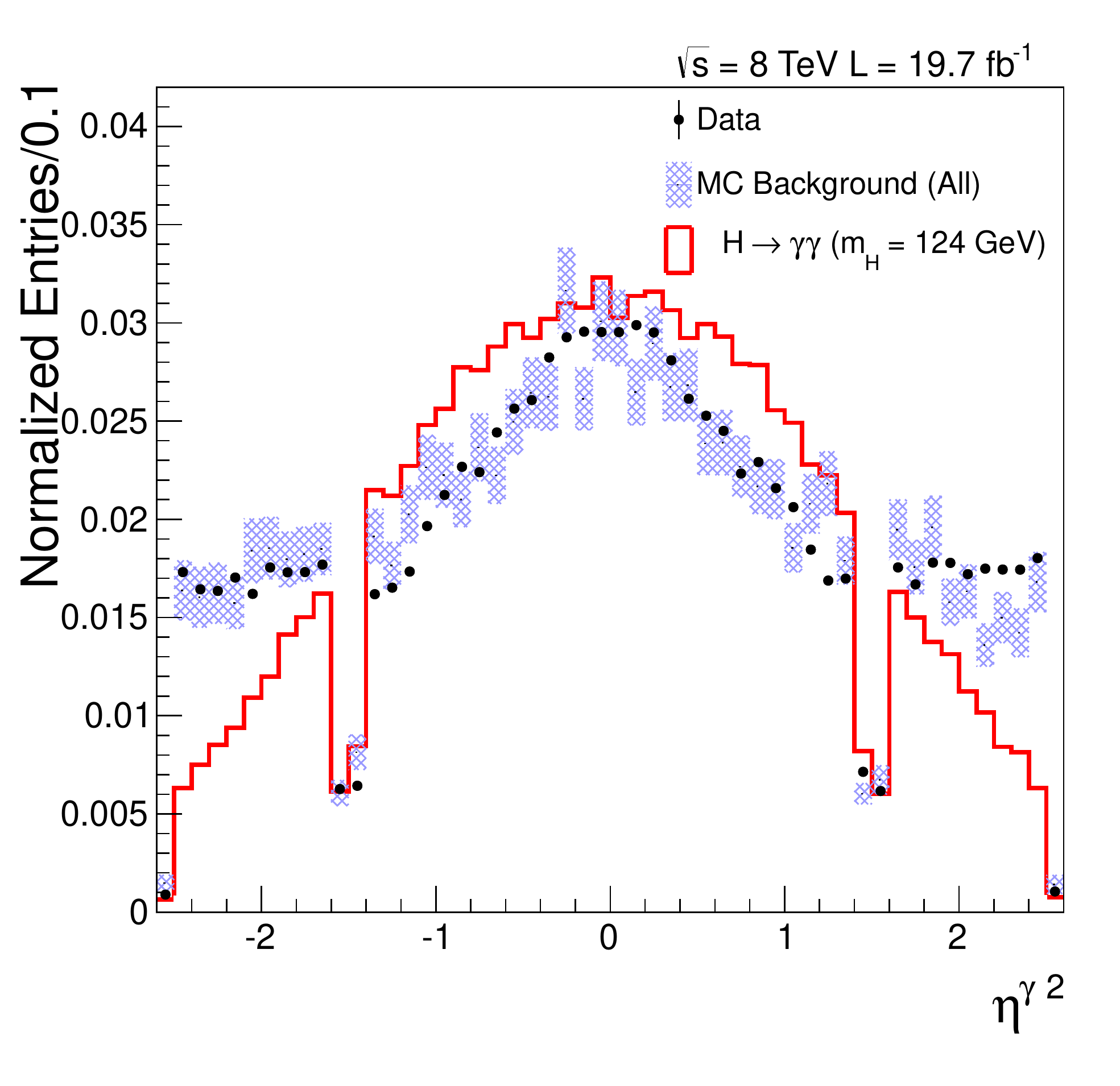}
    \includegraphics[width=0.37\textwidth]{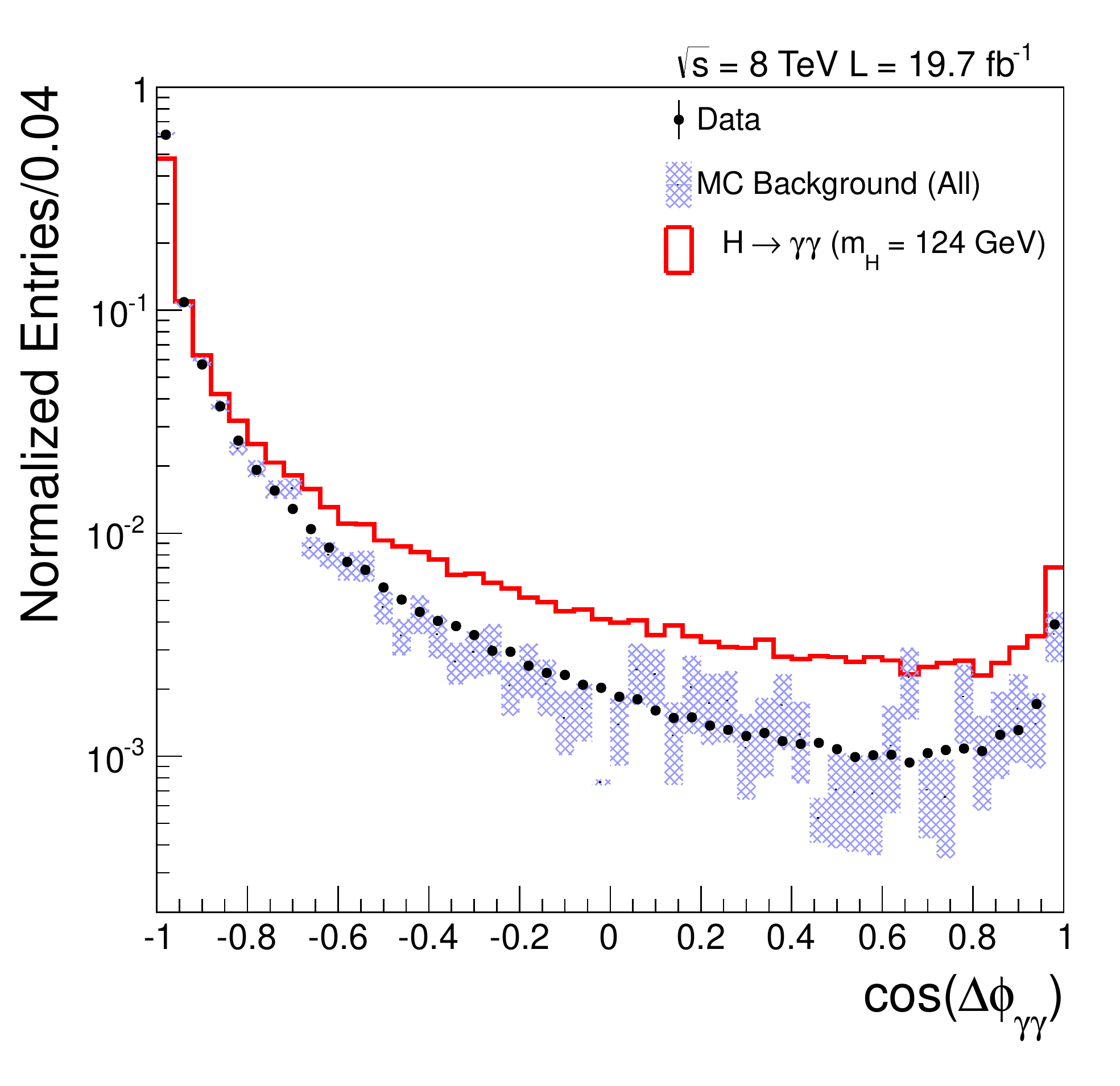}
  \end{center}
  \caption{The distributions of the diphoton BDT input variables $p_{T}^{\gamma1}/m_{\gamma\gamma}$ (top left), $p_{T}^{\gamma2}/m_{\gamma\gamma}$ (top right), $\eta^{\gamma1}$ (middle left), $\eta^{\gamma2}$ (middle right), and $cos(\Delta\phi_{\gamma\gamma})$ (bottom) for data (points), Monte Carlo background (histogram with blue band for statistical uncertainty) consisting of prompt diphoton, $\gamma$ + jet and dijet events weighted by cross section, and Monte Carlo signal (red line) consisting of $H\rightarrow \gamma\gamma$ events at a Higgs mass of 124 GeV with all four production processes weighted by cross section, at $8~\mathrm{TeV}$. The data and Monte Carlo background events in the signal region 120 GeV $<$ $m_{\gamma\gamma}$ $<$ 130 GeV are removed. All events pass the preselection with IDBDT $>$ $-$0.2 for both photons. Photon energy corrections are applied to both data and Monte Carlo events, and additional corrections are applied to Monte Carlo events including efficiency scaling and pileup reweighting.}
  \label{fig:diphotonmva input 2}
\end{figure}

The distributions of the input variables for data and Monte Carlo signal and background events at 8 TeV, which pass the preselection and IDBDT $>$ $-$0.2 for both photons, are shown in Figure \ref{fig:diphotonmva input 1} and Figure \ref{fig:diphotonmva input 2}. Photon energy corrections are applied to both data and Monte Carlo events, and additional corrections are applied to Monte Carlo events including efficiency scaling and pileup reweighting. The signal consists of $H\rightarrow \gamma\gamma$ events at a Higgs mass of 124 GeV, which is later used for the event classification optimization. The data and Monte Carlo background events in the signal region 120 GeV $<$ $m_{\gamma\gamma}$ $<$ 130 GeV are removed for the data and Monte Carlo background comparison. Good agreement between data and Monte Carlo simulation of background is shown. The remaining discrepancy makes the performance of the diphoton BDT sub-optimal but does not affect the correctness of the analysis as the background is evaluated from data for the final statistical analysis. 

\subsection{Output and Performance}
\label{sec:Output and Performance}
\begin{figure}[hbpt] 
  \begin{center}
    \includegraphics[width=0.49\textwidth]{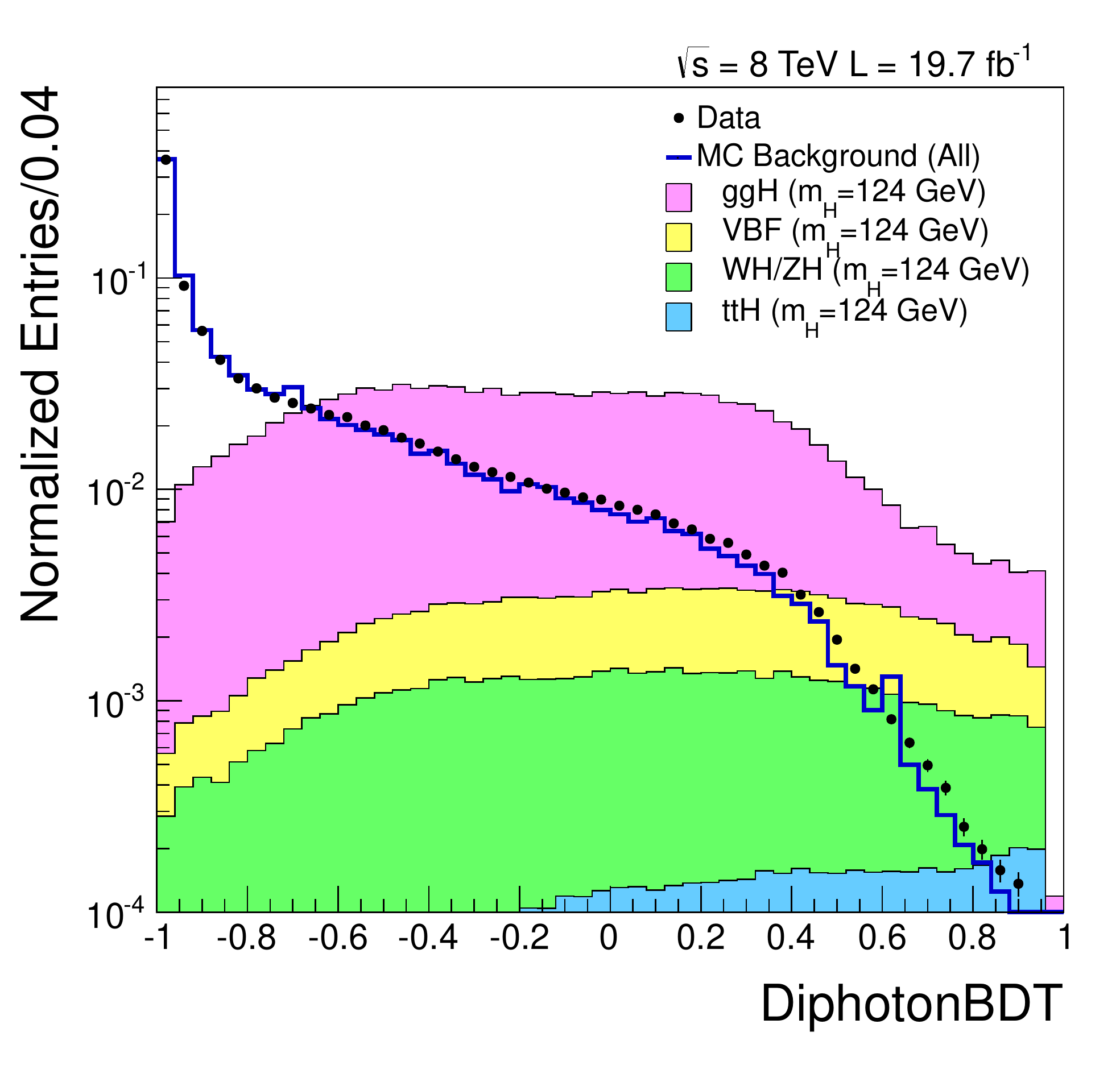}
    \includegraphics[width=0.49\textwidth]{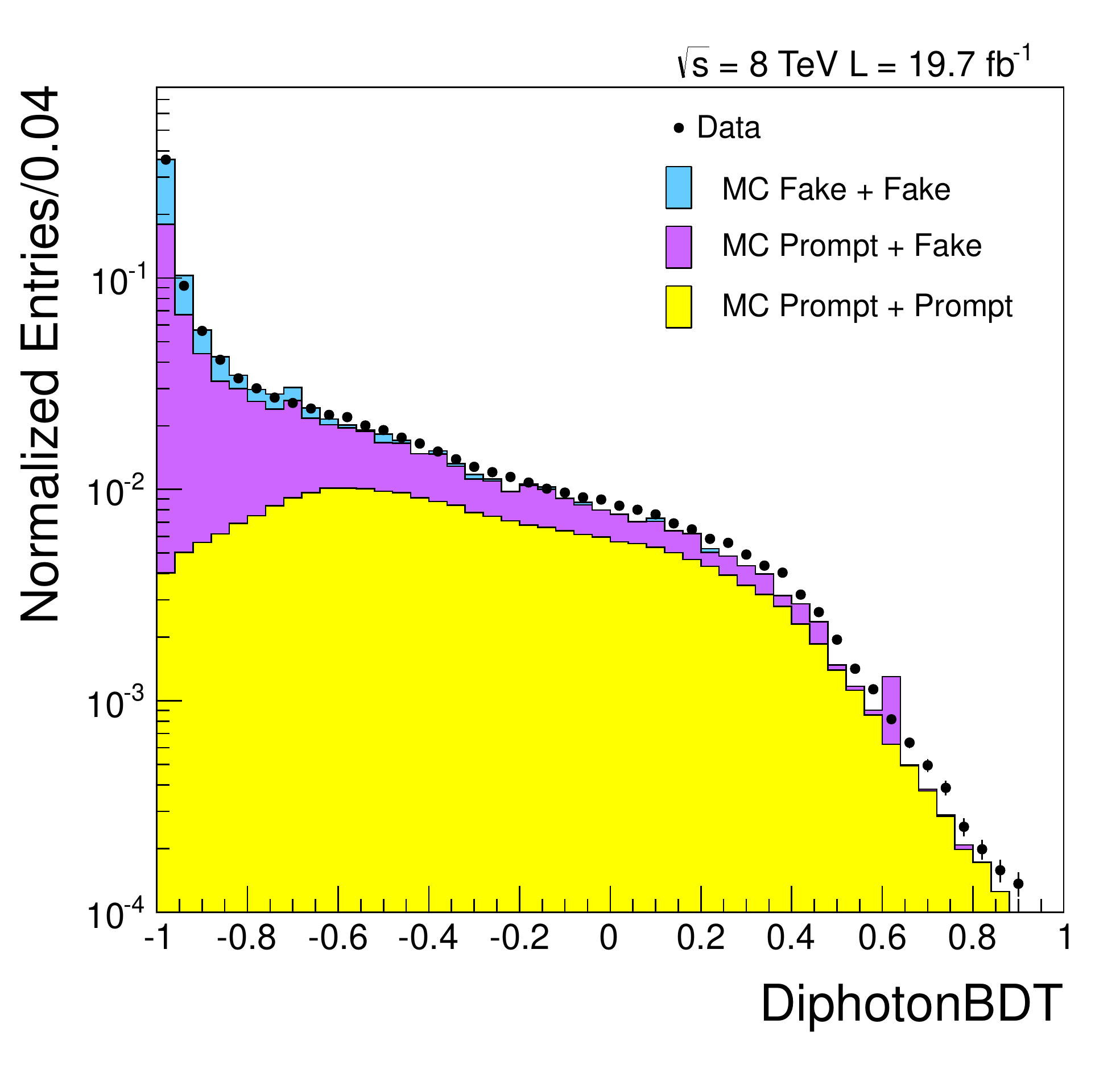}
  \end{center}
  \caption{Left: the distributions of DiphotonBDT for data (points), Monte Carlo background (blue line) consisting of prompt diphoton, $\gamma$ + jet and dijet events weighted by cross section, and Monte Carlo signal (stacked histogram) consisting of $H\rightarrow \gamma\gamma$ events at a Higgs mass of 124 GeV with all four production processes weighted by cross section, at $8~\mathrm{TeV}$. Right: the distributions of DiphotonBDT for data (points) and Monte Carlo background (stacked histogram) consisting of prompt diphoton, $\gamma$ + jet and dijet events weighted by cross section, at $8~\mathrm{TeV}$. The data and Monte Carlo background events in the signal region 120 GeV $<$ $m_{\gamma\gamma}$ $<$ 130 GeV are removed. All events pass the preselection with IDBDT $>$ $-$0.2 for both photons. Photon energy corrections are applied to both data and Monte Carlo events, and additional corrections are applied to Monte Carlo events including efficiency scaling and pileup reweighting.}
  \label{fig:diphotonmva output databkgsig}
\end{figure}

The distributions of the diphoton BDT output named DiphotonBDT for data, Monte Carlo signal and background events, which pass the preselection and IDBDT $>$ $-$0.2 for both photons, are shown on the left in Figure \ref{fig:diphotonmva output databkgsig}. The  DiphotonBDT ranges from $-$1 to 1. The number of $H\rightarrow \gamma\gamma$ events over the number of background events in each bin increases with the DiphotonBDT as expected. The $H\rightarrow \gamma\gamma$ events from the production processes \textit{VBF}, \textit{VH} and \textit{t$\overline{t}$H} tend to have higher score than the events from \textit{ggH} production process. This is due to the fact that \textit{VBF}, \textit{VH} and \textit{t$\overline{t}$H} events have higher Higgs $p_{T}$ and so higher cos$(\Delta\phi_{\gamma\gamma})$ on average than \textit{ggH} events. Also the BDT assigns higher score on the events with higher cos$(\Delta\phi_{\gamma\gamma})$ as the $H\rightarrow \gamma\gamma$ events have higher cos$(\Delta\phi_{\gamma\gamma})$ on average than the background events, as shown at the bottom plot in Figure \ref{fig:diphotonmva input 2}. The data and Monte Carlo background DiphotonBDT distributions are compared with signal region 120 GeV $<$ $m_{\gamma\gamma}$ $<$ 130 GeV removed. The Monte Carlo background in general describes the data well. The contributions from each background component are shown on the right in Figure \ref{fig:diphotonmva output databkgsig}. The average score increases with the number of prompt photons in the background as expected. The discrepancy between data and Monte Carlo background in the high score region is due to the discrepancy between the actual and simulated kinematics for the prompt diphoton background, but this does not influence the correctness of the analysis as explained above.

\begin{figure}[hbpt] 
  \begin{center}
    \includegraphics[width=0.49\textwidth]{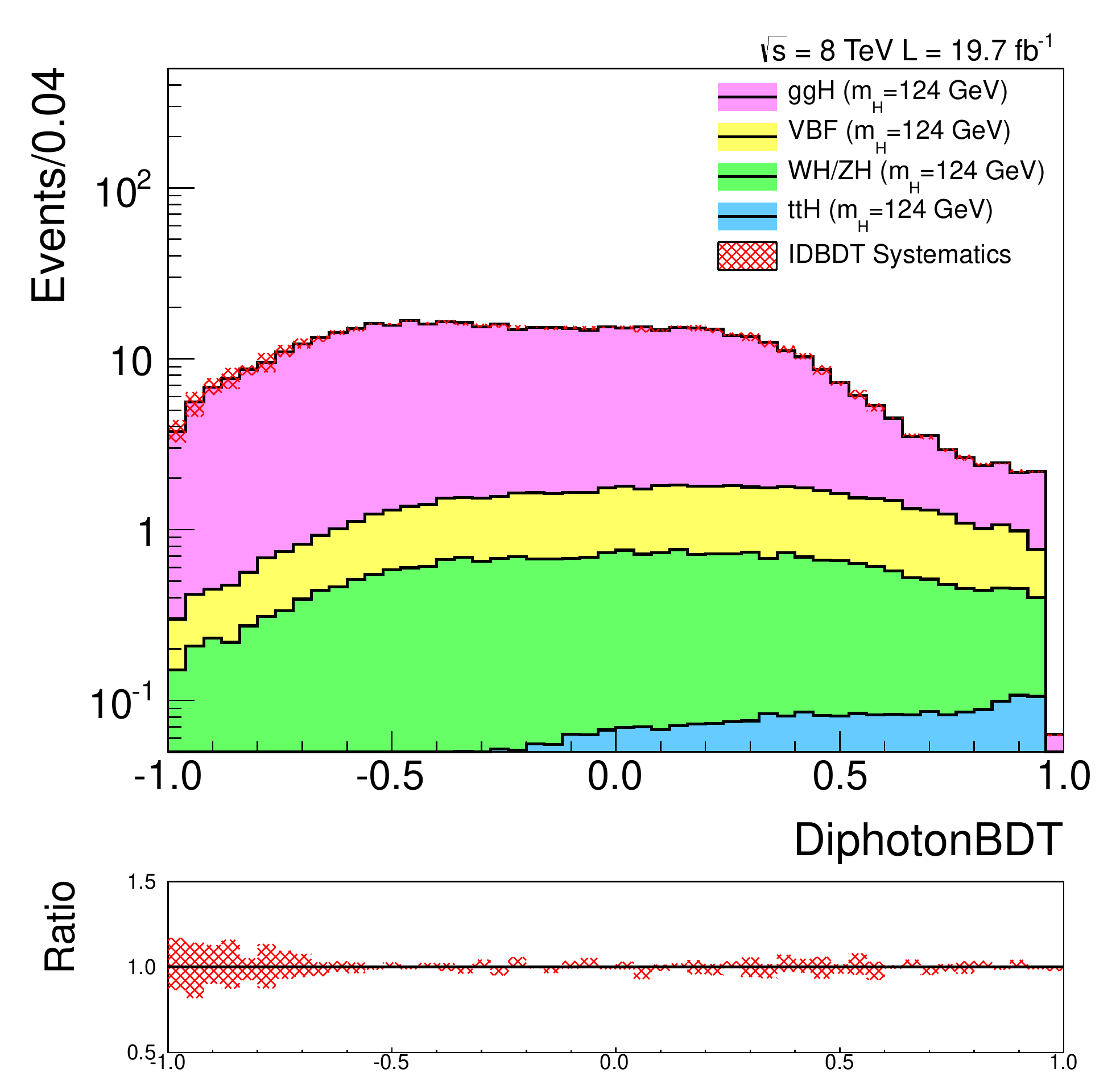}
    \includegraphics[width=0.49\textwidth]{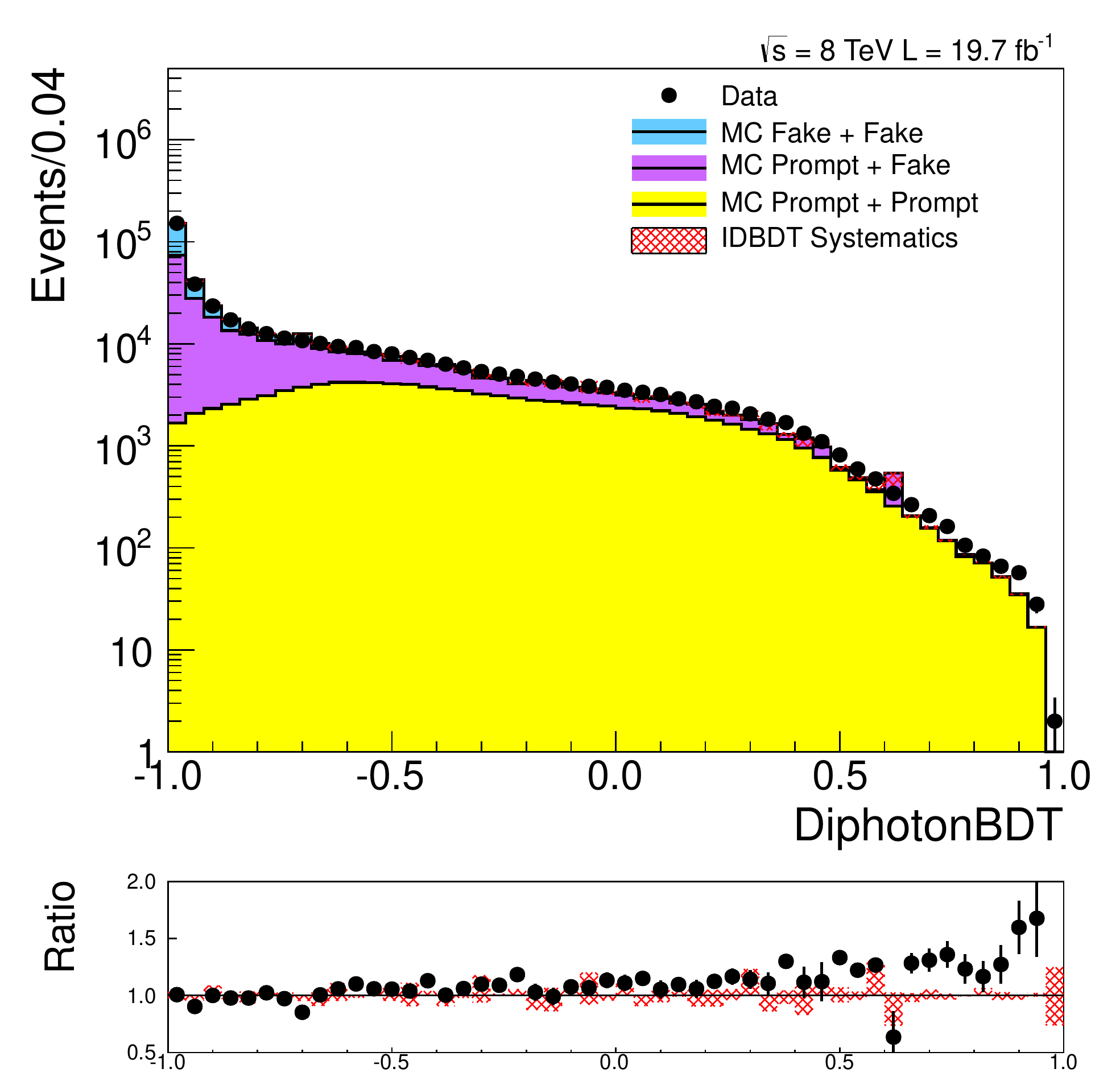}
  \end{center}
  \caption{Left: the distribution of DiphotonBDT for Monte Carlo signal (stacked histogram) consisting of $H\rightarrow \gamma\gamma$ events at a Higgs mass of 124 GeV with all four production processes weighted by cross section at 8 TeV. The variation of the DiphotonBDT distribution under the shift of IDBDT of both photons by $\pm$0.01 is shown as the red band on top of the stacked histogram. The corresponding ratios between the varied distributions and the original distribution are shown as the red band at the bottom. Right: the distributions of DiphotonBDT for data (points), Monte Carlo background (stacked histogram) consisting of prompt diphoton, $\gamma$ + jet and dijet events weighted by cross section at 8 TeV, and the variation of Monte Carlo background DiphotonBDT distribution under the shift of IDBDT of both photons by $\pm$0.01 (red band). At bottom, the ratio between the DiphotonBDT distributions of data and original Monte Carlo background (points), and the ratios between the varied Monte Carlo background distributions and the original Monte Carlo background distribution (red band) are shown. The data and Monte Carlo background events in the signal region 120 GeV $<$ $m_{\gamma\gamma}$ $<$ 130 GeV are removed. All events pass the preselection with IDBDT $>$ $-$0.2 for both photons. Photon energy corrections are applied to both data and Monte Carlo events, and additional corrections are applied to Monte Carlo events including efficiency scaling and pileup reweighting.}
  \label{fig:diphotonmva output classification idsys}
\end{figure}

\begin{figure}[hbpt] 
  \begin{center}
    \includegraphics[width=0.49\textwidth]{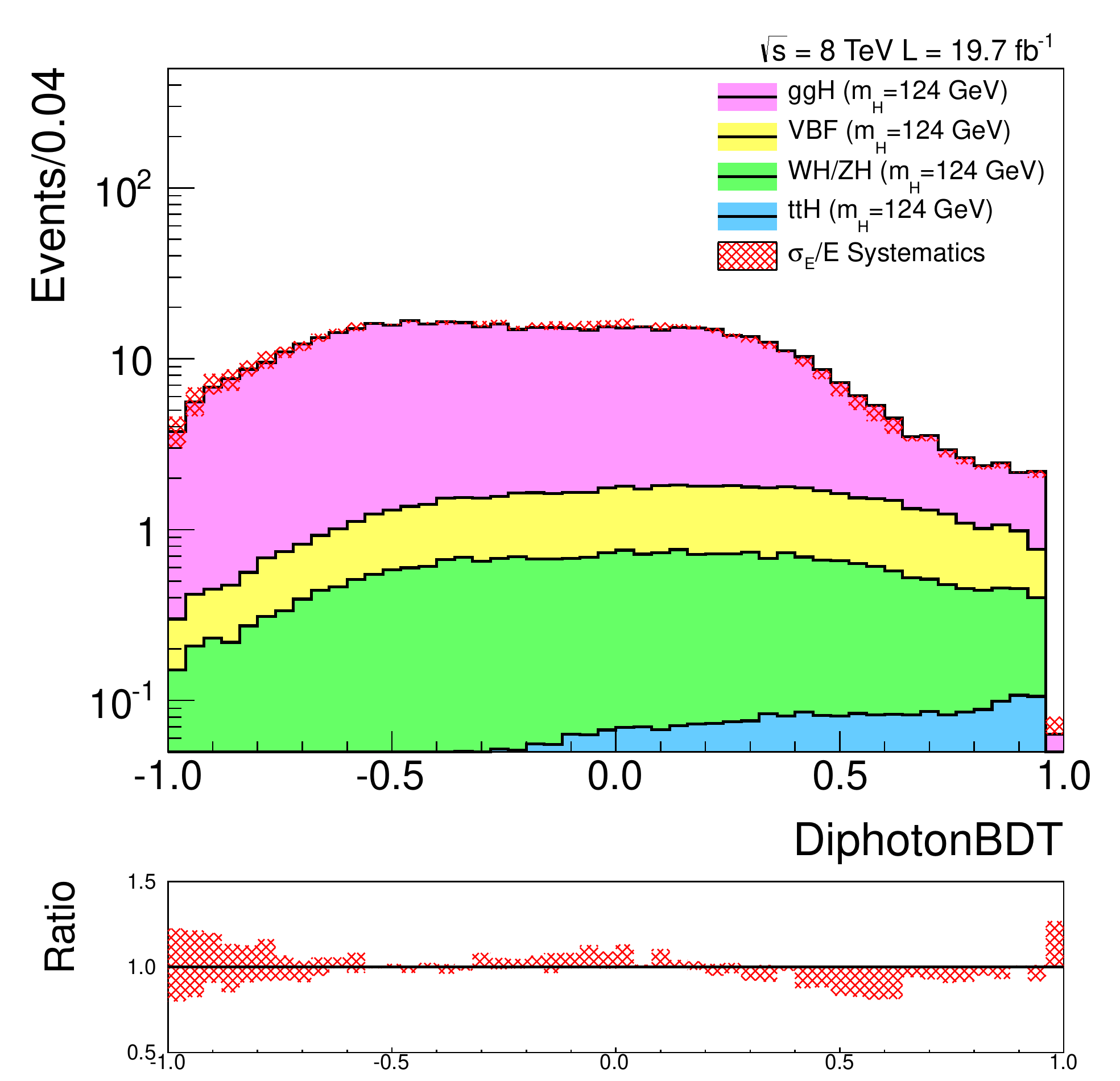}
    \includegraphics[width=0.49\textwidth]{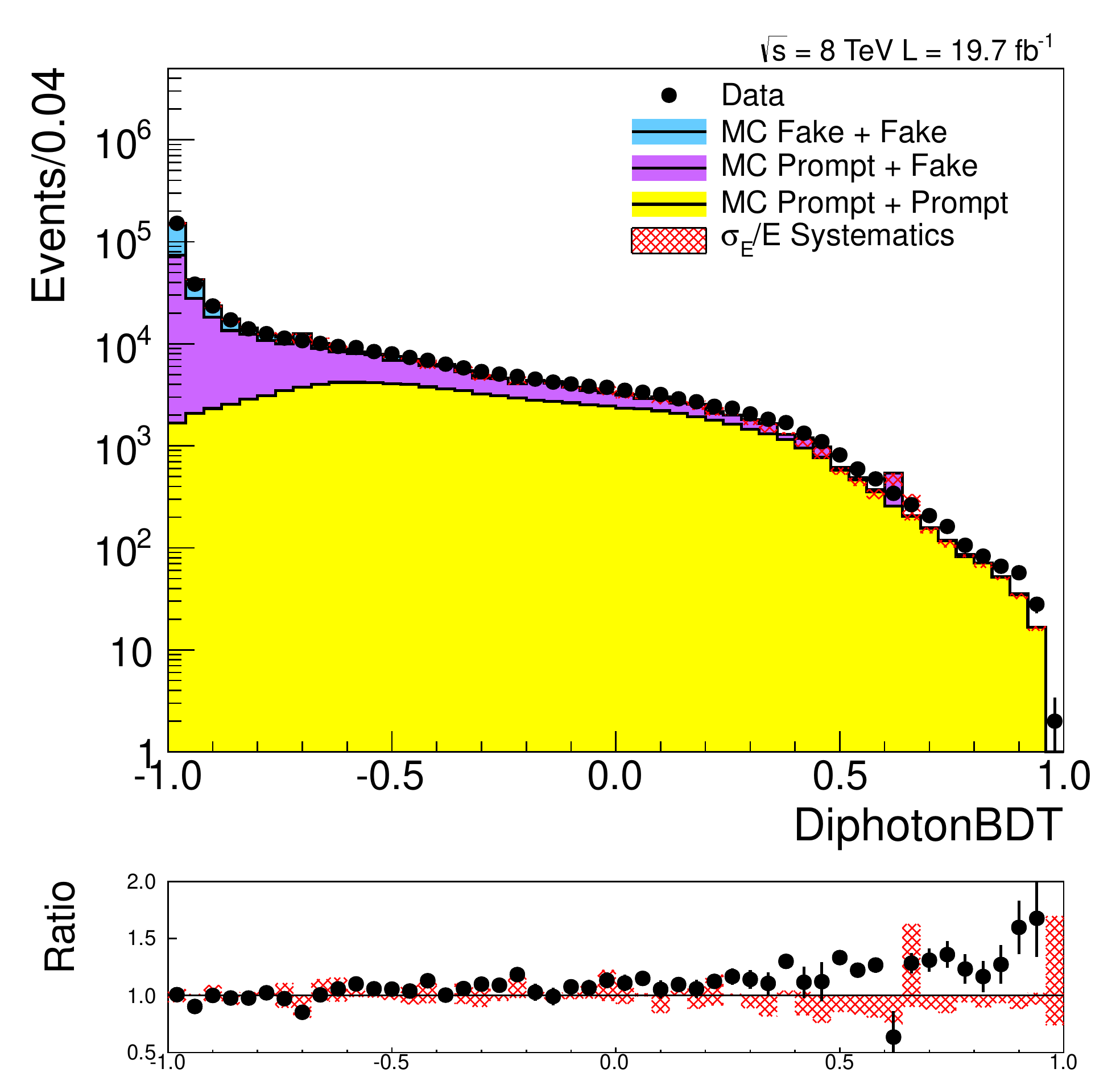}
  \end{center}
  \caption{The same with Figure \ref{fig:diphotonmva output classification idsys} except that the red bands represent the variation under the scaling of $\sigma_{E}/E$ of both photons by $\pm$10\% instead.}
  \label{fig:diphotonmva output classification sigsys}
\end{figure}

The uncertainty of IDBDT is propagated to DiphotonBDT by shifting the IDBDT of both photons by $\pm$0.01. The differences between the two varied DiphotonBDT distributions corresponding to the IDBDT shifts and the original DiphotonBDT distribution are shown as the red error bands for Monte Carlo signal and background on the left and right of Figure \ref{fig:diphotonmva output classification idsys}. The uncertainty of $\sigma_{E}/E$ is propagated to the DiphotonBDT by scaling the $\sigma_{E}/E$  of both photons by $\pm$10$\%$, and the corresponding error bands for Monte Carlo signal and background are shown on the left and the right in Figure \ref{fig:diphotonmva output classification sigsys}. The uncertainty of the DiphotonBDT due to diphoton kinematics is taken into account by varying the Higgs $p_{T}$ and rapidity by their theoretical uncertainties.   

%% file: ExclusiveTag.tex
\chapter{Tags of Higgs Production Processes}
\label{chapter:Tags for Higgs Boson Production Processes}
We assign tags of Higgs production processes to each diphoton event passing the preselection and IDBDT $>$ $-$0.2 for both photons. The tags are determined by identifying the signatures associated with \textit{VBF}, \textit{VH} and \textit{t$\overline{t}$H} processes, which are the presence of additional objects such as jets, leptons or transverse missing energy reconstructed following the descriptions in Section \ref{sec:Reconstruction}. Further energy correction and identification of these objects are provided in Section \ref{sec:Objects for Higgs Production Tagging}. The criteria for the \textit{VBF}, \textit{VH} and \textit{t$\overline{t}$H} tags based on these objects are given in Section \ref{sec:VBF Tag}, Section \ref{sec:VH Tag} and Section \ref{sec:ttH Tag}, respectively. The variables of these objects and photons used for the tagging are defined in Appendix \ref{chap:Variables For Higgs Production Tagging}. If none of the tagging criteria is satisfied, the diphoton event is labeled as ``untagged'', equivalent to the tag of \textit{ggH} process.   

\section{Objects for Higgs Production Tagging}
\label{sec:Objects for Higgs Production Tagging}

\subsection{Jets}
\label{sec:Jets}
For jets, their energies are corrected following a multi-step procedure. A set of selection cuts are applied afterwards to jets in the events at 8 TeV, in order to remove the ``fake jets'' due to clustering of random particles from pileup interactions, which are negligible for the events at 7 TeV. The corrections\cite{jetenergy,Cacciari:2008gn,Cacciari:2011ma,Cacciari:2007fd} are listed below, and the selection cuts are listed in Table \ref{tab:jetid}:
\begin{itemize}
\item The subtraction of the pileup contamination estimated by $\rho_{Event}$ times the jet area.
\item A relative scale correction to achieve a uniform jet energy response as a function of jet pseudorapidity $\eta^{j}$.
\item An absolute scale correction to match the original parton energy as a function of jet transverse momentum $p_{T}^{j}$  as derived from dijet, $\gamma$ + jet and $Z$ + jet events.
\item A residual scale correction between data and Monte Carlo simulation.
\end{itemize}  

\begin{table}[hbtp]
  \noindent
  \small\addtolength{\tabcolsep}{-6pt}
  \renewcommand{\arraystretch}{1.2}
  \caption{The selection for jet identification.}
  \begin{center}
    \setlength{\tabcolsep}{20pt}
    \begin{tabular}{|l|c|c|c|c|} 
      \hline
      $|\eta^{j}|$ & $\sum p_{T}^{TRK PU}/\sum p_{T}^{TRK}$& $\sum (({p_{T}^{PF}})^{2}\cdot{\Delta{R}}^{2})/\sum ({p_{T}^{PF}})^{2}$\\
      \hline
      \hline
      $<$ 2.5 & $<$ 0.2log($N_{Vtx}$$-$0.64) & $<$ 0.06\\
      \hline
      2.5 $<$ $|\eta^{j}|$ $<$ 2.75 & $<$ 0.3log($N_{Vtx}$$-$0.64)  & $<$ 0.05\\
      \hline
      2.75 $<$ $|\eta^{j}|$ $<$ 3 & - & $<$ 0.05\\
      \hline
      3 $<$ $|\eta^{j}|$ $<$ 4.7 & - & $<$ 0.055\\
      \hline
    \end{tabular}
    \label{tab:jetid}
  \end{center}
\end{table}

\subsection{Electrons}
\label{sec:Electrons}
For electrons, a set of selection cuts are applied to remove the electrons from jets or electrons from converted photons, as listed in the Table \ref{tab:eleid}:

\begin{table}[hbtp]
  \noindent
  \small\addtolength{\tabcolsep}{-6pt}
  \caption{The selection for electron identification.}
  \begin{center}
    \setlength{\tabcolsep}{20pt}
    \begin{tabular}{|l|c|} 
      \hline
       $d_{xy}^{e}$ &  $<$ 0.2 mm\\
      \hline
       $d_{z}^{e}$ &  $<$ 2 mm\\
      \hline
       $P_{ConvVtx}$ &   $<$ $10^{-6}$\\
      \hline
       $N_{Miss}$ &  $\leq$ 1\\
      \hline
       EleMVA&  $>$ 0.9\\
      \hline
       $\mathrm{ISO}_{RelPUCorrPFCombine03}$ &  $<$ 0.15\\
      \hline
    \end{tabular}
    \label{tab:eleid}
  \end{center}
\end{table}

\subsection{Muons}
\label{sec:Muons}
For muons, a set of selection cuts are applied to reject the background muons including muons from hadron decays, beam-halo muons induced by the accelerator and cosmic muons, as listed in Table \ref{tab:muonid}.

\begin{table}[hbtp]
  \noindent
  \small\addtolength{\tabcolsep}{-6pt}
  \caption{The selection for muon identification.}
  \begin{center}
    \setlength{\tabcolsep}{20pt}
    \begin{tabular}{|l|c|} 
      \hline
      $N_{Pixel}$ & $>$ 0\\
      \hline
      $N_{TRKLayper}$ & $>$ 5\\
      \hline
      $N_{MuonChamber}$ & $>$ 0\\
      \hline
      $N_{Matching}$ & $>$ 1\\
      \hline
      $d_{xy}^{\mu}$ & $<$ 2 mm\\
      \hline
      $d_{z}^{\mu}$ & $<$ 5 mm\\
      \hline
      $\chi^{2}/NDF$ & $<$ 10\\
      \hline
      ISO$_{RelBetaPuCorrPFCombine04}$ & $<$ 0.2\\
      \hline
    \end{tabular}
    \label{tab:muonid}
  \end{center}
\end{table}

\subsection{Transverse Missing Energy}
\label{sec:Transverse Missing Energy}
For transverse missing energy, the difference in the magnitude MET between data and Monte Carlo, due to the imperfect simulation of detector effects, is corrected by smearing the Monte Carlo jet energy to data in addition to the jet energy correction mentioned above. A further correction is applied to both data and Monte Carlo simulation to achieve a uniform distribution of the azimuthal angle of transverse missing energy.  

\section{\textit{VBF} Tag}
\label{sec:VBF Tag}
The criteria for the \textit{VBF} tag are based on the feature of two energetic jets with large separation in $\eta$. The \textit{VBF} candidates are first preselected from the diphoton events by applying a set of loose cuts on dijet kinematics. Each \textit{VBF} candidate is assigned a score from a so called combined BDT, measuring how likely it is a real \textit{VBF} event rather than a background event or a \textit{ggH} event with two jets from ISR. The \textit{VBF} tagged events are identified from the \textit{VBF} candidates and further classified according to the combined BDT score as described in Chapter \ref{chap:Event Classification}. To train the combined BDT, a dijet-diphoton kinematic BDT is trained beforehand, which provides a kinematic discriminator between the \textit{VBF} events and both the background events and the \textit{ggH} events. The combined BDT is then trained using the output of the dijet-diphoton kinematic BDT, along with the output of the diphoton BDT as a measure of the diphoton quality, and $p_{T}^{\gamma\gamma}$/$m_{\gamma\gamma}$ correlated to the outputs of both BDTs. The dijet preselection cuts for the \textit{VBF} candidates, and the details about the dijet-diphoton kinematic BDT and the combined BDT are provided below.

\subsection{Dijet Preselection}
\label{sec:Dijet Preselection}
The dijet kinematic cuts to select the \textit{VBF} candidates are summarized in Table \ref{tab:dijet preselection variables}. 

\begin{table}[hbtp]
  \noindent
  \small\addtolength{\tabcolsep}{-6pt}
  \renewcommand{\arraystretch}{1.2}
  \caption{The dijet kinematic cuts for the selection of \textit{VBF} candidates.}
  \begin{center}
    \setlength{\tabcolsep}{20pt}
    \begin{tabular}{|l|c|} 
      \hline
      $p_{T}^{j1}$ & $>$ 30 GeV\\
      \hline
      $p_{T}^{j2}$ & $>$ 20 GeV\\
      \hline
      $|{\eta}^{j1}|$ & $<$ 4.7\\
      \hline
      $|{\eta}^{j2}|$ & $<$ 4.7\\
      \hline
      $m_{jj}$ & $>$ 250 GeV \\
      \hline
    \end{tabular}
    \label{tab:dijet preselection variables}
  \end{center}
\end{table}

\subsection{Dijet-Diphoton Kinematic BDT}
\label{sec:Dijet-Diphoton Kinematic BDT}
The dijet-diphoton kinematic BDT is trained separately on Monte Carlo events at $7~\mathrm{TeV}$ and $8~\mathrm{TeV}$, which pass a looser diphoton and dijet preselection to increase the number of training events. The requirements on the transverse momenta and IDBDTs of photons as well as dijet kinematics are loosen as:
\begin{itemize}
\item $p_{T}^{\gamma1}/m_{\gamma\gamma}$ $>$ 1/4, $p_{T}^{\gamma2}/m_{\gamma\gamma}$ $>$ 1/5, IDBDT$^{\gamma1}$ $>$ $-$0.3, IDBDT$^{\gamma2}$ $>$ $-$0.3. 
\item $p_{T}^{j1}$ $>$ 15 GeV, $p_{T}^{j2}$ $>$ 10 GeV, $m_{jj}$ $>$ 75 GeV.
\end{itemize}
The signal sample consists of \textit{VBF} events at a Higgs mass of 123 GeV. The background sample consists of background diphoton, $\gamma$ + jet and dijet events weighted by cross section, and \textit{ggH} events at a Higgs mass of 123 GeV. The contribution of the \textit{ggH} events in the background sample is inflated by applying a weighting factor about 200. The weighting factor is chosen such that the BDT uses more features discriminating between the \textit{VBF} and \textit{ggH} events, while keeps good distinguishment between the Higgs and background events. The input variables are the following:
\begin{itemize}
\item All variables for the dijet kinematic cuts for \textit{VBF} candidate selection.
\item $p_{T}^{\gamma\gamma}/m_{\gamma\gamma}$: diphoton transverse momentum divided by the diphoton mass.   
\item $|\eta_{\gamma\gamma}-\frac{\eta^{j1}+\eta^{j2}}{2}|$: separation between the diphoton pseudorapidity and the average pseudorapidity of the dijet\cite{Rainwater:1996ud}. 
\item $\Delta\phi_{jj,\gamma\gamma}$: separation in the azimuthal angle between dijet and diphoton. The value is set as the maximum between $\Delta\phi_{jj,\gamma\gamma}$ and $\pi$$-$0.2 to avoid large theoretical uncertainty on the cross section of \textit{ggH} events with two jets from initial state radiation in the phase space where $\Delta\phi_{jj,\gamma\gamma}$ is close to $\pi$ \cite{LHCHiggsCrossSectionWorkingGroup3,Tackmann:2011}. 
\end{itemize}
The output is a score assigned to each event ranging from $-$1 to 1, which increases with the compatibility of the event kinematics to the \textit{VBF} kinematics.

\subsection{Combined BDT}
\label{sec:Combined BDT}
The combined BDT is trained separately for events at 7 TeV and 8 TeV. The signal sample is the same for the dijet-diphoton kinematic BDT. The background sample consists of the same background events for the dijet-diphoton kinematic BDT but not the \textit{ggH} events, to achieve good discrimination between the Higgs events and the background events. The training variables are the outputs of the dijet-diphoton kinematic BDT and the diphoton BDT, along with $p_{T}^{\gamma\gamma}$/$m_{\gamma\gamma}$. 

\begin{figure}[hbpt] 
  \begin{center}
    \includegraphics[width=0.49\textwidth]{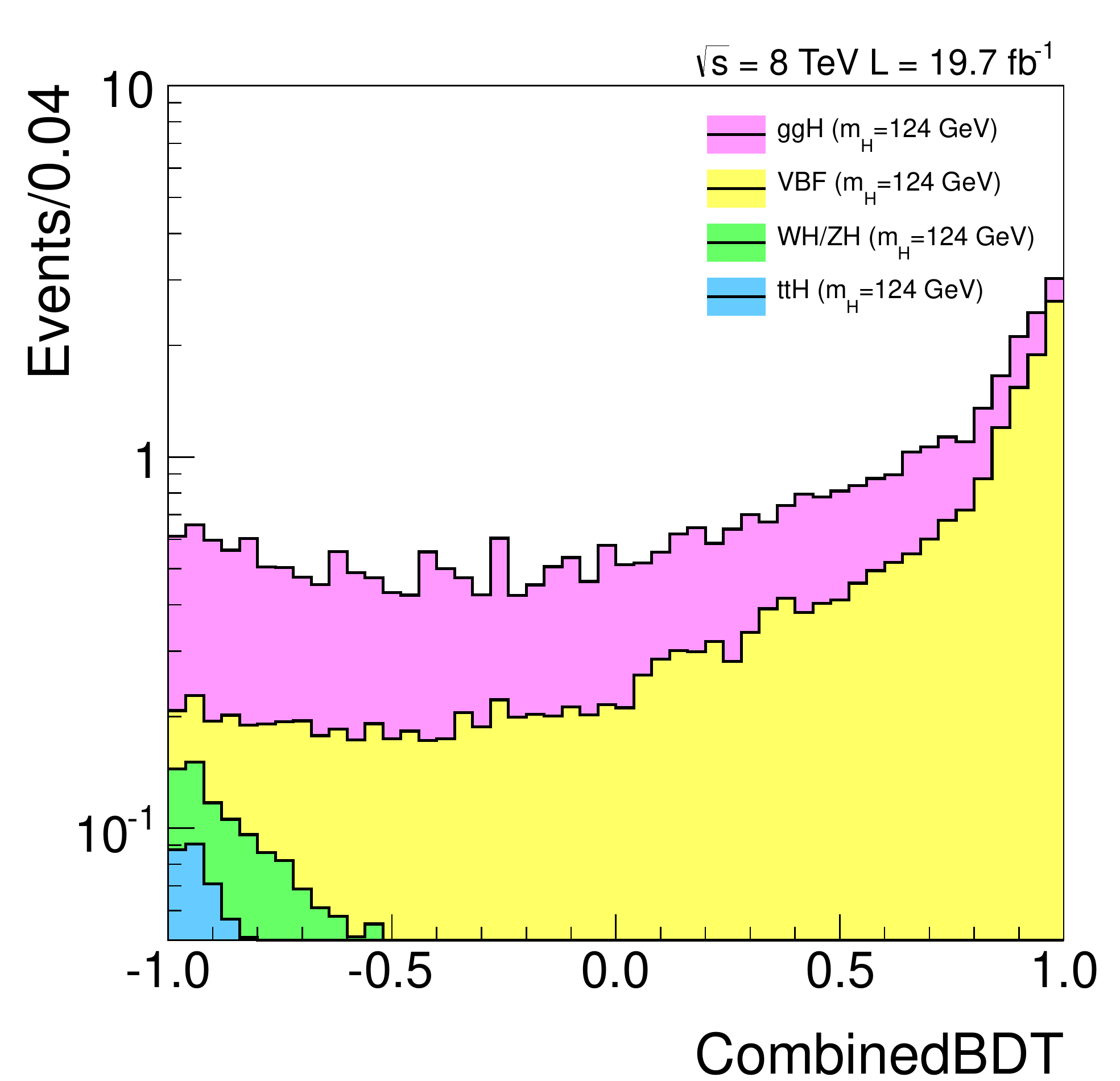}
    \includegraphics[width=0.49\textwidth]{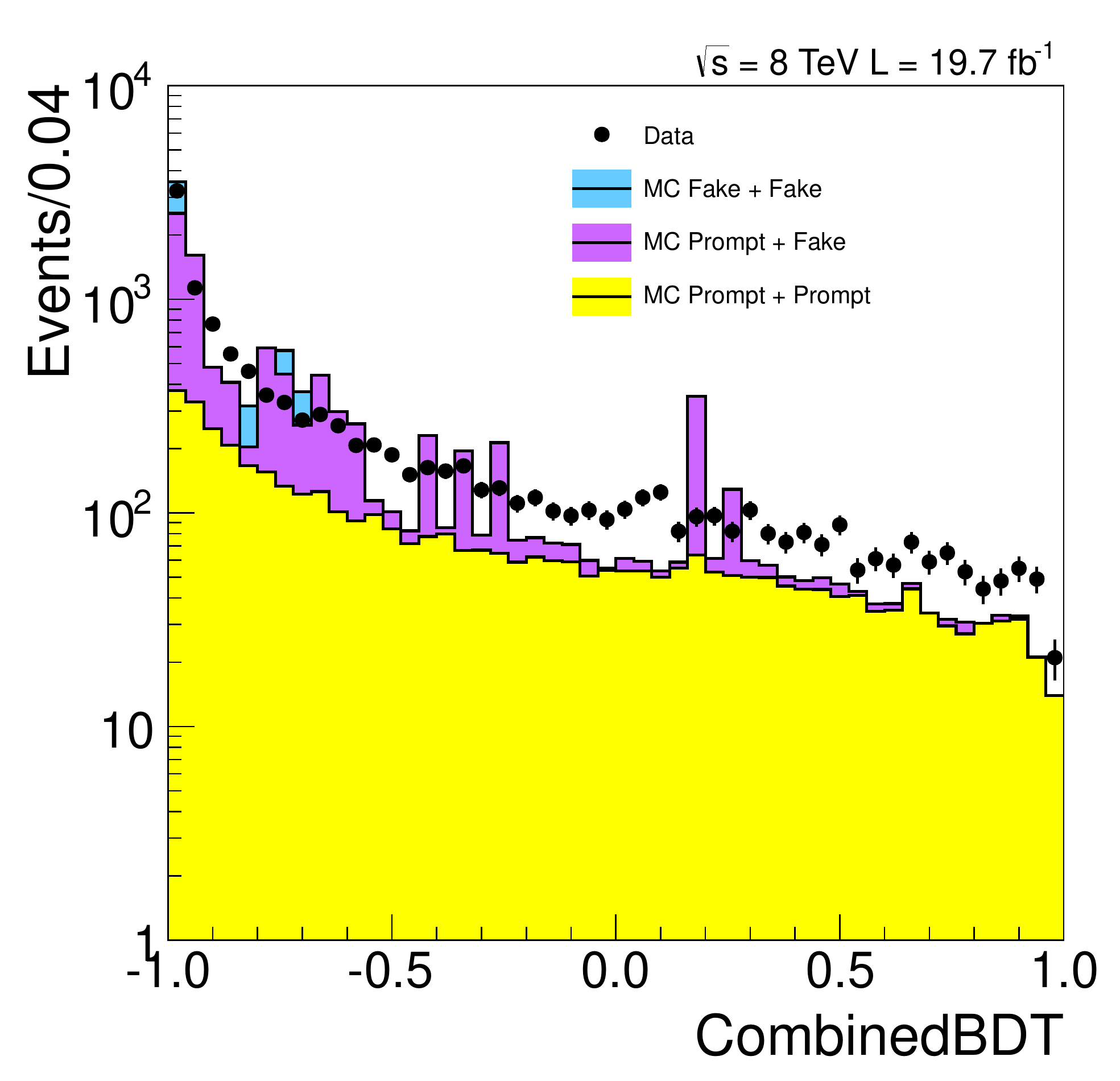}
  \end{center}
  \caption{Left: the distribution of CombinedBDT for Monte Carlo signal (stacked histogram) consisting of $H\rightarrow \gamma\gamma$ events at a Higgs mass of 124 GeV with all four production processes weighted by cross section at 8 TeV. Right: the distributions of CombinedBDT for data (points) and Monte Carlo background (stacked histogram) consisting of prompt diphoton, $\gamma$ + jet and dijet events weighted by cross section at 8 TeV. The data and Monte Carlo background events in the signal region 120 GeV $<$ $m_{\gamma\gamma}$ $<$ 130 GeV are removed. All events pass the preselection and IDBDT $>$ $-$0.2 for both photons. Photon and jet energy corrections are applied to both data and Monte Carlo events, and additional corrections are applied to Monte Carlo events including efficiency scaling and pileup reweighting.}
  \label{fig:combinedmva}
\end{figure}

The output of the combined BDT, named CombinedBDT, ranges from $-$1 to 1. The corresponding distributions for Monte Carlo $H\rightarrow \gamma\gamma$ events at a Higgs mass of $124~\mathrm{GeV}$ selected as \textit{VBF} candidates are shown on the left of Figure \ref{fig:combinedmva}, and the corresponding distributions for data versus Monte Carlo background events, with events in the signal region $120~\mathrm{GeV}$ $<$ $m_{\gamma\gamma}$ $<$ 130 GeV removed, are shown on the right. Photon and jet energy corrections are applied to both data and Monte Carlo events, and additional corrections are applied to Monte Carlo events including efficiency scaling and pileup reweighting. The number of \textit{VBF} events over the number of background events or the number of \textit{ggH} events in each bin increases with the CombinedBDT as expected. The background Monte Carlo in general describes the shape of the data. The granularity of the comparison is limited by the number of Monte Carlo $\gamma$ + jet and dijet events, and the spikes are due to the Monte Carlo events with large weight. As explained previously, though the discrepancy between the Monte Carlo background events and data makes the CombinedBDT sub-optimal, it does not influence the correctness of the analysis.        

\section{\textit{VH} Tag}
\label{sec:VH Tag}
The criteria for the \textit{VH} tag are based on the signatures from the decays of the $W$ or $Z$ boson. There are three sub-tags for different decay modes: 
\begin{itemize}
\item  Lepton (electron or muon) tag for the leptonic $W$ decay or $Z$ decay. 
\item  Dijet tag for the hadronic $W$ decay or $Z$ decay.
\item  MET tag for $Z$ decaying into two neutrinos, or leptonic $W$ decay with the lepton lost from reconstruction or outside of acceptance.
\end{itemize}
The tagging criteria optimized for the sensitivity of the \textit{VH} signal are introduced below.

\subsection{\textit{VH} Lepton Tag}
\label{sec:VH Lepton Tag}
The \textit{VH} lepton tag is further divided into tight and loose tags according to the number of leptons $N_{Lep}$ and MET as defined in Table \ref{tab:vh tag cats}. 
\begin{table}[h]
  \begin{center}
    \caption{The definition of tight and loose \textit{VH} lepton tags.}
    \begin{tabular}{|l|c|c|} 
      \hline
      & $N_{Lep}$  & MET\\
      \hline
      Tight & 2  & -\\
      \cline{2-3} 
      & 1  & $>$ 45 GeV \\
      \hline
      Loose & 1 & $\leq$ 45 GeV\\
      \hline
    \end{tabular}
    \label{tab:vh tag cats}
  \end{center}
\end{table}

 The requirements for these tags are summarized in Table \ref{vh lep selection}, which include:
\begin{itemize}
\item A set of kinematic cuts on leptons. The dilepton mass $m_{ll}$ is required to be close to the $Z$ mass since the leptons are supposed to come from the $Z$ decay.
\item Requirements for photons: 
  \begin{itemize}
  \item A $p_{T}$ cut on the leading photon higher than the preselection. This is due to the higher Higgs $p_{T}$ and so higher leading photon $p_{T}$ on average for the \textit{VH} events than for the \textit{ggH} events.
  \item Large enough distance in $\Delta R$ from the photon to electron, electron track and muon. This is to reject the photon from lepton radiation and electron faking photon, which reduces the dominant background from $W$ + $\gamma$ and $Z$ + $\gamma$.  
  \item The photon-electron mass $m_{\gamma,e}$ away from the $Z$ mass. This is to reject the electron faking photon from $Z\rightarrow e^{+}e^{-}$.  
  \end{itemize}
\item Less than three jets with $p_T^{j}>$ 20 GeV, $|\eta^{j}|$ $<$ 2.4 and $\Delta R$ $>$ 0.5 from any lepton or photon. This is to reject contamination from the \textit{t$\overline{t}$H} process. 
\end{itemize}

\begin{table}[h]
  \begin{center}
    \caption{The requirements for \textit{VH} lepton tag.}
    \begin{tabular}{|l|c|c|} 
      \hline
      $N_{Lep}$ & 1  & 2\\
      \hline
      \hline
      $p_{T}^{\mu}$ &  $>$ 20 GeV   &  $>$ 10 GeV \\
      \hline
      $p_{T}^{e}$ & $>$ 20 GeV  &  $>$ 10 GeV \\
      \hline
      $|\eta^{\mu}|$ & \multicolumn{2}{|c|}{$<$ 2.4}\\
      \hline
      $|\eta^{e}|$ & \multicolumn{2}{|c|}{$|\eta^{e}|<$ 1.4442 or 1.566$<|\eta^{e}|<$ 2.5}\\
      \hline
      $m_{ll}$ & - & 70 GeV $<$ $m_{ll}$ $<$ 110 GeV \\
      \hline
      \hline
      $\Delta R_{\gamma,\mu}$   & $>$ 1 & $>$ 0.5 \\
      \hline
      $\Delta R_{\gamma,e}$ & \multicolumn{2}{|c|}{$>$ 1} \\
      \hline
      $\Delta R_{\gamma,etrk}$ & \multicolumn{2}{|c|}{$>$ 1} \\
      \hline
      $|m_{\gamma,e}-M_{Z}|$ & \multicolumn{2}{|c|}{$>$ 10 GeV} \\
      \hline
      $p_{T}^{\gamma1}/m_{\gamma\gamma}$ & \multicolumn{2}{|c|}{$>$ 3/8}\\
      \hline
      \hline
      $N_{j}$ & $<$ 3 & - \\
      \hline
    \end{tabular}
    \label{vh lep selection}
  \end{center}
\end{table}

\subsection{\textit{VH} Dijet Tag}
\label{sec:VH Dijet Tag}
The \textit{VH} dijet tagged events are selected from the diphoton events with two jets, following the requirements summarized in Table \ref{vh jet selection}. For the events with more than two jets, the leading and sub-leading jets in $p_{T}$ are considered. Among the requirements, the dijet mass $m_{jj}$ is required to be close to the $Z$($W$) mass since the jets are supposed to come from the $Z$($W$) decay. The cosine of the angle $\theta^*$ between the diphoton momentum in the center-of-mass frame of diphoton-dijet and the total momentum of diphoton-dijet in the lab frame is used. Its distribution is flat for \textit{VH} events while peaking at $|cos(\theta^*)|$ $=$ 1 for background events. 
\begin{table}[h]
  \begin{center}
    \caption{The requirements for \textit{VH} dijet tag.}
    \begin{tabular}{|l|c|} 
      \hline
      $p_{T}^{j1}$ & $>$ 40 GeV \\
      \hline
      $p_{T}^{j2}$ & $>$ 40 GeV\\
      \hline
      $|\eta^{j1}|$ & $<$ 2.4 \\
      \hline
      $|\eta^{j2}|$ & $<$ 2.4 \\
      \hline
      $m_{jj}$ &  60 GeV $<$ $m_{jj}$ $<$ 120 GeV \\
      \hline
      $|cos(\theta^*)|$ & $<$ 0.5\\
      \hline
      \hline
      $p_{T}^{\gamma1}/m_{\gamma\gamma}$ & $>$ 1/2\\
      \hline
      $p_{T}^{\gamma\gamma}/m_{\gamma\gamma}$ & $>$ 13/12\\
      \hline
    \end{tabular}
    \label{vh jet selection}
  \end{center}
\end{table}

\subsection{\textit{VH} MET Tag}
\label{sec:VH MET Tag}
The \textit{VH} MET tagged events are selected from the diphoton events with large MET. The selection requirements are summarized in Table \ref{tab:vh MET selection}. Among the requirements, large separation in the azimuthal angle between the diphoton and MET $\phi_{\gamma\gamma,\textit{MET}}$ is required, because of the momentum balancing between the diphoton and MET in the \textit{VH} events with $Z$ decaying into two neutrinos, or $W$ leptonic decay with the lepton lost from reconstruction or outside of acceptance. An upper bound is put on the separation in the azimuthal angle between the diphoton and the leading jet, in order to reduce the contamination from the MET caused by the inaccurate measurement of jet energy when the jet and the diphoton is back to back.    
\begin{table}[h]
  \begin{center}
    \caption{The requirements for \textit{VH} MET tag.}
    \begin{tabular}{|l|c|} 
      \hline
      MET & $>$ 70 GeV\\
      \hline
      \hline
      $|\Delta \phi_{\gamma\gamma,\textit{MET}}|$ & $>$ 2.1\\
      \hline
      $|\Delta \phi_{\gamma\gamma,j1}|$ & $<$ 2.7\\ 
      \hline
      \hline
      $p_{T}^{\gamma1}/m_{\gamma\gamma}$ & $>$ 3/8\\
      \hline
    \end{tabular}
    \label{tab:vh MET selection}
  \end{center}
\end{table}

\section{\textit{t$\overline{\textbf{t}}$H} Tag}
\label{sec:ttH Tag}
The criteria for the \textit{t$\overline{t}$H} tag are based on the signatures from the decays of the $t\overline{t}$. There are two sub-tags for different decay modes: 
\begin{itemize}
\item  Lepton (electron or muon) tag for two or one leptonic $W$ decay.
\item  Multijet tag for two hadronic $W$ decays.
\end{itemize}
The tagging criteria optimized for the sensitivity of the \textit{t$\overline{t}$H} signal are introduced below.

\subsection{\textit{t$\overline{\textbf{t}}$H} Lepton Tag}
\label{sec:ttH Lepton Tag}
The \textit{t$\overline{t}$H} lepton tagged events are selected from the diphoton events with at least one lepton, following the requirements summarized in Table \ref{tab:tth lep}. Among the requirements, the $p_{T}$ cut on the leading photon is increased with respect to the preselection because of the higher Higgs $p_{T}$ and so the higher leading photon $p_{T}$ on average for the \textit{t$\overline{t}$H} events than for the \textit{ggH} events. For the requirement of the number of jets (b-jets), the jets (b-jets) with $p_{T}^{j}$ $>$ 25 GeV, $|\eta^{j}|$ $<$ 2.4, $\Delta R$ $>$ 0.5 from any lepton are counted. 
\begin{table}[h]
  \begin{center}
    \caption{The requirements for \textit{t$\overline{t}$H} lepton tag.}
    \begin{tabular}{|l|c|} 
      \hline
      $p_{T}^{\mu}$ &  $>$ 20 GeV \\
      \hline
      $p_{T}^{e}$ & $>$ 20 GeV  \\
      \hline
      $|\eta^{\mu}|$ & $<$ 2.4\\
      \hline
      $|\eta^{e}|$ & $|\eta^{e}|<$ 1.4442 or 1.566$<|\eta^{e}|<$ 2.5\\
      \hline
      \hline
      $\Delta R_{\gamma,\mu}$   & $>$ 0.5\\
      \hline
      $\Delta R_{\gamma,e}$ & $>$ 1\\
      \hline
      $\Delta R_{\gamma,etrk}$ & $>$ 1\\
      \hline
      $p_{T}^{\gamma1}/m_{\gamma\gamma}$ & $>$ 1/2\\
      \hline
      \hline
      $N_{j}$ & $>$ 1\\
      \hline
      $N_{B-j}$ & $>$ 0\\
      \hline
    \end{tabular}
    \label{tab:tth lep}
  \end{center}
\end{table}

\subsection{\textit{t$\overline{\textbf{t}}$H} Multijet Tag}
\label{sec:ttH Multijet Tag}
The \textit{t$\overline{t}$H} Multijet tagged events are selected from the diphoton events with multiple jets, following the requirements summarized in Table \ref{tab:tth multijet}. For the requirement of the number of jets (b-jets), the jets (b-jets) with $p_{T}^{j}$ $>$ 25 GeV, $|\eta^{j}|$ $<$ 2.4 are counted.

\begin{table}[h]
  \begin{center}
    \caption{The requirements for \textit{t$\overline{t}$H} multijet tag.}
    \begin{tabular}{|l|c|} 
      \hline
      $N_{j}$ & $>$ 4\\
      \hline
      $N_{B-j}$ & $>$ 0\\
      \hline
      \hline
      $p_{T}^{\gamma1}/m_{\gamma\gamma}$ & $>$ 1/2\\
      \hline
    \end{tabular}
    \label{tab:tth multijet}
  \end{center}
\end{table}

%% file: Class.tex
\chapter{Event Classification}
\label{chap:Event Classification}
We classify the diphoton events passing the preselection and IDBDT $>$ $-$0.2 for both photons into the tagged and the untagged Higgs production process classes. The events in \textit{VBF} tagged classes are selected from \textit{VBF} candidates passing the dijet kinematic selection, and are classified in terms of the CombinedBDT. The corresponding class boundaries are chosen to minimize the expected uncertainty of the signal strength for the \textit{VBF}+\textit{VH} processes $\mu_{VBF,VH}$, sensitive to the Higgs coupling to bosons. The events in the \textit{VH} and \textit{t$\overline{t}$H} tagged classes are the \textit{VH} and \textit{t$\overline{t}$H} tagged events, which pass the additional DiphotonBDT cuts to improve the \textit{VH} and \textit{t$\overline{t}$H} sensitivity. The untagged events are classified into the untagged classes in terms of the DiphotonBDT, and the corresponding class boundaries are chosen to minimize the expected uncertainty of the overall signal strength $\mu_{H}$. The optimization of the class boundaries for the \textit{VBF} tagged classes and the untagged classes is described in Section \ref{sec:Boundary Optimization for VBF Tagged and Untagged Classes}. The final tagged and untagged event classes are summarized in Section \ref{sec:Final Event Classes}.  

\section{Boundary Optimization for VBF Tagged Classes and Untagged Classes}
\label{sec:Boundary Optimization for VBF Tagged and Untagged Classes}
The boundaries on CombinedBDT and DiphotonBDT for the \textit{VBF} tagged classes and the untagged classes are optimized separately for events at 7 TeV and 8 TeV, using Monte Carlo diphoton events passing the preselection and IDBDT $>$ $-$0.2 for both photons, independent from the training sample. The signal sample consists of $H\rightarrow \gamma\gamma$ events for a Higgs mass of 124 GeV (121 GeV) with all four production processes weighted by cross section at $8~\mathrm{TeV}$ ($7~\mathrm{TeV}$). The background sample consists of prompt diphoton, $\gamma$ + jet and dijet events not used for BDT training and weighted by cross section. The total number of Monte Carlo events are weighted to match the luminosity in data, and corrections on the Monte Carlo simulation including photon and jet energy corrections, efficiency scaling and pileup reweighting are applied. The background CombinedBDT and DiphotonBDT distributions are smoothed using adaptive Gaussian kernel estimations\cite{gassuiankernal}. 

\subsection{VBF Tagged Class Optimization}
\label{sec:VBF Tagged Class Optimization}
The CombinedBDT boundaries are first optimized on events passing \textit{VBF} dijet kinematic selection. The number of boundaries and the corresponding values for the boundaries are adjusted interactively until the decrease of the expected uncertainty is less than 1$\%$. The evaluation of the expected uncertainty is based on the profile likelihood fit on the diphoton mass spectra cross all the classes, which follows the procedure as described in Chapter \ref{chap:Statistical Analysis} with a simplified signal model and background model. For each class, the histogram of the diphoton mass for Monte Carlo Higgs events is used as the signal model while a power law funtion fitted from the Monte Carlo background events is used as the background model. The variation of the signal model due to systematic uncertainties is not considered for simplicity. 

\begin{figure}[hbpt] 
  \begin{center}
    \includegraphics[width=0.495\textwidth]{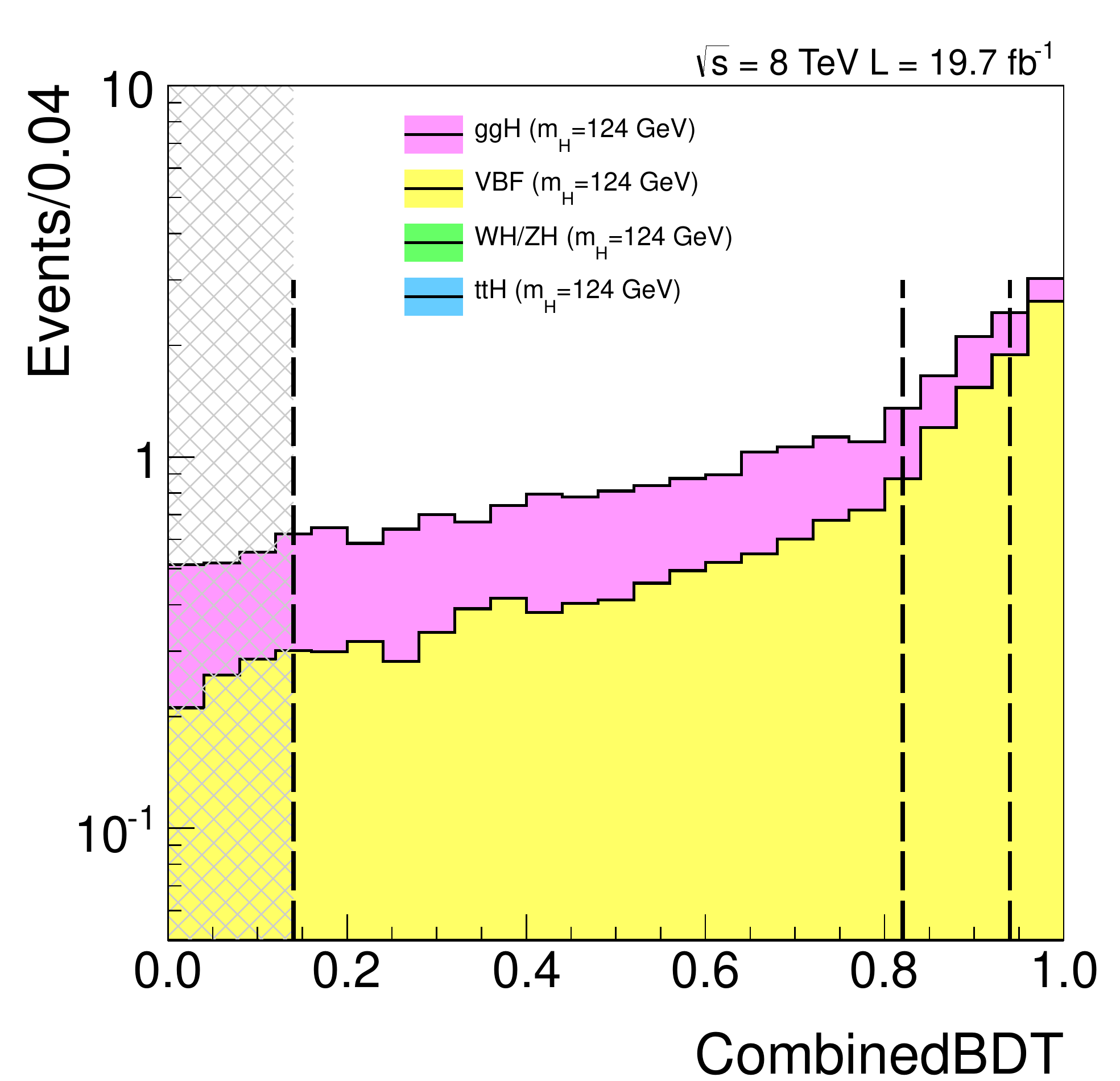}
    \includegraphics[width=0.495\textwidth]{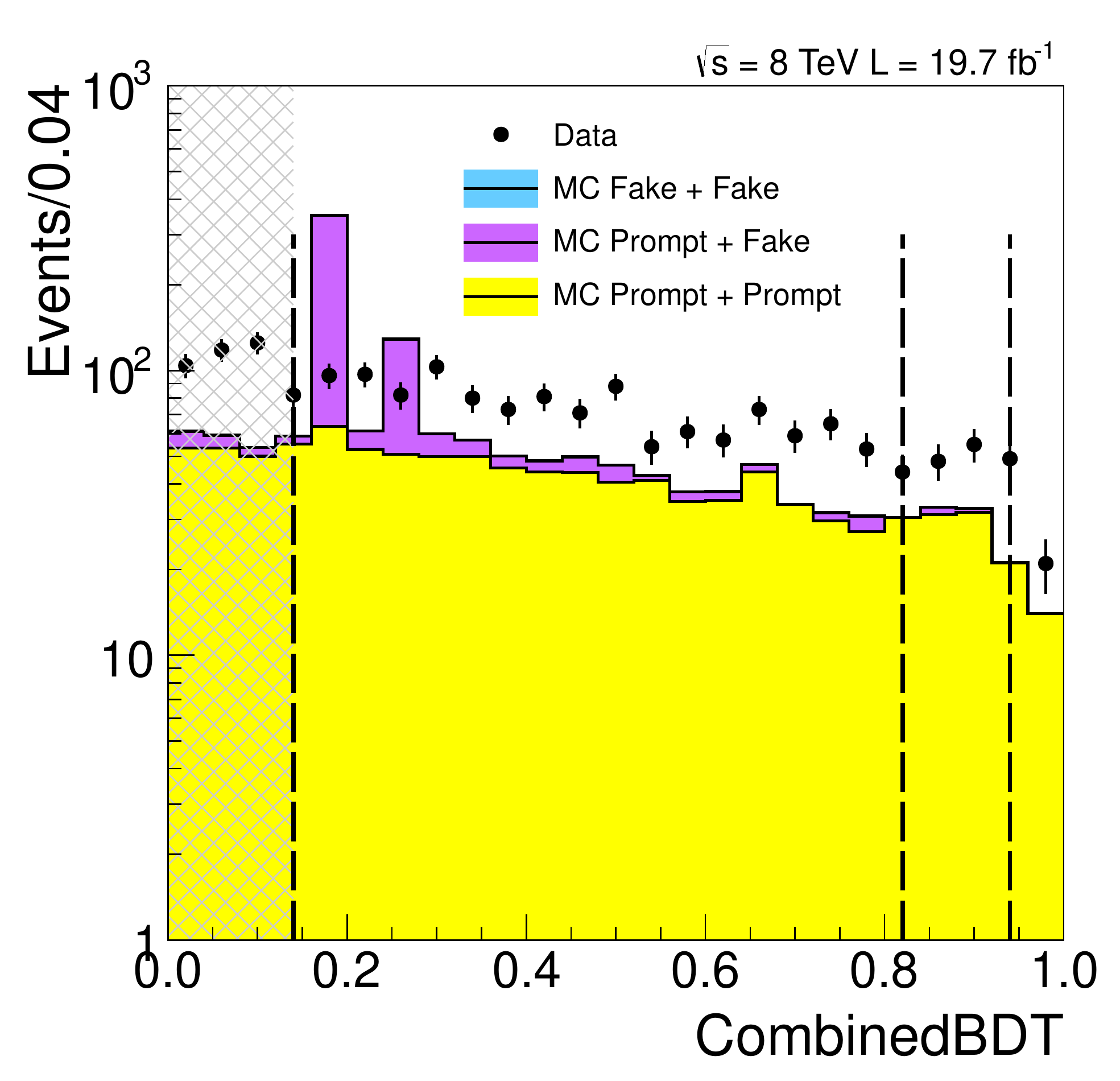}
  \end{center}
  \caption{Left: the distribution of CombinedBDT in the range of CombinedBDT $\geq$ 0 for Monte Carlo signal (stacked histogram) consisting of $H\rightarrow \gamma\gamma$ events at a Higgs mass of 124 GeV with all four production processes weighted by cross section at 8 TeV. Right: the distributions of CombinedBDT in the range of CombinedBDT $\geq$ 0 for data (points) and Monte Carlo background (stacked histogram) consisting of prompt diphoton, $\gamma$ + jet and dijet events weighted by cross section at 8 TeV. The data and Monte Carlo background events in the signal region 120 GeV $<$ $m_{\gamma\gamma}$ $<$ 130 GeV are removed. All events pass the preselection with IDBDT $>$ $-$0.2 for both photons and dijet kinematic cuts. Photon and jet energy corrections are applied to both data and Monte Carlo events, and additional corrections are applied to Monte Carlo events including efficiency scaling and pileup reweighting. The class boundaries are labeled as dashed lines. The events in the shaded region below the dashed line with the lowest CombinedBDT value are taken away from the \textit{VBF} tagged classes and used for the selection for the rest of classes.}
 \label{fig:combinedmva output classification}
\end{figure}

For the \textit{VBF} tagged classes, 3 classes for events at 8 TeV and 2 classes for events at 7 TeV are determined. Due to the limited number of Monte Carlo prompt diphoton background events passing the dijet kinematic selection, the \textit{VBF} tagged classes for events at 7 TeV are determined by matching the acceptance times efficiency for \textit{VBF} events to those of the \textit{VBF} tagged classes of events at 8 TeV, instead of using the optimization procedure as described above. Figure \ref{fig:combinedmva output classification} shows the CombinedBDT distributions, along with the class boundaries (dashed lines), in the range of CombinedBDT $\geq$ 0, of Monte Carlo signal events (left), Monte Carlo background events and data (right) at 8 TeV, passing the preselection and IDBDT $>$ $-$0.2 for both photons, and dijet kinematic cuts. The events in the shaded region below the dashed line with the lowest CombinedBDT value is taken away from the \textit{VBF} tagged classes, and used for the selection for the rest of classes. 

\subsection{Untagged Class Optimization}
\label{sec:Untagged Class Optimization}
The events with CombinedBDT below the lowest boundary and the events not passing the dijet kinematic cuts are used for the optimization of the DiphotonBDT boundaries of the untagged classes. The procedure is the same as for the determination of the boundaries of the CombinedBDT. The events with DiphotonBDT below the lowest boundary are removed, which includes few signal but large number of  background events. The removal of these events causes a negligible loss in the sensitivity for the Higgs signal and largely simplifies the final statistical analysis. 

\begin{figure}[hbpt] 
 \begin{center}
    \includegraphics[width=0.495\textwidth]{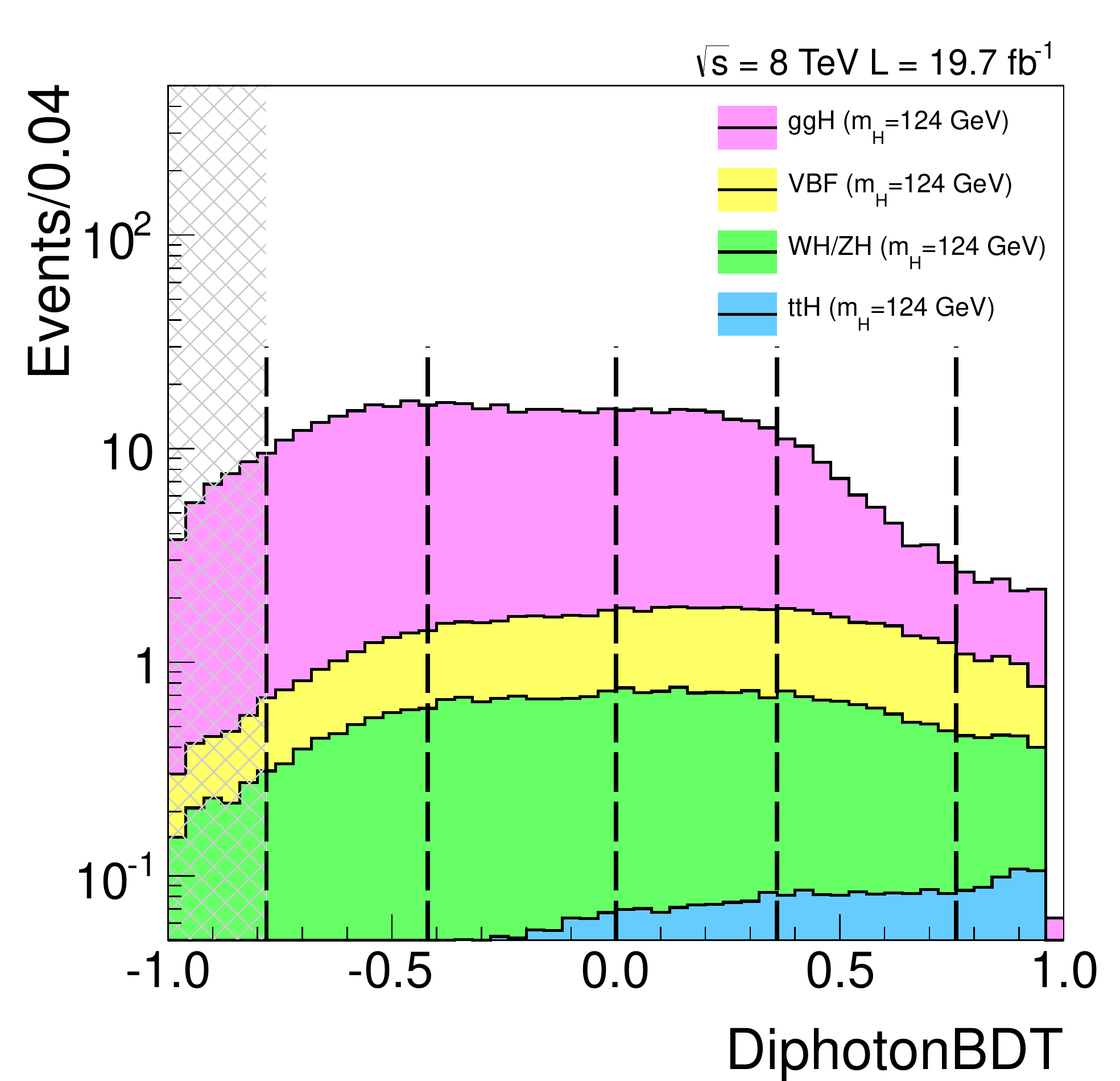}
    \includegraphics[width=0.495\textwidth]{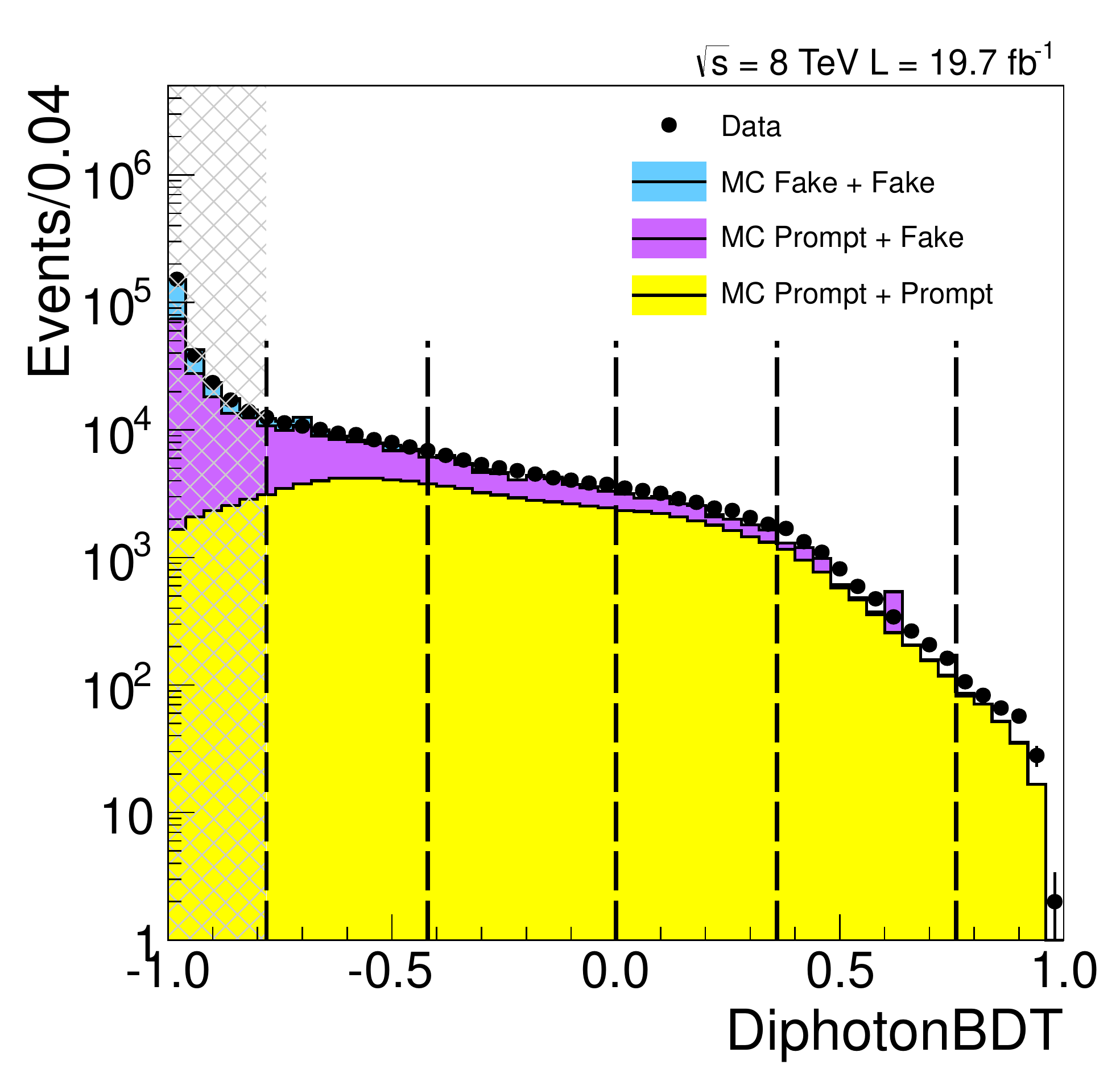}
  \end{center}
  \caption{Left: the distribution of DiphotonBDT for Monte Carlo signal (stacked histogram) consisting of $H\rightarrow \gamma\gamma$ events at a Higgs mass of 124 GeV with all four production processes weighted by cross section at 8 TeV. Right: the distributions of DiphotonBDT for data (points), Monte Carlo background (stacked histogram) consisting of prompt diphoton, $\gamma$ + jet and dijet events weighted by cross section at 8 TeV. The data and Monte Carlo background events in the signal region 120 GeV $<$ $m_{\gamma\gamma}$ $<$ 130 GeV are removed. All events pass the preselection with IDBDT $>$ $-$0.2 for both photons. Photon energy corrections are applied to both data and Monte Carlo events, and additional corrections are applied to Monte Carlo events including efficiency scaling and pileup reweighting. The class boundaries are labeled as dashed lines. The events in the shaded region below the dashed line with the lowest DiphotonBDT value are dropped.}
 \label{fig:diphotonmva output classification idsysclass}
\end{figure}

For the untagged classes, 4 classes for events at 7 TeV and 5 classes for events at $8~\mathrm{TeV}$ are determined. Figure \ref{fig:diphotonmva output classification idsysclass} shows the DiphotonBDT distributions, along with the class boundaries (dashed lines), of Monte Carlo signal events (left), Monte Carlo background events and data (right) at 8 TeV, passing the preselection and IDBDT $>$ $-$0.2 for both photons. The region below the dashed line with the lowest DiphotonBDT value is removed.

\section{Final Event Classes}
\label{sec:Final Event Classes}
The diphoton events, passing the preselection and IDBDT $>$ $-$0.2 for both photons, are first selected into the tagged classes and the rest are selected into the untagged classes. The classes are mutually exclusive. In the case that an event satisfies the criteria of more than one tagged classes, the class with higher fraction of events from the corresponding tagged production process among the selected signal events is chosen in general. The final event classes including 11 classes for events at 7 TeV and 14 classes for events at 8 TeV are summarized in Table \ref{tab:eveclass}. There are very few \textit{t$\overline{t}$H} lepton tagged events and multijet tagged events at 7 TeV, and so these events are combined into a single class.   

\begin{table}[h]
  \newcolumntype{L}[1]{>{\raggedright\let\newline\\\arraybackslash\hspace{0pt}}m{#1}}
  \newcolumntype{C}[1]{>{\centering\let\newline\\\arraybackslash\hspace{0pt}}m{#1}}
  \begin{center}
    \caption{The event classes listed in the event selection order. The events for each class are selected from the preselected diphoton events with IDBDT $>$ $-$0.2 for both photons.}
    \begin{tabular}{|L{0.2cm}|L{3cm}|c|c|c|} 
      \hline
      \multicolumn{2}{|c|}{Event classes} & Tag & DiphotonBDT & CombinedBDT\\
      \hline
      \multirow{12}{*}{\begin{sideways}{7 TeV 5.1 fb$^{-1}$}\end{sideways}}&\textit{t$\overline{t}$H} Lepton + Multijet & \textit{t$\overline{t}$H} Lepton  & $>$ 0.6 & -\\
      && or \textit{t$\overline{t}$H} Multijet  & & \\
      \cline{2-5}
      &\textit{VH} Lepton Tight & \textit{VH} Lepton Tight   & $>$ 0.1 & -\\
      \cline{2-5}
      &\textit{VH} Lepton Loose & \textit{VH} Lepton Loose  & $>$ 0.1 & -\\
      \cline{2-5}
      &\textit{VBF} Dijet 0  & \textit{VBF} Candidate & - & $>$ 0.995\\
      \cline{2-5}
      &\textit{VBF} Dijet 1  & \textit{VBF} Candidate & - & $>$ 0.917 \&\& $\leq$ 0.995\\
      \cline{2-5}
      &\textit{VH} MET  & \textit{VH} MET  & $>$ 0.8 & -\\
      \cline{2-5}
      &\textit{VH} Dijet  & \textit{VH} Dijet  & $>$ 0.6 & -\\
      \cline{2-5}
      &Untagged 0  & - & $>$ 0.93 & -\\
      \cline{2-5}
      &Untagged 1  & - & $>$ 0.85 \&\& $\leq$ 0.93  & -\\
      \cline{2-5}
      &Untagged 2  & - & $>$ 0.7 \&\& $\leq$ 0.85  & -\\
      \cline{2-5}
      &Untagged 3  & - & $>$ 0.19 \&\& $\leq$ 0.7  & -\\
      \cline{2-5}
      \hline
      \hline
       \multirow{14}{*}{\begin{sideways}{8 TeV 19.7 fb$^{-1}$}\end{sideways}}&\textit{t$\overline{t}$H} Lepton & \textit{t$\overline{t}$H} Lepton  & $>$ $-$0.6  & -\\
      \cline{2-5}
      &\textit{VH} Lepton Tight& \textit{VH} Lepton Tight  & $>$ $-$0.6 & -\\
      \cline{2-5}
      &\textit{VH} Lepton Loose& \textit{VH} Lepton Loose & $>$ $-$0.6 & -\\
      \cline{2-5}
      &\textit{VBF} Dijet 0  & \textit{VBF} Candidate & - & $>$ 0.94\\
      \cline{2-5}
      &\textit{VBF} Dijet 1  & \textit{VBF} Candidate & - & $>$ 0.82 \&\& $\leq$ 0.94\\
      \cline{2-5}
      &\textit{VBF} Dijet 2  & \textit{VBF} Candidate & - & $>$ 0.14 \&\& $\leq$ 0.82\\
      \cline{2-5}
      &\textit{VH} MET  & \textit{VH} MET  & $>$ 0  & -\\
      \cline{2-5}
      &\textit{t$\overline{t}$H} Multijet & \textit{t$\overline{t}$H} Multijet  & $>$ $-$0.2 & -\\
      \cline{2-5}
      &\textit{VH} Dijet  & \textit{VH} Dijet  & $>$ 0.2  & -\\
      \cline{2-5}
      &Untagged 0  & - & $>$ 0.76 & -\\
      \cline{2-5}
      &Untagged 1  & - & $>$  0.36 \&\& $\leq$  0.76  & -\\
      \cline{2-5}
      &Untagged 2  & - & $>$  0 \&\& $\leq$  0.36  & -\\
      \cline{2-5}
      &Untagged 3  & - & $>$ $-$0.42 \&\& $\leq$ 0   & -\\
      \cline{2-5}
      &Untagged 4  & - & $>$ $ -$0.78 \&\& $\leq$ $-$0.42    & -\\
      \hline
    \end{tabular}
    \label{tab:eveclass}
  \end{center}
 \end{table}

%% file: StatAna.tex
\chapter{Statistical Procedure for the Extraction of the Higgs Signal}
\label{chap:Statistical Analysis}
We extract the signal of Higgs boson from the observed diphoton mass spectra of all the event classes. For each event class, the models of the expected diphoton mass spectrum of the Higgs signal events and that of background events are constructed. The signal model is built for each Higgs mass hypothesis $m_{H}$ in the search range [115,135] GeV, using parametric functions fitted from Monte Carlo simulated events, as described in Section \ref{sec:Signal Model}. The background model is built using a set of parametric functions fitted directly from data, and the uncertainty due to the limited knowledge of the true background function is taken into account by profiling the choice of functions in the signal extraction, as described in Section \ref{sec:Treatment of Background for the Signal Extraction}. The main systematic uncertainties related to the signal model are discussed in Section \ref{sec:Systematic Uncertainties}. The statistical procedure for the final Higgs signal extraction, based on the simultaneous likelihood fit to the diphoton mass spectra over all event classes, is described in Section \ref{sec:Higgs Signal Extraction Procedure}.
    
\section{Signal Model}
\label{sec:Signal Model}
For each event class, there are three steps in the signal model construction. First, the signal models of the Standard Model (SM) Higgs boson are built for five reference Higgs mass hypotheses separated by a 5 GeV step, $m_{H}$$\in$\{115,120,125,130,135\} GeV, on Monte Carlo simulations with the resolution correction, preselection efficiency scale factors and trigger efficiency applied to match data. Second, the signal model as a function of Higgs mass is built through interpolation between the neighboring reference masses of Monte Carlo models. Finally, the variations of the signal model are constructed based on the SM model with signal strengths or coupling strengths included as free parameters (their values equal to one for the SM Higgs boson), which are used in the final signal extraction. These three steps are described below.
  
\subsection{Signal Model for a Reference Higgs Mass}
\label{sec:Signal Model at A Reference Higgs Mass}    
For a reference Higgs mass $m_{H}^{'}$, models for the four Higgs production processes are first built respectively, and then combined according to their cross sections. The model for a particular Higgs production process \textit{XH}, any of \textit{ggH}, \textit{VBF}, \textit{VH} and \textit{t$\overline{t}$H}, is described as follow. The combined model is described afterwards.
    
\subsubsection{Model for a Higgs Production Process}
\label{sec:Model of Individual Higgs Production Process}
The model for a Higgs production process \textit{XH} consists of the expected yield and the diphoton mass distribution. 

The expected yield, $N_{\mathit{XH}}^{SM}(m_{H}^{'})$, is the product of luminosity, $L$, SM Higgs production cross section for the process, $\sigma_{\mathit{XH}}^{SM}(m_{H}^{'})$, SM branching ratio of the Higgs decaying to two photons, $B_{H\rightarrow\gamma\gamma}^{SM}(m_{H}^{'})$, acceptance, A($m_{H}^{'}$), and efficiency, $\epsilon(m_{H}^{'})$:
  \begin{equation}
    N_{\mathit{XH}}^{SM}(m_{H}^{'}) = L \cdot  \sigma_{\mathit{XH}}^{SM}(m_{H}^{'})\cdot  B_{H\rightarrow\gamma\gamma}^{SM}(m_{H}^{'}) \cdot  A(m_{H}^{'}) \cdot \epsilon(m_{H}^{'}).
    \label{eqn:processyield}
  \end{equation}
  The luminosity is taken from the experimental measurement described in References\cite{CMS-PAS-SMP-12-008,CMS-PAS-LUM-13-001}. The cross section and branching ratio are taken from the LHC Higgs boson Cross Section Working Group \cite{LHCHiggsCrossSectionWorkingGroup3}. The acceptance times efficiency is evaluated on the Higgs Monte Carlo sample for the process \textit{XH} with all the corrections applied, which is the fraction of events passing all the selection.

The expected diphoton mass distribution is modeled by an empirical parametric function, which consists of two components, one for the events with right vertex selected while the other for the events with wrong vertex selected. The right (wrong) vertex component, $P_{\mathit{XH}}^{R(W)}(m_{\gamma\gamma}|\overrightarrow{x}_{\mathit{XH}}^{R(W)}(m_{H}^{'}))$, with the set of parameters, $\overrightarrow{x}_{\mathit{XH}}^{R(W)}(m_{H}^{'})$, is parametrized as a Gaussian distribution, or a sum of two Gaussians, with the set of parameters $\overrightarrow{x}_{\mathit{XH}}^{R(W)}(m_{H}^{'})$ determined by maximum likelihood fit to the Monte Carlo events:  
  \begin{multline}
    P_{\mathit{XH}}^{R(W)}(m_{\gamma\gamma}|\overrightarrow{x}_{\mathit{XH}}^{R(W)}(m_{H}^{'})) = \\ \sum_{i}f_{i}^{R(W)}(m_{H}^{'})\cdot G_{i}(m_{\gamma\gamma}|\mu_{mi}^{R(W)}(m_{H}^{'})=m_{H}^{'}+\Delta m_{i}^{R(W)}(m_{H}^{'}), \sigma_{mi}^{R(W)}(m_{H}^{'})),
    \label{eqn:Gaussian2}
  \end{multline}
where $G_{i}(m_{\gamma\gamma}|\mu_{mi}^{R(W)}(m_{H}^{'})=m_{H}^{'}+\Delta m_{i}^{R(W)}(m_{H}^{'}), \sigma_{mi}^{R(W)}(m_{H}^{'}))$ represents the $i_{th}$ Gaussian with mean, $\mu_{mi}^{R(W)}(m_{H}^{'})$, and standard deviation, $\sigma_{mi}^{R(W)}(m_{H}^{'})$, $\Delta m_{i}^{R(W)}(m_{H}^{'})$ represents the shift of the mean with respect to the nominal Higgs mass, $m_{H}^{'}$, and $f_{i}^{R(W)}(m_{H}^{'})$ represents the fraction coefficient for the  $i_{th}$ Gaussian. The diphoton mass distribution for the production process, $P_{\mathit{XH}}(m_{\gamma\gamma}|\overrightarrow{x}_{\mathit{XH}}(m_{H}^{'}))$, with the set of parameters, $\overrightarrow{x}_{\mathit{XH}}(m_{H}^{'})$, is then the sum of right and wrong components, weighted according to the vertex efficiency $\epsilon_{R}(m_{H}^{'})$ calculated from the Monte Carlo events:
\begin{multline}
  P_{\mathit{XH}}(m_{\gamma\gamma}|\overrightarrow{x}_{\mathit{XH}}(m_{H}^{'})) =  \epsilon_{R}(m_{H}^{'})\cdot P_{\mathit{XH}}^R(m_{\gamma\gamma}|\overrightarrow{x}_{\mathit{XH}}^R(m_{H}^{'}))+\\(1- \epsilon_{R}(m_{H}^{'}))\cdot P_{\mathit{XH}}^W(m_{\gamma\gamma}|\overrightarrow{x}_{\mathit{XH}}^W(m_{H}^{'})).
  \label{eqn:processpdf}
\end{multline}

The complete model for the process, $S_{\mathit{XH}}(m_{\gamma\gamma}|m_{H}^{'})$, is then the product of the expected yield and the diphoton mass distribution:
  \begin{equation}
    S_{\mathit{XH}}(m_{\gamma\gamma}|m_{H}^{'}) = N_{\mathit{XH}}^{SM}(m_{H}^{'})\cdot P_{\mathit{XH}}(m_{\gamma\gamma}|\overrightarrow{x}_{\mathit{XH}}(m_{H}^{'}))
    \label{eqn:xhsigmodel}
  \end{equation} 

\subsubsection{Combined Model}
\label{sec:Combined Model}
 The model for the SM Higgs boson, $S(m_{\gamma\gamma}|m_{H}^{'})$, is constructed as the sum of the models of all processes:
\begin{equation}
  S(m_{\gamma\gamma}|m_{H}^{'}) = \sum_{\mathit{XH} \in\{\mathrm{\mathit{ggH, VBF, VH, t\overline{t}H\}}}} S_{\mathit{XH}}(m_{\gamma\gamma}|m_{H}^{'})
  \label{eqn:totalsigmodel}
\end{equation}   

\begin{figure}[hbpt] 
  \begin{center}
  \includegraphics[width=0.5\textwidth]{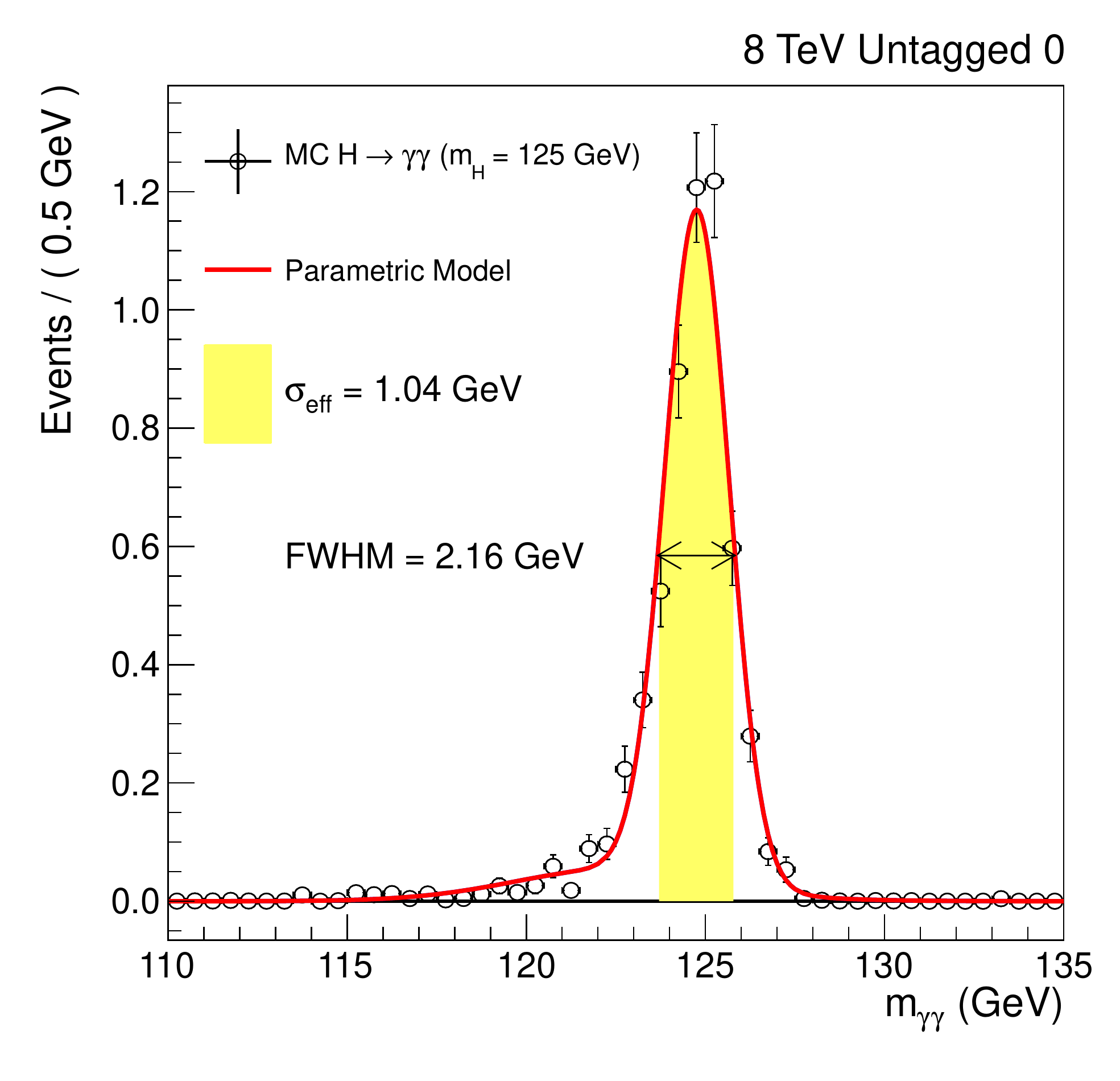}
  \end{center}
  \caption{The 8 TeV untagged 0 class's diphoton mass spectrum (points) and the fitted distribution (red line) of Monte Carlo $H\rightarrow \gamma\gamma$ events at a Higgs mass of 125 GeV.}
  \label{fig:sigmod untaggged0 8TeV}
\end{figure}

Figure \ref{fig:sigmod untaggged0 8TeV} shows the Monte Carlo diphoton mass spectrum, along with the fitted distribution (red line), for $H\rightarrow \gamma\gamma$ at $m_{H}$ = 125 GeV in 8 TeV untagged 0 class, which has the best resolution among all the event classes. The measures of the resolution, the half of the narrowest mass interval containing 68.3$\%$ of the area under the distribution (yellow), $\sigma_{eff}$, and the full width half maximum, FWHM, are 1.04 GeV and 2.16 GeV for this class. The corresponding figures for all the other event classes are provided in Appendix \ref{chap:Figures of Signal Model}. 

Table \ref{tab:sigbkg} shows the total expected yield, the fraction of contribution from each production process (the contribution less than 0.1 \% is ignored) and $\sigma_{eff}$ for $H\rightarrow \gamma\gamma$ at $m_{H}$ = 125 GeV for each event class. The expected number of the selected signal events is 475.9 (96.1) at 8 TeV (7 TeV), corresponding to the acceptance times efficiency 48\% (48\%). The contribution from the corresponding tagged process is dominant in the tagged classes, while the contribution from the \textit{ggH} process is dominant in the untagged class as expected. The resolution $\sigma_{eff}$ increases with the class number of the untagged classes also as expected.

\begin{table}[h]
  \renewcommand{\arraystretch}{1.1}
  \newcolumntype{L}[1]{>{\raggedright\let\newline\\\arraybackslash\hspace{0pt}}m{#1}}
  \newcolumntype{C}[1]{>{\centering\let\newline\\\arraybackslash\hspace{0pt}}m{#1}}
 \begin{center}
   \caption{The expected yield $S$, the fraction of each production process $f_{ggH}$, $f_{\textit{VBF}}$, $f_{\textit{VH}}$, $f_{t\overline{t}H}$ and the resolution $\sigma_{eff}$ for $H\rightarrow \gamma\gamma$ at $m_{H}$ = 125 GeV, along with the number of background events per GeV at 125 GeV $dB/dm_{\gamma\gamma}$, $S/B$ and $S/\sqrt{B}$ for each event class. The number of background events under the signal peak $B$ is estimated as  $dB/dm_{\gamma\gamma}$ multiplied by 4 $\sigma_{eff}$.}
    \begin{tabular}{|L{0.1495 cm}|L{2.221 cm}|C{0.9407 cm}|C{0.7342 cm}|C{0.8539 cm}|C{0.7342 cm}|C{0.7342 cm}|C{1.1384 cm}|C{1.6905 cm}|C{0.8 cm}|C{1.1486 cm}|} 
      \hline
      \multicolumn{2}{|c|}{\multirow{3}{*}{Event classes}} & \multicolumn{6}{c|}{Expected Higgs Boson at $m_{H}$ = 125 GeV}  & $dB/dm_{\gamma\gamma}$ & $S/B$ & $S/\sqrt{B}$\\
      \cline{3-8}
      \multicolumn{2}{|c|}{}& $S$ & f$_{ggH}$ & f$_{\textit{VBF}}$ & f$_{\textit{VH}}$ & f$_{t\overline{t}H}$ & $\sigma_{eff}$&  & &\\
      \multicolumn{2}{|c|}{} & & (\%) & (\%) & (\%) & (\%) & (GeV) & (GeV$^{-1}$ ) & &\\
      \hline
      \hline
      \multirow{11}{*}{\begin{sideways}{7 TeV 5.1 fb$^{-1}$}\end{sideways}} & $t\overline{t}$H Lepton + Multijet & 0.2 & 2.9 & 1.1 & 3.5 & 92.5 & 1.38 & 0.2 & 0.18&0.19\\
      \cline{2-11}
      & VH Lepton Tight & 0.3  & - & - & 97.7 & 2.3 & 1.59 & 0.1 & 0.47&0.38\\ 
      \cline{2-11}
      & VH Lepton Loose & 0.2 & 3.0 & 1.1 & 94.9 & 1 & 1.62 & 0.2  & 0.15&0.18\\ 
      \cline{2-11}
      & VBF Dijet 0  & 1.6 & 18.1 & 81.4 & 0.5 & - & 1.43 & 0.4 & 0.70&1.06\\ 
      \cline{2-11}
      & VBF Dijet 1  & 3.0 & 38.1 & 59.5 & 1.9 & 0.5 & 1.64 & 3.3 & 0.14&0.64\\ 
      \cline{2-11}
      & VH MET  & 0.3 & 5.7 & 1 & 85 & 8.3 & 1.52 & 0.2 & 0.25&0.27\\
      \cline{2-11}
      & VH Dijet  & 0.4 & 28.7 & 2.8 & 66.4 & 2.1 & 1.55 & 0.5 & 0.13&0.23\\ 
      \cline{2-11}
      & Untagged 0  & 5.9 & 79.7 & 10.0 & 9.6 & 0.7 & 1.12 & 11.0 & 0.12&0.84\\
      \cline{2-11}
      & Untagged 1  & 23.0 & 91.9 & 4.2 & 3.7 & 0.2 & 1.26 & 69.2 & 0.07&1.23\\
      \cline{2-11}
      & Untagged 2  & 27.2 & 91.9 & 4.2 & 3.8 & 0.1 & 1.76 & 133.5 & 0.03&0.89\\
      \cline{2-11}
      & Untagged 3  & 34.0  & 92.0 & 4.1 & 3.7 & 0.2 & 2.32 & 311.8 & 0.01&0.63\\
      \hline
      \hline
      \multirow{14}{*}{\begin{sideways}{8 TeV 19.7 fb$^{-1}$}\end{sideways}} & $t\overline{t}$H Lepton  & 0.5 & - & - & 2.8 & 97.2 & 1.32 & 0.1 & 0.95&0.69\\
      \cline{2-11}
      & VH Lepton Tight & 1.4 & - & 0.2 & 96.0 & 3.8 & 1.60 & 0.4 & 0.55&0.88\\ 
      \cline{2-11}
      & VH Lepton Loose & 0.9 & - & 1.3 & 97.1 & 1.6 & 1.56 & 1.1 & 0.13&0.34\\ 
      \cline{2-11}
      & VBF Dijet 0  & 4.4 & 17.3 & 82.3 & 0.3 & 0.1 & 1.27 & 0.7 & 1.24&2.33\\ 
      \cline{2-11}
      & VBF Dijet 1  & 5.4 & 26.0 & 73.0 & 0.8 & 0.2 & 1.44 & 2.7 & 0.35&1.37\\ 
      \cline{2-11}
      & VBF Dijet 2  & 13.7 & 44.0 & 53.1 & 2.2 & 0.7 & 1.56 & 21.9 & 0.10&1.17\\ 
      \cline{2-11}
      & VH MET & 1.7 & 12.0 & 2.3 & 74.0 & 11.7 & 1.58 & 1.2 & 0.22&0.62\\ 
      \cline{2-11}
      & $t\overline{t}$H Multijet  & 0.6 & 7.5 & 1.0 & 1.7 & 89.8 & 1.41 & 0.5 & 0.21&0.36\\
      \cline{2-11}
      & VH Dijet & 1.6 & 31.1 & 3.0 & 63.3 & 2.6 & 1.33 & 1.0 & 0.3&0.7\\ 
      \cline{2-11}
      & Untagged 0  & 5.8 & 74.7 & 12.3 & 11.0 & 2.0 & 1.04 & 4.3 & 0.32&1.37\\
      \cline{2-11}
      & Untagged 1  & 50.5 & 85.1 & 7.8 & 6.5 & 0.6 & 1.18 & 117.9 & 0.09&2.14\\
      \cline{2-11}
      & Untagged 2  & 116.5 & 91.1 & 4.8 & 3.9 & 0.2 & 1.43 & 410.9 & 0.05&2.40\\
      \cline{2-11}
      & Untagged 3  & 151.7  & 91.5 & 4.4 & 3.8 & 0.3 & 1.99 & 856.9 & 0.02&1.84\\
      \cline{2-11}
      & Untagged 4  & 121.2  & 93.2 & 3.6 & 3.1 & 0.1 & 2.56 & 1395.0 & 0.01&1.01\\
      \hline
    \end{tabular}
    \label{tab:sigbkg}
 \end{center}
\end{table}

\subsection{Signal Model as a Function of Higgs Mass}
\label{sec:Signal Model as A function of Higgs Mass} 
After building the models at the five Higgs mass hypotheses from Monte Carlo simulations, the final signal model for the SM Higgs boson as a function of Higgs mass hypothesis, $S(m_{\gamma\gamma}|m_{H})$, is constructed by interpolating between the five masses of the Monte Carlo models. The distribution uses the same functional form as the Monte Carlo models. The parameters associated with each process, ${\overrightarrow{Y}}_{\mathit{XH}}({m}_{H}) =\{N_{\mathit{XH}}^{SM}(m_{H}), \overrightarrow{x}_{\mathit{XH}}(m_{H})\}$, are piecewise linear functions in $m_{H}$:
\begin{align}
  {\overrightarrow{Y}}_{\mathit{XH}}(m_{H}) &= {\overrightarrow{Y}}_{\mathit{XH}}(m_{i})+\frac{{\overrightarrow{Y}}_{\mathit{XH}}(m_{i}+5 \:\text{GeV})-{\overrightarrow{Y}}_{\mathit{XH}}(m_{i})}{5}\cdot (m_{H}-m_{i}),\nonumber \\ 
   i \in \{1,2,3,4\}, m_{i} &\in \{115,120,125,130\} \text{GeV}, m_{i} \leq m_{H} \leq m_{i} + 5 \: \text{GeV}.
    \label{eqn:sigmodelfuncmass}
\end{align}

\subsection{Variations of Signal Model}
\label{sec:Variations of Signal Model} 
The variations of the signal model, with signal strengths or coupling strengths included as free parameters, are constructed by modifying the Higgs cross section times its branching ratio to two photons $\sigma_{\mathit{XH}}(m_{H})\cdot  B_{H\rightarrow\gamma\gamma}(m_{H})$. These parameters, along with the Higgs mass $m_{H}$, are later measured to quantify the compatibility of the SM Higgs model with respect to data. The varied models are summarized as following\cite{combined7tev,LHCHiggsCrossSectionWorkingGroup3}:
\begin{itemize}  
\item  $S(m_{\gamma\gamma}|\mu_{H},m_{H})$: signal model with total signal strength, $\mu_{H}$, and Higgs mass, $m_{H}$, as free parameters of interest, in which\\
  $\sigma_{\mathit{XH}}(m_{H})\cdot  B_{H\rightarrow\gamma\gamma}(m_{H})$ = $\mu_{H} \cdot \sigma_{\mathit{XH}}^{SM}(m_{H})\cdot B_{H\rightarrow\gamma\gamma}^{SM}(m_{H})$.
\item  $S(m_{\gamma\gamma}|\mu_{ggH,t\overline{t}H},\mu_{\textit{VBF,VH}},m_{H})$: signal model with signal strength for \textit{ggH} and \textit{t$\overline{t}$H} processes, $\mu_{ggH,t\overline{t}H}$ (sensitive to Higgs coupling strength to fermions), signal strength for \textit{VBF} and \textit{VH} processes, $\mu_{\textit{VBF,VH}}$ (sensitive to Higgs coupling strength to bosons), and Higgs mass, $m_{H}$, as free parameters of interest, in which\\
  $\sigma_{ggH(t\overline{t}H)}(m_{H})\cdot  B_{H\rightarrow\gamma\gamma}(m_{H})$ = $\mu_{ggH,t\overline{t}H} \cdot \sigma_{ggH(t\overline{t}H)}^{SM}(m_{H})\cdot  B_{H\rightarrow\gamma\gamma}^{SM}(m_{H})$,\\ 
  $\sigma_{\textit{VBF(VH)}}(m_{H})\cdot  B_{H\rightarrow\gamma\gamma}(m_{H})$ = $\mu_{\textit{VBF,VH}} \cdot \sigma_{\textit{VBF(VH)}}^{SM}(m_{H})\cdot  B_{H\rightarrow\gamma\gamma}^{SM}(m_{H})$.
\item $S(m_{\gamma\gamma}|\kappa_{V},\kappa_{f},m_{H})$: signal model with Higgs coupling strength to bosons, $\kappa_{V}$, and  Higgs coupling strength to fermions, $\kappa_{f}$---benchmark parameterization defined in Reference \cite{LHCHiggsCrossSectionWorkingGroup3}---and Higgs mass, $m_{H}$, as free parameters of interest.
\item $S(m_{\gamma\gamma}|\kappa_{\gamma},\kappa_{g},m_{H})$: signal model with effective Higgs coupling strength to photon, $\kappa_{\gamma}$, effective Higgs coupling strength to gluon, $\kappa_{g}$---benchmark parameterization defined in Reference \cite{LHCHiggsCrossSectionWorkingGroup3}---and  Higgs mass, $m_{H}$, as free parameters of interest. 
\end{itemize}  

\section{Treatment of Background for the Signal Extraction}
\label{sec:Treatment of Background for the Signal Extraction}
For each event class, the model of the background diphoton spectrum, the product of the expected yield and the diphoton mass distribution, is constructed using parametric functions fitted to data. The functional forms are chosen to describe the continuously falling character of the expected background spectrum. The fit range is 100 GeV $<m_{\gamma\gamma}<$ 180 GeV, such that the background under an emerging narrow peak, for any given Higgs mass hypotheses within 115 GeV $\leq$ $m_{H}$ $\leq$ 135 GeV, gets constraint from sufficient events in the sidebands of the signal region. The differences between the chosen background functions and the unknown true function lead to an uncertainty in the extraction of Higgs signal. In our previous $H\rightarrow \gamma\gamma$ analysis\cite{longhgg}, a single background function is chosen for each event class, following the criterion that the potential bias of the Higgs results is negligible with respect to the statistical uncertainty, at the price of increasing the number of parameters of the function and inflating the statistical uncertainty. An updated method\cite{bkgenv1,bkgenv2} is used in this analysis, which incorporates the uncertainty due to the choice of the functional form into the total uncertainty of the Higgs results, and thus avoids the inflation of the statistical uncertainty. 

The basic idea of the updated method is: first, choose a set of background functions describing the data well and that are generic enough to cover the true function. Second, build a negative log-likelihood function of Higgs parameter of interest, e.g. signal strength $\mu_{H}$, for each background function, with a correction term penalizing the increase of the number of parameters. Third, construct the envelope negative log-likelihood function by taking the minimum value of the individual functions at each $\mu_{H}$, from which the best fit $\hat{\mu}_{H}$ and the associated confidence interval are obtained. The uncertainty of the background function choice is taken into account in the confidence interval as a result of profiling background functions. The implementation of the method is described below, and the performance of the method, in terms of the bias of results and the coverage of confidence interval, is discussed afterwards.        

\subsection{Selection of the Set of Background Functions}
\label{sec:Construction of the Background Function Set}
For each event class, a set of background functions, $\{B_{1}(m_{\gamma\gamma}|\theta_{B_{1}}),...,B_{n}(m_{\gamma\gamma}|\theta_{B_{n}})\}$ with $\theta_{B_{i}}$ representing the set of parameters for the $i_{th}$ background function, are chosen from the following four function families:  
\begin{itemize}  
\item $N_{th}$ order Bernstein polynomial  
  \begin{equation}
    \mathrm{NBer}(m_{\gamma\gamma}) = \sum_{i=0}^{N}{\beta_i^2}{N \choose i}(\bar{m})^i(1-{\bar{m}})^{N-i}, \bar{m}=\frac{m_{\gamma\gamma}-100}{80} 
    \label{eqn:nlau}
  \end{equation}
\item $N_{th}$ order exponential sum  
  \begin{equation}
    \mathrm{NExp}(m_{\gamma\gamma}) = \sum_{i=1}^{N} \beta_i e^{\alpha_i m_{\gamma\gamma}}
    \label{eqn:nexp}
  \end{equation}
\item $N_{th}$ order power sum 
  \begin{equation}
    \mathrm{NPow}(m_{\gamma\gamma}) = \sum_{i=1}^{N} \beta_i m_{\gamma\gamma}^{\alpha_i}
    \label{eqn:npow}
  \end{equation}
\item $N_{th}$ order Laurent series 
  \begin{equation}
    \mathrm{NLau}(m_{\gamma\gamma}) = \sum_{i=1}^{N} \beta_i m_{\gamma\gamma}^{(-4+ \sum_{j=1}^{i}(-1)^{(j-1)}(j-1))}
    \label{eqn:nlau}
  \end{equation}
\end{itemize}   

For each function family, starting from the function order $N=1$, except for the Laurent series which starts from $N=2$ because $N=1$ corresponds to a trivial power law function $\beta_1 m_{\gamma\gamma}^{-4}$, background only fits are performed on data with increasing order $N$. The goodness of the fits is measured using $\chi^{2}$, and the so called p-value, the probability of getting a result as compatible or less to data than the observed one given that the function under consideration is true\cite{Glen}. If the loose criterion of the fit quality p-value $>$ 0.01 is satisfied, the function is included into the function set. This process keeps going for the $(N+1)_{th}$ order function until the higher order function is no longer significantly favored by data, quantified by: 
\begin{equation}
  P(\chi^{2}>(-2\mathrm{ln}\frac{\mathcal{L}_{N}}{\mathcal{L}_{N+1}})_{obs})\geq0.1.
  \label{eqn:bkgpval}
\end{equation}
In the equation above, $\mathcal{L}_{N}$ is the maximum likelihood for the $N_{th}$ order function; $ P(\chi^{2}>(-2\mathrm{ln}\frac{\mathcal{L}_{N}}{\mathcal{L}_{N+1}})_{obs})$ is the p-value of the observed $-2\mathrm{ln}\frac{\mathcal{L}_{N}}{\mathcal{L}_{N+1}}$ for a $\chi^{2}$ distribution, with the degree of freedom as the difference in the number of parameters between the $(N+1)_{th}$ and $N_{th}$ order function, which is the distribution of $-2\mathrm{ln}\frac{\mathcal{L}_{N}}{\mathcal{L}_{N+1}}$ in the case that $N_{th}$ order function is the true function and sufficient number of events is available for fitting. The highest $(N+1)_{th}$ order function satisfying $P(\chi^{2}>(-2\mathrm{ln}\frac{\mathcal{L}_{N}}{\mathcal{L}_{N+1}})_{obs})<0.05$ is automatically included into the function set without the requirement of goodness of the fit. This function corresponds to the true function for the pseudo-experiments used to study the potential biases associated with different background functions in the previous $H\rightarrow \gamma\gamma$ analysis\cite{longhgg}, and is used as the true background function to study the bias and coverage of confidence interval of the function set for the update method. The orders of the final input background functions for each event class are listed in Table \ref{tab:bkg input}.     

\begin{table}[h]
  \renewcommand{\arraystretch}{1.1}
  \caption{The orders of the input background functions for all event classes.}
  \begin{center}
    \begin{tabular}{|l|c|c|c|c|c|} 
    \hline
     \multicolumn{2}{|c|}{Event classes} &  NBer & NExp & NPow &  NLau\\
    \hline
      \multirow{11}{*}{\begin{sideways}{7 TeV 5.1 fb$^{-1}$}\end{sideways}} & Untagged 0         & 1 2 3 & 1  & 1    & 2 \\
     \cline{2-6}
      & Untagged 1         & 3 & 1 2   & 1   & 2   \\
     \cline{2-6}
      & Untagged 2         & 2 3 & 1 2 & 1   & 2  \\
     \cline{2-6}
      & Untagged 3         & 3 4 5  & 1 2 3  & 1 & 2    \\
     \cline{2-6}
      & VBF Dijet 0         & 1  & 1  &  1    & 2   \\
     \cline{2-6}
      & VBF Dijet 1         & 1 2 3 & 1  & 1     & 2   \\
     \cline{2-6}
      & VH Lepton Tight        & 1 & 1 &  1   & 2   \\
     \cline{2-6}
      & VH Lepton Loose        & 1 & 1 &  1   & 2   \\
     \cline{2-6}
      & VH MET       & 1 &  1 &  1   & 2  \\
     \cline{2-6}
      & VH Dijet       & 1 2 & 1 & 1  & 2   \\
     \cline{2-6}
      & $t\overline{t}$H Lepton + Multijet       & 1 & 1 &  1   & 2   \\
     \hline
     \hline
      \multirow{14}{*}{\begin{sideways}{8 TeV 19.7 fb$^{-1}$}\end{sideways}} & Untagged 0         &  1 2 3 & 1  & 1   &  2 \\
     \cline{2-6}
      & Untagged 1         &  2 3 & 1   & 1    & 2   \\
     \cline{2-6}
      & Untagged 2         & 3 4 & 1 2  & 1    & 2   \\
     \cline{2-6}
      & Untagged 3         & 4 5 & 2 &  1   & 2   \\
     \cline{2-6}
      & Untagged 4         & 4 5 & 2 &  1 2   & 2   \\
     \cline{2-6}
      & VBF Dijet 0         & 1 & 1 &  1   & 2   \\
     \cline{2-6}
      & VBF Dijet 1         & 1 2 & 1 &  1   & 2  \\
     \cline{2-6}
      & VBF Dijet 2         & 2 3 & 1 &  1   &  2  \\
     \cline{2-6}
      & VH Lepton Tight       & 1 & 1 &  1   & 2    \\
     \cline{2-6}
      & VH Lepton Loose        & 1 2 & 1 &  1   & 2  \\
     \cline{2-6}
      & VH MET        & 1 & 1 &  1   & 2   \\
     \cline{2-6}
      & VH Dijet        & 1 2 & 1 & 1    & 2   \\
     \cline{2-6}
      & $t\overline{t}$H Lepton     & 1 2 & 1 &  1   & 2   \\
     \cline{2-6}
      & $t\overline{t}$H Multijet       & 1 & 1 &  1   & 2  \\
     \hline
    \end{tabular}
    \label{tab:bkg input}
  \end{center}
\end{table}

\subsection{Construction of Envelope Negative Log-Likelihood Function}
\label{sec:Construction of Likelihood Function}        
After selecting the set of background functions, to extract a Higgs parameter under interest, e.g. total signal strength $\mu_{H}$ at a given Higgs mass hypothesis $m_{H}^{'}$, the so called envelop negative log-likelihood function, the envelope function, of the parameter is constructed, with signal plus background model on the observed diphoton mass spectrum. The data is binned in 320 bins of the diphoton mass with 250 MeV per bin---this choice permits a relatively quick extraction process while preserving the precision.

To construct the envelope function, the likelihood function for individual background function, e.g. the likelihood function for the $i_{th}$ background function $\mathcal{L}_{i}(\mu_{H},m_{H}^{'},\theta_{B_{i}})$, is first built as a product of Poisson distributions:
\begin{equation}
  \mathcal{L}_{i}(\mu_{H},m_{H}^{'},\theta_{B_{i}}) = \prod\limits_{j=1}^{320} \operatorname{Poisson}(n_{j}|s_{j}(\mu_{H},m_{H}^{'})+b_{j,i}(\theta_{B_{i}})),
  \label{eqn:bkglikelihood}
\end{equation}
where $n_{j}$ is the observed number of events in the $j_{th}$ bin of the data, $s_{j}(\mu_{H},m_{H}^{'})$ is the expected number of signal events in the $j_{th}$ bin under the $m_{H}^{'}$ mass hypothesis, which is obtained by integrating $S(m_{\gamma\gamma}|\mu_{H},m_{H}^{'})$ over the bin, and $b_{j,i}(\theta_{B_{i}})$ is the expected number of background events in the $j_{th}$ bin obtained by integrating $B_{i}(m_{\gamma\gamma}|\theta_{B_{i}})$. 

The envelope function $-2\mathrm{ln}\mathcal{L}_{E}(\mu_{H},m_{H}^{'})$ is then constructed as:
\begin{equation}
  -2\mathrm{ln}\mathcal{L}_{E}(\mu_{H},m_{H}^{'})=\min_{\forall i \in {1,...,n}}\{-2\mathrm{ln}\mathcal{L}_{i}(\mu_{H},m_{H}^{'}, \hat{\theta}_{B_{i},\mu_{H},m_{H}^{'}})+l_{B_{i}}\},
  \label{eqn:overall}
\end{equation} 
where $\hat{\theta}_{B_{i},\mu_{H},m_{H}^{'}}$ represents the set of values of the background parameters maximizing the $i_{th}$ likelihood function at $\mu_{H}$ and a given $m_{H}^{'}$, and $l_{B_{i}}$ represents the number of parameters of the $i_{th}$ background function, acting as a correction term penalizing the increase of number of parameters. For two background functions from the same function family $B_{n}$ and $B_{m}$, with $B_{n}$ having larger number of parameters than $B_{m}$, the penalty works in the way that the two times negative log-likelihood value after correction for $B_{n}$ roughly equals to that for $B_{m}$, if the $\chi^{2}$ p-values associated with $B_{n}$ and $B_{m}$ are the same. This correction reduces the statistical uncertainty, while keeps a small bias and a good coverage of confidence interval of fitted signal strength. 

The best fit $\hat{\mu}_{H}$ is then the $\mu_{H}$ minimizing $-2\mathrm{ln}\mathcal{L}_{E}(\mu_{H},m_{H}^{'})$. The confidence intervals are determined from the likelihood ratio $-2\Delta \mathrm{ln}\mathcal{L}_{E}(\mu_{H},m_{H}^{'})$:
\begin{equation}
  -2\Delta \mathrm{ln}\mathcal{L}_{E}(\mu_{H},m_{H}^{'})= -2\mathrm{ln}\frac{\mathcal{L}_{E}(\mu_{H},m_{H}^{'})}{\mathcal{L}_{E}(\hat{\mu}_{H},m_{H}^{'})}.
  \label{eqn:deltanll}
\end{equation}
For example, the boundary points for the 68.3$\%$ confidence interval $[\mu_{H}^{68.3\%-},\mu_{H}^{68.3\%+}]$ correspond to: 
\begin{equation}
  -2\Delta\mathrm{ln}\mathcal{L}_{E}(\mu_{H}^{68.3\%-},m_{H}^{'})=-2\Delta\mathrm{ln}\mathcal{L}_{E}(\mu_{H}^{68.3\%+},m_{H}^{'})= 1,
  \label{eqn:bkgmu68.3}
\end{equation}
for which the uncertainty of the background function choice is taken into account as a result of profiling the background functions.

\subsection{Performance}
\label{sec:Performance of The Method}      
For each event class, the bias of the best fit $\hat{\mu}_{H}$, defined as the median difference between the measured and true $\mu_{H}$ relative to the uncertainty, and the coverage of the confidence interval are evaluated on toy datasets, which are generated from signal plus background model for each background truth function as mentioned above. For untagged classes and production process tagged classes with sufficient large samples, the bias and the deviation of the confidence interval coverage from the nominal value are within 14$\%$ and 1$\%$ respectively, or slightly above, in most cases for the signal region 115 GeV $\leq$ $m_{H}$ $\leq$ 135 GeV, which are considered as neglegible. For tagged classes with few events, the bias and the deviation of the confidence interval coverage are in general larger, with the maximum value about 30$\%$ and 10$\%$ respectively; and the expected background functions are not so well constrained by the data in the sidebands. The influence from these classes is negligible since the final Higgs results are extracted by simultaneous fitting over all classes.

\section{Systematic Uncertainties Associated with the Signal Model}
\label{sec:Systematic Uncertainties}
The systematic uncertainties associated with the signal model are considered for the final Higgs signal extraction. There are two types of uncertainties. One leads to the variations of the expected signal yield, and dominates the systematic uncertainty of the signal strength. The other leads to the variations of the signal shape, and dominates the systematic uncertainty of the Higgs mass. These uncertainties are summarized in the following, more descriptions are available in Reference\cite{hggfinalpaper}. The statistical procedure to incorporate the corresponding signal variations into the Higgs signal extraction is described afterwards.    

\subsection{Systematic Uncertainties Related to the Signal Yield} 
\label{sec:Systematic Uncertainties Related to the Signal Yield}  
There are two kinds of systematic uncertainties influencing the signal yield. The first kind causes 100\% correlated variations of yields of all the event classes under influence. The second kind causes migrations of events among classes and so $-$100\% correlated variations of yields of the classes the events migrating between. These two kinds of uncertainties are introduced below respectively.

\subsubsection{Uncertainties Causing 100\% Correlated Variations of Yields} 
\label{sec:Systematic Uncertainties Causing Fully Correlated Variations of Yields}  
The systematic uncertainties causing 100\% correlated variations of yields of the event classes under influence are summarized in Table \ref{tab:yield1}. The systematic sources are listed in the first column, and their corresponding uncertainties are listed in the second column. Among the uncertainties, the cross section uncertainty of each Higgs production process and the branching ratio uncertainty of Higgs decaying to two photons are associated with theoretical calculations. The former consists of two components: one is from the uncertainty of the Parton Distribution Functions (PDF); the other is from the effect of missing higher order correction terms, evaluated by varying the factorization scale and the renormalization scale (scale). For this analysis, the events from \textit{WH} and \textit{ZH} processes are considered together as events from \textit{VH}, and the larger uncertainty of \textit{WH} and \textit{ZH} is taken. The rest of the uncertainties are associated with experimental measurements. The theoretical uncertainties, especially the cross section uncertainty of the \textit{ggH} process, drive the uncertainty of the expected total signal yield, and thus the uncertainty of the signal strength. 
\begin{table}[bthp]
  \newcolumntype{L}[1]{>{\raggedright\let\newline\\\arraybackslash\hspace{0pt}}m{#1}}
  \newcolumntype{R}[1]{>{\raggedleft\let\newline\\\arraybackslash\hspace{0pt}}m{#1}}
  \newcolumntype{C}[1]{>{\centering\let\newline\\\arraybackslash\hspace{0pt}}m{#1}}
  \caption{The systematic uncertainties causing 100\% correlated variations of yields of all the event classes under influence.}
  \begin{center}
    \begin{tabular}{|L{2.1cm}|L{2.1cm}L{5.3cm}L{5.3cm}|}
      \hline
      \multicolumn{2}{|L{4.2cm}|}{Source} &  \multicolumn{2}{L{10.6cm}|}{Uncertainty} \\ 
      \hline
      \hline
      \multicolumn{2}{|L{4.2cm}|}{Cross Section} &  \multicolumn{1}{L{5.3cm}}{PDF 8 TeV (7 TeV)} &  \multicolumn{1}{L{5.3cm}|}{Scale 8 TeV (7 TeV)}\\ 
      \multicolumn{2}{|L{4.2cm}|}{\textit{ggH}}& \multicolumn{1}{L{5.3cm}}{$+7.5\%$$-6.9\%$ ($+7.6\%$$-7.1\%$)}& \multicolumn{1}{L{5.3cm}|}{$+7.2\%$$-7.8\%$ ($+7.1\%$$-7.8\%$)}\\
      \multicolumn{2}{|L{4.2cm}|}{\textit{VBF}}& \multicolumn{1}{L{5.3cm}}{$+2.6\%$$-2.8\%$ ($+2.5\%$$-2.1\%$)}& \multicolumn{1}{L{5.3cm}|}{$+0.2\%$$-0.2\%$ ($+0.3\%$$-0.3\%$)}\\
      \multicolumn{2}{|L{4.2cm}|}{\textit{WH}}& \multicolumn{1}{L{5.3cm}}{$+2.3\%$$-2.3\%$ ($+2.6\%$$-2.6\%$)}& \multicolumn{1}{L{5.3cm}|}{$+1.0\%$$-1.0\%$ ($+0.9\%$$-0.9\%$)}\\
      \multicolumn{2}{|L{4.2cm}|}{\textit{ZH}}& \multicolumn{1}{L{5.3cm}}{$+2.5\%$$-2.5\%$ ($+2.7\%$$-2.7\%$)}& \multicolumn{1}{L{5.3cm}|}{$+3.1\%$$-3.1\%$ ($+2.9\%$$-2.9\%$)}\\ 
      \multicolumn{2}{|L{4.2cm}|}{\textit{$t\overline{t}H$}}& \multicolumn{1}{L{5.3cm}}{$+8.1\%$$-8.1\%$ ($+8.1\%$$-8.1\%$)}& \multicolumn{1}{L{5.3cm}|}{$+3.8\%$$-9.3\%$ ($+3.2\%$$-9.3\%$)}\\
      \hline
      \multicolumn{2}{|L{4.2cm}|}{Branching Ratio $H\rightarrow \gamma\gamma$} & \multicolumn{2}{L{10.6cm}|}{$+5.0\%$/$-4.9\%$}   \\  
      \hline
      \multicolumn{2}{|L{4.2cm}|}{Integrated Luminosity} & \multicolumn{2}{L{10.6cm}|}{$2.6\%$ ($2.2\%$) 8 TeV (7 TeV) }   \\  
       \hline
       \multicolumn{2}{|L{4.2cm}|}{Trigger Efficiency} & \multicolumn{2}{L{10.6cm}|}{$1.0\%$}   \\  
       \hline
       \multicolumn{2}{|L{4.2cm}|}{Preselection Efficiency Per Photon} & \multicolumn{1}{L{5.3cm}}{$1.0\%$ ($2.6\%$) Barrel (Endcap)} & \\  
       \hline
    \end{tabular}
    \label{tab:yield1}
  \end{center}
\end{table}

\subsubsection{Uncertainties Causing Migration of Events} 
\label{sec:Systematic Uncertainties Causing Migrations of Yields}  
The systematic uncertainties causing migration of yields among classes are further divided into two groups. One group is related to the DiphotonBDT, and mainly causes the events to migrate among the untagged classes, or to migrate into/out of the selection range of the analysis which is DiphotonBDT $>$ $-0.78$ (0.19) for events at 8 TeV (7 TeV). The other group is related to the tags of the Higgs production processes and causes the events to migrate among the tagged classes, or to migrate between the tagged classes and the untagged classes.

The systematic uncertainties related to the DiphotonBDT are summarized in Table \ref{tab:yield2}. The uncertainty of each source is propagated to the variation of the DiphotonBDT distribution as already described in Section \ref{sec:Output and Performance}. The resulting relative yield uncertainty of any event class is evaluated as the change of the yield due to the variation, and the maximum uncertainty is shown.  
\begin{table}[bthp]
   \newcolumntype{L}[1]{>{\raggedright\let\newline\\\arraybackslash\hspace{0pt}}m{#1}}
   \newcolumntype{C}[1]{>{\centering\let\newline\\\arraybackslash\hspace{0pt}}m{#1}}
   \caption{The systematic uncertainties related to the DiphotonBDT.}
   \begin{center}
     \begin{tabular}{|L{4cm}|L{5.5cm}|L{2cm}L{2cm}|}
       \hline
       \multicolumn{2}{|L{9.5cm}|}{Source} & \multicolumn{2}{L{3.5cm}|}{Yield Uncertainty Per Event Class (Up To)}\\
       \hline
       \hline
       \multicolumn{1}{|L{4cm}}{IDBDT}&\multicolumn{1}{L{5.5cm}|}{\textit{Shifting 0.01}} & \multicolumn{2}{L{4cm}|}{\textit{$\sim$5\%}}\\
       \hline
       \multicolumn{1}{|L{4cm}}{$\sigma_{E}/{E}$}&\multicolumn{1}{L{5.5cm}|}{\textit{Scaling $10\%$}} & \multicolumn{2}{L{4cm}|}{\textit{$\sim$16\%}}\\
       \hline
       \multicolumn{1}{|L{4cm}}{Diphoton Kinematics}&\multicolumn{1}{L{5.5cm}|}{\textit{Varying Higgs $p_{T}$ and rapidity}} & \multicolumn{2}{L{4cm}|}{\textit{$\sim$20\%}}\\ 
       \hline
     \end{tabular}
     \label{tab:yield2}
   \end{center}
 \end{table}

The main systematic uncertainties related to the tags of the Higgs production processes are summarized in Table \ref{tab:yield3}. For each source, the tagged classes under event migration and the corresponding migration mode, either among the tagged classes or between the tagged and the untagged classes, are shown in the second column. The maximum relative yield uncertainty of each relevant Higgs production process for a type of classes, e.g. \textit{VBF} Dijet classes, is shown in the third column. Among all the sources, the uncertainty related to the production of additional jets in the events from \textit{ggH} process has the largest effect on the event migration. This contributes to 30\% \textit{ggH} yield uncertainty for all the \textit{VBF} Dijet classes and for the $t\overline{t}H$ Multijet class, through the \textit{ggH} event migration between these classes and the untagged classes, and up to 14\% additional \textit{ggH} yield uncertainty for the \textit{VBF} Dijet classes, through the event migration among themselves. 
 \begin{table}[bthp]
   \newcolumntype{L}[1]{>{\raggedright\let\newline\\\arraybackslash\hspace{0pt}}m{#1}}
   \newcolumntype{C}[1]{>{\centering\let\newline\\\arraybackslash\hspace{0pt}}m{#1}}
   \newcolumntype{R}[1]{>{\raggedleft\let\newline\\\arraybackslash\hspace{0pt}}m{#1}}
   \caption{The systematic uncertainties related to the tags of the Higgs production processes.}
   \begin{center}
     \begin{tabular}{|L{3.15cm}|L{2.3cm}|L{3.45cm}|R{0.9cm}L{0.9cm}R{0.68cm}L{1.52cm}|}
       \hline
       \multicolumn{1}{|L{3.15cm}}{Source}&\multicolumn{1}{|L{2.3cm}}{Class}&\multicolumn{1}{L{3.45cm}|}{(from/to Class)}&\multicolumn{4}{L{3.92cm}|}{Yield Uncertainty Per Event Class  \:\:\:\:\:\:(Up To)}\\
       \hline
       \hline
       \multicolumn{1}{|L{3.15cm}}{Production of}&\multicolumn{1}{|L{2.3cm}}{\textit{VBF} Dijet}&\multicolumn{1}{L{3.45cm}|}{(Untagged)}&\multicolumn{1}{R{0.9cm}}{30\%}&\multicolumn{1}{L{0.9cm}}{\textit{ggH}}&&\\ 
       \multicolumn{1}{|L{3.15cm}}{Additional Jets}&\multicolumn{1}{|L{2.3cm}}{\textit{VBF} Dijet}&\multicolumn{1}{L{3.45cm}|}{(Other \textit{VBF} Dijet)}&\multicolumn{1}{R{0.9cm}}{14\%}&\multicolumn{1}{L{0.9cm}}{\textit{ggH}}&&\\ 
       \multicolumn{1}{|L{3.15cm}}{in \textit{ggH}}&\multicolumn{1}{|L{2.3cm}}{$t\overline{t}H$ Multijet}&\multicolumn{1}{L{3.45cm}|}{(Untagged)}&\multicolumn{1}{R{0.9cm}}{30\%}&\multicolumn{1}{L{0.9cm}}{\textit{ggH}}&&\\ 
       \hline
       \multicolumn{1}{|L{3.15cm}}{Jet Energy Scale}&\multicolumn{1}{|L{2.3cm}}{\textit{VBF} Dijet}&\multicolumn{1}{L{3.45cm}|}{(Untagged)}&\multicolumn{1}{R{0.9cm}}{10\%}&\multicolumn{1}{L{0.9cm}}{\textit{ggH}}&\multicolumn{1}{R{0.68cm}}{~4\%}&\multicolumn{1}{L{1.55cm}|}{\textit{VBF}}\\ 
       \multicolumn{1}{|L{3.15cm}}{and Resolution}&\multicolumn{1}{|L{2.3cm}}{\textit{VBF} Dijet}&\multicolumn{1}{L{3.45cm}|}{(Other \textit{VBF} Dijet)}&\multicolumn{1}{R{0.9cm}}{6\%}&\multicolumn{1}{L{0.9cm}}{\textit{ggH}}&\multicolumn{1}{R{0.68cm}}{~1\%}&\multicolumn{1}{L{1.55cm}|}{\textit{VBF}}\\ 
       \hline
       \multicolumn{1}{|L{3.15cm}}{Muon Selection}&\multicolumn{1}{|L{2.3cm}}{\textit{VH} Lepton}&\multicolumn{1}{L{3.45cm}|}{(Untagged)}&\multicolumn{1}{R{0.9cm}}{0.4\%}&\multicolumn{1}{L{0.9cm}}{\textit{VH}}&&\\ 
       \multicolumn{1}{|L{3.15cm}}{}&\multicolumn{1}{|L{2.3cm}}{$t\overline{t}H$ Lepton}&\multicolumn{1}{L{3.45cm}|}{(Untagged)}&\multicolumn{1}{R{0.9cm}}{0.2\%}&\multicolumn{1}{L{0.9cm}}{$t\overline{t}H$}&&\\ 
       \hline
       \multicolumn{1}{|L{3.3cm}}{Electron Selection}&\multicolumn{1}{|L{2.3cm}}{\textit{VH} Lepton}&\multicolumn{1}{L{3.45cm}|}{(Untagged)}&\multicolumn{1}{R{0.9cm}}{0.4\%}&\multicolumn{1}{L{0.9cm}}{\textit{VH}}&&\\ 
       \multicolumn{1}{|L{3.3cm}}{}&\multicolumn{1}{|L{2.3cm}}{$t\overline{t}H$ Lepton}&\multicolumn{1}{L{3.45cm}|}{(Untagged)}&\multicolumn{1}{R{0.9cm}}{0.2\%}&\multicolumn{1}{L{1cm}}{$t\overline{t}H$}&&\\ 
       \hline
       \multicolumn{1}{|L{3.15cm}}{MET Selection}&\multicolumn{1}{|L{2.3cm}}{\textit{VH} MET}&\multicolumn{1}{L{3.45cm}|}{(Untagged)}&\multicolumn{1}{R{0.9cm}}{3\%}&\multicolumn{1}{L{1.cm}}{\textit{VH}}&\multicolumn{1}{R{0.68cm}}{4\%}&\multicolumn{1}{L{1.55cm}|}{Non \textit{VH}}\\ 
      
       \hline
       \multicolumn{1}{|L{3.15cm}}{B-jet Selection}&\multicolumn{1}{|L{2.3cm}}{$t\overline{t}H$ Multijet}&\multicolumn{1}{L{3.45cm}|}{(Untagged)}&\multicolumn{1}{R{0.9cm}}{1\%}&\multicolumn{1}{L{0.9cm}}{$t\overline{t}H$}&\multicolumn{1}{R{0.68cm}}{2\%}&\multicolumn{1}{L{1.55cm}|}{\textit{ggH}}\\ 
       \multicolumn{1}{|L{3.15cm}}{}&\multicolumn{1}{|L{2.3cm}}{$t\overline{t}H$ Lepton}&\multicolumn{1}{L{3.45cm}|}{(Untagged)}&\multicolumn{1}{R{0.9cm}}{1\%}&\multicolumn{1}{L{1.cm}}{$t\overline{t}H$}&&\\ 
       \hline
     \end{tabular}
     \label{tab:yield3}
   \end{center}
 \end{table}

\subsection{Systematic Uncertainties Related to the Signal Shape} 
\label{sec:Systematics Uncertainties Related to the Signal Shape}   
The systematic uncertainties related to the signal shape include the uncertainties associated with photon energy scale and resolution, and the uncertainty of vertex efficiency. The former influences the mean and width of both the right vertex and wrong vertex components of the signal shape, while the latter influences the relative contributions of these two components. The types of systematic sources are listed in the first column of Table \ref{tab:shape}, and the number of corresponding sources, if more than one, are denoted in parenthesis. For each type, the largest relative uncertainty of the signal shape parameters for an event class due to a single source is shown. A brief description of these systematic uncertainties is provided below.   

\subsubsection{Uncertainties Associated With Photon Energy Scale And Resolution}
The systematic uncertainties associated with photon energy scale and resolution originate from the imperfect energy correction between data and Monte Carlo simulation using $Z\rightarrow e^{+}e^{-}$ events, which are due to three factors.

The first factor is the different effects on the photon and electron energy reconstructions from the imperfect Monte Carlo simulations, the photon/electron differences. The main differences come from the deficits of material simulation in the regions before ECAL, effectively about 10\% deficit in the region $|\eta|$ $<$ $1$ and 20\% deficit in the region $|\eta|$ $>$ $1$ from esimations, and contribute up to 0.2\% relative uncertainty of the mean value of the signal shape for an event class as shown in Table \ref{tab:shape}. The rest of the differences come from the imperfect simulations of the electromanetic shower, and the variation of collection rate of scintilation lights with their emission location along the longitudinal direction of the crystal, the light collection nonuniformity.   

The second factor is the variation of the energy scale difference between data and the Monte Carlo simulation as a function of the particle energy, the energy scale nonlinearity. The average electron energy used for the derivation of energy scale correction is lower than the average photon energy from the Higgs decay, since the $Z$ boson mass is 91.2 GeV while the Higgs mass in the search range is from 115 GeV to 135 GeV. This energy difference contributes up to 0.2\% relative uncertainty of the mean value of the signal shape for an event class. 

The third factor is the imperfect method for energy scale and resolution correction between data and Monte Carlo simulation, the energy correction method. This leads to the independent energy scale uncertainty and energy resolution uncertainty of each category of photons classified according to the photon location (barrel or endcap) and the shower shape $R_{9}$ ($>$ 0.94 or $\leq$ 0.94). For photons in the barrel from events at 8 TeV, there are two uncertainties associated with the resolution, one for the constant smearing term and the other for the energy dependent term. All together, there are 10 (8) independent single photon energy uncertainties for events at 8 TeV (7 TeV). The largest relative uncertainty of the mean value, or of the width, of the signal shape for an event class due to a single photon energy uncertainty is $0.04\%$, or $3\%$. There is an additional source of uncertainty associated with the imperfect simulation of the intrinsic distribution of $Z\rightarrow e^{+}e^{-}$, the $Z\rightarrow e^{+}e^{-}$ line-shape, which contributes to $0.01\%$ relative uncertainty on the mean value.

\subsubsection{Uncertainty of Vertex Efficiency}
The vertex efficiency is corrected between data and Monte Carlo simulation using $Z\rightarrow \mu^{+}\mu^{-}$ events. The uncertainty associated with the correction is 1.5$\%$ of the right vertex component fraction of the signal shape for an event class. 

 \begin{table}[bthp]
   \newcolumntype{L}[1]{>{\raggedright\let\newline\\\arraybackslash\hspace{0pt}}m{#1}}
   \newcolumntype{C}[1]{>{\centering\let\newline\\\arraybackslash\hspace{0pt}}m{#1}}
   \newcolumntype{R}[1]{>{\raggedleft\let\newline\\\arraybackslash\hspace{0pt}}m{#1}}
   \caption{The systematic uncertainties related to the signal shape.}
   \begin{center}
     \begin{tabular}{|L{4.99cm}|L{5.9cm}|R{1.17cm}L{2.4cm}|}
       \hline
       \multicolumn{2}{|L{10.89cm}|}{Source}&\multicolumn{2}{|L{3.57cm}|}{Shape Uncertainty Per Event Class (Up To)}\\
       \hline
       \multicolumn{2}{|L{10.89cm}|}{Photon/Electron Differences}&\multicolumn{1}{|R{1.17cm}}{}&\multicolumn{1}{L{2.4cm}|}{}\\
       \multicolumn{2}{|L{10.89cm}|}{\textit{\:\:\:\:\:\:\:\:\:\:\:\:\:\:\:Material Before ECAL (2)}}&\multicolumn{1}{|R{1.17cm}}{$0.2\%$}&\multicolumn{1}{L{2.4cm}|}{Mean}\\
       \multicolumn{2}{|L{10.89cm}|}{\textit{\:\:\:\:\:\:\:\:\:\:\:\:\:\:\:Light Collection Nonuniformity}}&\multicolumn{1}{|R{1.17cm}}{$0.02\%$}&\multicolumn{1}{L{2.4cm}|}{Mean}\\
       \multicolumn{2}{|L{10.89cm}|}{\textit{\:\:\:\:\:\:\:\:\:\:\:\:\:\:\:Electromagnetic Shower}}&\multicolumn{1}{|R{1.17cm}}{$0.05\%$}&\multicolumn{1}{L{2.4cm}|}{Mean}\\
       \hline
       \multicolumn{2}{|L{10.89cm}|}{Energy Scale Nonlinearity}&\multicolumn{1}{|R{1.17cm}}{$0.2\%$}&\multicolumn{1}{L{2.4cm}|}{Mean}\\
       \hline
       \multicolumn{2}{|L{10.89cm}|}{Energy Correction Method}&\multicolumn{1}{|R{1.17cm}}{}&\multicolumn{1}{L{2.4cm}|}{}\\
       \multicolumn{2}{|L{10.89cm}|}{\textit{\:\:\:\:\:\:\:\:\:\:\:\:\:\:\:Single photon energy scale/resolution}}&\multicolumn{1}{|R{1.17cm}}{$0.04\%$}&\multicolumn{1}{L{2.4cm}|}{Mean}\\
       \multicolumn{2}{|L{10.89cm}|}{\textit{\:\:\:\:\:\:\:\:\:\:\:\:\:\:\:(8 for 7 TeV, 10 for 8 TeV)}}&\multicolumn{1}{|R{1.17cm}}{$3\%$}&\multicolumn{1}{L{2.4cm}|}{Width}\\
       \multicolumn{2}{|L{10.89cm}|}{\textit{\:\:\:\:\:\:\:\:\:\:\:\:\:\:\:} $Z\rightarrow e^{+}e^{-}$ \textit{line-shape}}&\multicolumn{1}{|R{1.17cm}}{$0.01\%$}&\multicolumn{1}{L{2.4cm}|}{Mean}\\
       \hline
       \multicolumn{2}{|L{10.89cm}|}{Vertex Efficiency}&\multicolumn{1}{|R{1.17cm}}{~~$1.5\%$}&\multicolumn{1}{L{2.4cm}|}{Right Vertex Fraction}\\
       \hline
     \end{tabular}
     \label{tab:shape}
   \end{center}
 \end{table}

\subsection{Correlation of Uncertainties Among Event Classes} 
\label{sec:Systematic Uncertainties Correlation}  
The systematic uncertainties due to different sources are independent. For the systematic uncertainties due to the same source, the uncertainties related to the signal yield are 100\% or $-$100\% correlated among the 7 TeV and 8 TeV event classes under influence. For the signal shape, the uncertainties associated with the photon/electron differences, the $Z\rightarrow e^{+}e^{-}$ line-shape and the vertex efficiency are 100\% correlated among the 7 TeV and 8 TeV event classes. The uncertainties assocaited with the energy nonlinerity and the effect of energy correction method on single photon energy scale and resolution are 100\% correlated within $7~\mathrm{TeV}$ classes or 8 TeV classes. These uncertainties are not 100\% correlated between the 7 TeV and the $8~\mathrm{TeV}$ event classes since they are sensitive to the independent energy calibrations, regressions and the differences in the energy correction procedures of $7~\mathrm{TeV}$ and 8 TeV events. There are 20$\%$ and 50$\%$ correlations assigned to the uncertainties associated with the energy nonlinearity and the effect of energy correction method on single photon energy scale between 7 TeV and 8 TeV classes, respectively, and no correlation assigned to the uncertainties associated with the effect of energy correction method on single photon energy resolution.
     
\subsection{Procedure to Incorporate Systematic Uncertainties}  
\label{sec:Procedure to Incorporate Systematic Uncertainties}  
The signal model as introduced in Section \ref{sec:Signal Model} and the corresponding likelihood function for each event class as introduced in Section \ref{sec:Treatment of Background for the Signal Extraction} are modified to incorporate the signal yield and shape uncertainties through nuisance parameters, each associated with a particular source of systematic uncertainty. The procedure follows the description in References \cite{LHC-HCG-Report,combined7tev} and is introduced below.   

\subsubsection{Modification of Signal Yield}      
The expected yield of a Higgs production process is modified as:
\begin{equation}
  N_{\mathit{XH}}^{SM}(m_{H},\theta_{N}) = N_{\mathit{XH}}^{SM}(m_{H})\cdot\prod\limits_{k=1}^{n({\theta_{N}})} e^{{\theta}_{N}^{k}\cdot \mathrm{ln}(1+\delta_{N}^{k})},
  \label{eqn:nuisanceyields}
\end{equation}
where $\theta_{N}$ represents the set of nuisance parameters associated with the sources of the signal yield uncertainties, $n({\theta_{N}})$ represents the number of nuisance parameters, $\theta_{N}^{k}$ represents the nuisance parameter associated with the $k_{th}$ source and $\delta_{N}^{k}$ represents the corresponding relative yield uncertainty of the process in the event class.

\subsubsection{Modification of Signal Shape} 
The mean for the $i_{th}$ Gaussian component of the signal shape is modified as:
\begin{multline}
  \mu_{mi}^{R(W)}(m_{H},\theta_{\mu},\theta_{\mu}(\sqrt{s})) = \mu_{mi}^{R(W)}(m_{H})\{1+\sum\limits_{k=1}^{n({\theta_{\mu}})}\theta_{\mu}^{k}\cdot\delta_{\mu}^{k}\\+\sum\limits_{k=1}^{n({\theta_{\mu}(\sqrt{s})})}(\sqrt{1-{c_{k}^{2}(\sqrt{s})}}\cdot\theta_{\mu}^{k}(8)+c_{k}(\sqrt{s})\cdot\theta_{\mu}^{k}(7))\cdot\delta_{\mu}^{k}(\sqrt{s})\},
  \label{eqn:nuisancemu}
\end{multline}
where $\theta_{\mu}$ represents the set of nuisance parameters associated with the sources of the mean uncertainties 100\% correlated between 7 TeV and 8 TeV classes, $\theta_{\mu}(\sqrt{s})$ represents the set of nuisance parameters independent for $\sqrt{s}$=7 TeV or $\sqrt{s}$=8 TeV, $n(\theta_{\mu})$ and $n({\theta_{\mu}(\sqrt{s})})$ represent the numbers of corresponding nuisance parameters, $\theta_{\mu}^{k}$ represents the nuisance parameter associated with the $k_{th}$ source of 100\% correlated uncertainties, $\delta_{\mu}^{k}$ represents the corresponding relative uncertainty on the mean, $\theta_{\mu}^{k}(7(8))$ represents the nuisance parameter for 7 TeV (8 TeV) associated with the $k_{th}$ source of partially correlated uncertainties, $\delta_{\mu}^{k}(\sqrt{s})$ represents the corresponding relative uncertainty on the mean, $c_{k}(\sqrt{s})$ represents the coefficient assoicated with the correaltion which is 0.2 (0.5) for uncertainties realted to energy nonlinearity (effect of energy correction method on single photon energy scale) at$\sqrt{s}$=8 TeV and is 1(1) at $\sqrt{s}$=7 TeV.   
 
The standard deviation for the $i_{th}$ Gaussian component of the signal shape is modified as:
\begin{equation}
  \sigma_{mi}^{R(W)}(m_{H},\theta_{\sigma}(\sqrt{s})) = \sigma_{mi}^{R(W)}(m_{H})\{1+{\sqrt{\sum\limits_{k=1}^{n({\theta_{\sigma}(\sqrt{s})})}(\theta_{\sigma}^{k}(\sqrt{s})\cdot\delta_{\sigma}^{k}(\sqrt{s}))^{2}}}\}, 
  \label{eqn:nuisancesigma}
\end{equation}
where $\theta_{\sigma}(\sqrt{s})$ represents the set of nuisance parameters associated with the sources of the width uncertainties for $\sqrt{s}$=7 TeV or $\sqrt{s}$=8 TeV, $n({\theta_{\sigma}(\sqrt{s})})$ represents the number of nuisance parameters, $\theta_{\sigma}^{k}(\sqrt{s})$ represents the nuisance parameter associated with the $k_{th}$ source and $\delta_{\sigma}^{k}(\sqrt{s})$ represents the corresponding relative uncertainty on the width. 

The vertex selection efficiency is modified as:
\begin{equation}
  \epsilon_{R}(m_{H},\theta_{V}) = \epsilon_{R}(m_{H})\cdot\min\{(1+\theta_{V}\cdot\delta\epsilon_{R}),1\},
  \label{eqn:nuisancevtx}
\end{equation}
where $\theta_{V}$ represents the nuisance parameter associated with the vertex selection efficiency uncertainty and $\delta\epsilon_{R}$ represents the corresponding relative uncertainty.

\subsubsection{Modification of Likelihood Function} 
A likelihood function, chosen as the standard Gaussian distribution, is assigned to each nuisance parameter. The likelihood function for the $i_{th}$ input background function defined in Equation \ref{eqn:bkglikelihood} is modified accordingly as:
\begin{equation}
  \mathcal{L}_{i}(\mu_{H},m_{H}^{'},\theta_{B_{i}},\theta_{S}) = \mathcal{L}_{i}(\mu_{H},m_{H}^{'},\theta_{B_{i}})\cdot\rho(\theta_{S}),
  \label{eqn:bkglikelihoodsys}
\end{equation}
where $\theta_{S}$ represents the set of signal nuisance parameters, and $\rho(\theta_{S})$ represents the product of the likelihood functions of the nuisance parameters. The envelope function defined in Equation \ref{eqn:overall} is modified accordingly as:
\begin{equation}
  -2\mathrm{ln}\mathcal{L}_{E}^{*}(\mu_{H},m_{H}^{'})=\min_{\forall i \in {1,...,n}}\{-2\mathrm{ln}\mathcal{L}_{i}(\mu_{H},m_{H}^{'}, \hat{\theta}_{B_{i},\mu_{H},m_{H}^{'}},\hat{\theta}_{S,i,\mu_{H},m_{H}^{'}})+l_{B_{i}}\},
  \label{eqn:overallsys}
\end{equation}
where $\hat{\theta}_{S,i,\mu_{H},m_{H}^{'}}$ represents the set of values of the signal nuisance parameters maximizing the likelihood function with the $i_{th}$ background function at $\mu_{H}$ and a given $m_{H}^{'}$.
The generalized envelop function $-2\mathrm{ln}\mathcal{L}_{E}^{*}(p_{H})$ for any signal model $S(m_{\gamma\gamma}|p_{H})$ with Higgs parameters $p_{H}$ is then defined as:
 \begin{equation}
   -2\mathrm{ln}\mathcal{L}_{E}^{*}(p_{H})=\min_{\forall i \in {1,...,n}}\{-2\mathrm{ln}\mathcal{L}_{i}(p_{H}, \hat{\theta}_{B_{i},p_{H}},\hat{\theta}_{S,i,p_{H}})+l_{B_{i}}\},
  \label{eqn:overallsys_general}
\end{equation}    
where $\hat{\theta}_{B_{i},p_{H}}$ and $\hat{\theta}_{S,i,p_{H}}$ represent the set of values of the background parameters and the set of values of the signal nuisance parameters maximizing the likelihood function with the $i_{th}$ background function at $p_{H}$.
 
 \section{Higgs Signal Extraction Procedure}
 \label{sec:Higgs Signal Extraction Procedure}
The Higgs signal is finally extracted by performing simultaneous profile likelihood fits to the observed diphoton mass spectra of all the 25 event classes, including 11 classes for $7~\mathrm{TeV}$ data and 14 classes for 8 TeV data. The existence of a signal is demonstrated by a background only hypothesis test. The properties of the signal and its compatibility with the SM Higgs boson are quantified by measuring various Higgs parameters. The statistical method used is described in  References\cite{LHC-HCG-Report,combined7tev,AsymptoticCLsFormulae} and introduced as below.

For fitting the Higgs parameters, $p_{H}$, with the associated signal model, $S(m_{\gamma\gamma}|p_{H})$, the parameters $p_{H}$ and the signal systematic nuisance parameters are varied simultaneously across all the event classes, while the background nuisance parameters are varied independently for each event class. The total envelope function, $-2\mathrm{ln}\mathcal{L}_{Tot}(p_{H})$, is constructed as: 
 \begin{equation}
  -2\mathrm{ln}\mathcal{L}_{Tot}(p_{H}) = \sum_{i=1}^{25} -2\mathrm{ln}\mathcal{L}_{Ei}^{*}(p_{H}),
  \label{totlikelihood}
\end{equation}     
where $-2\mathrm{ln}\mathcal{L}_{Ei}^{*}(p_{H})$ is the envelop function for the $i_{th}$ class. From the total envelope function, the best fit $\hat{p}_{H}$, the values of $p_{H}$ minimizing the function, and the associated confidence interval or region are extracted. For extracting the confidence interval or region for a subset of $p_{H}$, $p_{H}^{I}$, the remaining parameters of $p_{H}$, $p_{H}^{0}$, are profiled as nuisance parameters, and the resulting likelihood ratio function $q_{s}(p_{H}^I)$ is used:
 \begin{equation}
   q_{s}(p_{H}^I) = -2\mathrm{ln}\frac{\mathcal{L}_{Tot}(p_{H}^{I},{\hat{p}}_{H,p_{H}^{I}}^{0})}{\mathcal{L}_{Tot}(\hat{p}_{H})},  
  \label{deltatotlikelihood}
\end{equation}     
where ${\hat{p}}_{H,p_{H}^{I}}^{0}$ represents the values of $p_{H}^{0}$ maximizing $\mathcal{L}_{Tot}(p_{H})$ at a given $p_{H}^{I}$.
 
For testing the background-only hypothesis against the existence of a signal at Higgs mass hypothesis $m_{H}^{'}$, in the presence of an excess of events above the background-only expectation, the test statistic $q_{b}(m_{H}^{'})$ is constructed as:
 \begin{equation}
   q_{b}(m_{H}^{'}) = -2\mathrm{ln}\frac{\mathcal{L}_{Tot}(\mu_{H}=0,m_{H}^{'})}{\mathcal{L}_{Tot}(\hat{\mu}_{H},m_{H}^{'})},\:\hat{\mu}_{H}\ge0 \:\:\text{or}\:\:  q_{b}(m_{H}^{'}) = 0, \:\hat{\mu}_{H}<0,
  \label{qb}
\end{equation}   
where $\hat{\mu}_{H}$ is the $\mu_{H}$ maximizing the likelihood $\mathcal{L}_{Tot}(\mu_{H},m_{H}^{'})$ at a given $m_{H}^{'}$. The probability of the test statistic under the background only hypothesis, $p(q_{b}(m_{H}^{'})|\mu_{H}=0)$, is 0.5 for $q_{b}(m_{H}^{'}) = 0$, and follows 0.5 times the $\chi^{2}$ distribution with one degree of freedom for $q_{b}(m_{H}^{'}) > 0$ in the limit of large number of events. The probability for observing equal or larger excess as the observed one under the background only hypothesis is then quantified by the local p-value, and is translated into the local significance $\sigma_{local}$ through the standard Gaussian distribution g(x):  
 \begin{equation}
   \text{local p-value} = \int_{q_{b}^{obs}(m_{H}^{'})}^\infty p(q_{b}(m_{H}^{'})|\mu_{H}=0)\, \mathrm{d}q_{b}(m_{H}^{'}) =  \int_{\sigma_{local}}^\infty g(x)\, \mathrm{d}x, 
  \label{pvalue}
\end{equation}  
where $q_{b}^{obs}(m_{H}^{'})$ is the value of the test statistic observed from the data and $\sigma_{local}$ is $\sqrt{q_{b}^{obs}(m_{H}^{'})}$ in the limit of large number of events.     

In the end of this analysis, the Higgs parameters introduced in Section \ref{sec:Signal Model} are measured. The total signal strength $\mu_{H}$ is measured using the signal model $S(m_{\gamma\gamma}|\mu_{H},m_{H})$ with $m_{H}$ treated as a nuisance parameter. The Higgs mass, $m_{H}$, the signal strength for \textit{ggH} and \textit{$t\overline{t}$H} processes, $\mu_{ggH,t\overline{t}H}$, and the signal strength for \textit{VBF} and \textit{VH} processes, $\mu_{\textit{VBF,VH}}$ are measured using the signal model  $S(m_{\gamma\gamma}|\mu_{ggH,t\overline{t}H},\mu_{\textit{VBF,VH}},m_{H})$. For the measurement of each of the parameters, the rest two are treated as nuisance parameters. The Higgs coupling strengths to bosons and to fermions, $\kappa_{V}$ and $\kappa_{f}$, and the effective Higgs coupling strengths to photon and to gluon, $\kappa_{\gamma}$ and $\kappa_{g}$, are measured using the signal models $S(m_{\gamma\gamma}|\kappa_{V},\kappa_{f},m_{H})$ and $S(m_{\gamma\gamma}|\kappa_{\gamma},\kappa_{g},m_{H})$ respectively, at the measured $m_{H}$.

%% file: Results.tex
\chapter{Results of Higgs Search from CMS $H\rightarrow\gamma\gamma$ Channel}
\label{chap:Results}

\section{Diphoton Mass Spectra and Fits}
\begin{figure}[hbpt] 
   \begin{center}
     \includegraphics[width=0.415\textwidth]{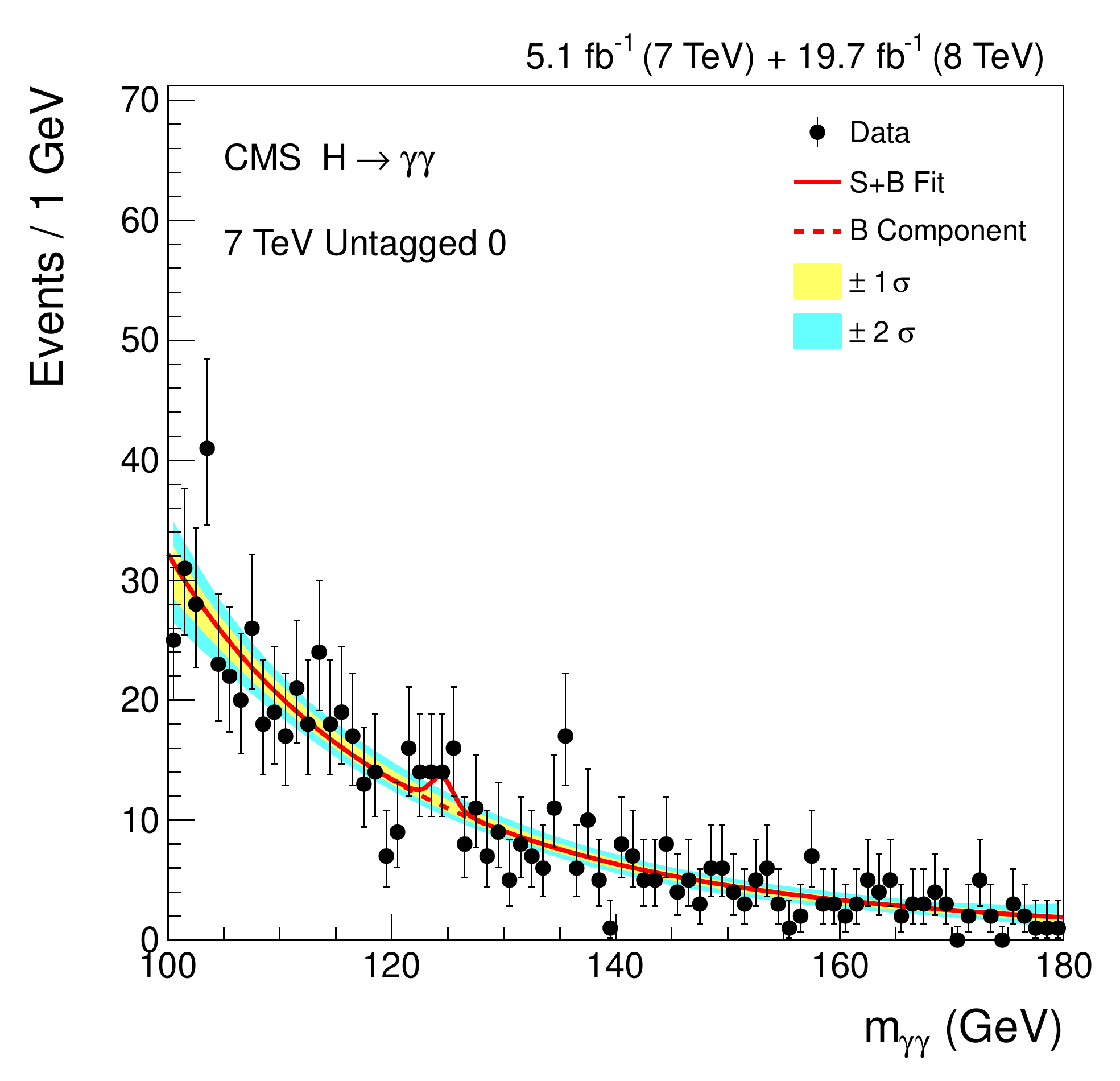}
     \includegraphics[width=0.415\textwidth]{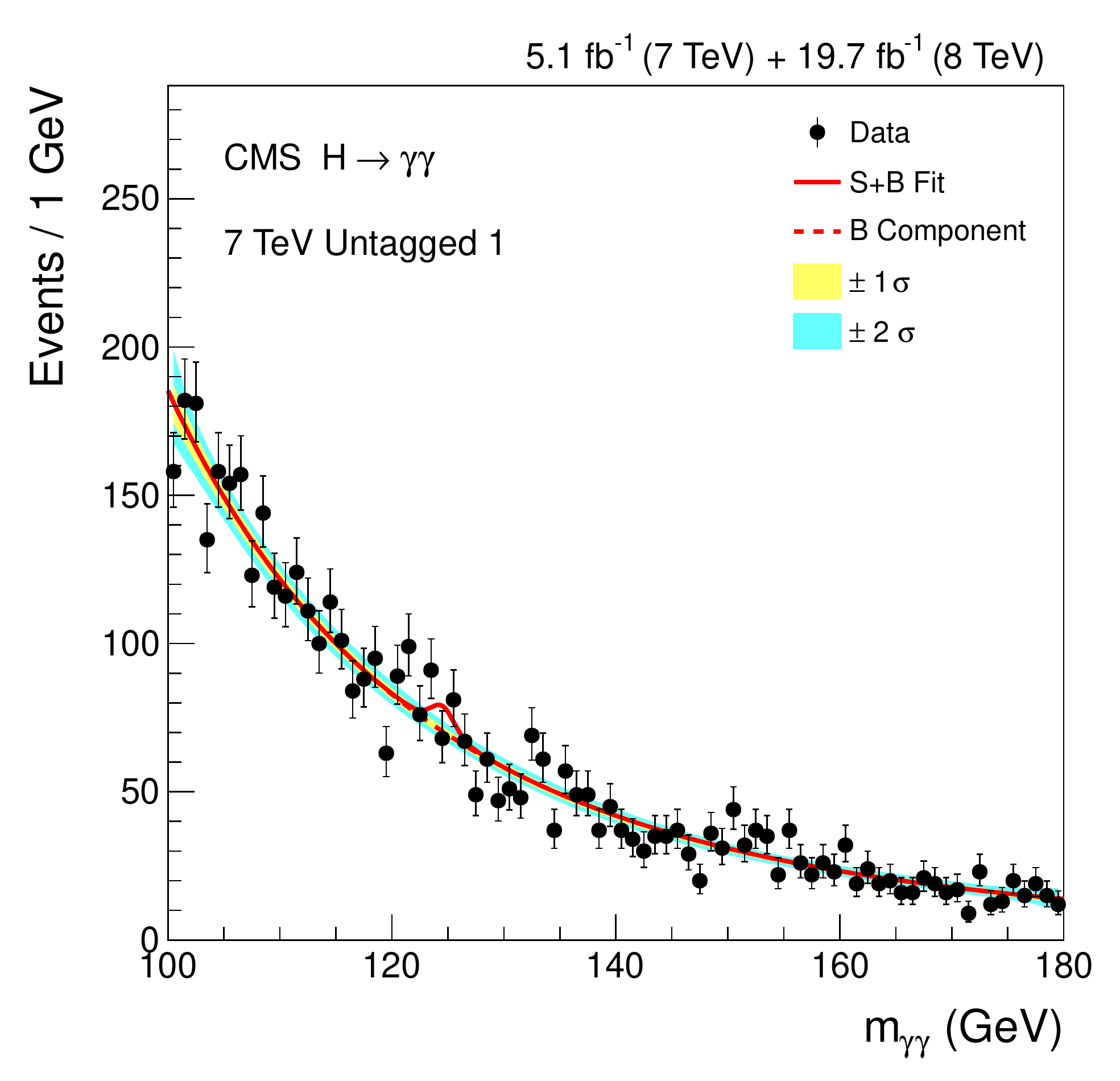}
     \includegraphics[width=0.415\textwidth]{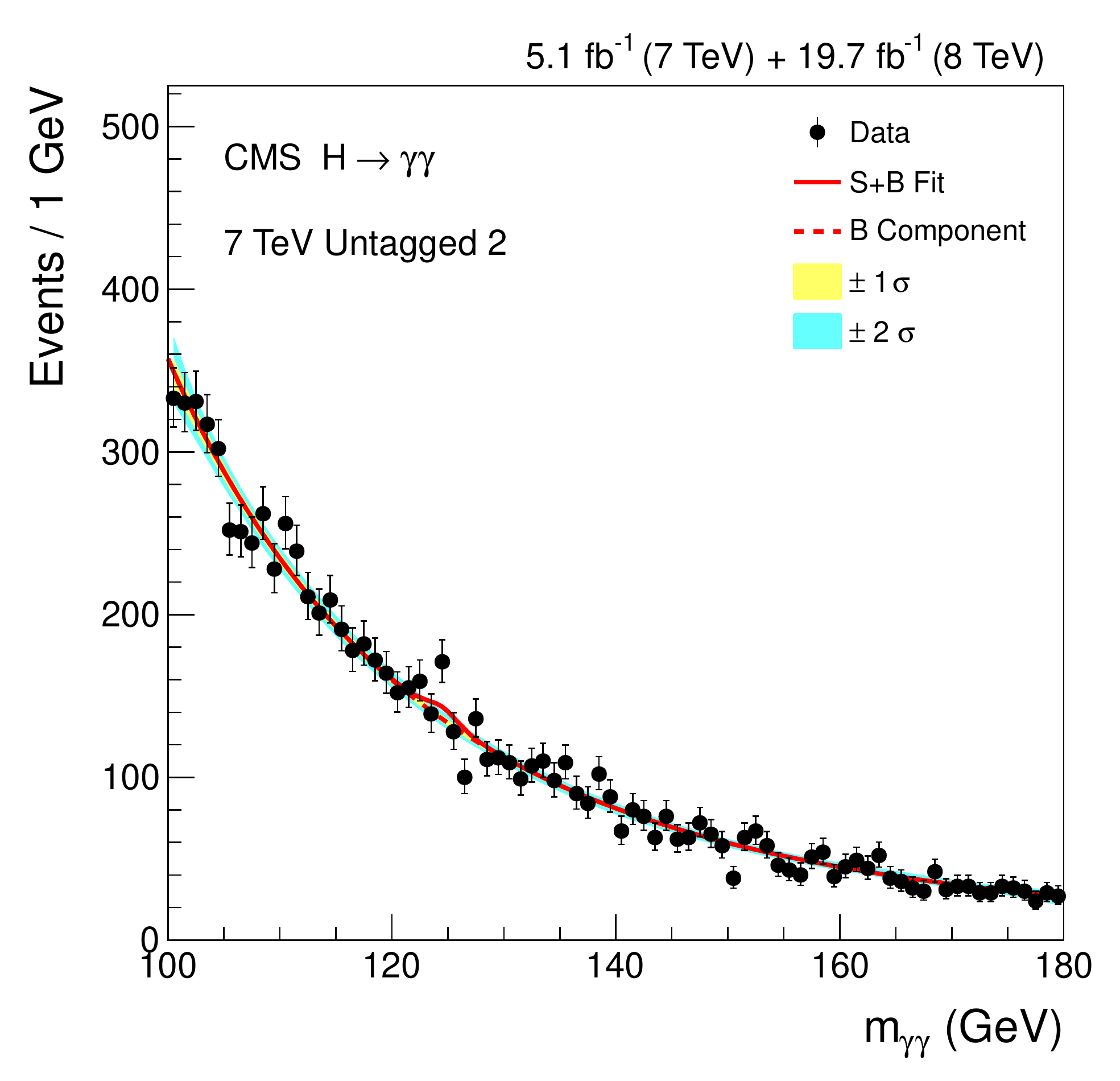}
     \includegraphics[width=0.415\textwidth]{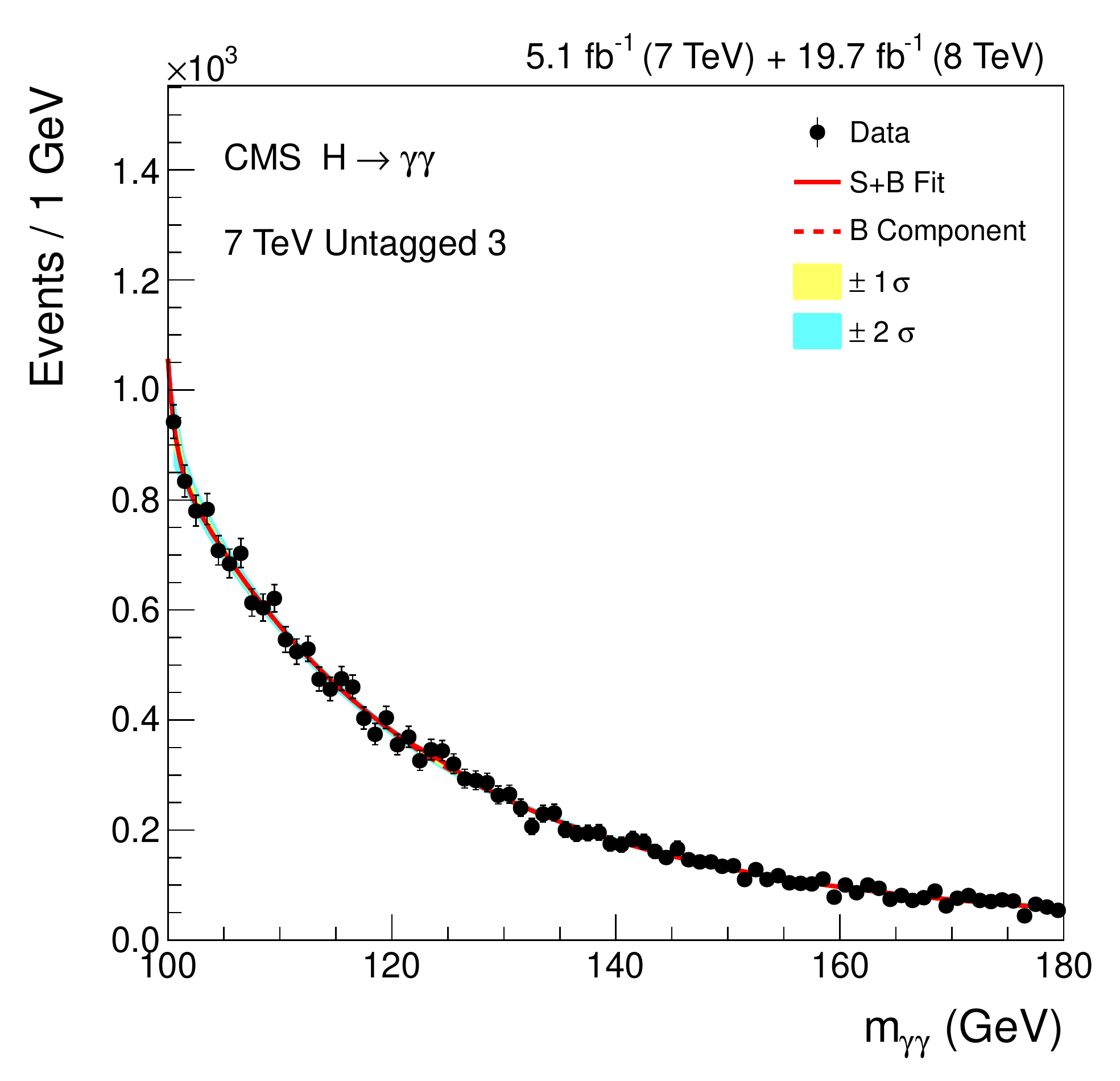}
   \end{center}
   \caption{The observed diphoton mass spectra of the untagged classes for the $7~\mathrm{TeV}$ dataset (points) binned in 1 GeV steps. For each class, the signal plus background model (solid red line), at the best-fit $\hat{\mu}_{H}$ = 1.12 and $\hat{m}_H$ = $\mathrm{124.72~GeV}$ associated with the signal model $S(m_{\gamma\gamma}|\mu_{H},m_{H})$ for the combined $\mathrm{7~TeV}$ and $\mathrm{8~TeV}$ datasets, is shown. The background component for the fit (dashed red line), the 68.3$\%$ (1 $\sigma$) confidence band (yellow) and the 95.4$\%$ (2 $\sigma$) confidence band (cyan) are also shown.}
   \label{fig:sbmass inclusive 7TeV}
 \end{figure}

\begin{figure}[hbpt] 
   \begin{center}
     \includegraphics[width=0.415\textwidth]{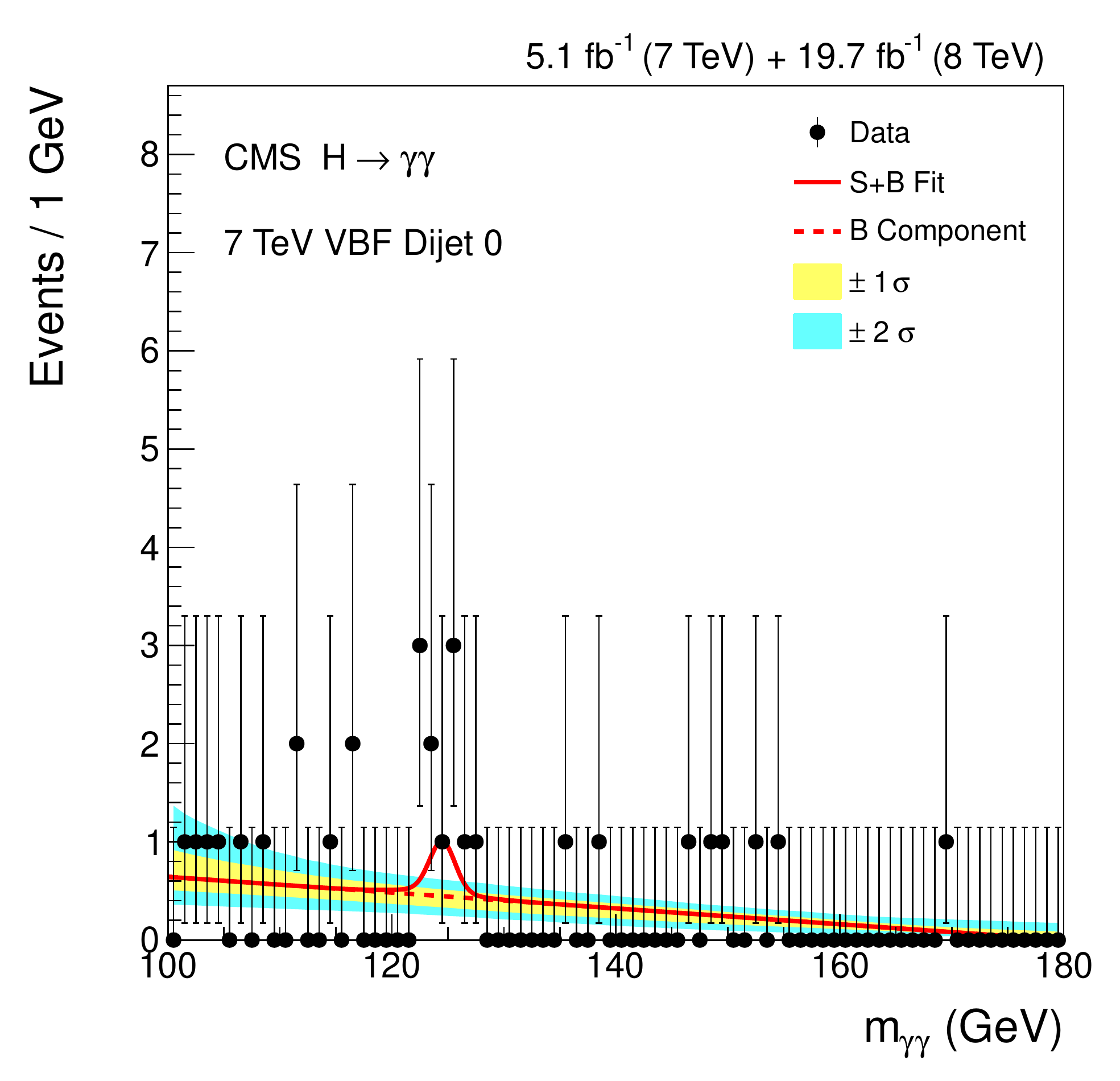}
     \includegraphics[width=0.415\textwidth]{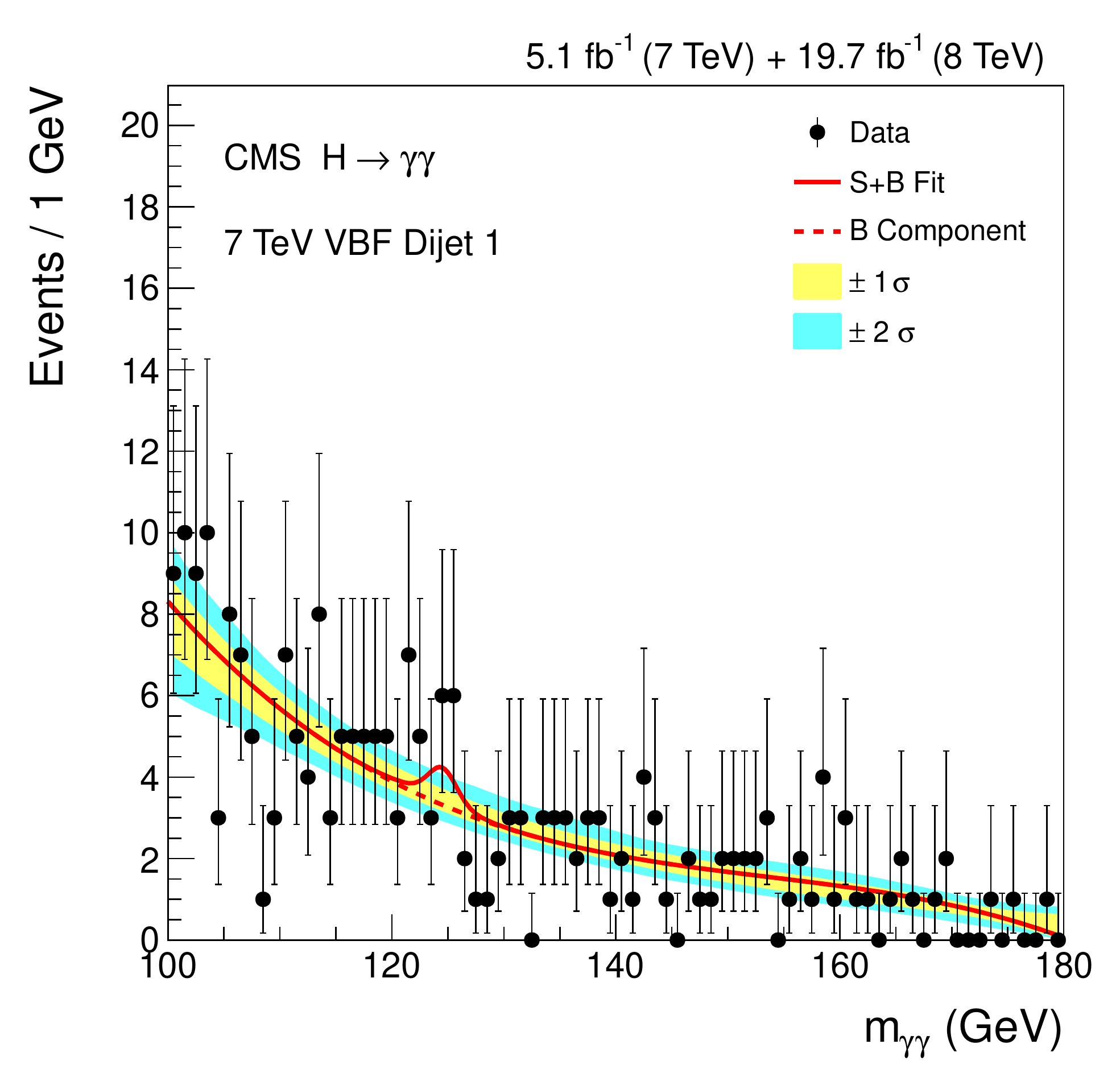}
   \end{center}
   \caption{The observed diphoton mass spectra of the \textit{VBF} tagged classes for the $7~\mathrm{TeV}$ dataset (points) binned in 1 GeV steps. For each class, the signal plus background model (solid red line), at the best-fit $\hat{\mu}_{H}$ = 1.12 and $\hat{m}_H$ = $\mathrm{124.72~GeV}$ associated with the signal model $S(m_{\gamma\gamma}|\mu_{H},m_{H})$ for the combined $\mathrm{7~TeV}$ and $\mathrm{8~TeV}$ datasets, is shown. The background component for the fit (dashed red line), the 68.3$\%$ (1 $\sigma$) confidence band (yellow) and the 95.4$\%$ (2 $\sigma$) confidence band (cyan) are also shown.}
   \label{fig:sbmass vbf 7TeV}
 \end{figure}

\begin{figure}[hbpt] 
   \begin{center}
     \includegraphics[width=0.415\textwidth]{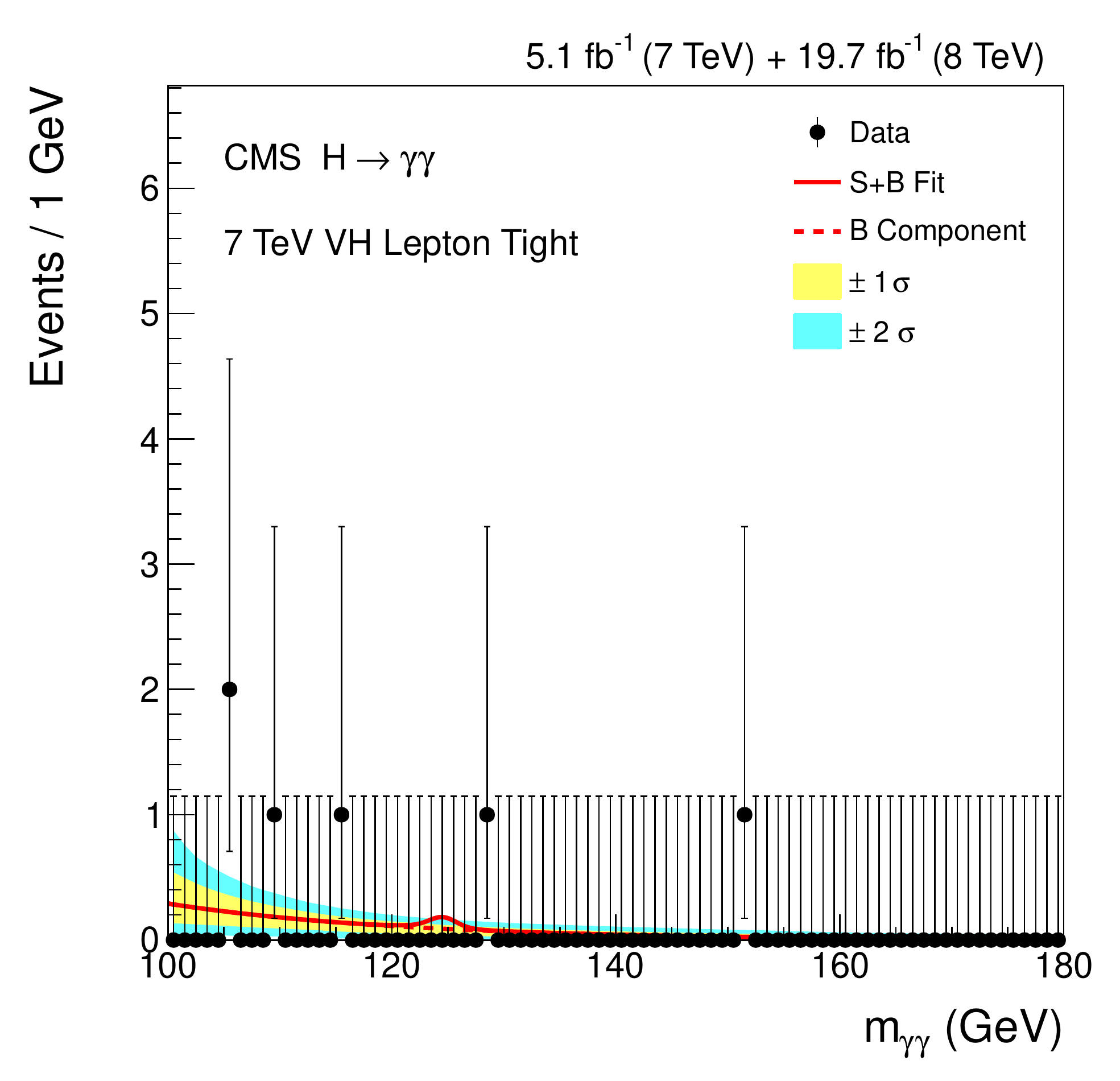}
     \includegraphics[width=0.415\textwidth]{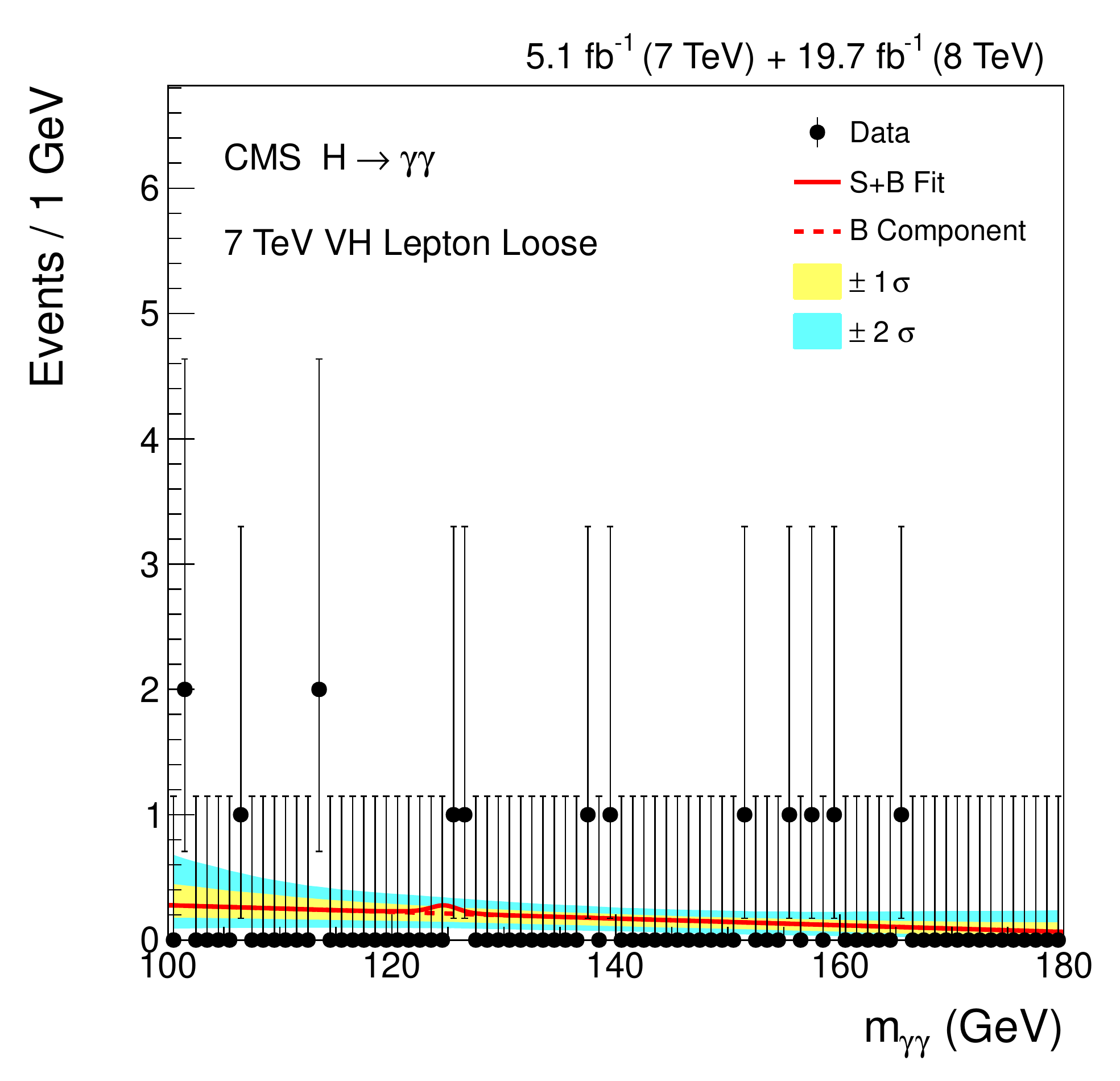}
     \includegraphics[width=0.415\textwidth]{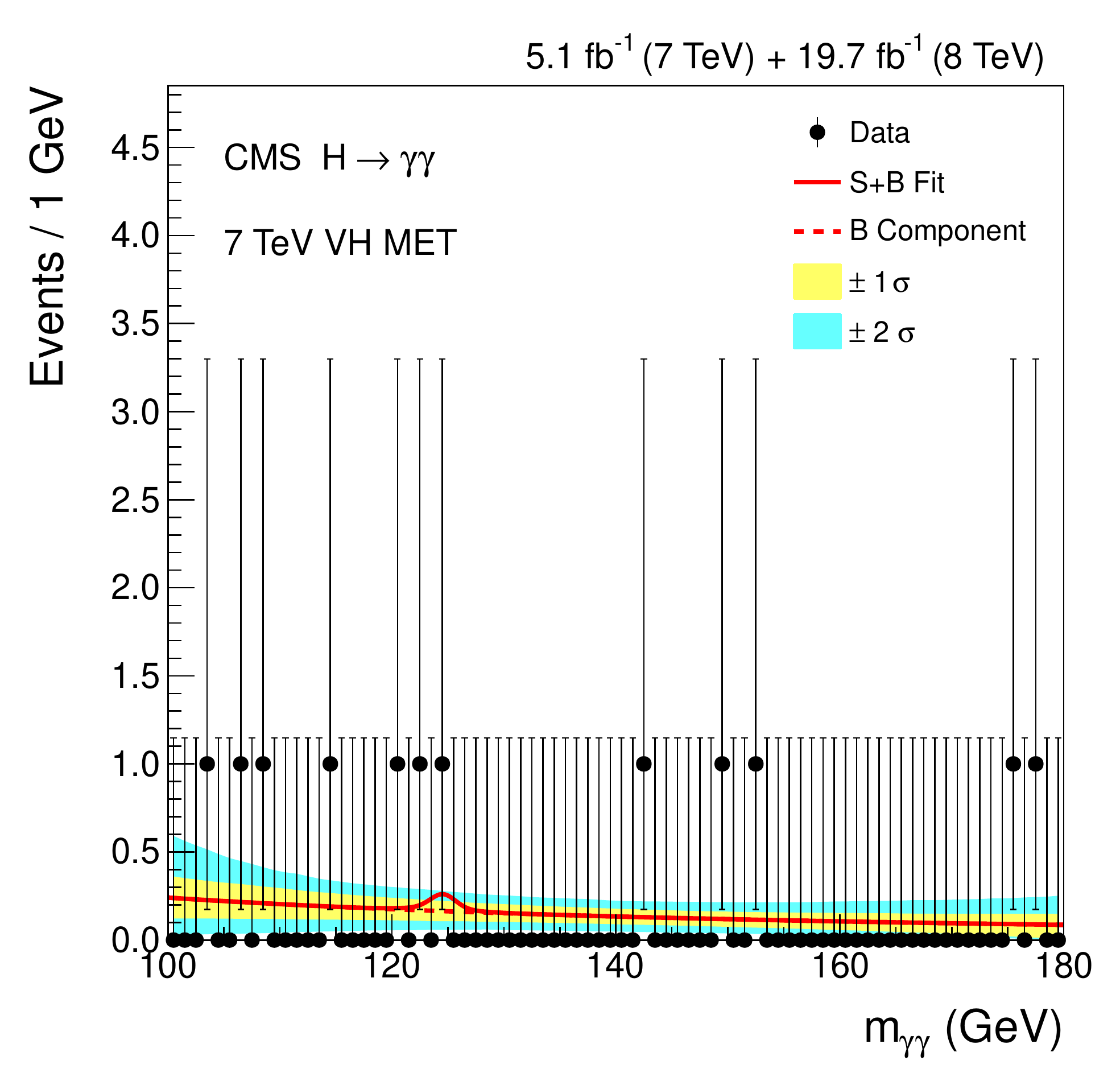}
     \includegraphics[width=0.415\textwidth]{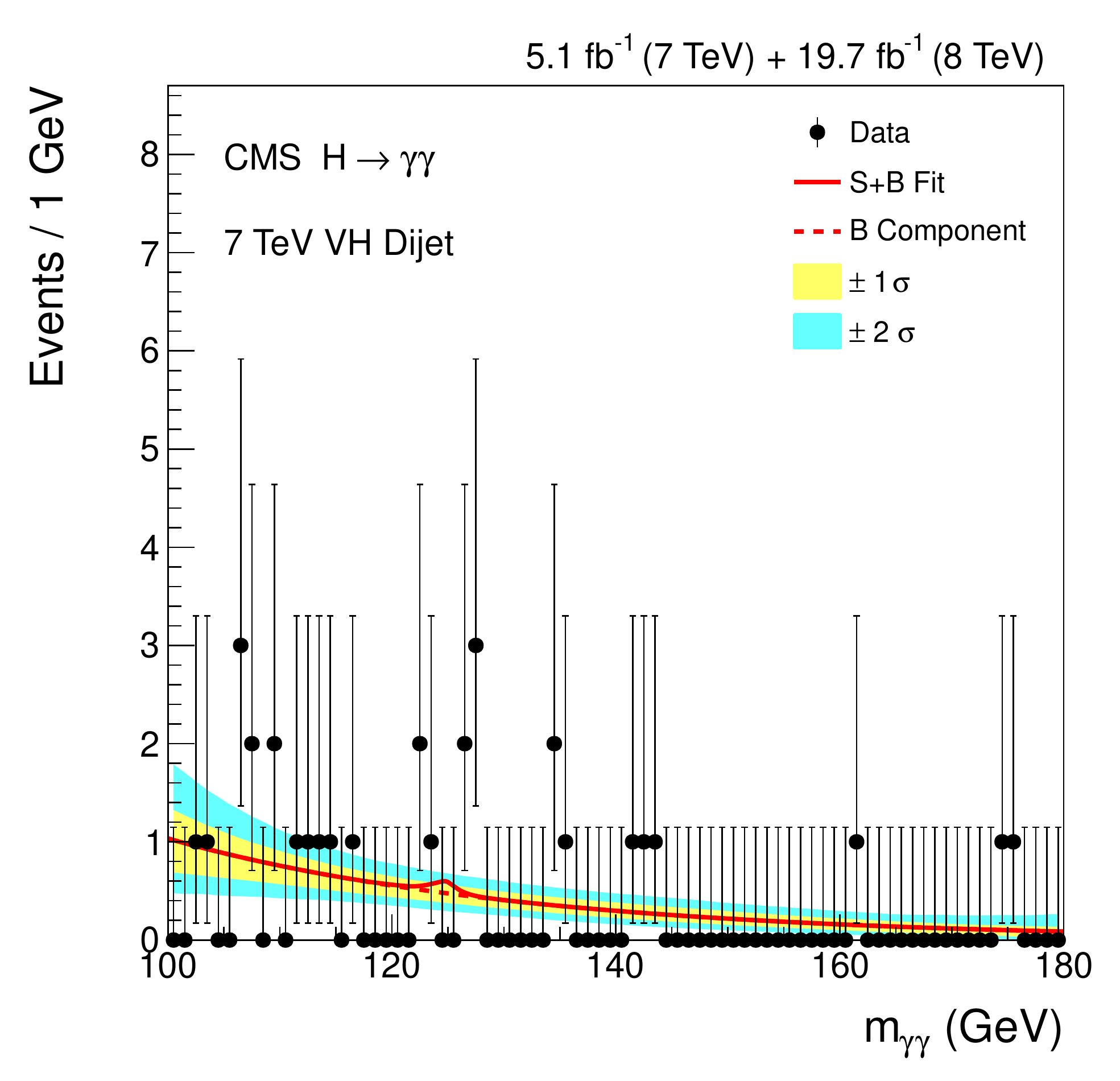}
     \includegraphics[width=0.415\textwidth]{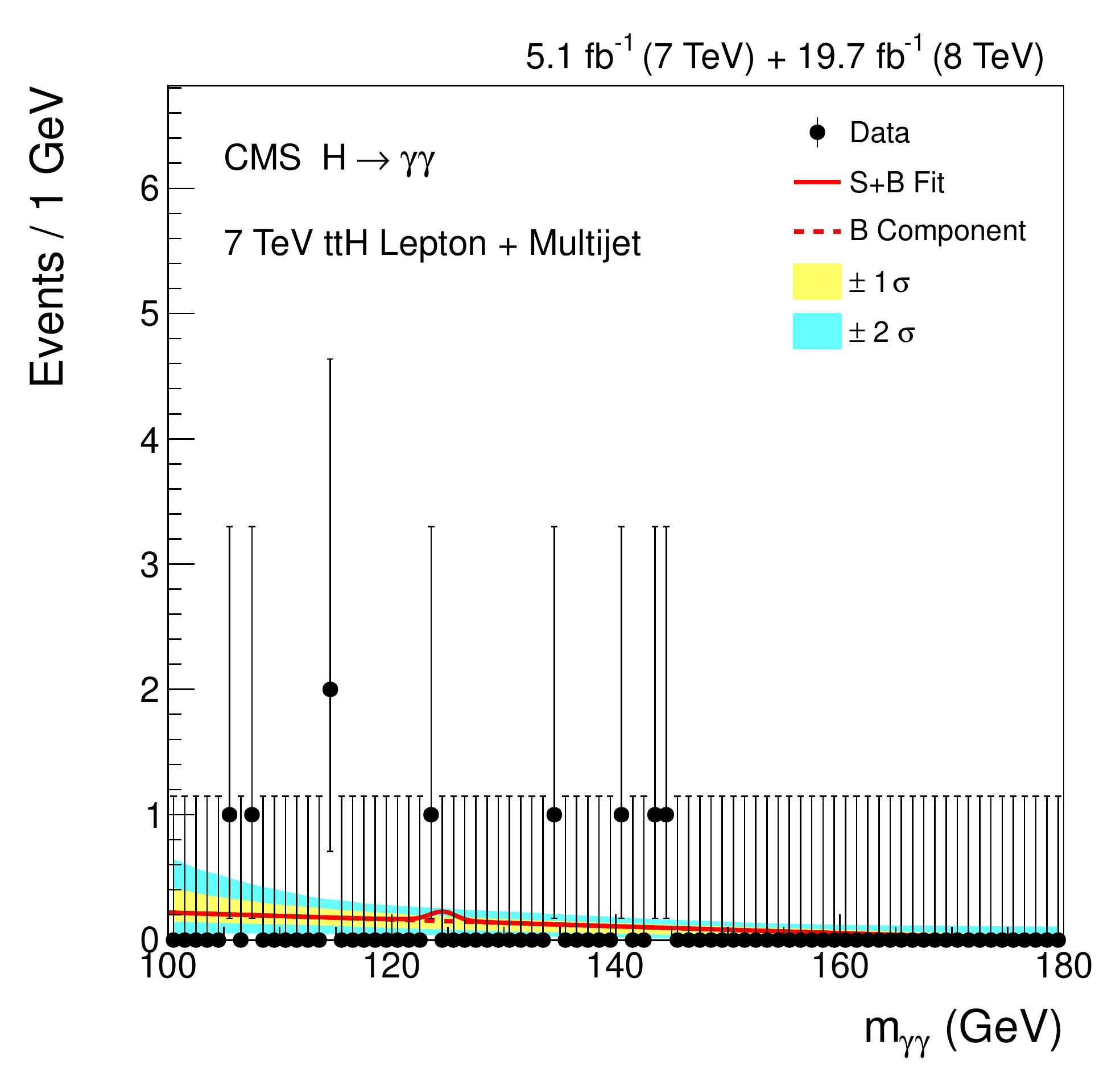}
   \end{center}
   \caption{The observed diphoton mass spectra of the \textit{VH} and \textit{t$\overline{t}$H} tagged classes for the $7~\mathrm{TeV}$ dataset (points) binned in 1 GeV steps. For each class, the signal plus background model (solid red line), at the best-fit $\hat{\mu}_{H}$ = 1.12 and $\hat{m}_H$ = $\mathrm{124.72~GeV}$ associated with the signal model $S(m_{\gamma\gamma}|\mu_{H},m_{H})$ for the combined $\mathrm{7~TeV}$ and $\mathrm{8~TeV}$ datasets, is shown. The background component for the fit (dashed red line), the 68.3$\%$ (1 $\sigma$) confidence band (yellow) and the 95.4$\%$ (2 $\sigma$) confidence band (cyan) are also shown.}
   \label{fig:sbmass vh tth 7TeV}
 \end{figure}

\begin{figure}[hbpt] 
   \begin{center}
     \includegraphics[width=0.415\textwidth]{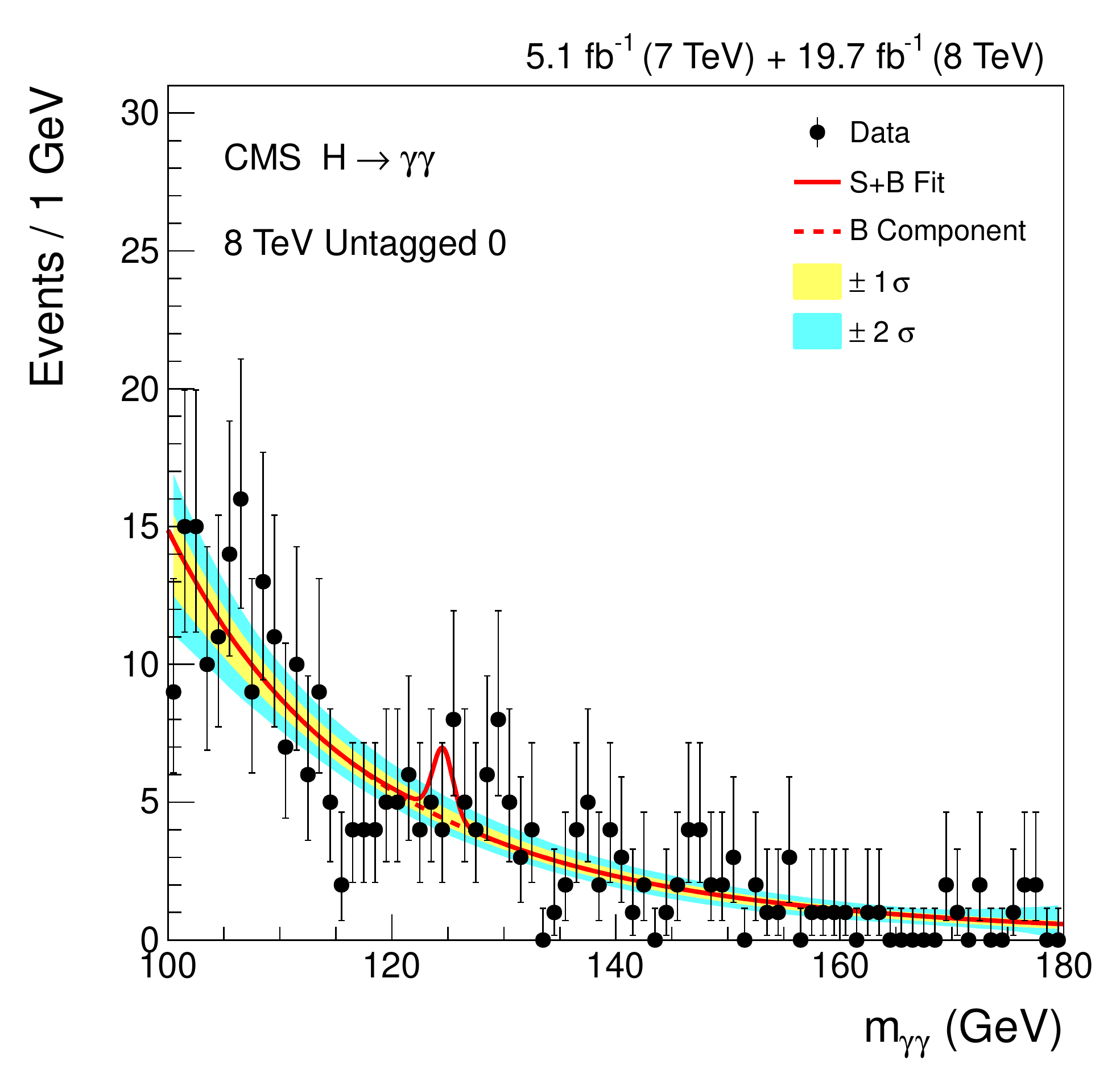}
     \includegraphics[width=0.415\textwidth]{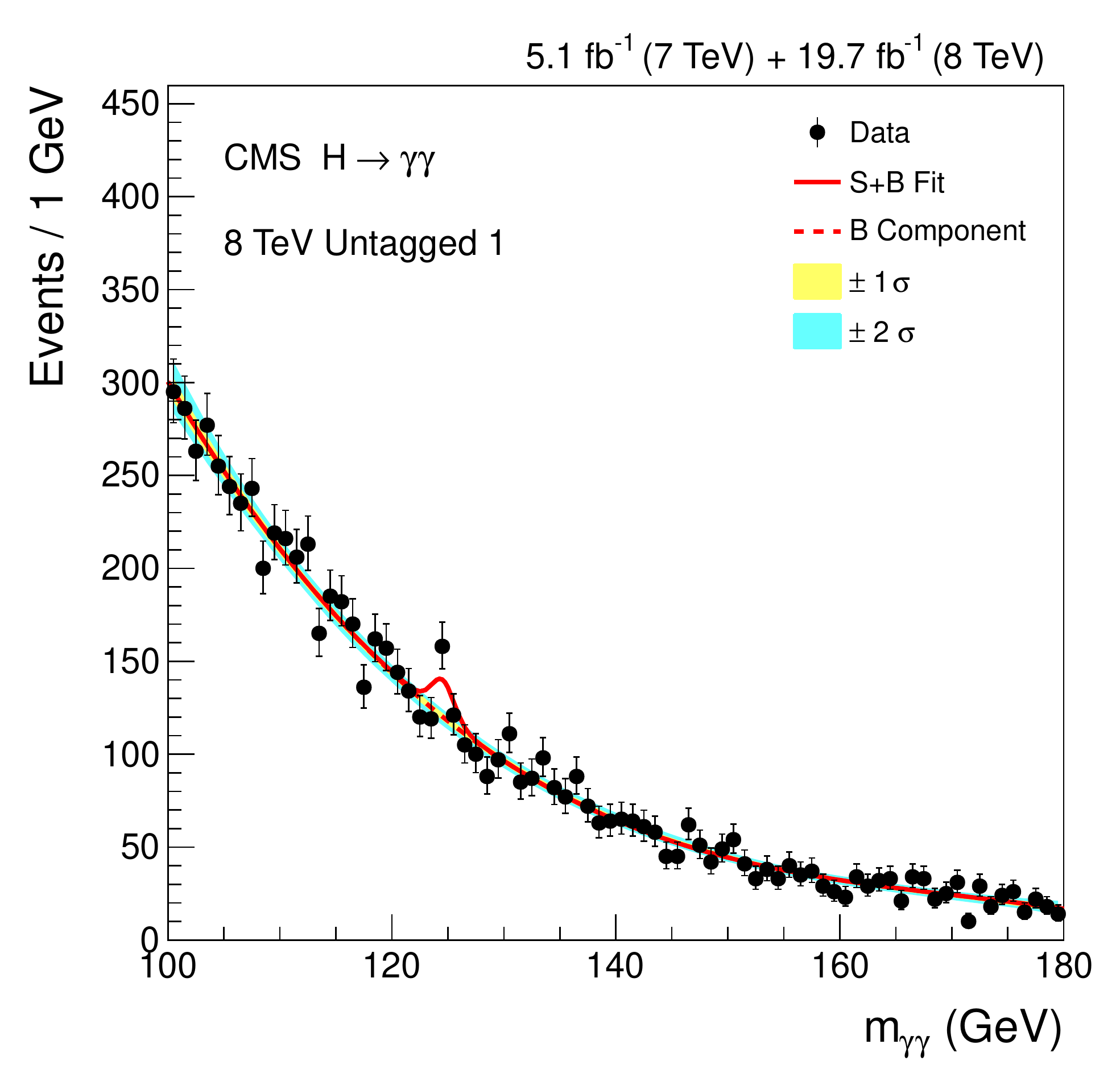}
     \includegraphics[width=0.415\textwidth]{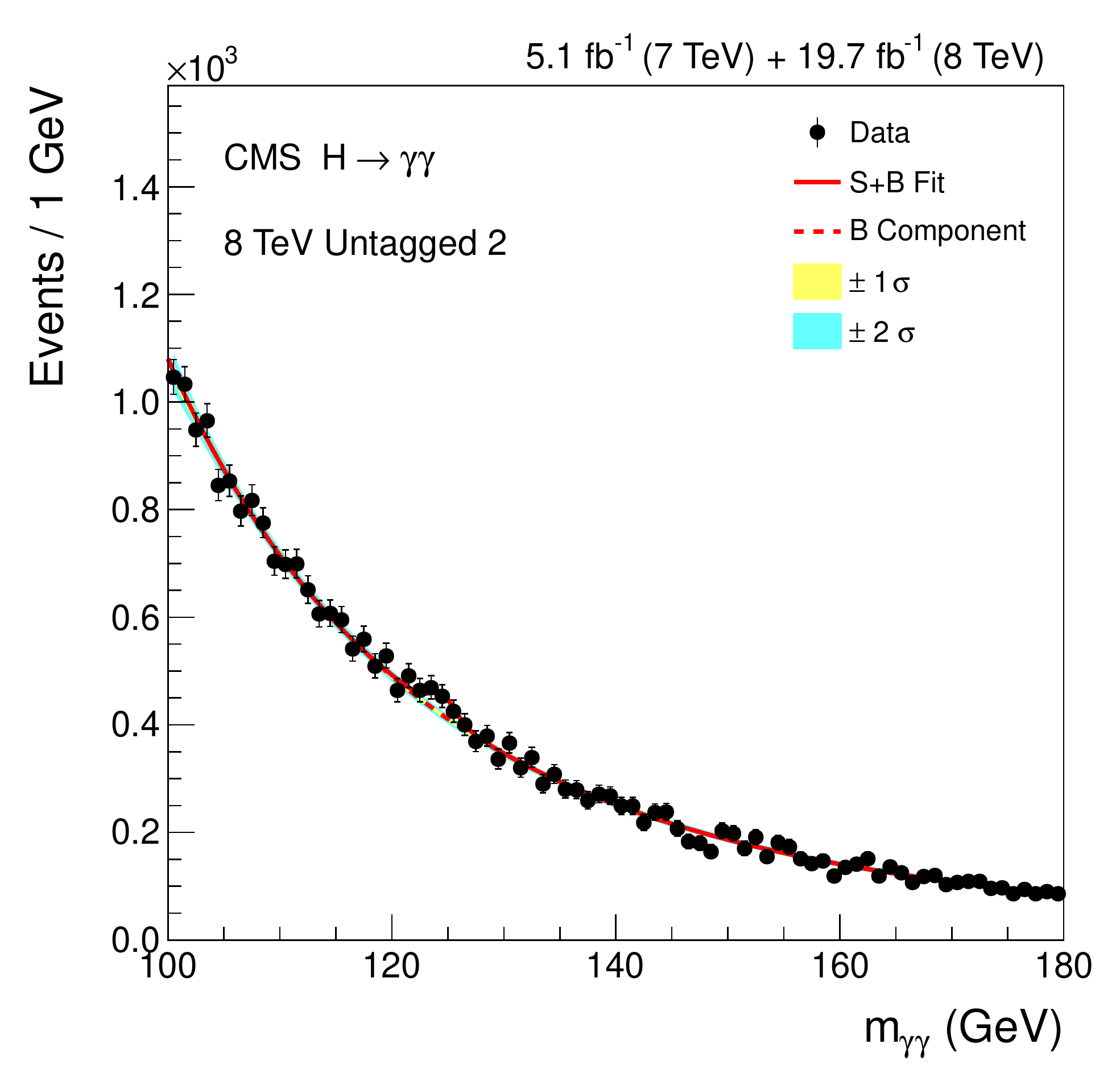}
     \includegraphics[width=0.415\textwidth]{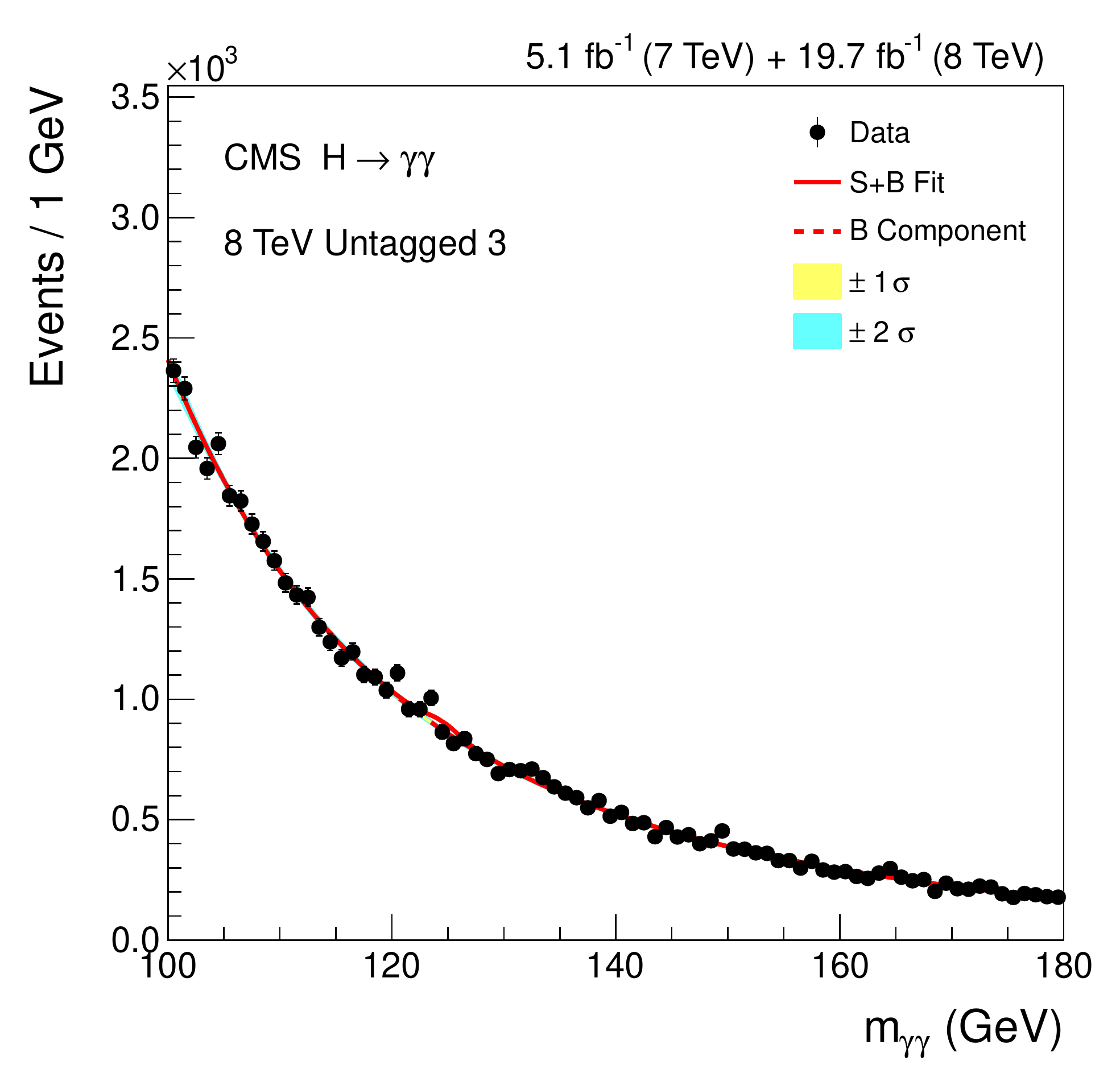}
     \includegraphics[width=0.415\textwidth]{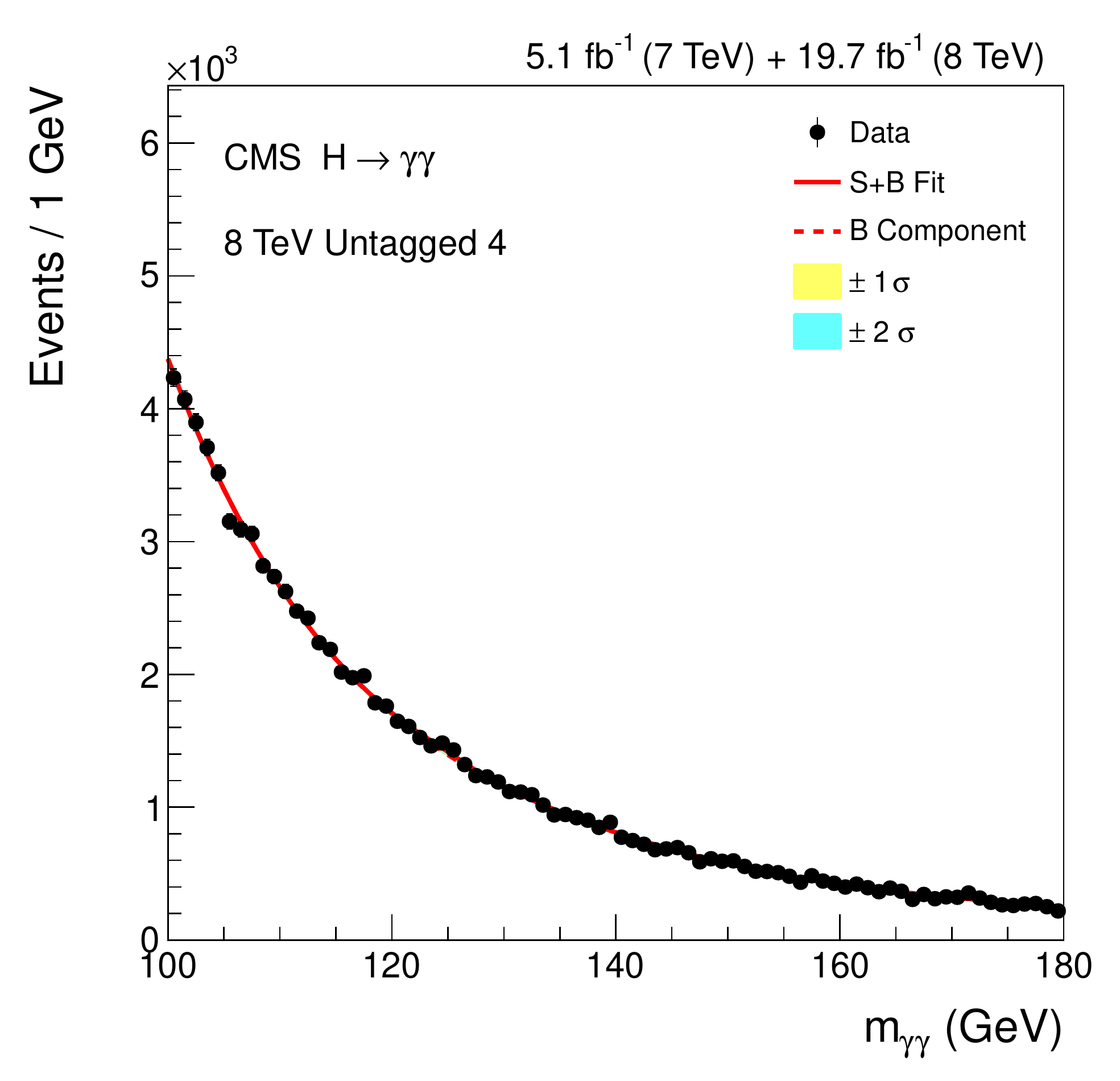}
   \end{center}
   \caption{The observed diphoton mass spectra of the untagged classes for the $8~\mathrm{TeV}$ dataset (points) binned in 1 GeV steps. For each class, the signal plus background model (solid red line), at the best-fit $\hat{\mu}_{H}$ = 1.12 and $\hat{m}_H$ = $\mathrm{124.72~GeV}$ associated with the signal model $S(m_{\gamma\gamma}|\mu_{H},m_{H})$ for the combined $\mathrm{7~TeV}$ and $\mathrm{8~TeV}$ datasets, is shown. The background component for the fit (dashed red line), the 68.3$\%$ (1 $\sigma$) confidence band (yellow) and the 95.4$\%$ (2 $\sigma$) confidence band (cyan) are also shown.}
   \label{fig:sbmass inclusive 8TeV}
 \end{figure}

\begin{figure}[hbpt] 
   \begin{center}
     \includegraphics[width=0.415\textwidth]{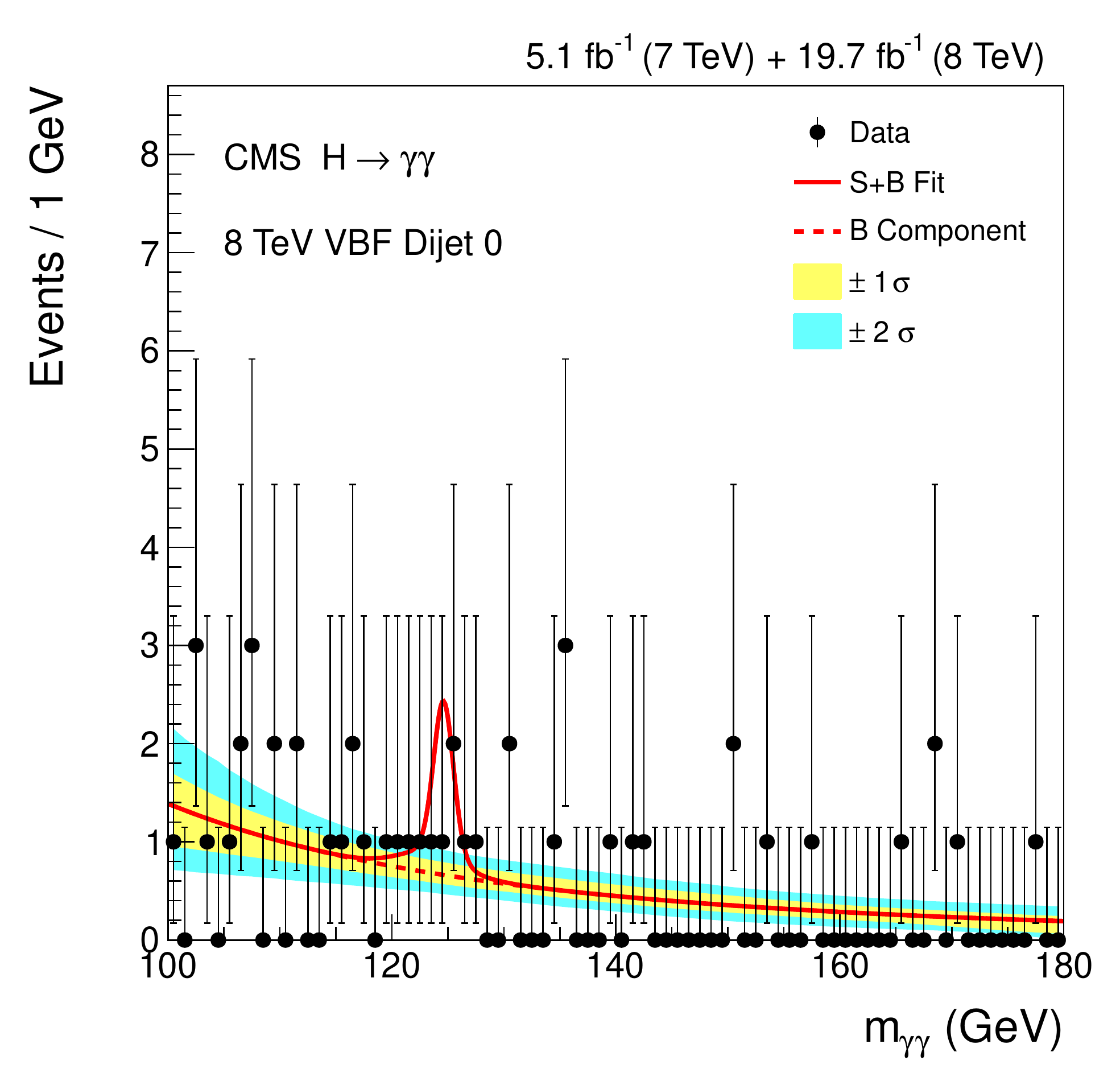}
     \includegraphics[width=0.415\textwidth]{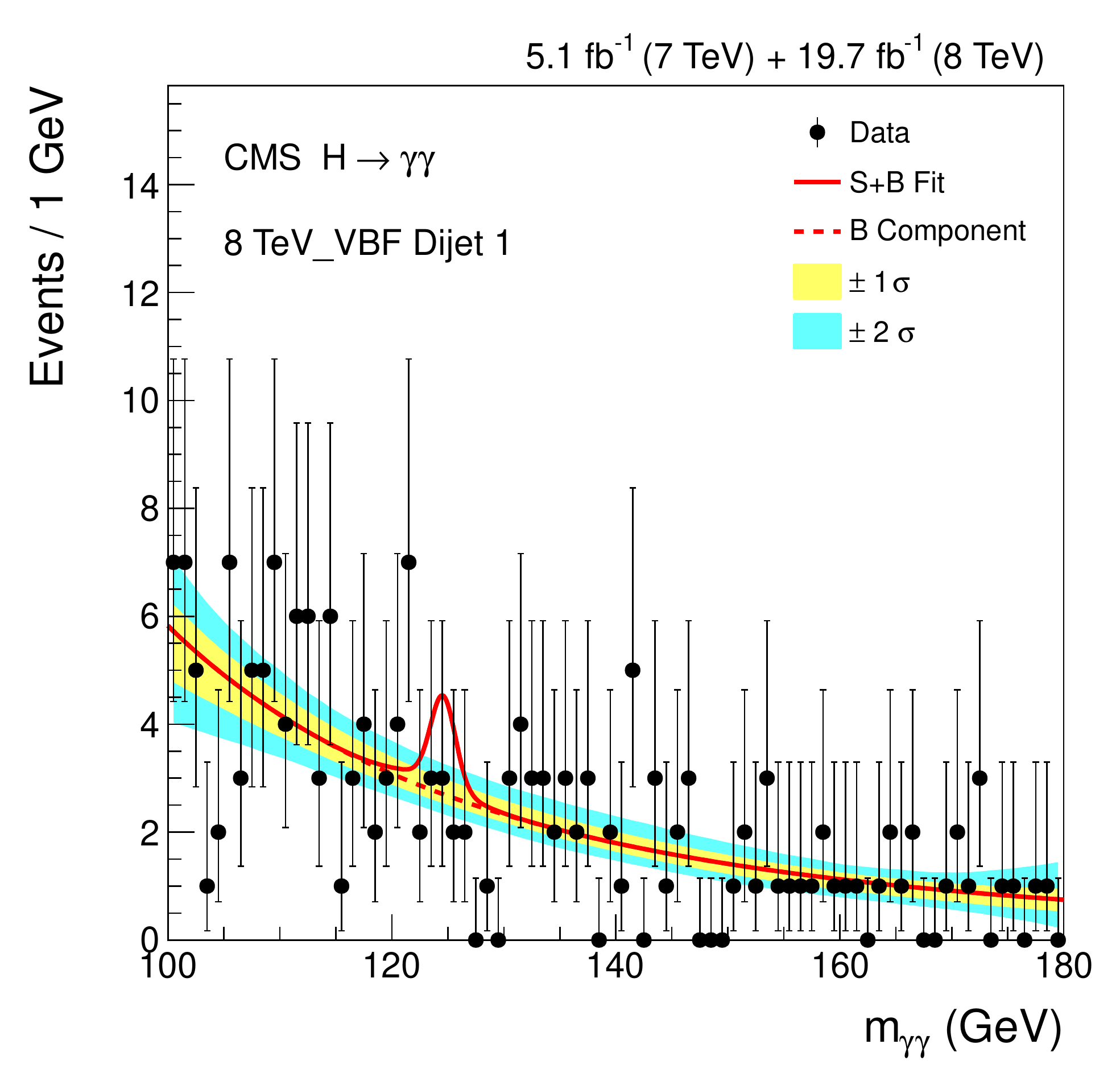}
     \includegraphics[width=0.415\textwidth]{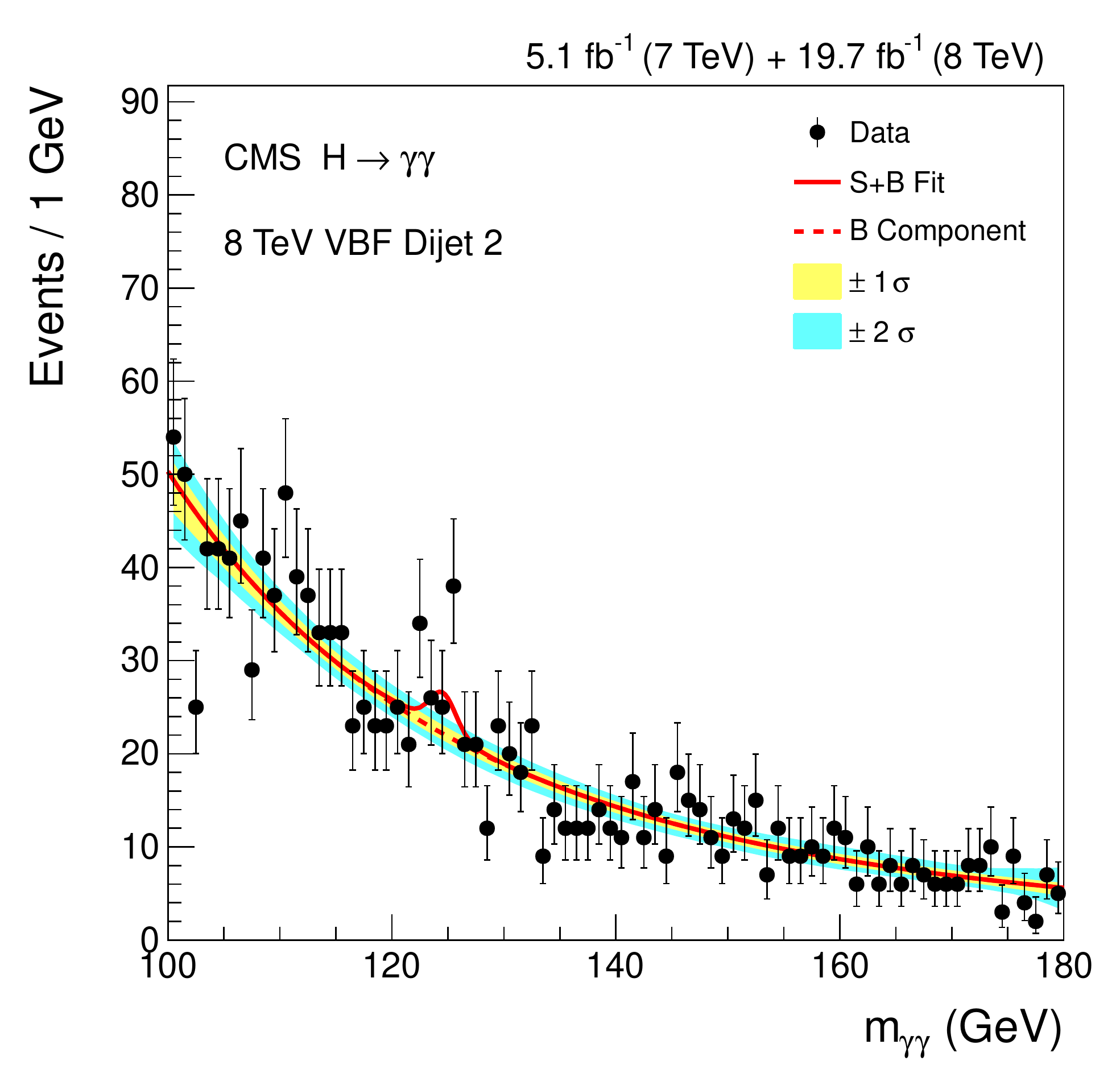}
   \end{center}
   \caption{The observed diphoton mass spectra of the \textit{VBF} tagged classes for the $8~\mathrm{TeV}$ dataset (points) binned in 1 GeV steps. For each class, the signal plus background model (solid red line), at the best-fit $\hat{\mu}_{H}$ = 1.12 and $\hat{m}_H$ = $\mathrm{124.72~GeV}$ associated with the signal model $S(m_{\gamma\gamma}|\mu_{H},m_{H})$ for the combined $\mathrm{7~TeV}$ and $\mathrm{8~TeV}$ datasets, is shown. The background component for the fit (dashed red line), the 68.3$\%$ (1 $\sigma$) confidence band (yellow) and the 95.4$\%$ (2 $\sigma$) confidence band (cyan) are also shown.}
   \label{fig:sbmass vbf 8TeV}
 \end{figure}

\begin{figure}[hbpt] 
   \begin{center}
     \includegraphics[width=0.415\textwidth]{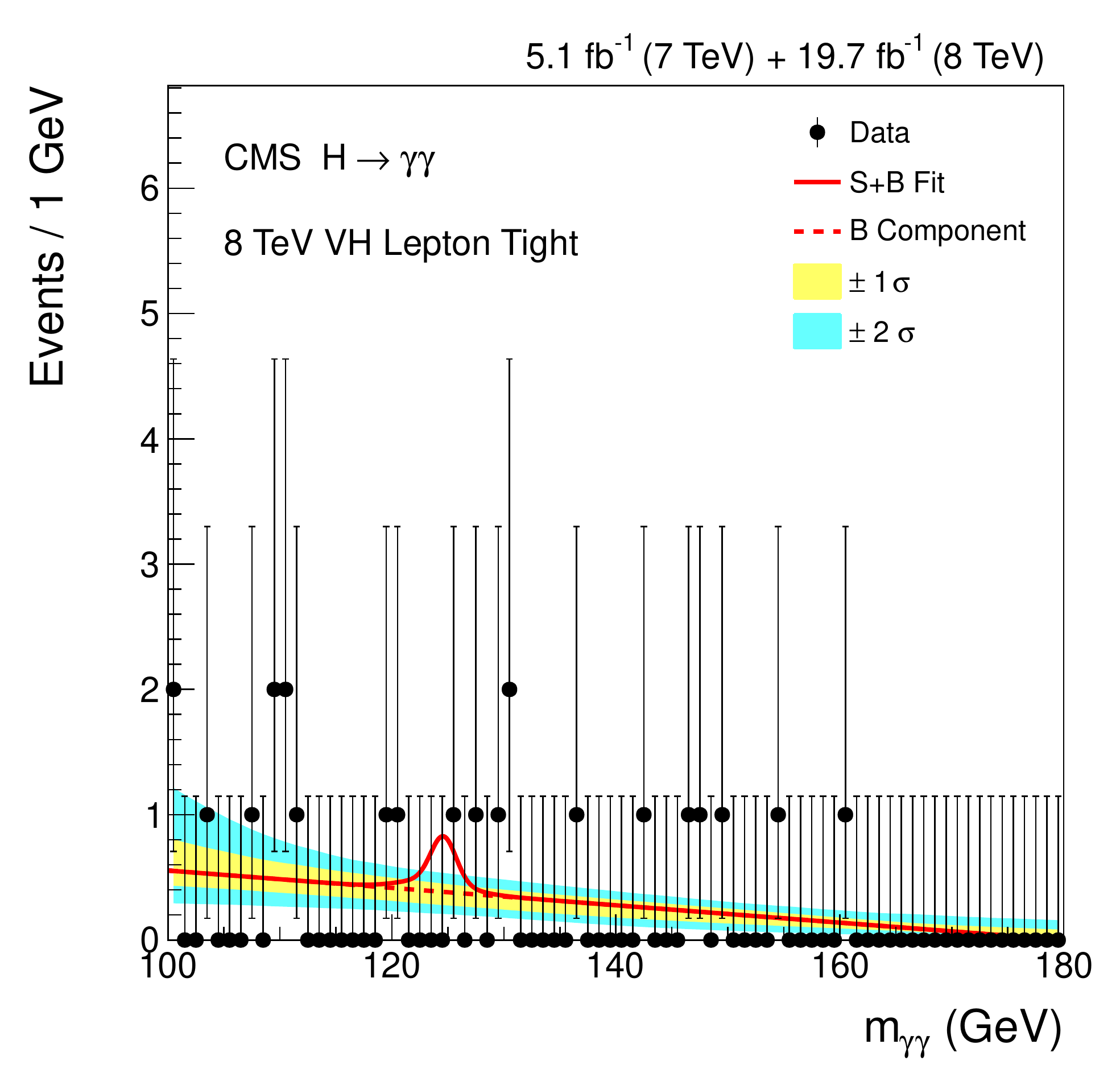}
     \includegraphics[width=0.415\textwidth]{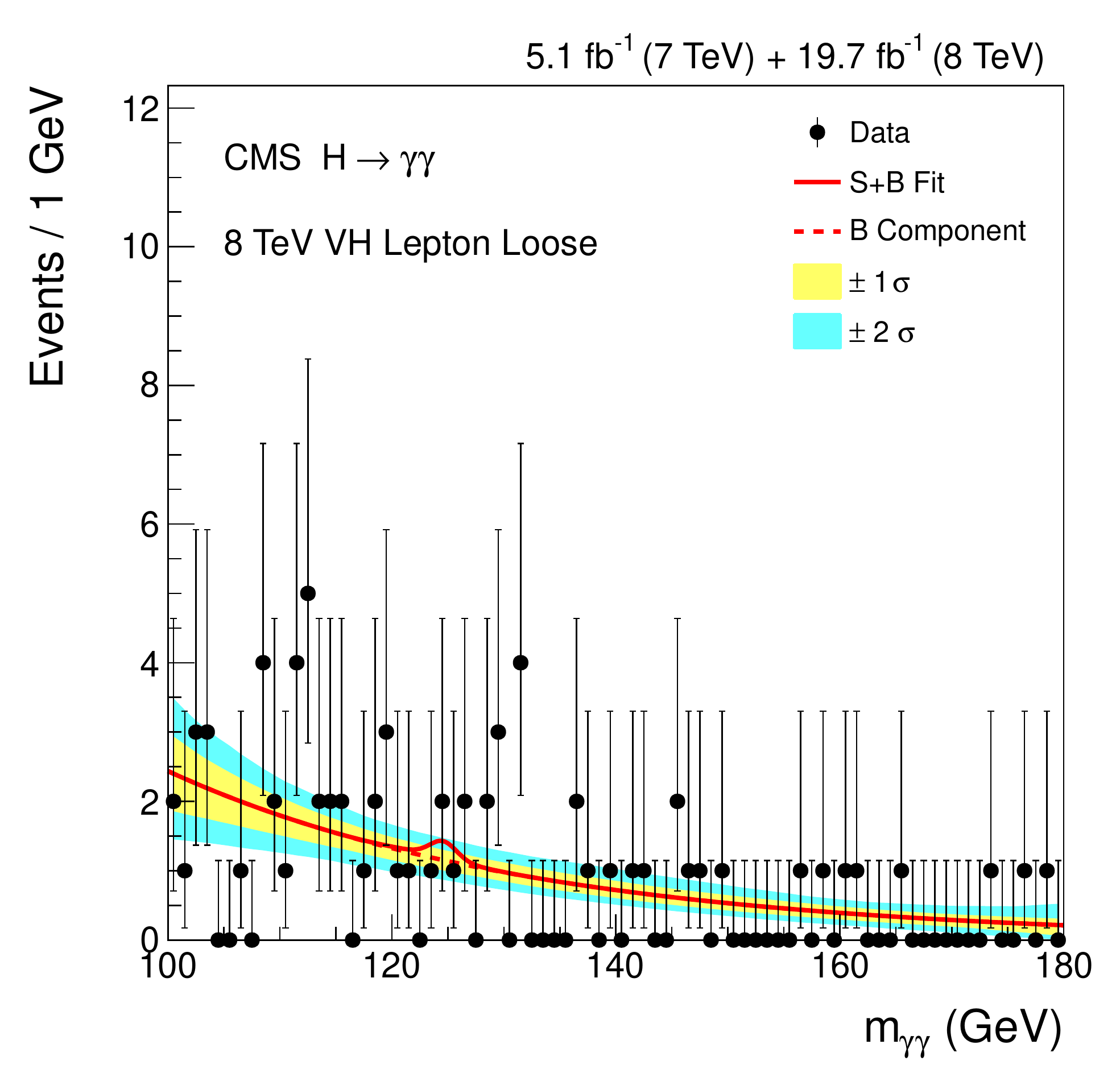}
     \includegraphics[width=0.415\textwidth]{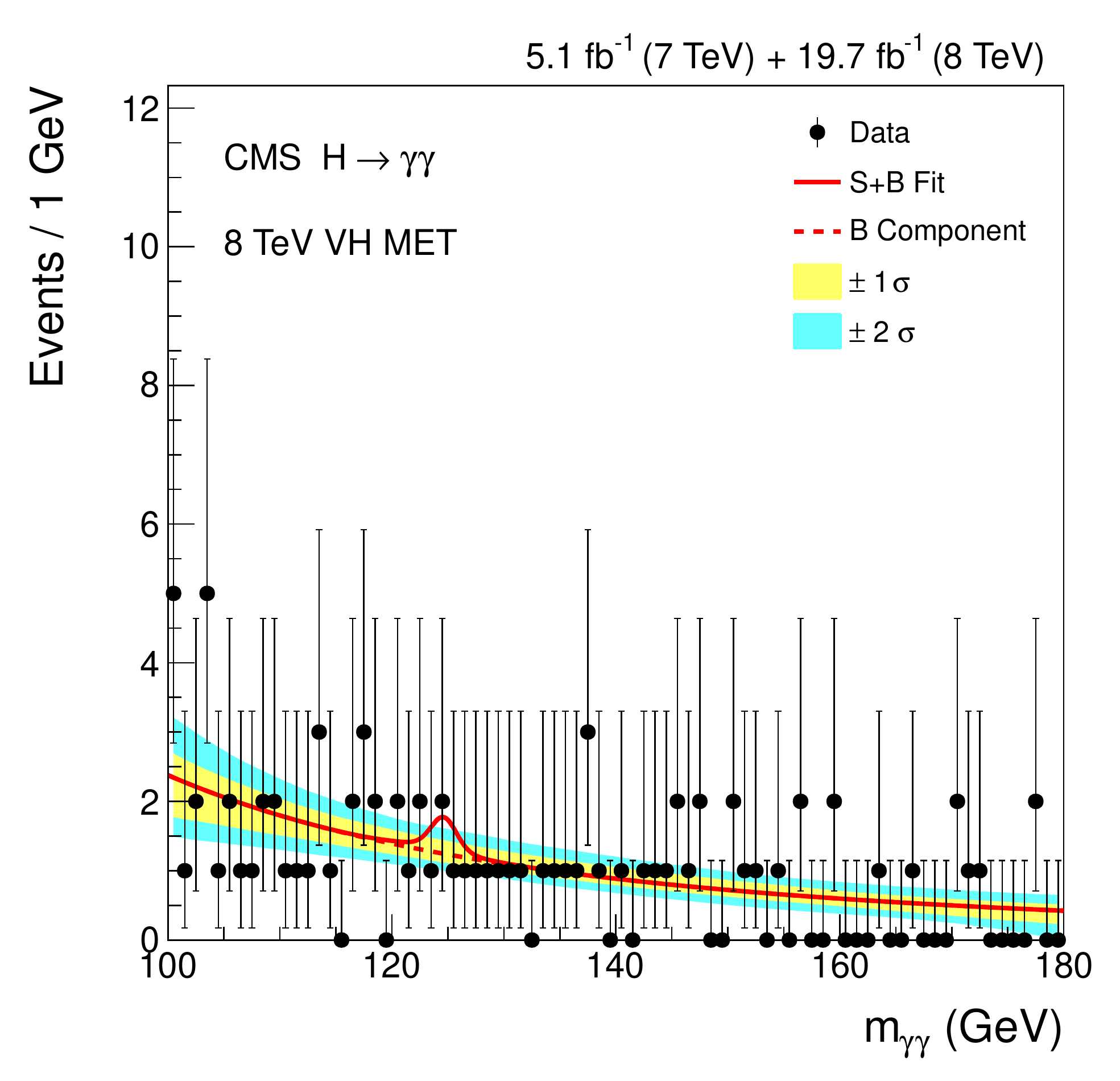}
     \includegraphics[width=0.415\textwidth]{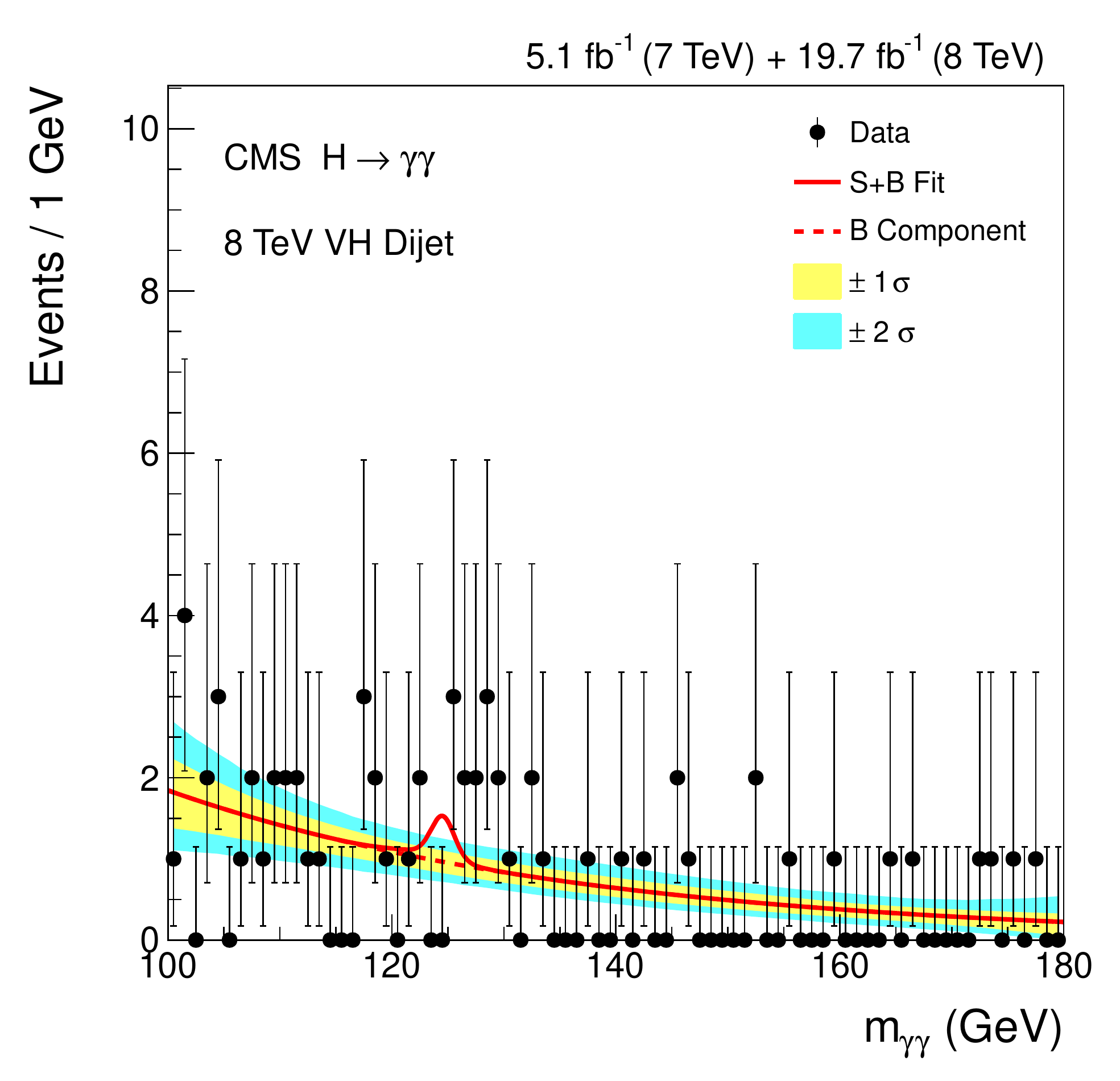}
     \includegraphics[width=0.415\textwidth]{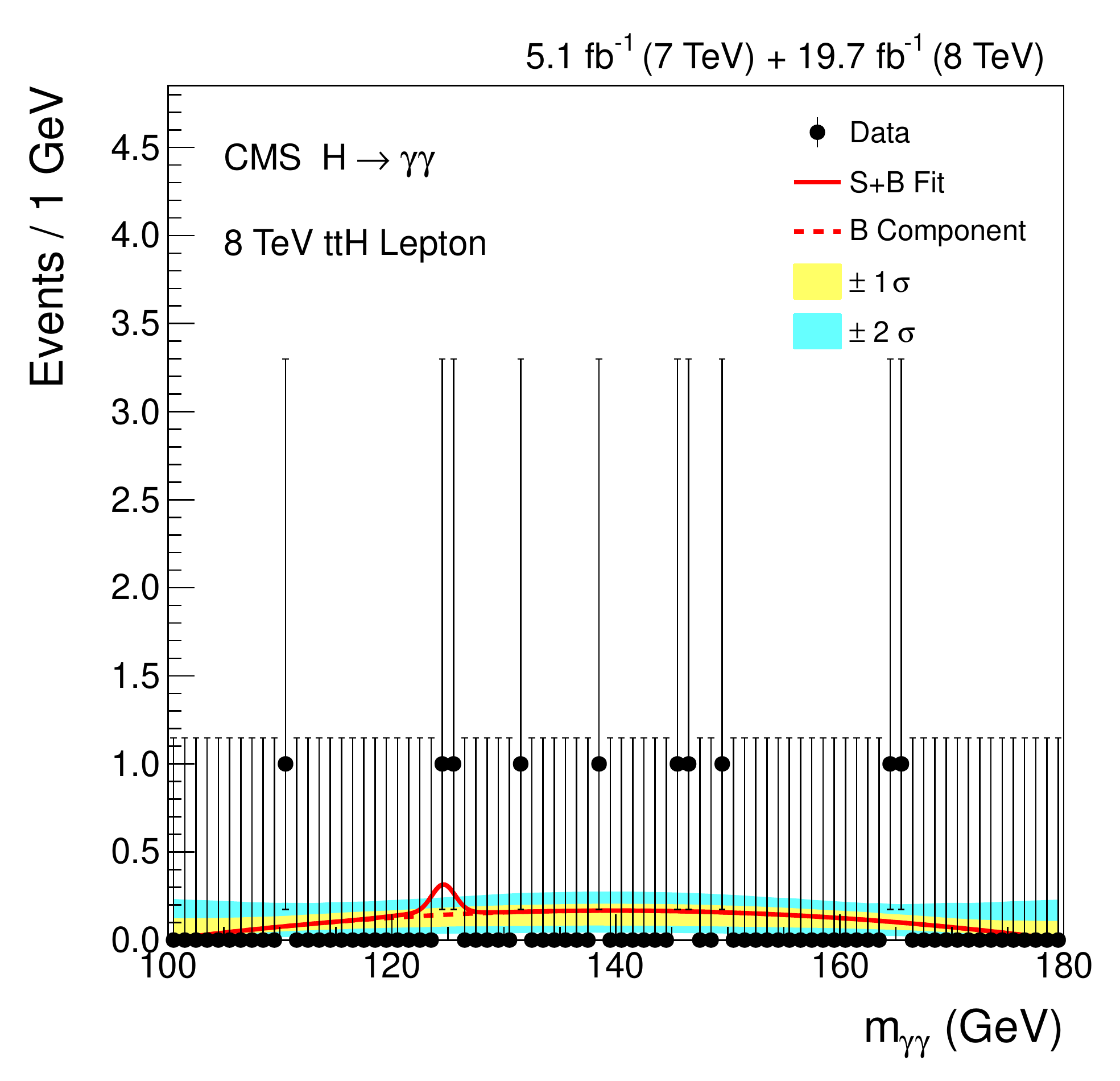}
     \includegraphics[width=0.415\textwidth]{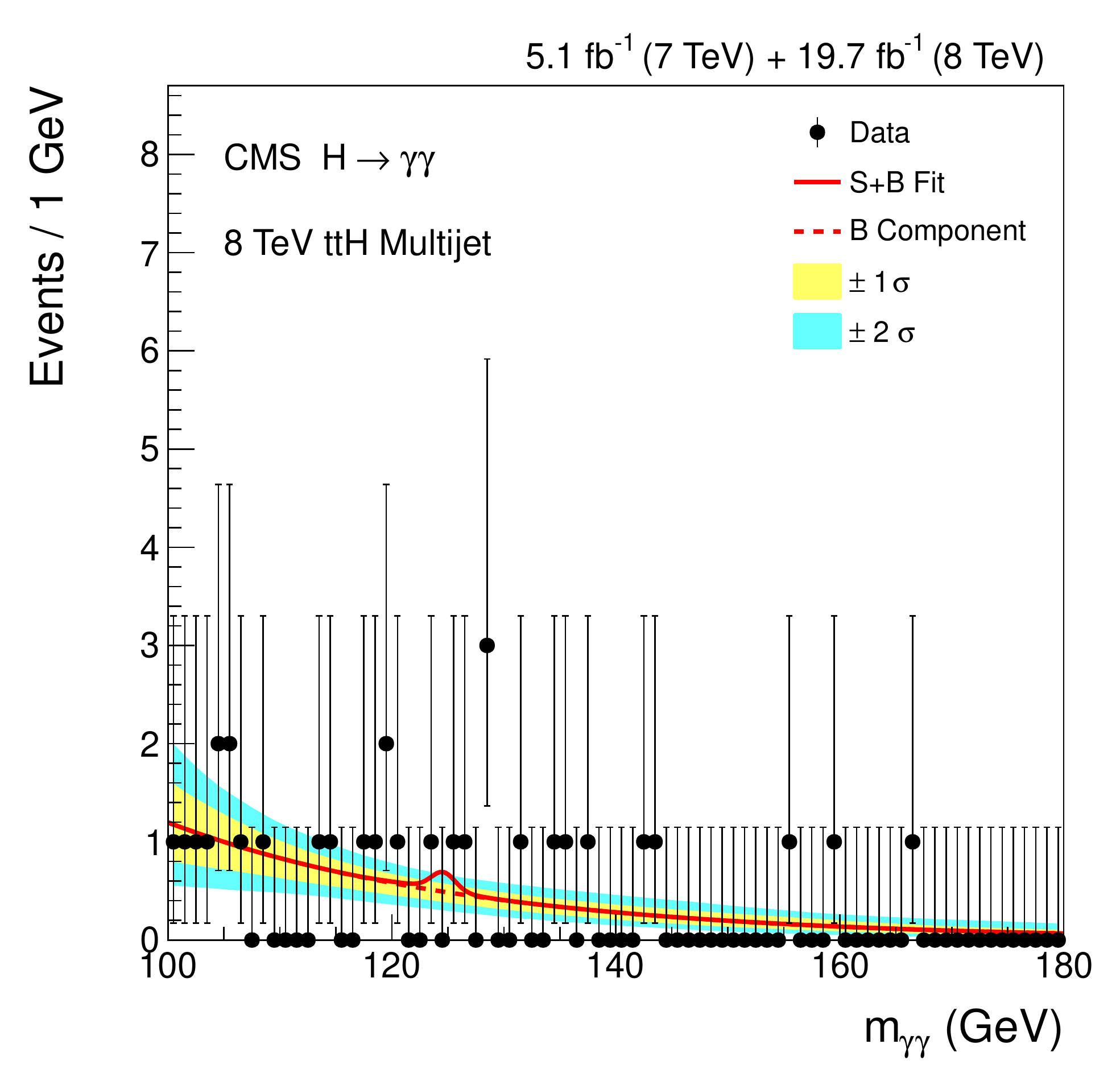}
   \end{center}
   \caption{The observed diphoton mass spectra of the \textit{VH} and \textit{t$\overline{t}$H} tagged classes for the $8~\mathrm{TeV}$ dataset (points) binned in 1 GeV steps. For each class, the signal plus background model (solid red line), at the best-fit $\hat{\mu}_{H}$ = 1.12 and $\hat{m}_H$ = $\mathrm{124.72~GeV}$ associated with the signal model $S(m_{\gamma\gamma}|\mu_{H},m_{H})$ for the combined $\mathrm{7~TeV}$ and $\mathrm{8~TeV}$ datasets, is shown. The background component for the fit (dashed red line), the 68.3$\%$ (1 $\sigma$) confidence band (yellow) and the 95.4$\%$ (2 $\sigma$) confidence band (cyan) are also shown.}
  \label{fig:sbmass vh tth 8TeV}
 \end{figure}

The observed diphoton mass spectra are shown in Figure \ref{fig:sbmass inclusive 7TeV}, Figure \ref{fig:sbmass vbf 7TeV} and Figure \ref{fig:sbmass vh tth 7TeV} for the $7~\mathrm{TeV}$ classes, and in Figure \ref{fig:sbmass inclusive 8TeV}, Figure \ref{fig:sbmass vbf 8TeV} and Figure \ref{fig:sbmass vh tth 8TeV} for the $8~\mathrm{TeV}$ classes. A Higgs signal-like excess is observed and quantified through the simultaneous signal plus background fit, using the signal model $S(m_{\gamma\gamma}|\mu_{H},m_{H})$, to the diphoton mass spectra over all event classes. The corresponding best-fit values of the signal strength and the Higgs mass are $\hat{\mu}_{H}$ = 1.12 and $\hat{m}_H$ = 124.72 GeV. For each event class, the signal plus background model at the best-fit (solid red line) is shown. The background component for the fit (dashed red line), along with the 68.3$\%$ (1 $\sigma$) confidence band (yellow) and the 95.4$\%$ (2 $\sigma$) confidence band (cyan) for the expected number of background events from the fit, is shown as well. 

More information for each event class, including the expected number of background events $\mathrm{per~GeV}$ ($dB/dm_{\gamma\gamma}$) at 125 GeV, and the expected $S/B$ and $S/\sqrt{B}$ at $m_{H}$ = $\mathrm{125~GeV}$, is presented in Table \ref{tab:sigbkg}, where the number of background events under the signal peak, $B$, is estimated as $dB/dm_{\gamma\gamma}$ at 125 GeV multiplied by 4 $\sigma_{eff}$. The $S/B$ is higher for the tagged classes than for the untagged classes in general, and decreases with the increase of the class number for the untagged classes, as expected. The $S/\sqrt{B}$ provides a measure of the signal sensitivity of each event class, according to which the 8 TeV untagged 2 class is the most sensitive class though not the one with the highest $S/B$, as a result of its relatively large signal yield.         

\begin{figure}[hbpt] 
   \begin{center}
     \includegraphics[width=0.6\textwidth]{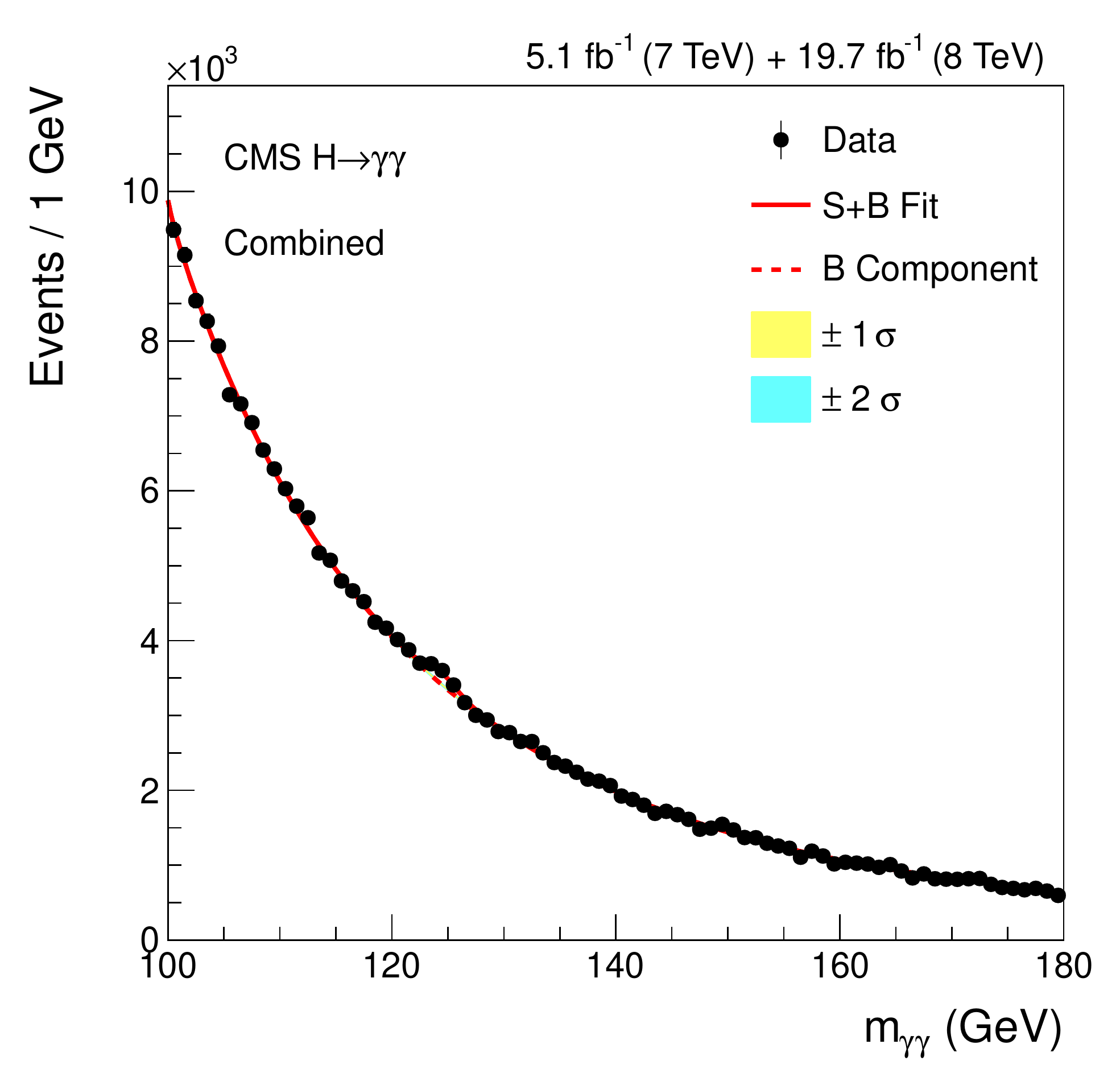}
   \end{center}
   \caption{The sum of the observed diphoton mass spectra of all the event classes for the $7~\mathrm{TeV}$ and $8~\mathrm{TeV}$ datasets (points) binned in 1 GeV steps. The corresponding signal plus background model (solid red line), obtained by summing the signal plus background models of all the event classes according to their fractions of the total number of events, is shown. The models correspond to the best-fit $\hat{\mu}_{H}$ = 1.12 and $\hat{m}_H$ = $\mathrm{124.72~GeV}$ associated with the signal model $S(m_{\gamma\gamma}|\mu_{H},m_{H})$ for the combined $\mathrm{7~TeV}$ and $\mathrm{8~TeV}$ datasets. The background component for the combined model (dashed red line), the 68.3$\%$ (1 $\sigma$) confidence band (yellow) and the 95.4$\%$ (2 $\sigma$) confidence band (cyan) are also shown.}
  \label{fig:sbmass combined}
 \end{figure}

The combined diphoton mass spectrum, with the corresponding signal plus background model, of all the $7~\mathrm{TeV}$ and $8~\mathrm{TeV}$ event classes is shown in Figure \ref{fig:sbmass combined}. The combined signal plus background model is obtained by summing the best-fit signal plus background models of all the event classes according to their fractions of the total number of events. The signal peak is not obvious because the signals in the high $S/B$ classes are submerged by mixing with large number of background events from the low $S/B$ classes. This is the reason that we classify events according to $S/B$ and extract the signal by simultaneous fit to the diphoton mass spectra over all event classes, instead of fitting an combined diphoton mass spectrum, in order to achieve the best signal sensitivity.   

\begin{figure}[hbpt] 
   \begin{center}
     \includegraphics[width=1\textwidth]{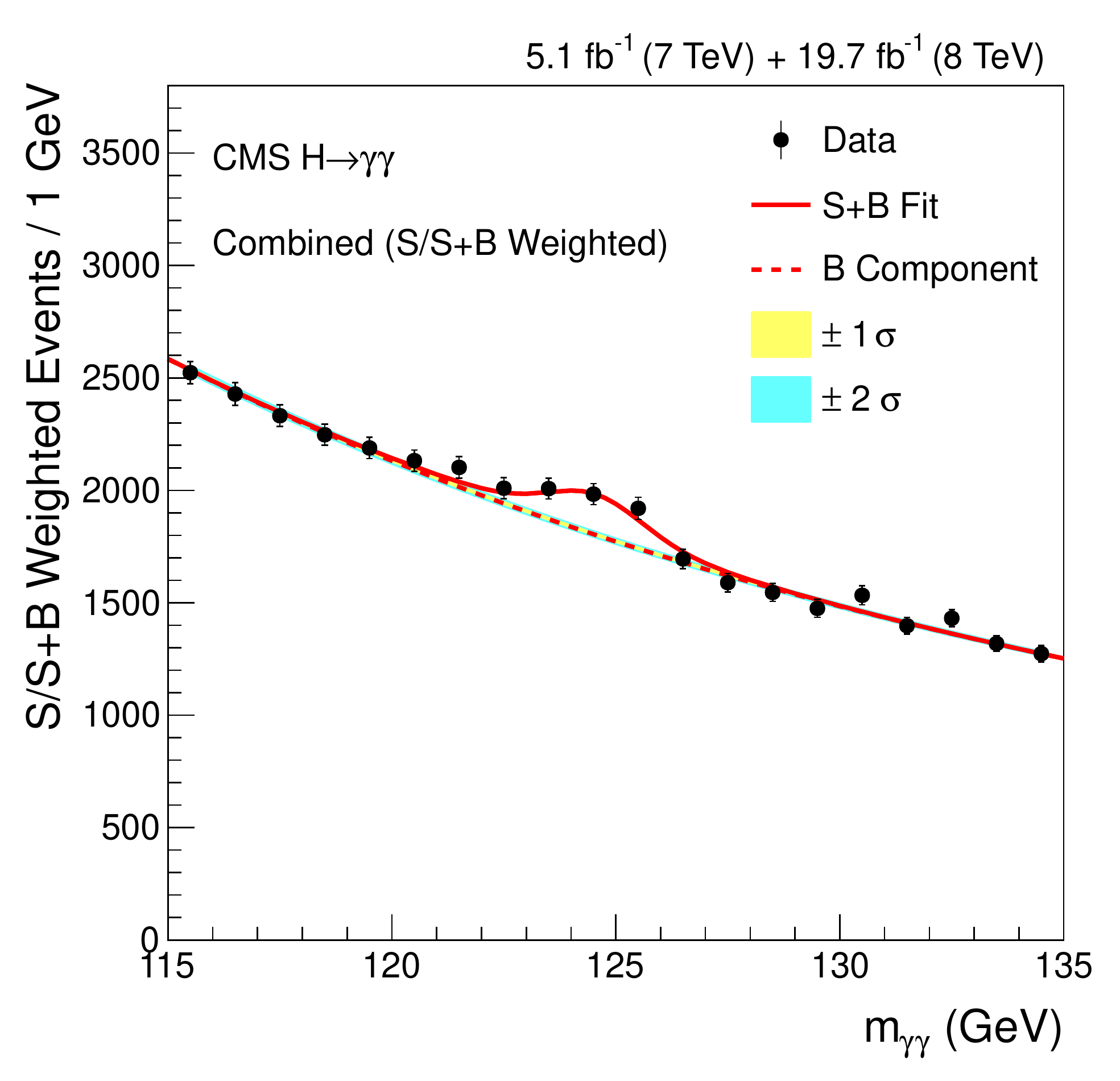}
   \end{center}
   \caption{The $S/S+B$ weighted sum of the observed diphoton mass spectra of all the event classes for the $7~\mathrm{TeV}$ and $8~\mathrm{TeV}$ datasets (points) binned in 1 GeV steps. The corresponding signal plus background model (solid red line), obtained by summing the signal plus background models of all the event classes according to their weighted fractions of the total number of events, is shown. The models correspond to the best-fit $\hat{\mu}_{H}$ = 1.12 and $\hat{m}_H$ = 124.72 GeV associated with the signal model $S(m_{\gamma\gamma}|\mu_{H},m_{H})$ for the combined $\mathrm{7~TeV}$ and $\mathrm{8~TeV}$ datasets. The background component for the weighted model (dashed red line), the 68.3$\%$ (1 $\sigma$) confidence band (yellow) and the 95.4$\%$ (2 $\sigma$) confidence band (cyan) are also shown.}
  \label{fig:sbmass combined weighted} 
 \end{figure}

The weighted version of the combined mass spectrum, with the corresponding signal plus background model, is shown in Figure \ref{fig:sbmass combined weighted}, which provides a better view of the observed signal-like excess. The data for each individual class is weighted by the ratio $S/(S+B)$, which is evaluated using the values of the signal model and background model at the best-fit signal strength and the Higgs mass. A normalization factor is applied such that the total number of fitted signal events keeps unchanged after the weighting. The signal plus background curve shown in the figure for the weighted spectrum is obtained by summing the best-fit signal plus background models of all the event classes according to their weighted fractions of the total number of events. The weighting is chosen according to the optimal signal extraction by fitting to the weighted diphoton mass spectrum\cite{Barlow:1986ek}. This weighting procedure estimates and visualizes the contribution of each event class in the simultaneous diphoton mass fit, though the fitting to the weighted diphoton mass spectrum is still not as optimal as the simultaneous fit used in this analysis for the signal extraction. 

\section{Local P-Value and Significance}
\begin{figure}[hbpt] 
  \begin{center}
    \includegraphics[width=1\textwidth]{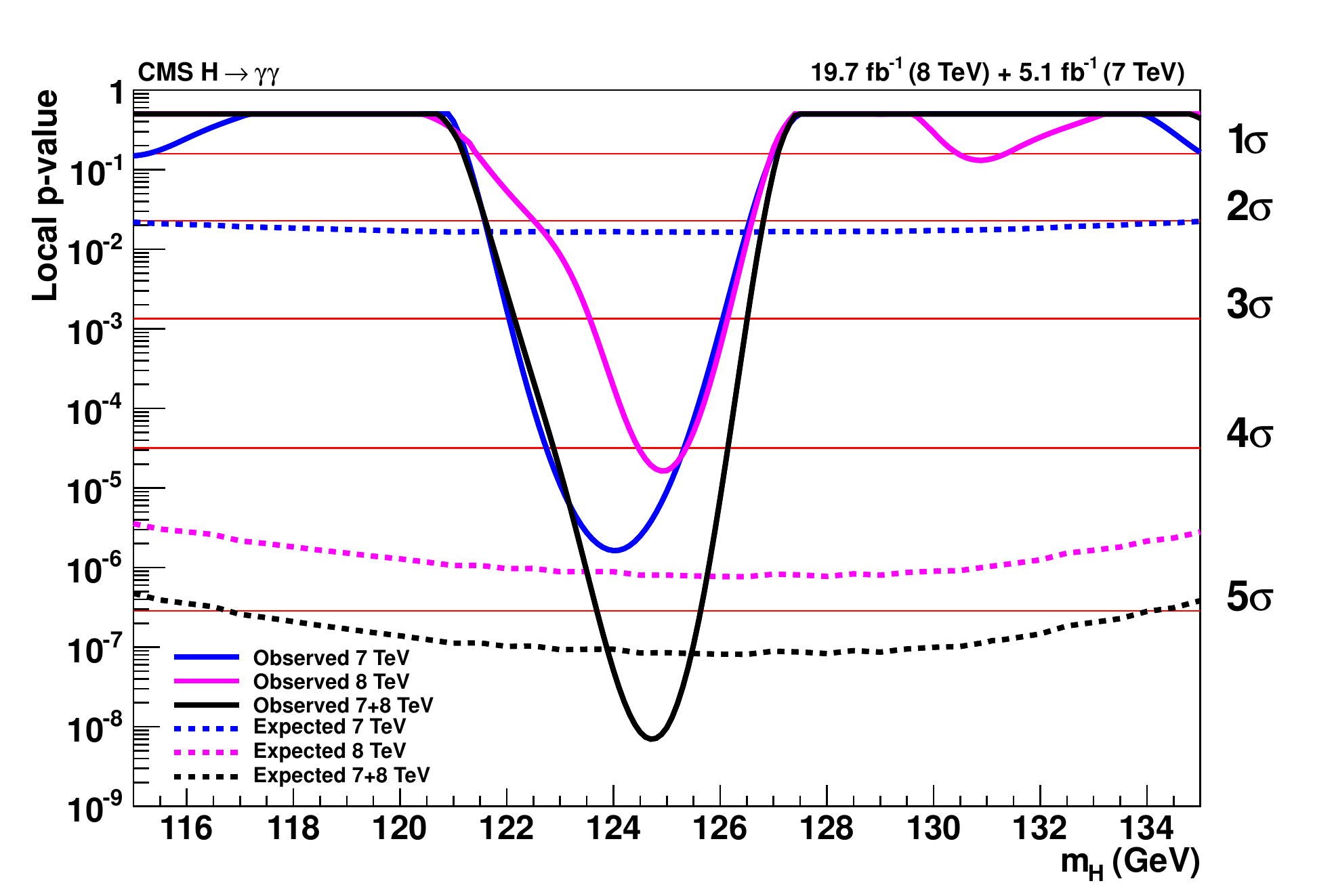}
   \end{center}
  \caption{The local p-value (left axis) of the background only hypothesis and the corresponding significance (right axis) of the excess against the Higgs hypotheses in the range 115 GeV $\leq$ $m_{H}$ $\leq$ 135 GeV. The observed values from the combined $7~\mathrm{TeV}$ and $8~\mathrm{TeV}$ datasets (solid black line), and the ones from separate $7~\mathrm{TeV}$ (solid blue line) and $8~\mathrm{TeV}$ (solid magenta line) datasets are shown. The corresponding expected values (dashed line) are also shown. The excess corresponds to a significance of 5.7 standard deviations.}
  \label{fig:significance}
\end{figure}

The local p-value of the background only hypothesis is scanned against the Higgs hypotheses in the range 115 GeV $\leq$ $m_{H}$ $\leq$ 135 GeV, in steps of 0.1 GeV. The observed local p-value and the corresponding significance of the excess as a function of $m_{H}$ for the combined $7~\mathrm{TeV}$ and $8~\mathrm{TeV}$ datasets (solid black line), and the ones for the separate $7~\mathrm{TeV}$ (solid blue line) and $8~\mathrm{TeV}$ (solid magenta line) datasets are shown in Figure \ref{fig:significance}. The corresponding expected local p-value and local significance (dashed lines) under the SM Higgs hypotheses are also shown. The expected values at each $m_{H}$ are evaluated on an Asimov dataset\cite{AsymptoticCLsFormulae}, a representative dataset following the expected distribution of the corresponding signal plus background model with $\mu_{H}=$ 1. For the generation of the Asimov dataset, the background model at the best-fit $\hat{\mu}_{H}$ and $\hat{m}_{H}$ are used and the systematic nuisance parameters for the signal model are also set to the values at the best-fit. 

The minimum observed local p-value from the combined $7~\mathrm{TeV}$ and $8~\mathrm{TeV}$ datasets is $7.0\cdot10^{-9}$ at $m_{H}=$ 124.7 GeV, which corresponds to an excess with a local significance of 5.7 standard deviations. This result, strongly disfavoring the background only hypothesis, leads to the observation of a new diphoton resonance---the conventional threshold for an observation in particle physics is 5.0 standard deviations. The expected local p-value for the Higgs at $m_{H}=$ 124.7 GeV is $8.5\cdot10^{-8}$, corresponding to an excess with a local significance of 5.2 standard deviations.

The observed and expected local significance at $m_{H}=$ 124.7 GeV for the $7~\mathrm{TeV}$, $8~\mathrm{TeV}$ and combined $7~\mathrm{TeV}$ and $8~\mathrm{TeV}$ datasets are summarized in Table \ref{tab:significance}.              

\begin{table}[hbtp]
  \noindent
  \small\addtolength{\tabcolsep}{-6pt}
  \caption{The observed and expected local significance $\sigma_{local}$ at $m_{H}$ $=$ 124.7 GeV.}
  \begin{center}
    \setlength{\tabcolsep}{20pt}
    \begin{tabular}{|l|c|c|c|} 
      \hline
      $\sigma_{local}$   & $7~\mathrm{TeV}$ & $8~\mathrm{TeV}$ & $7~\mathrm{TeV}$ + $8~\mathrm{TeV}$\\
      \hline
      Observed & 4.5 $\sigma$ & 4.1 $\sigma$ & 5.7 $\sigma$\\
      \hline
      Expected & 2.1 $\sigma$ & 4.8 $\sigma$ & 5.2 $\sigma$\\
      \hline
    \end{tabular}
    \label{tab:significance}
  \end{center}
\end{table}

\section{Overall Higgs Signal Strength}
The overall signal strength extracted from the combined $7~\mathrm{TeV}$ and $8~\mathrm{TeV}$ datasets is $\hat{\mu}_{H}$ $=$ $1.12_{-0.23}^{+0.26}$ at $\hat{m}_{H}$ $=$ 124.72 GeV, where the upper and lower uncertainties are the differences between the best-fit and the boundary points of the 68.3$\%$ confidence interval. This obtained signal strength is consistent with the SM Higgs expectation within the uncertainty.  

\begin{figure}[hbpt] 
  \begin{center}
    \includegraphics[width=0.496\textwidth]{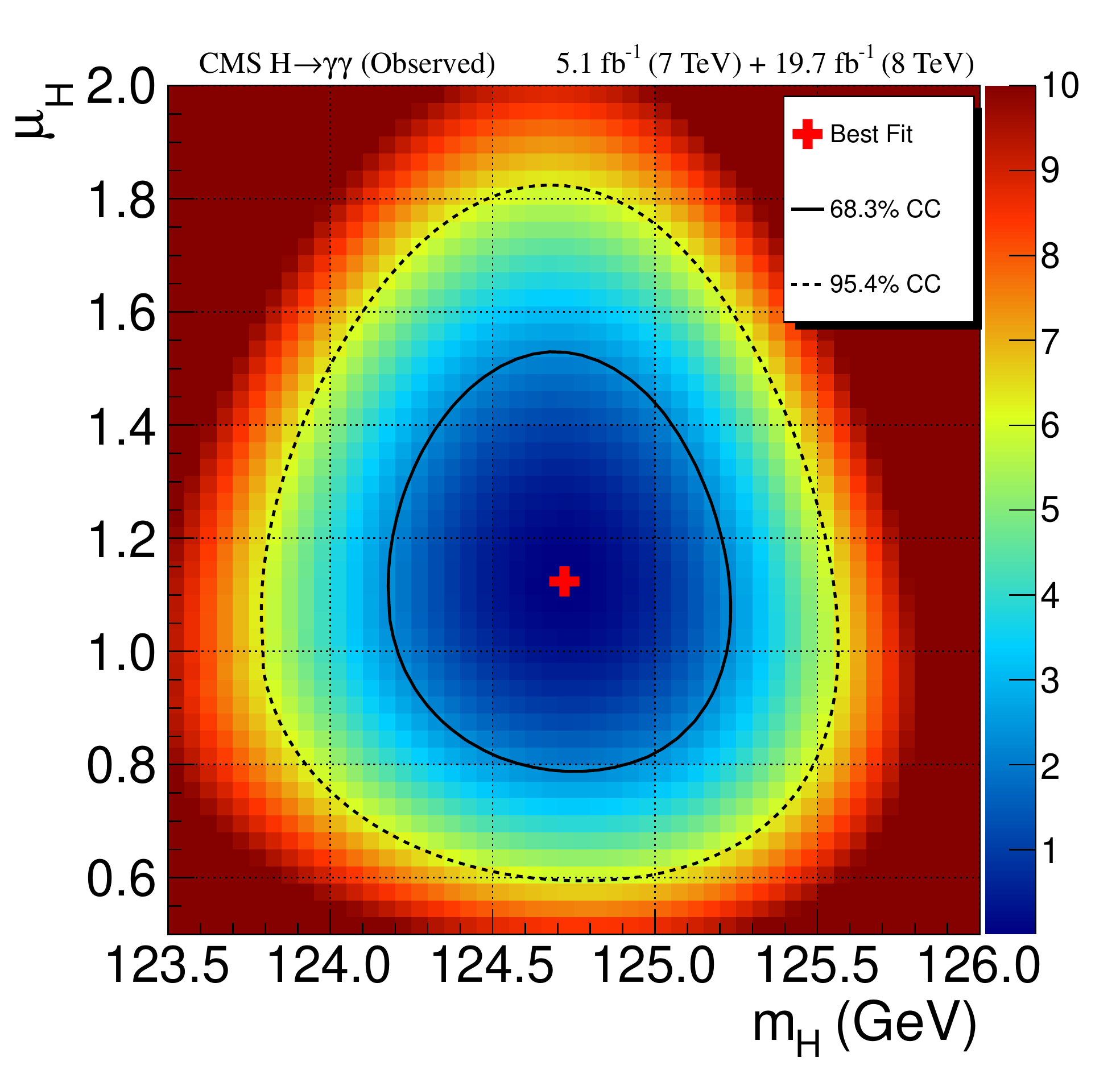}
    \includegraphics[width=0.496\textwidth]{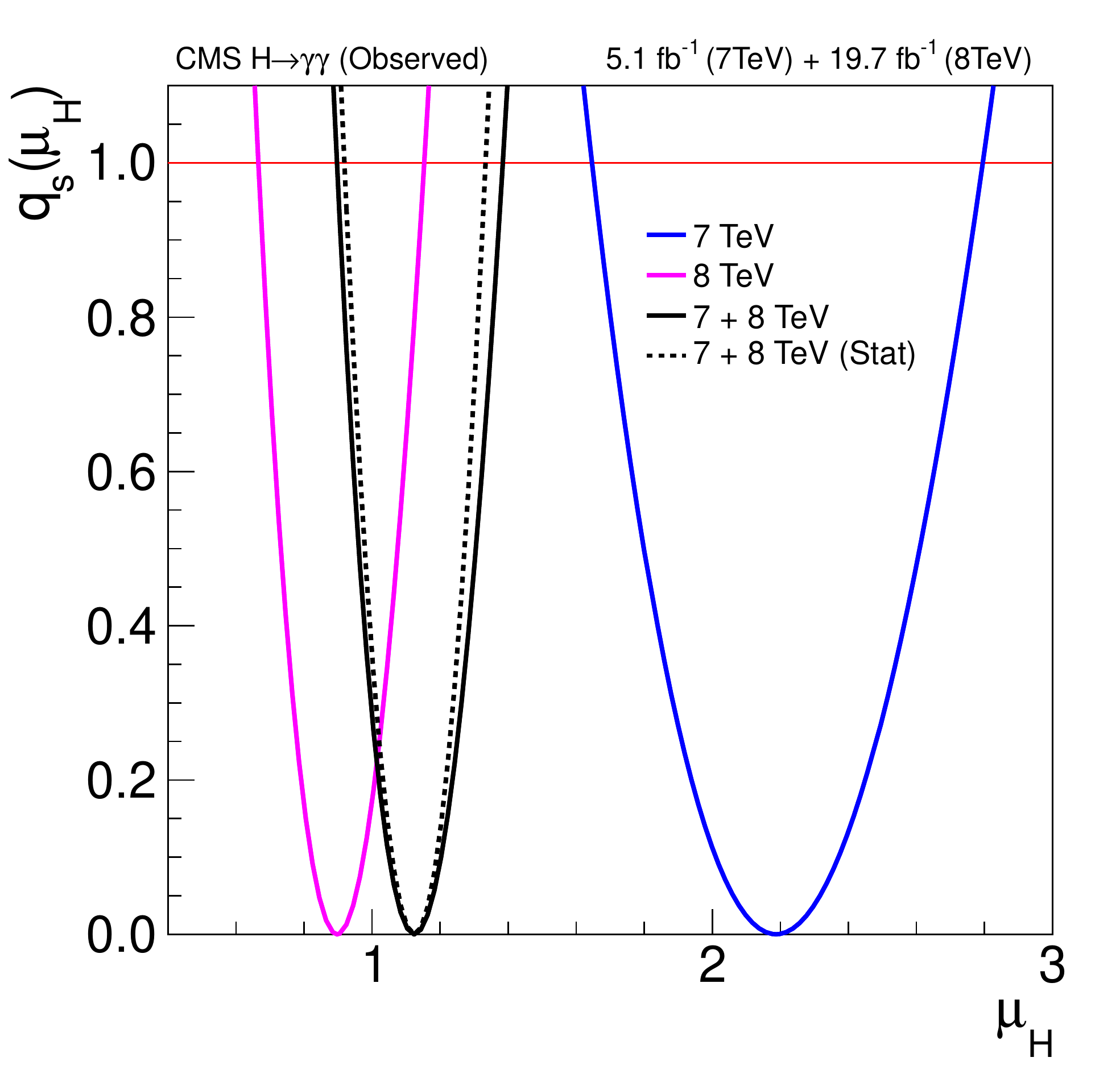}
    \end{center}
  \caption{The observed likelihood ratio $q_{s}(\mu_{H},m_{H})$ and $q_{s}(\mu_{H})$. On the left, the observed likelihood ratio $q_{s}(\mu_{H},m_{H})$ from the combined $7~\mathrm{TeV}$ and $8~\mathrm{TeV}$ datasets is shown as a contour plot. The best-fit (red cross) is $\hat{\mu}_{H}=$ 1.12 and $\hat{m}_{H}=$ $124.72~\mathrm{GeV}$. The 68.3$\%$ confidence contour (solid black line) and the 95.4$\%$ confidence contour (dashed black line) are shown. On the right, the observed likelihood ratio $q_{s}(\mu_{H})$ from the combined $7~\mathrm{TeV}$ and $8~\mathrm{TeV}$ dataset (solid black line), and the ones from the separate $7~\mathrm{TeV}$ (solid blue line) and $8~\mathrm{TeV}$ (solid magenta line) datasets are shown. The $q_{s}(\mu_{H})$ with the signal systematic nuisance parameters fixed to the best-fit values from the combined $7~\mathrm{TeV}$ and $8~\mathrm{TeV}$ dataset (dashed black line) is shown as well. The total uncertainty of the extracted signal strength from the combined $7~\mathrm{TeV}$ and $8~\mathrm{TeV}$ dataset is +0.26/$-$0.23, which consists of the statistical uncertainty +0.21/$-$0.21 and the systematic uncertainty +0.15/$-$0.09.}
  \label{fig:mumh}
\end{figure}

The observed contour plot of likelihood ratio $q_{s}(\mu_{H},m_{H})$ is shown on the left of Figure \ref{fig:mumh}. The best-fit (red cross), and the 68.3$\%$ (solid black line) and 95.4$\%$ (dashed black line) confidence contours, correponding to $q_{s}(\mu_{H},m_{H})$ = 2.3 and  $q_{s}(\mu_{H},m_{H})$ = 6.17 respectively, are also shown. The corresponding likelihood ratio $q_{s}(\mu_{H})$ treating $m_{H}$ as a nuisance parameter obtained from the combined $7~\mathrm{TeV}$ and $8~\mathrm{TeV}$ (solid black line) datasets, and the ones obtained from the separate $7~\mathrm{TeV}$ (solid blue line) and $8~\mathrm{TeV}$ (solid magenta line) datasets are shown on the right of Figure \ref{fig:mumh}. The boundary points for the 68.3$\%$ confidence interval of $\mu_{H}$ correspond to $q_{s}(\mu_{H})$ $=$ 1. In order to quantify separately the statistical uncertainty, including the uncertainty associated with the background model, and the systematic uncertainty, the $q_{s}(\mu_{H})$ with the signal systematic nuisance parameters fixed to the best-fit values for the combined $7~\mathrm{TeV}$ and $8~\mathrm{TeV}$ datasets (dashed black line) is obtained, from which the statistical upper and lower uncertainties are evaluated as +0.21/$-$0.21. The systematic upper and lower uncertainties are computed by subtracting the corresponding statistical uncertainties from the overall uncertainties in quadrature, which are +0.15/$-$0.09. 

The observed $\hat{\mu}_{H}$ and the corresponding $\hat{m}_{H}$ for the $7~\mathrm{TeV}$, $8~\mathrm{TeV}$ and combined $7~\mathrm{TeV}$ and $8~\mathrm{TeV}$ datasets are summarized in Table \ref{tab:mu}.  

\begin{table}[hbtp]
  \renewcommand{\arraystretch}{1.5}
  \noindent
  \small\addtolength{\tabcolsep}{-6pt}
  \caption{The observed signal strength $\hat{\mu}_{H}$ and the corresponding mass $\hat{m}_{H}$.}
  \begin{center}
    \setlength{\tabcolsep}{20pt}
    \begin{tabular}{|l|c|c|} 
      \hline
      & $\hat{\mu}_{H}$ & $\hat{m}_{H}$\\
      \hline
      $7~\mathrm{TeV}$ Observed   & 2.19$_{-0.54}^{+0.61}$ & 124.03 GeV\\
      \hline
      $8~\mathrm{TeV}$ Observed   & 0.90$_{-0.23}^{+0.26}$ & 124.93 GeV\\
      \hline
      $7~\mathrm{TeV}$ + $8~\mathrm{TeV}$ Observed   & $1.12_{-0.23}^{+0.26}$ = 1.12$_{-0.21}^{+0.21}$(stat)$_{-0.09}^{+0.15}$(syst) & 124.72 GeV\\
      \hline
    \end{tabular}
    \label{tab:mu}
  \end{center}
\end{table}  

\section{Mass}
The mass of the observed signal is extracted using the signal model with $\mu_{ggH,t\overline{t}H}$ and $\mu_{\textit{VBF,VH}}$ treated as nuisance parameters. The measured mass from the combined $7~\mathrm{TeV}$ and $8~\mathrm{TeV}$ datasets is $\hat{m}_{H} = 124.72_{-0.36}^{+0.35}$ GeV. 

\begin{figure}[hbpt] 
  \begin{center}
    \includegraphics[width=1\textwidth]{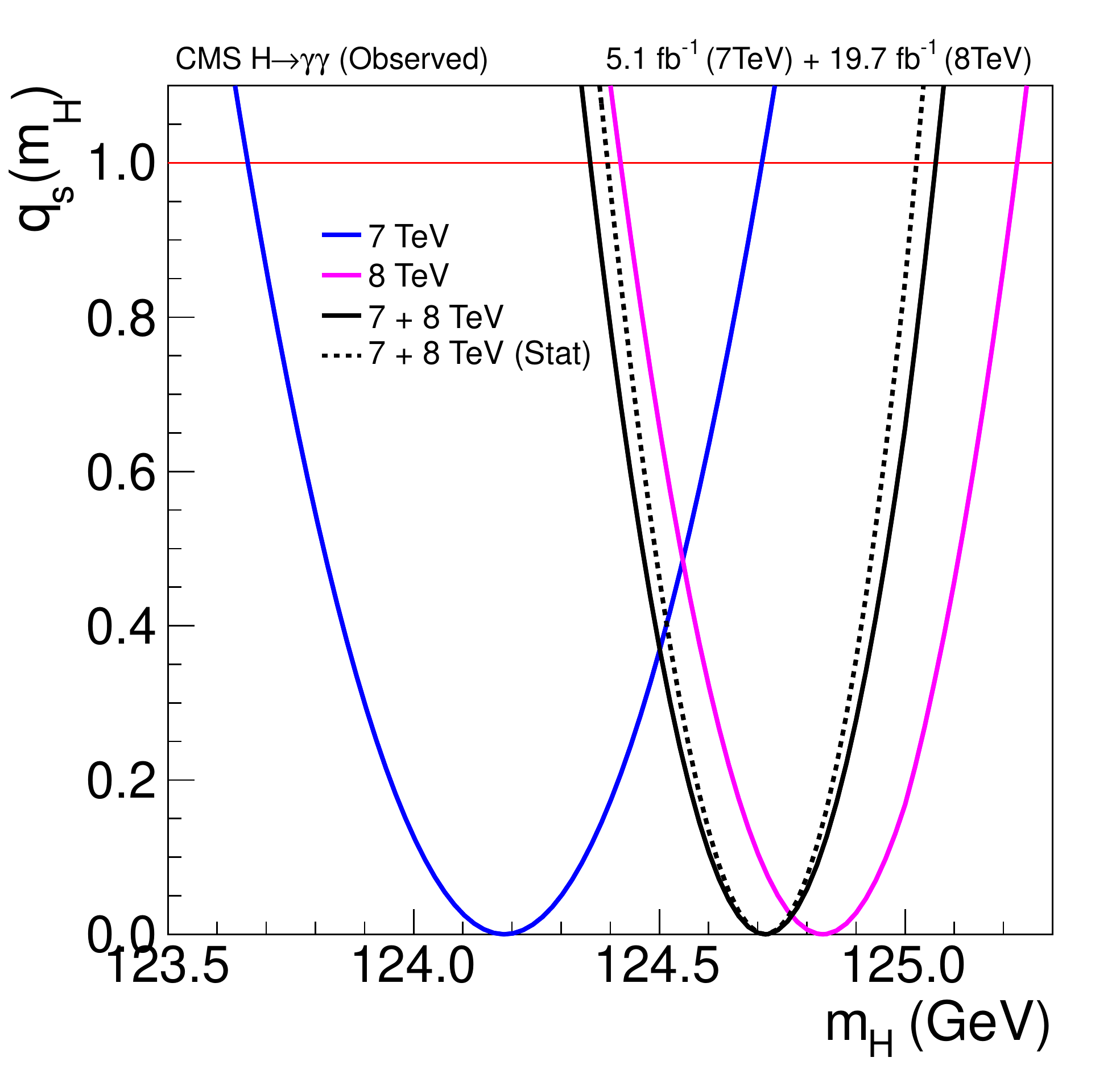}
  \end{center}
  \caption{The observed likelihood ratio $q_{s}(m_{H})$ with $\mu_{ggH,t\overline{t}H}$ and $\mu_{\textit{VBF,VH}}$ treated as nuisance parameters. The $q_{s}(m_{H})$ from the combined $7~\mathrm{TeV}$ and $8~\mathrm{TeV}$ datasets (solid black line), and the ones from the separate $7~\mathrm{TeV}$ (solid blue line) and $8~\mathrm{TeV}$ (solid magenta line) datasets are shown. The $q_{s}(m_{H})$ with the signal systematic nuisance parameters fixed to the best-fit values from the combined $7~\mathrm{TeV}$ and $8~\mathrm{TeV}$ datasets (dashed black line) is shown as well. The best-fit is $\hat{m}_{H}=$ 124.72 GeV. The total uncertainty of the measured mass is +0.35/$-0.36~\mathrm{GeV}$, which consists of the statistical uncertainty +0.31/$-$0.32 GeV and the systematic uncertainty +0.16/$-0.16~\mathrm{GeV}$.}
  \label{fig:massrvrf}
\end{figure}

The corresponding likelihood ratio $q_{s}(m_{H})$ obtained from the combined $7~\mathrm{TeV}$ and $8~\mathrm{TeV}$ datasets (solid black line), and the ones obtained from the separate $7~\mathrm{TeV}$ (solid blue line) and $8~\mathrm{TeV}$ (solid magenta line) datasets are shown in Figure \ref{fig:massrvrf}. For the combined $7~\mathrm{TeV}$ and $8~\mathrm{TeV}$ datasets, the $q_{s}(m_{H})$ with the signal systematic nuisance parameters fixed to the best-fit values (dashed black line) is also shown, from which the statistical uncertainties are evaluated as +0.31/$-$0.32 GeV. The corresponding systematic uncertainties are +0.16/$-$$\mathrm{0.16~GeV}$. 

The observed $\hat{m}_{H}$ for the $7~\mathrm{TeV}$, $8~\mathrm{TeV}$, and combined $7~\mathrm{TeV}$ and $8~\mathrm{TeV}$ datasets are summarized in Table \ref{tab:mhrvrf}.   

\begin{table}[hbtp]
  \renewcommand{\arraystretch}{1.5}
  \noindent
  \small\addtolength{\tabcolsep}{-6pt}
  \caption{The results for the measurement of mass $\hat{m}_{H}$.}
  \begin{center}
    \setlength{\tabcolsep}{20pt}
    \begin{tabular}{|l|c|c|} 
      \hline
      & $\hat{m}_{H}$\\
      \hline
      $7~\mathrm{TeV}$ Observed   & $124.19_{-0.52}^{+0.52}$ GeV\\
      \hline
      $8~\mathrm{TeV}$ Observed   & $124.83_{-0.41}^{+0.40}$ GeV\\
      \hline
      $7~\mathrm{TeV}$ + $8~\mathrm{TeV}$ Observed   & $124.72_{-0.36}^{+0.35}$ GeV  = 124.72$_{-0.32}^{+0.31}$(stat)$_{-0.16}^{+0.16}$(syst) GeV\\
      \hline
    \end{tabular}
    \label{tab:mhrvrf}
  \end{center}
\end{table}  

\section{Signal Strengths for Separate Higgs Production Processes}
The signal strength for \textit{ggH} and \textit{t$\overline{t}$H} processes extracted from the combined $7~\mathrm{TeV}$ and $8~\mathrm{TeV}$ datasets is $\hat{\mu}_{ggH,t\overline{t}H} = 1.14_{-0.31}^{+0.36}$, while the signal strength for \textit{VBF} and \textit{VH} processes is $\hat{\mu}_{\textit{VBF,VH}} = 1.08_{-0.56}^{+0.62}$. Both obtained signal strengths are consistent with the SM Higgs expectation within the uncertainty. 

\begin{figure}[hbpt] 
  \begin{center}
    \includegraphics[width=1\textwidth]{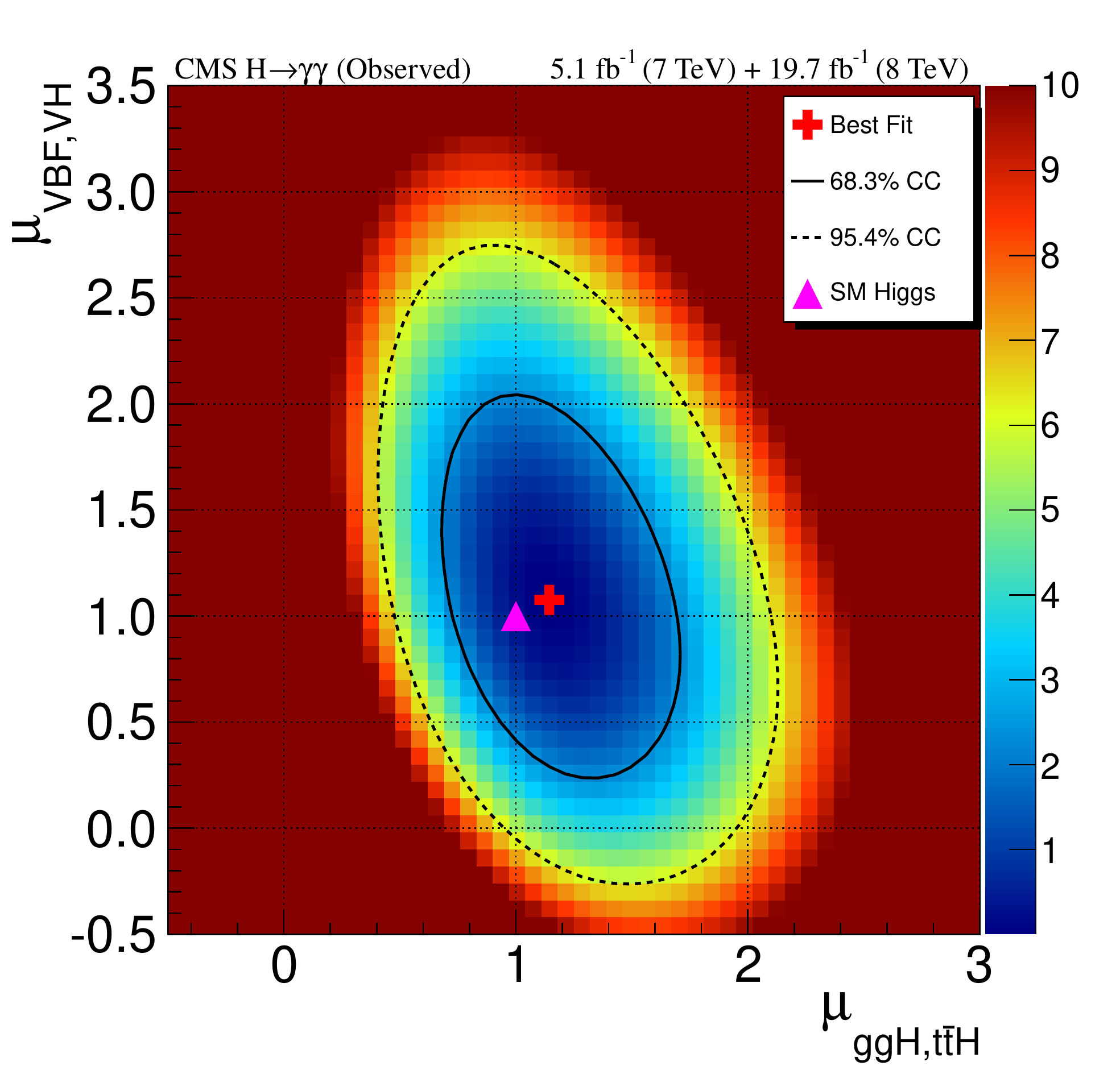}
  \end{center}
  \caption{The observed likelihood ratio $q_{s}(\mu_{ggH,t\overline{t}H},\mu_{\textit{VBF,VH}})$ with $m_{H}$ treated as a nuisance parameter from the combined $7~\mathrm{TeV}$ and $8~\mathrm{TeV}$ datasets. The best-fit (red cross) is  $\hat{\mu}_{ggH,t\overline{t}H} = 1.14$ and $\hat{\mu}_{\textit{VBF,VH}} = 1.08$. The 68.3$\%$ confidence contour (solid black line) and the 95.4$\%$ confidence contour (dashed black line) are shown. The point (magenta triangle) corresponding to the SM Higgs expectation $\mu_{ggH,t\overline{t}H} =$ 1 and $\mu_{\textit{VBF,VH}} =$ 1 is within the $68.3\%$ confidence contour.}
  \label{fig:2dRVRF}
\end{figure}

\begin{figure}[hbpt] 
  \begin{center}
    \includegraphics[width=0.496\textwidth]{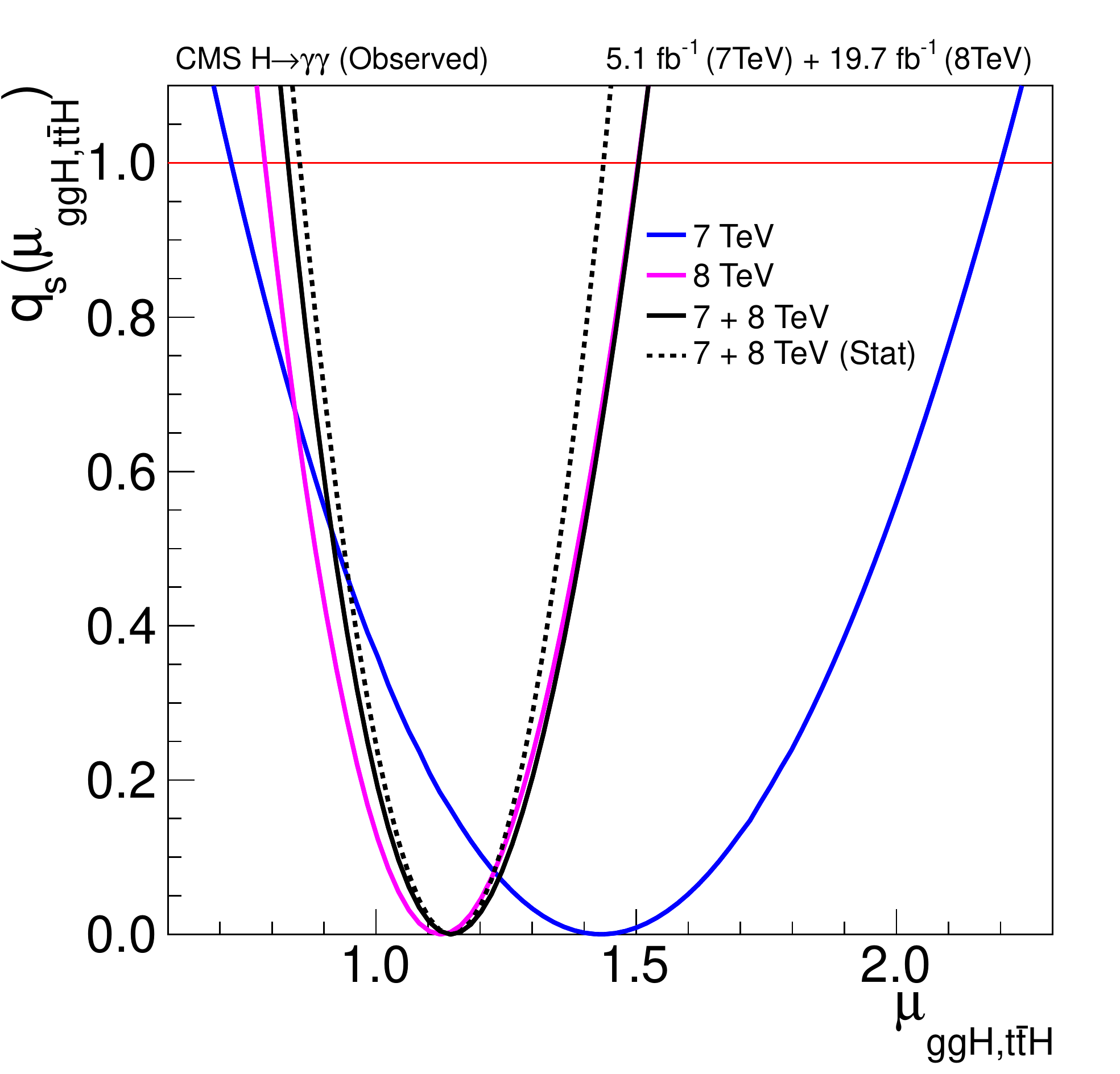}
    \includegraphics[width=0.496\textwidth]{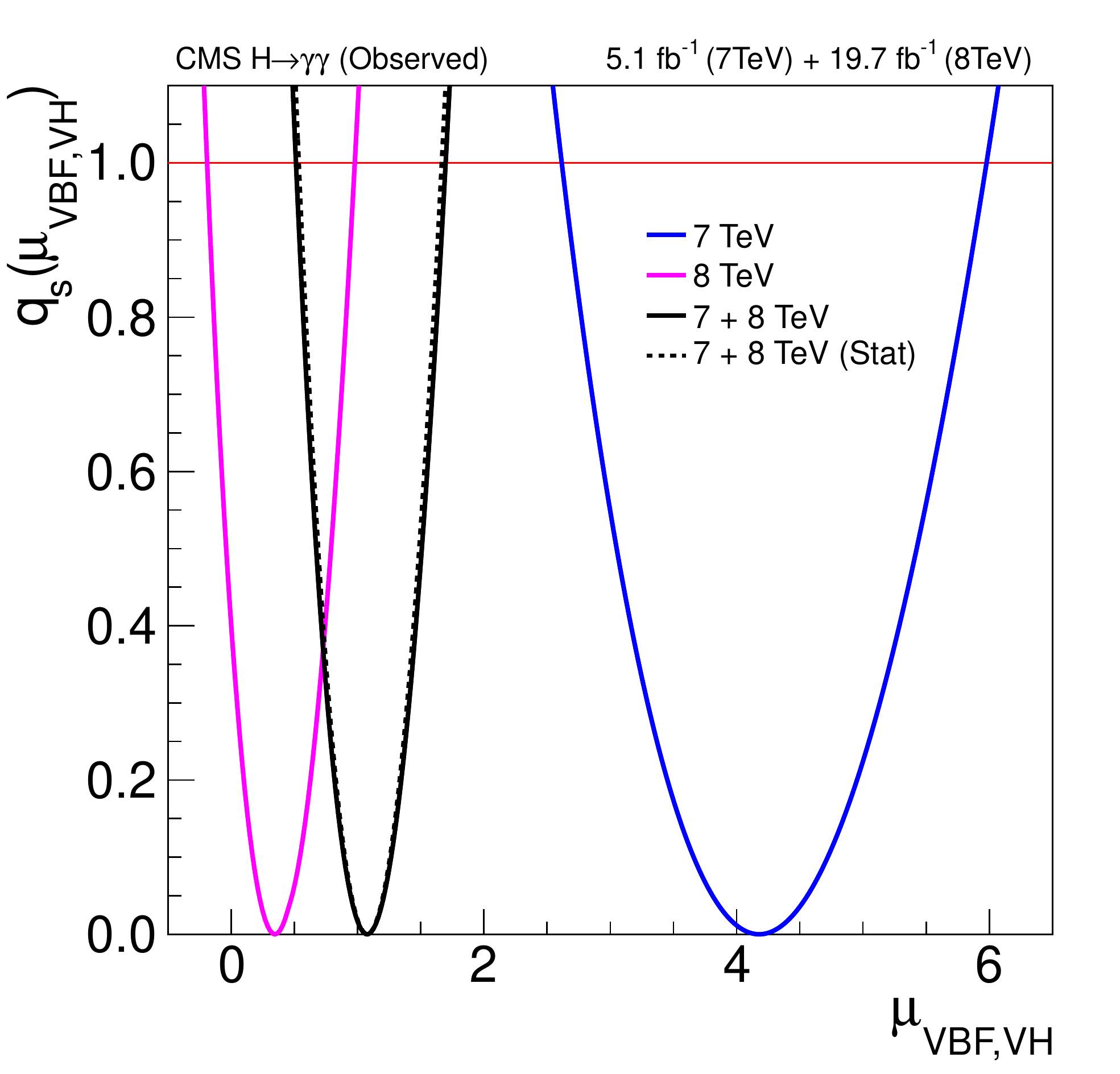}
  \end{center}
  \caption{The observed likelihood ratio $q_{s}(\mu_{ggH,t\overline{t}H})$ ($q_{s}(\mu_{\textit{VBF,VH}})$) with $m_{H}$ and $\mu_{\textit{VBF,VH}}$ ($\mu_{ggH,t\overline{t}H}$) treated as nuisance parameters. On the left (right), the observed likelihood ratio $q_{s}(\mu_{ggH,t\overline{t}H})$ ($q_{s}(\mu_{\textit{VBF,VH}})$) from the combined $7~\mathrm{TeV}$ and $8~\mathrm{TeV}$ datasets (solid black line), and the ones from the separate $7~\mathrm{TeV}$ (solid blue line) and $8~\mathrm{TeV}$ (solid magenta line) datasets are shown. The $q_{s}(\mu_{ggH,t\overline{t}H})$ ($q_{s}(\mu_{\textit{VBF,VH}})$) with the signal systematic nuisance parameters fixed to the best-fit values from the combined $7~\mathrm{TeV}$ and $8~\mathrm{TeV}$ datasets (dashed black line) is also shown. The best-fit is $\hat{\mu}_{ggH,t\overline{t}H} = 1.14$ ($\hat{\mu}_{\textit{VBF,VH}} = 1.08$). The uncertainty of the extracted signal strength from the combined $7~\mathrm{TeV}$ and $8~\mathrm{TeV}$ datasets is +0.36/$-$0.31 (+0.62/$-$0.56).}
  \label{fig:RVRF}
\end{figure}

The observed likelihood ratio $q_{s}(\mu_{ggH,t\overline{t}H},\mu_{\textit{VBF,VH}})$ with $m_{H}$ treated as a nuisance parameter is shown in Figure \ref{fig:2dRVRF}. The best-fit (red cross), the 68.3$\%$ (solid black line) and 95.4$\%$ (dashed black line) confidence contours are also shown. The point (magenta triangle) corresponding to the SM Higgs expectation $\mu_{ggH,t\overline{t}H} =$ 1 and $\mu_{\textit{VBF,VH}} =$ 1 is within the $68.3\%$ confidence contour. The corresponding $q_{s}(\mu_{ggH,t\overline{t}H})$, with $m_{H}$ and $\mu_{\textit{VBF,VH}}$ treated as nuisance parameters, and $q_{s}(\mu_{\textit{VBF,VH}})$, with $m_{H}$ and $\mu_{ggH,t\overline{t}H}$ treated as nuisance parameters, obtained from the combined $7~\mathrm{TeV}$ and $8~\mathrm{TeV}$ (solid black line) datasets, along with the ones obtained from the separate $7~\mathrm{TeV}$ (solid blue line) and $8~\mathrm{TeV}$ (solid magenta line) datasets, are shown on the left and right of Figure \ref{fig:RVRF}, respectively. For the combined $7~\mathrm{TeV}$ and $8~\mathrm{TeV}$ datasets, the $q_{s}(\mu_{ggH,t\overline{t}H})$ and $q_{s}(\mu_{\textit{VBF,VH}})$ with the signal systematic nuisance parameters fixed to the best-fit values (dashed black line) are also shown, whose corresponding statistical uncertainties dominate the overall uncertainties. 

The observed $\hat{\mu}_{ggH,t\overline{t}H}$, $\hat{\mu}_{\textit{VBF,VH}}$ and the corresponding $\hat{m}_{H}$ for the $7~\mathrm{TeV}$, $8~\mathrm{TeV}$ and combined $7~\mathrm{TeV}$ and $8~\mathrm{TeV}$ datasets are summarized in Table \ref{tab:rvrf}.

\begin{table}[hbtp]
  \renewcommand{\arraystretch}{1.5}
  \noindent
  \small\addtolength{\tabcolsep}{-6pt}
  \caption{The observed \textit{ggH} and \textit{t$\overline{t}$H} signal strength $\hat{\mu}_{ggH,t\overline{t}H}$ and the \textit{VBF} and \textit{VH} signal strength $\hat{\mu}_{\textit{VBF,VH}}$ along with the corresponding mass $\hat{m}_{H}$.}
  \begin{center}
    \setlength{\tabcolsep}{20pt}
    \begin{tabular}{|l|c|c|c|} 
      \hline
      & $\hat{\mu}_{ggH,t\overline{t}H}$ & $\hat{\mu}_{\textit{VBF,VH}}$ & $\hat{m}_{H}$\\
      \hline
      $7~\mathrm{TeV}$ Observed   & $1.43_{-0.71}^{+0.77}$ & $4.18_{-1.56}^{+1.80}$  & 124.19 GeV \\
      \hline
      $8~\mathrm{TeV}$ Observed   & $1.13_{-0.34}^{+0.38}$ & $0.35_{-0.54}^{+0.63}$  & 124.83 GeV \\   
      \hline
      $7~\mathrm{TeV}$ + $8~\mathrm{TeV}$ Observed   & $1.14_{-0.31}^{+0.36}$ & $1.08_{-0.56}^{+0.62}$ & 124.72 GeV\\  
      \hline
    \end{tabular}
    \label{tab:rvrf}
  \end{center}
\end{table}   

To have a further look, the separate signal strengths for all the four production processes $\mu_{ggH}$, $\mu_\textit{VBF}$, $\mu_\textit{VH}$ and $\mu_{t\bar{t}H}$ are also extracted. For the determination of each signal strength, the other three signal strengths and $m_{H}$ are treated as nuisance parameters. Since the dominant event classes for \textit{VH} and \textit{$t\bar{t}$H} processes have low statistics, the accuracy of the obtained $\hat{\mu}_\textit{VH}$ and $\hat{\mu}_{t\bar{t}H}$ along with their uncertainties suffer from the background estimation as mentioned in Section \ref{sec:Performance of The Method}. Instead of providing the most accurate evaluations of individual signal strength, the results provide an overall estimation of the compatibility with the SM Higgs boson. The results are listed in Table \ref{tab:fourproc}. The largest deviation from the expectation of the Higgs boson is the signal strength of $\textit{VH}$ production process, which is still compatible with the expectation within 2 standard deviations.     
\begin{table}[hbtp]
  \renewcommand{\arraystretch}{1.5}
  \noindent
  \small\addtolength{\tabcolsep}{-6pt}
  \caption{The observed signal strengths for all the four production processes $\hat{\mu}_{ggH}$, $\hat{\mu}_\textit{VBF}$, $\hat{\mu}_\textit{VH}$ and $\hat{\mu}_{t\bar{t}H}$ along with the corresponding mass $\hat{m}_{H}$.}
  \begin{center}
    \setlength{\tabcolsep}{8pt}
    \begin{tabular}{|l|c|c|c|c|c|} 
      \hline
      & $\hat{\mu}_{ggH}$ & $\hat{\mu}_\textit{VBF}$ & $\hat{\mu}_\textit{VH}$ & $\hat{\mu}_{t\bar{t}H}$ & $\hat{m}_{H}$\\
      \hline
      $7~\mathrm{TeV}$ + $8~\mathrm{TeV}$ Observed  & 1.15$_{-0.32}^{+0.37}$ & 1.51$_{-0.68}^{+0.77}$ & $-$0.35$_{-1.02}^{+1.19}$ & 2.56$_{-1.79}^{+2.50}$ & 124.60 GeV\\
      \hline
    \end{tabular}
    \label{tab:fourproc}
  \end{center}
\end{table}       

\section{Higgs Coupling Strengths}
\begin{figure}[hbpt] 
  \begin{center}
    \includegraphics[width=0.496\textwidth]{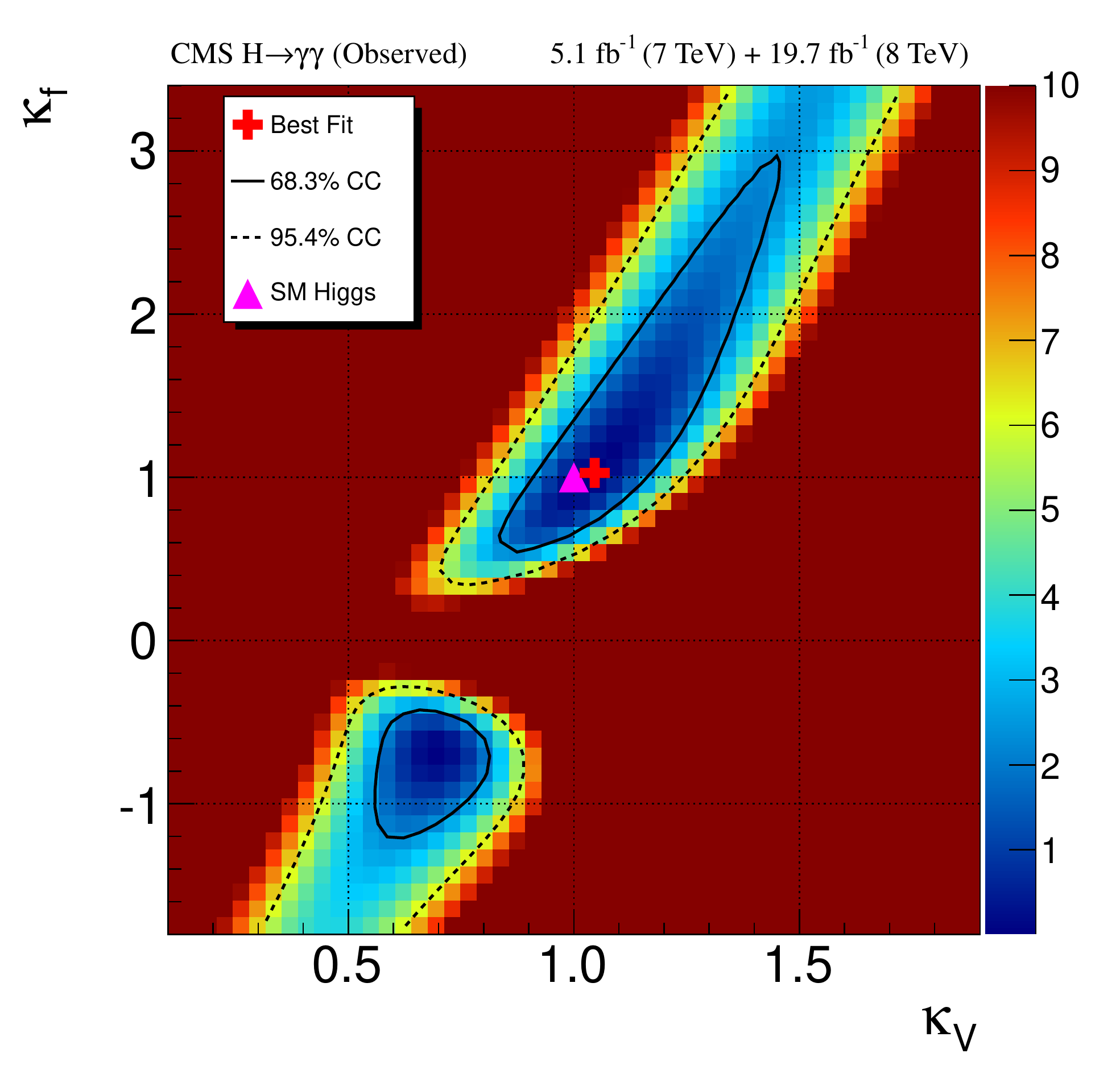}
    \includegraphics[width=0.496\textwidth]{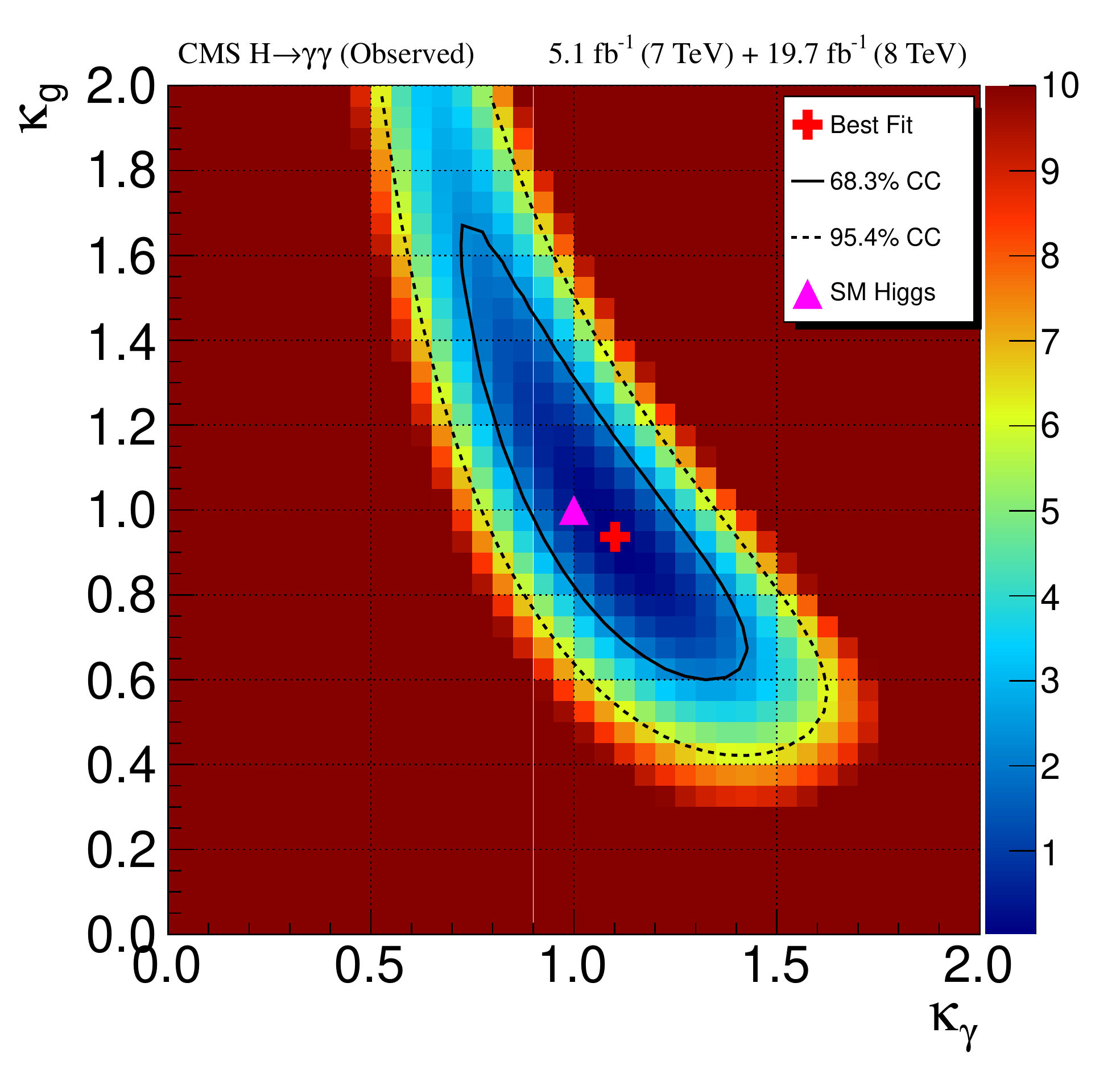}
  \end{center}
  \caption{The observed likelihood ratios $q_{s}(\kappa_{V},\kappa_{f})$ and $q_{s}(\kappa_{\gamma},\kappa_{g})$ at $m_{H} = $ 124.72 GeV from the combined $7~\mathrm{TeV}$ and $8~\mathrm{TeV}$ datasets shown on the left and right. The best-fit (red cross) for the Higgs coupling strengths to bosons and to fermions is $\hat{\kappa}_{V} =$ 1.05 and $\hat{\kappa}_{f} =$ 1.03. The best-fit (red cross) for the effective Higgs coupling strengths to photon and to gluon is $\hat{\kappa}_{\gamma} =$ 1.10 and $\hat{\kappa}_{g} =$ 0.94. The associated 68.3$\%$ (solid black line) confidence contours and 95.4$\%$ (dashed black line) confidence contours are also shown. The points (magenta triangle) corresponding to the SM Higgs expectation $\kappa_{V} =$ 1 and $\kappa_{f} =$ 1, and $\kappa_{\gamma} =$ 1 and $\kappa_{g} =$ 1 are within the $68.3\%$ confidence contours of the best-fits.} 
  \label{fig:2dCVCFhiggsloop}
\end{figure}

The likelihood ratios $q_{s}(\kappa_{V},\kappa_{f})$ and $q_{s}(\kappa_{\gamma},\kappa_{g})$ scanned at $m_{H} = $ $\mathrm{124.72~GeV}$ from the combined $7~\mathrm{TeV}$ and $8~\mathrm{TeV}$ datasets, along with the corresponding best-fits (red cross), the 68.3$\%$ (solid black line) and 95.4$\%$ (dashed black line) confidence contours, are shown on the left and right of Figure \ref{fig:2dCVCFhiggsloop}, respectively.

For the likelihood scan of $\kappa_{V}$ and $\kappa_{f}$, it assumes $\kappa_{V}$ $>$ 0 as only the relative sign between $\kappa_{V}$ and $\kappa_{f}$ is measurable. The best-fit is $\hat{\kappa}_{V} =$ 1.05 and $\hat{\kappa}_{f} =$ 1.03, which supports the same sign scenario and is consistent with the SM Higgs expectation $\hat{\kappa}_{V} =$ 1 and $\hat{\kappa}_{f} =$ 1 (magenta triangle) at 68.3$\%$ confidence level. The opposite sign scenario is not excluded though, and the local mimimum in the region $\kappa_{f}$ $<$ 0 is within the  68.3$\%$ contour. The 68.3$\%$ confidence interval (CL) for $\kappa_{V}$ is [0.61, 0.77] $\cup$ [0.90, 1.24], and that for $\kappa_{f}$ is [$-$0.95, $-$0.50] $\cup$ [0.69, 1.75].    

For $\kappa_{\gamma}$ and $\kappa_{g}$, the extracted values are $\hat{\kappa}_{\gamma} = 1.10_{-0.23}^{+0.21}$ and $\hat{\kappa}_{g} = 0.94_{-0.23}^{+0.38}$, consistent with the SM Higgs expectation.

%% file: OtherResults.tex
\chapter{Other CMS and ATLAS Higgs Results}
\label{chap:OtherResults}
To provide an overall picture of the Higgs searches at LHC, other main results from CMS and ATLAS experiments using LHC Run I data are briefly summarized below. 

\section{Signal Significance, Mass and Compatibility with SM Higgs in Terms of Signal and Coupling Strengths}

\subsection{CMS Results}

For the Higgs searches at CMS experiment through main decay channels, the $H\rightarrow ZZ \rightarrow 4\ell$ channel \cite{Hzz} reports the observation of a narrow resonance with a local significance of 6.8 standard deviations. The measured mass is $\hat{m}_{H} = \mathrm{125.6\pm0.4(stat)\pm0.2(syst)}$ GeV---compatible with the measured mass from the $H\rightarrow\gamma\gamma$ channel within 2 standard deviations, and the best-fit overall Higgs signal strength is $\hat{\mu}_{H} = \mathrm{0.93_{-0.23}^{+0.26}(stat)_{-0.09}^{+0.13}(syst)}$, consistent with the SM Higgs expectation. The $H\rightarrow W^{+}W^{-} \rightarrow 2\ell 2\nu$ channel \cite{Hww} reports an excess of events above background with a local significance of 4.3 standard deviations at the Higgs mass of $\mathrm{125.6~GeV}$ measured from the $H\rightarrow ZZ \rightarrow 4\ell$ channel, and the corresponding best-fit signal strength $\hat{\mu}_{H} = \mathrm{0.72_{-0.18}^{+0.20}}$, consistent with the SM Higgs expectation as well. Besides the bosonic decay channels, the two fermionic decay channels $H\rightarrow \tau^{+}\tau^{-}$ \cite{Htautau} and $H \rightarrow b\overline{b}$ \cite{Hbb} report an excess with a local significance of 3.2 and 2.1 standard deviations for a Higgs mass of $\mathrm{125~GeV}$, and the corresponding best-fit signal strength $\hat{\mu}_{H} = \mathrm{0.78\pm 0.27}$ and $\hat{\mu}_{H}=1.0\pm 0.5$, respectively. The combination of these two channels \cite{Hleptoncombine} leads to the strong evidence for the $\mathrm{125~GeV}$ Higgs decaying into down-type fermions with a local significance of 3.8 standard deviations, for which the corresponding best-fit signal strength is $\hat{\mu}_{H} = 0.83\pm 0.24$. To test the direct Higgs coupling to up-type top quark, a search for \textit{t$\overline{t}$H} production \cite{Htt} is performed by analyzing the events from the above decay channels and the two photon decay channel tagged according to the \textit{t$\overline{t}$H} signature, assuming a Higgs mass of 125.6 GeV. An excess with a local significance of 3.4 standard deviations is observed, and the best-fit signal strength is $\hat{\mu}_{t\bar{t}H} = 2.8_{-0.9}^{+1.0}$, which is compatible with the SM Higgs expectation at 2 standard deviations level.   

In addition, searches are performed through $H\rightarrow \mu^{+}\mu^{-}$ and $H\rightarrow e^{+}e^{-}$ (analysis only performed on events at 8 TeV for $e^{+}e^{-}$) channels \cite{Hmue} as well, despite their very small branching ratios and low sensitivity. The observed (expected) 95$\%$ CL upper limits on their branching ratio for a Higgs mass of 125 GeV---assuming the SM cross section---are 0.0016 and 0.0019, corresponding to 7.4($6.5_{-1.9}^{+2.8}$) and $3.7\times 10^{5}$ times the SM value, respectively. Since the result from $H\rightarrow \tau^{+}\tau^{-}$ is consistent with the SM Higgs expectation with a branching ratio $0.0632\pm 0.0036$ larger than the limits for $\mu^{+}\mu^{-}$ and $e^{+}e^{-}$, the leptonic couplings of the Higgs are shown as not flavour-universal as expected by the SM. Furthermore, a search is performed for the Higgs decaying into particles not interacting with the detector---the invisible decays ($H\rightarrow \mathrm{invisible}$) \cite{hinvisible}, targeting the non-SM decay particles such as dark matter candidates. The observed data is consistent with the SM background expectation, and the observed (expected) 95$\%$ CL upper limit on the invisible branching ratio for the $\mathrm{125~GeV}$ Higgs is 0.58(0.44).   

For the combined CMS results \cite{cmshiggscombinefinal}, the Higgs mass measured through both the $H\rightarrow\gamma\gamma$ and $H\rightarrow ZZ \rightarrow 4\ell$ channels is $\hat{m}_{H} = \mathrm{125.02_{-0.27}^{+0.26}(stat)_{-0.15}^{+0.14}(syst)~GeV}$. The overall Higgs signal strength---the relative Higgs production cross section with respect to the SM expectation---as well as the signal strengths for different production processes are extracted at this mass combining the main decay channels, $H\rightarrow \gamma\gamma$, $H\rightarrow ZZ \rightarrow 4\ell$, $H\rightarrow W^{+}W^{-} \rightarrow 2\ell 2\nu$, $H \rightarrow b\overline{b}$ and $H\rightarrow \tau^{+}\tau^{-}$, with multiple Higgs production tags explored. The best-fit overall signal strength is $\hat{\mu}_{H} = \mathrm{1.00\pm 0.09(stat)_{-0.07}^{+0.08}(theo)\pm 0.07(syst)}$, where the systematic uncertainty is further decomposed into the theoretical related component (theo) and the rest (syst). The best-fit signal strengths for the individual production processes are $\hat{\mu}_{ggH} = 0.85_{-0.16}^{+0.19}$, $\hat{\mu}_\textit{VBF} = 1.16_{-0.34}^{+0.37}$, $\hat{\mu}_\textit{VH} = 0.92_{-0.36}^{+0.38}$ and $\hat{\mu}_{t\bar{t}H} = 2.90_{-0.94}^{+1.08}$. Both the overall signal strength and the individual production signal strengths are compatible with the expectations of the SM Higgs---for the \textit{t$\overline{t}$H} signal strength, agreeing with the result from the dedicated \textit{t$\overline{t}$H} search as mentioned above, the compatibility is at about 2 standard deviations level. Furthermore, various Higgs coupling strengths are probed under different physics scenarios, using the inputs from the main decay channels as well as the $H\rightarrow \mu^{+}\mu^{-}$ and $H\rightarrow \mathrm{invisible}$ channels. For the benchmark scenarios of Reference \cite{LHCHiggsCrossSectionWorkingGroup3} assuming no invisible or undetectable Higgs decays, the best-fits for the Higgs coupling strengths to bosons and fermions are $\hat{\kappa}_{V} = 1.01\pm 0.07$ and $\hat{\kappa}_{f} = 0.87_{-0.13}^{+0.14}$, which supports the same sign scenario between ${\kappa}_{V}$ and ${\kappa}_{f}$ as expected by the SM. The data excludes the opposite sign scenario at the 95\% CL while not at the 99.7\% CL. For the effective Higgs coupling strengths to photon and to gluon, the best-fits are $\hat{\kappa}_{\gamma} = 1.14_{-0.13}^{+0.12}$ and $\hat{\kappa}_{g} = 0.89_{-0.10}^{+0.11}$. The above results from combination are consistent with the results from $H\rightarrow\gamma\gamma$ channel alone, and with smaller uncertainties due to the extra constraints from the other decay channels. The full combined results are provided in Reference \cite{cmshiggscombinefinal}, which are all compatible with the SM Higgs expectation.

\subsection{ATLAS Results}
For the corresponding results from ATLAS experiment \cite{atlashiggsCombined,compatlashiggs1}, all the main bosonic decay channels report observation of excess with significance beyond 5 standard deviations---5.2 standard deviations from $H\rightarrow\gamma\gamma$. Strong evidence for the Higgs coupling to down-type fermions is obtained with a significance of 4.5 standard deviations. The measured Higgs mass from the $H\rightarrow\gamma\gamma$ and $H\rightarrow ZZ \rightarrow 4\ell$ channels is $\hat{m}_{H} = \mathrm{125.36\pm 0.37(stat)\pm 0.18(syst)~GeV}$. At this mass, the best-fit overall signal strength from the $H\rightarrow\gamma\gamma$ channel is $\hat{\mu}_{H} = \mathrm{1.17 \pm 0.27}$, which agrees with the result from the CMS $H\rightarrow\gamma\gamma$ channel. Combining all the main decay channels, together with the $H\rightarrow Z\gamma$ and  $H\rightarrow \mu\mu$ channels, the best-fit overall signal strength is $\hat{\mu}_{H} = 1.18\pm 0.10\mathrm{(stat)}_{-0.07}^{+0.08}\mathrm{(theo)}\pm 0.07 \mathrm{(syst)}$. The best-fit signal strengths for the individual production processes are $\hat{\mu}_{ggH} = 1.23_{-0.20}^{+0.23}$, $\hat{\mu}_\textit{VBF} = 1.23\pm 0.32$, $\hat{\mu}_\textit{VH} = 0.80\pm 0.36$ and $\hat{\mu}_{t\bar{t}H} = 1.81\pm 0.80$. Various Higgs coupling strengths are probed as well. In particular, the best-fits for the Higgs coupling strengths to bosons and fermions are $\hat{\kappa}_{V} = 1.09\pm 0.07$ and $\hat{\kappa}_{f} = 1.11_{-0.15}^{+0.17}$. The best-fits for the effective Higgs coupling strengths to photon and to gluon are $\hat{\kappa}_{\gamma} = 1.00\pm 0.12$ and $\hat{\kappa}_{g} = 1.12\pm 0.12$ with the effective Higgs coupling strength to $Z\gamma$, $\kappa_{Z\gamma}$, profiled. These combined results are summarized in the right column of Table \ref{tab:cmsatlascomp}, and are compared with the CMS combined results summarized in the left column. 

ATLAS and CMS, with different detector design, independent analysis methods and similar luminosities for the analyzed data, obtain results compatible with each other, which lead to the observation of a new particle with the signal and coupling strengths consistent with the Standard Model Higgs boson.  

\begin{table}[hbtp]
  \renewcommand{\arraystretch}{1.5}
  \newcolumntype{C}[1]{>{\centering\let\newline\\\arraybackslash\hspace{0pt}}m{#1}}
  \noindent
  \small\addtolength{\tabcolsep}{-6pt}
  \caption{The comparison between combined CMS results (left) and ATLAS results (right).}
  \begin{center}
   \setlength{\tabcolsep}{1pt}
    \begin{tabular}{|l|C{7cm}|C{7 cm}|} 
      \hline
      & CMS & ATLAS \\
      \hline
      $\hat{m}_{H}$ (GeV) & $\mathrm{125.02_{-0.27}^{+0.26}(stat)_{-0.15}^{+0.14}(syst)}$ & $\mathrm{125.36\pm 0.37(stat)\pm 0.18(syst)}$\\
      \hline
      $\hat{\mu}_{H}$ & $\mathrm{1.00\pm 0.09(stat)_{-0.07}^{+0.08}(theo)\pm 0.07(syst)}$ & $1.18\pm 0.10\mathrm{(stat)}_{-0.07}^{+0.08}\mathrm{(theo)}\pm 0.07 \mathrm{(syst)}$\\
      \hline
      $\hat{\mu}_{ggH}$ & $0.85_{-0.16}^{+0.19}$ & $1.23_{-0.20}^{+0.23}$\\
      $\hat{\mu}_\textit{VBF}$ & $1.16_{-0.34}^{+0.37}$ & $1.23\pm 0.32$\\
      $\hat{\mu}_\textit{VH}$ & $0.92_{-0.36}^{+0.38}$ & $0.80\pm 0.36$\\
      $\hat{\mu}_{t\bar{t}H}$ & $2.90_{-0.94}^{+1.08}$ & $1.81\pm 0.80$\\
      \hline
      $\hat{\kappa}_{V}$  & $1.01\pm 0.07$ & $1.09\pm 0.07$\\
      $\hat{\kappa}_{f}$  & $0.87_{-0.13}^{+0.14}$ & $1.11_{-0.15}^{+0.17}$\\
      \hline
      $\hat{\kappa}_{\gamma}$ & $1.14_{-0.13}^{+0.12}$ & $1.00\pm 0.12$\\
      $\hat{\kappa}_{g}$  & $0.89_{-0.10}^{+0.11}$ & $1.12\pm 0.12$\\
      \hline
    \end{tabular}
    \label{tab:cmsatlascomp}
  \end{center}
\end{table}   

\section{Spin and Parity}
The new particle is identified as a boson since it is observed through the $H\rightarrow\gamma\gamma$ and $H\rightarrow ZZ \rightarrow 4\ell$ channels. Its observation through the $H\rightarrow\gamma\gamma$ channel further indicates that its spin is not equal to 1 \cite{Landau,Yang} and its charge conjugation is positive. All observations are in favor of the SM Higgs spin-parity hypothesis with spin-0 and even parity, while disfavor opposite parity under spin-0 hypothesis, spin-1 hypothesis and several models under spin-2 hypothesis tested so far \cite{hggfinalpaper,spin1,spin2,spin3,spin4,spin5}.

%% file: Conclusion.tex
\chapter{Conclusion}
\label{Conclusion}
\textbf{Passed by mornings and nights, bright and dark, we are now at the end of this Odyssey, searching for the Higgs boson through its decay into two photons at the CMS experiment at CERN's Large Hadron Collider.} This thesis concludes here with our final results concerning the observation of a new particle and the measurements of its properties from the refined and extended analysis, using the advanced multivariate analysis techniques, that we have developed since 2011, on the full LHC ``Run I'' data collected by the CMS detector during 2011 and 2012, consisting of proton-proton collision events at $\sqrt{s}$ $=$ $7~\mathrm{TeV}$ with $L$ $=$ $5.1~\mathrm{fb^{-1}}$ and at $\sqrt{s}$ $=$ $8~\mathrm{TeV}$ with $L$ $=$ $19.7~\mathrm{fb^{-1}}$, with the final calibration.

\vspace{\baselineskip}
An excess of events above the background expectation is observed, with a local significance of 5.7 standard deviations at a mass of 124.7 GeV. This result confirms our observation of an excess of events, with a local significance of 4.1 standard deviations near 125 GeV in 2012, which provided the strongest evidence among all the Higgs search channels for the observation of a new particle from the CMS experiment\cite{cmsdiscover,longhgg}. This result further constitutes the standalone observation of the new particle through the two photon decay channel.   
\vspace{\baselineskip}

A further measurement provides the precise mass of this new particle as
\begin{center} $\hat{m}_{H}$ = $124.72_{-0.36}^{+0.35}$ GeV  = 124.72$_{-0.32}^{+0.31}$(stat)$_{-0.16}^{+0.16}$(syst) GeV,  \end{center} 
with a relative total uncertainty less than 0.3\% dominated by the statistical uncertainty.  

The production cross section times the two photon decay branching ratio of this new particle relative to that of the Standard Model Higgs boson, the signal strength, for all the Higgs production processes combined, is extracted as 
\begin{center}  $\hat{\mu}_{H}$ = $1.12_{-0.23}^{+0.26}$ = 1.12$_{-0.21}^{+0.21}$(stat)$_{-0.09}^{+0.15}$(syst).  \end{center}
The relative uncertainty is about 20\% dominated by the statistical uncertainty. This result is compatible with the Standard Model Higgs boson expectation within the uncertainty.
 
\vspace{\baselineskip}
The separate signal strengths for \textit{VBF} and \textit{VH} production processes, sensitive to Higgs couplings to bosons, and for the \textit{ggH} and \textit{t$\overline{t}$H} production processes, sensitive to Higgs couplings to fermions, are further extracted as
\begin{center} $\hat{\mu}_{\textit{VBF,VH}} = 1.08_{-0.56}^{+0.62}$,      \end{center} 
\begin{center} $\hat{\mu}_{ggH,t\overline{t}H}$ = $1.14_{-0.31}^{+0.36}$, \end{center}
which have large uncertainties and are consistent with the Standard Model Higgs boson. 
 
\vspace{\baselineskip}
The signal strengths for individual production processes are also extracted as
\begin{center} $\hat{\mu}_{ggH}$ = 1.15$_{-0.32}^{+0.37}$,   \end{center} 
\begin{center} $\hat{\mu}_\textit{VBF}$ = 1.51$_{-0.68}^{+0.77}$,\end{center} 
\begin{center} $\hat{\mu}_\textit{VH}$ =  $-$0.35$_{-1.02}^{+1.19}$,   \end{center} 
\begin{center} $\hat{\mu}_{t\bar{t}H}$ = 2.56$_{-1.79}^{+2.50}$. \end{center} 
These results, especially for $\textit{VH}$ and $t\bar{t}H$, are limited by the large statistical uncertainties. The largest deviation from the expectation of the Higgs boson is the signal strength of $\textit{VH}$ production process, which is still compatible with the expectation within 2 standard deviations. 

\clearpage
The couplings of this new particle to bosons and to fermions relative to the key predictions from the Standard Model about those of the Higgs boson, proportional to boson mass squared and to fermion mass, respectively, assuming the existence of the Yukawa interactions between the Higgs boson and fermions, are further extracted as
\begin{center} $\hat{\kappa}_{V} =$ 1.05 with 68.3$\%$ confidence interval [0.61, 0.77] $\cup$ [0.90, 1.24], \end{center} 
\begin{center} $\hat{\kappa}_{f} =$ 1.03 with 68.3$\%$ confidence interval [$-$0.95, $-$0.50] $\cup$ [0.69, 1.75]. \end{center}
The extracted ${\kappa}_{V}$ shows that the coupling of the new particle to bosons is compatible with the Standard Model prediction at 68.3$\%$ confidence level. The extracted $\kappa_{f}$ supports the existence of the interaction between the new particle and fermions, and further shows that the coupling of the new particle to fermions is compatible with the Standard Model prediction at 68.3$\%$ confidence level.   

\vspace{\baselineskip}
The effective couplings of the particle to photon and to gluon relative to the Standard Model Higgs boson are extracted as
\begin{center} $\hat{\kappa}_{\gamma} = 1.10_{-0.23}^{+0.21}$, \end{center}
\begin{center} $\hat{\kappa}_{g} = 0.94_{-0.23}^{+0.38}$. \end{center}
These results are also compatible with the Standard Model Higgs boson expectation and provide no evidence for the existence of new heavy particles in the loops given the current precision.

\clearpage
The observation of a new particle from the $H\rightarrow\gamma\gamma$ channel is supported by the final search results from the other main Higgs decay channels at the CMS experiment on the LHC ``Run I'' data, including the standalone observation of a new particle from the $H\rightarrow ZZ \rightarrow 4\ell$ channel\cite{Hzz}, and strong evidences from the $H\rightarrow W^{+}W^{-} \rightarrow 2\ell 2\nu$ channel \cite{Hww} and also from the combination of fermionic decay channels $H\rightarrow \tau^{+}\tau^{-}$ and $H \rightarrow b\overline{b}$ \cite{Hleptoncombine}. These results confirm the observation of a new particle from the CMS experiment in 2012\cite{cmsdiscover}. Combining the $H\rightarrow\gamma\gamma$ and $H\rightarrow ZZ \rightarrow 4\ell$ channels, the mass of the new particle is measured precisely as $\mathrm{125.02_{-0.27}^{+0.26}(stat)_{-0.15}^{+0.14}(syst)~GeV}$. Combining all main decay channels, the total production cross section of this new particle relative to that of the Standard Model Higgs boson is extracted as $\mathrm{1.00\pm 0.09(stat)_{-0.07}^{+0.08}(theo)\pm 0.07(syst)}$, with the relative uncertainty reduced to 10\% with respect to the result from the $H\rightarrow\gamma\gamma$ channel alone, and is compatible with the Standard Model Higgs boson expectation. All the other CMS combined results on the relative cross sections for separate Higgs prodcution processes and couplings are compatible with the Standard Model Higgs boson expectations as well. In particular, the coupling to bosons and that to fermions relative to those of the Higgs boson are extracted as $1.01\pm 0.07$, with an uncertainty within 10\%, and $0.87_{-0.13}^{+0.14}$, with an uncertainty of about 15\%, respectively\cite{cmshiggscombinefinal}.   

\vspace{\baselineskip}
The above observation and measurements of a new particle from the CMS experiment are confirmed by the results from the ATLAS experiment also at LHC---with different design of detector, independent analysis methods, and similar luminosity of analyzed data---in the $H\rightarrow\gamma\gamma$ channel and all the Higgs decay channels combined\cite{atlashiggsCombined,compatlashiggs1}. 

\vspace{\baselineskip}
The new particle is identified as a boson since it is observed through the $H\rightarrow\gamma\gamma$ and $H\rightarrow ZZ \rightarrow 4\ell$ channels. Its observation through $H\rightarrow\gamma\gamma$ channel further indicates that its spin is not equal to 1 \cite{Landau,Yang} and its charge conjugation is positive. All the studies regarding the spin and parity of this new particle are in favor of the SM Higgs boson hypothesis with spin-0 and even parity\cite{hggfinalpaper,spin1,spin2,spin3,spin4,spin5}.     
 
\clearpage
\subsubsection{Final Remarks}
Standing at the end of this Odyssey of searching for the Higgs boson at the Large Hadron Collider, we have a list of results on our hands, which points to a new particle looking very similar to the Higgs boson in terms of the production rate, couplings, and spin and parity. 

\vspace{\baselineskip}
What is the ultimate reality behind it? 

\vspace{\baselineskip}
Is this particle the quantum of the scalar field, slowing down the particles with masses such that they could get together to form the structures in the universe including ourselves, as described by the Standard Model of particle physics? Does it relate to the phenomena beyond the description of the Standard Model such as dark matter? Further measurements of this particle from LHC ``Run II'', with the center-of-mass energy increasing to 13 TeV and total luminosity about $\mathrm{100~fb^{-1}}$,  and from other experiments in the future would provide more information to tell. 

\vspace{\baselineskip}
What is sure for the moment ------

\vspace{\baselineskip}
Searching for the Higgs boson does bring us together to experience a series of events in that space and time, becoming a fundamental part of our existence and a monumental part of human history. All the information we having obtained from the proton-proton collisions at the Large Hadron Collider, along with the epic efforts of generations of physicists and engineers from all over the world, all the sleepless nights, all the memorable moments, all the collisions among ourselves, all the beautiful minds and hearts, all the emotions, and all the stories, are folded into the results sent towards the future, passing through layers of time, as pairs of photons passing through layers of nights, as what we have received from our predecessors. 

%% file: appa.tex
\chapter{Figures of Signal Model}
\label{chap:Figures of Signal Model}

\begin{figure}[hbpt] 
  \begin{center}
    \includegraphics[width=0.4\textwidth]{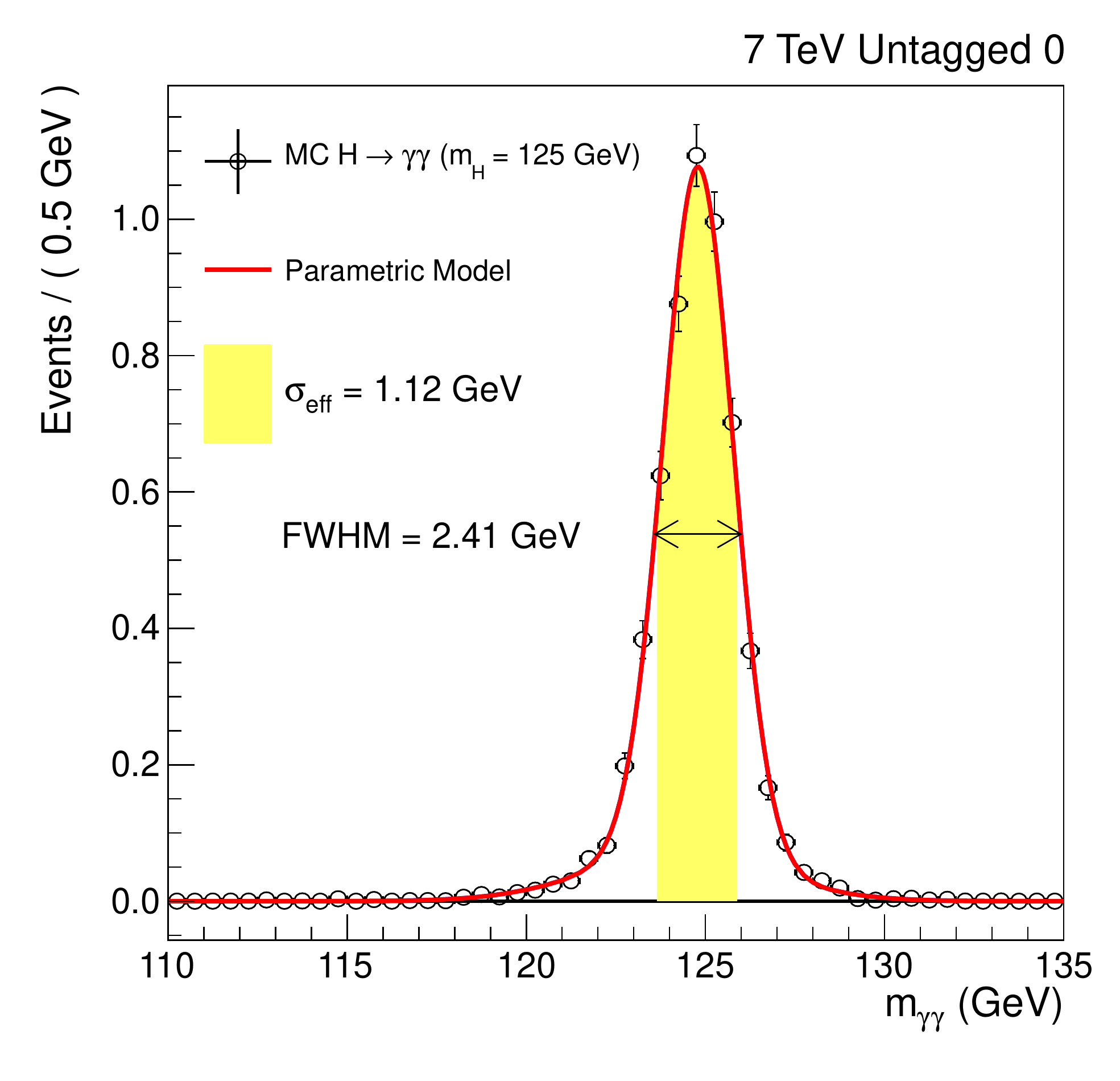}
    \includegraphics[width=0.4\textwidth]{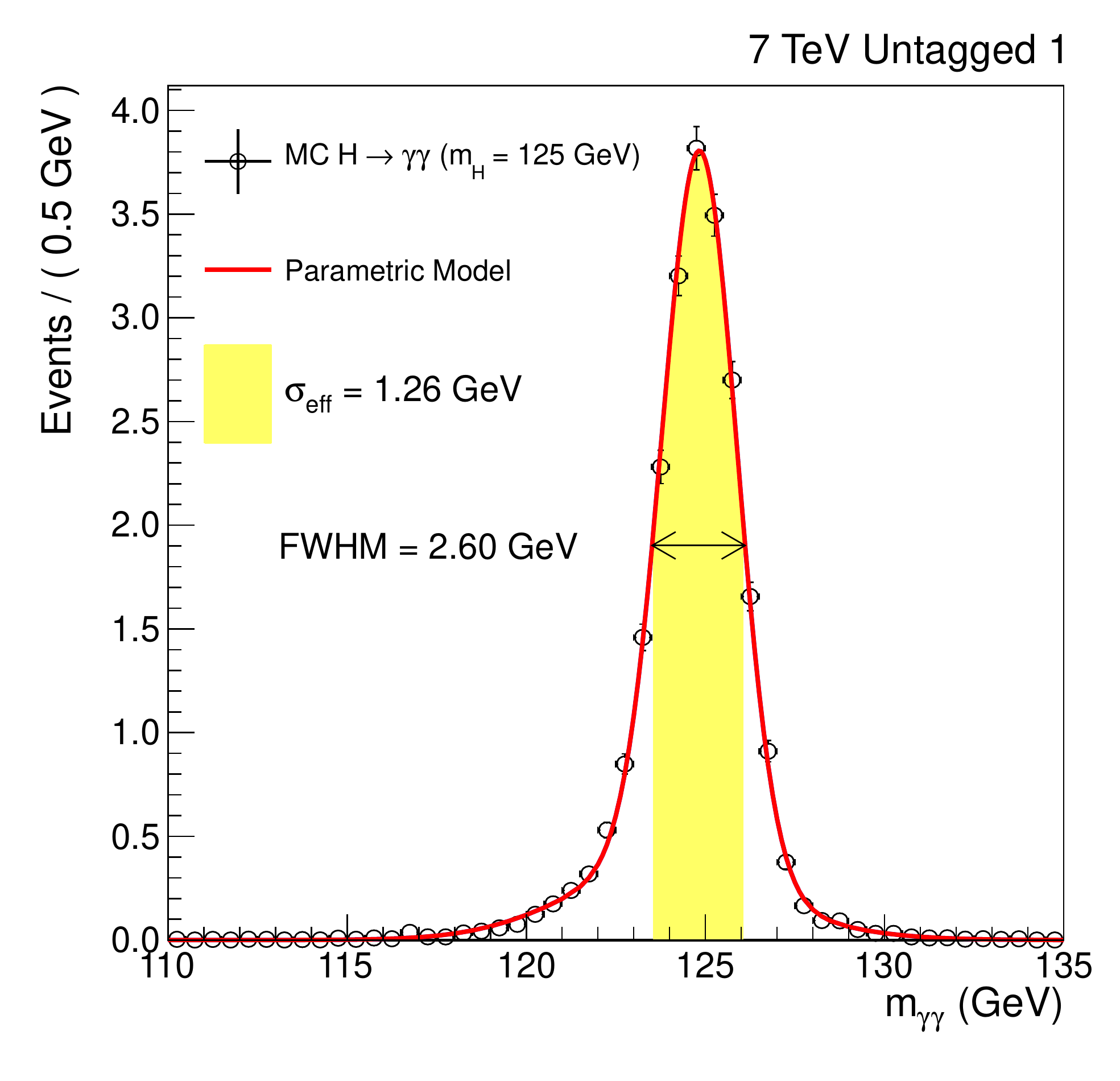}
    \includegraphics[width=0.4\textwidth]{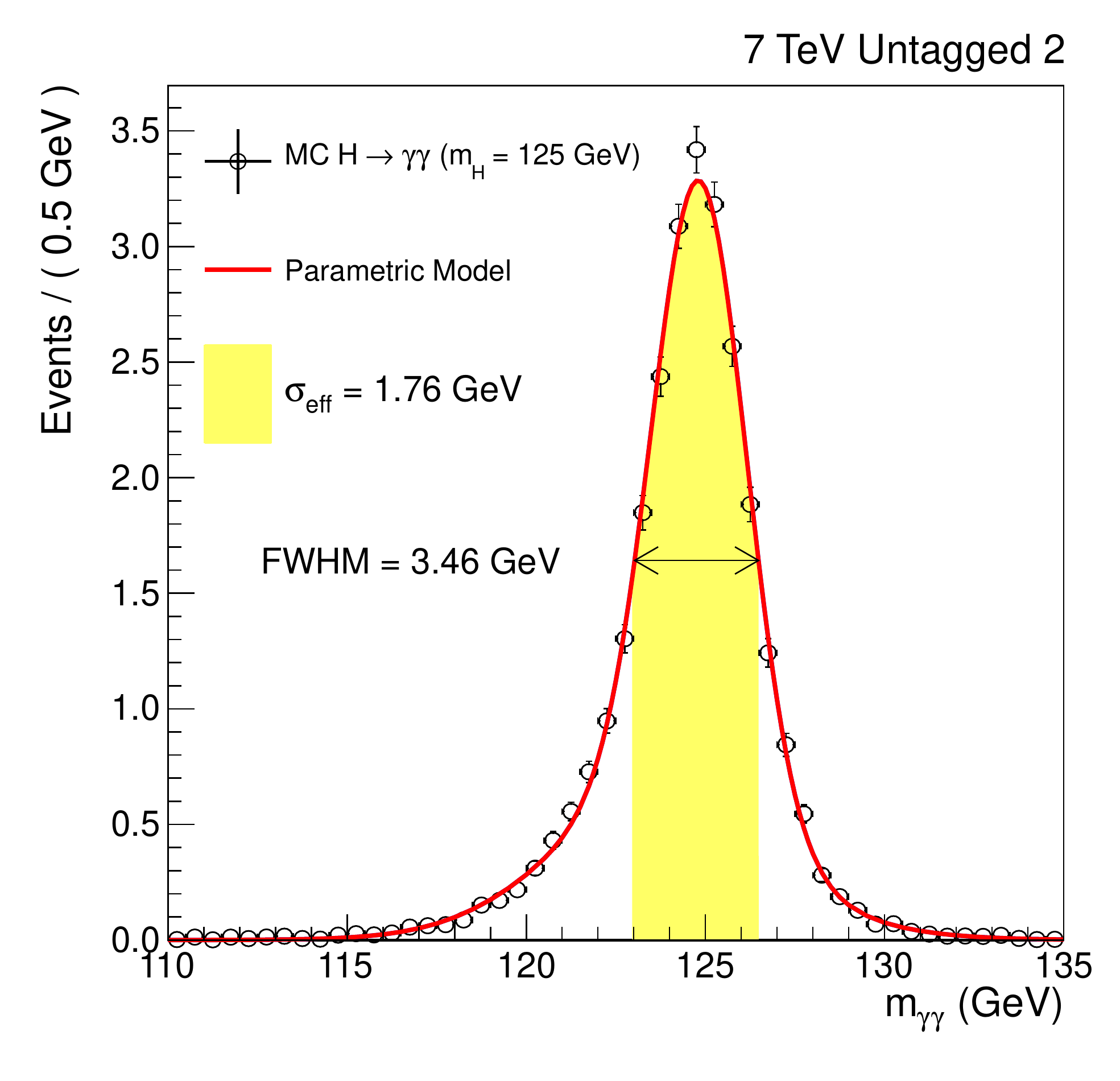}
    \includegraphics[width=0.4\textwidth]{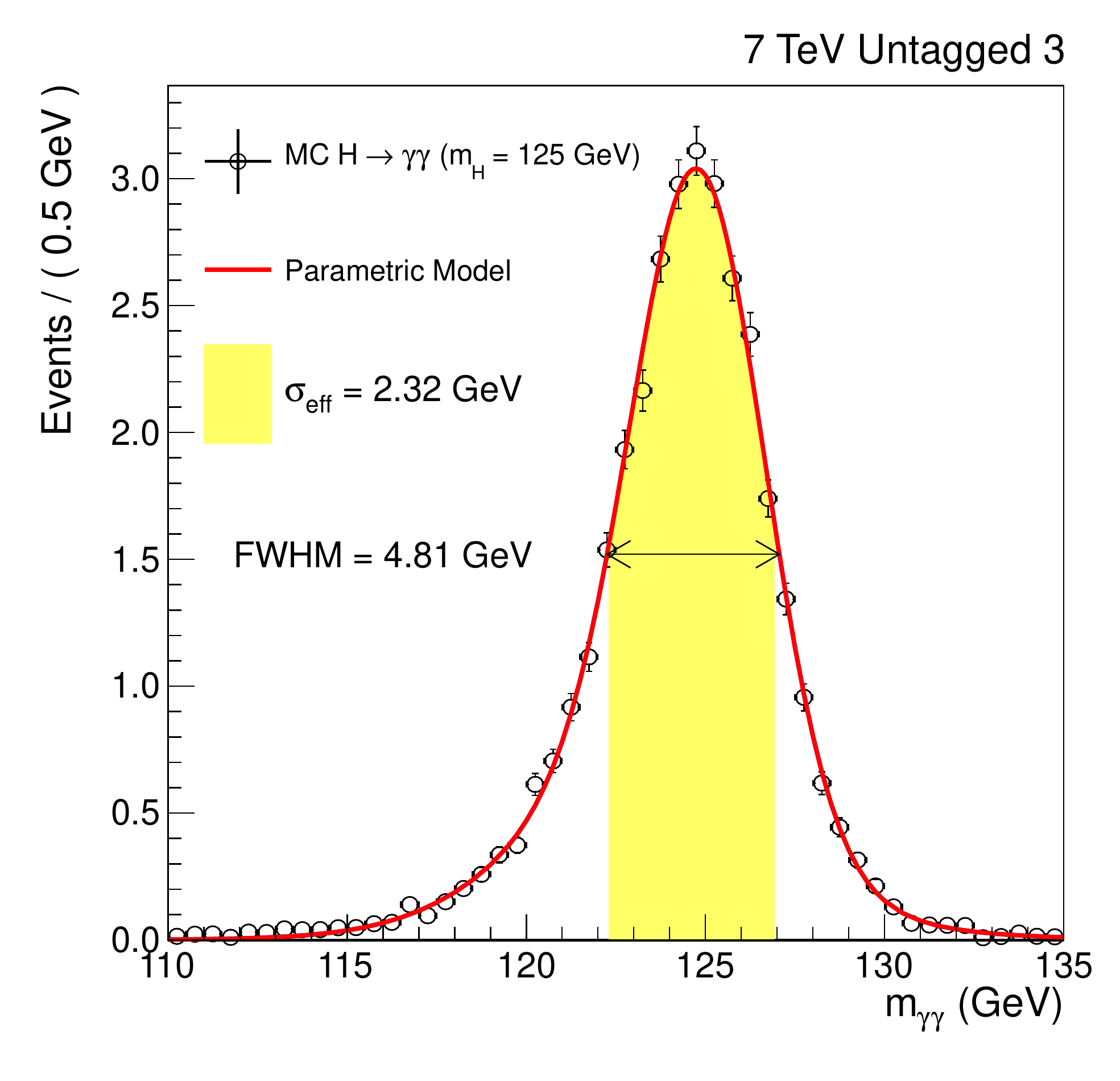}
  \end{center}
  \caption{The 7 TeV untagged classes's diphoton mass spectra (points) and the fitted distributions (red lines) of Monte Carlo $H\rightarrow \gamma\gamma$ events at a Higgs mass of 125 GeV.}
  \label{fig:sigmod inclusive 7TeV}
\end{figure}
\begin{figure}[hbpt] 
  \begin{center}
    \includegraphics[width=0.4\textwidth]{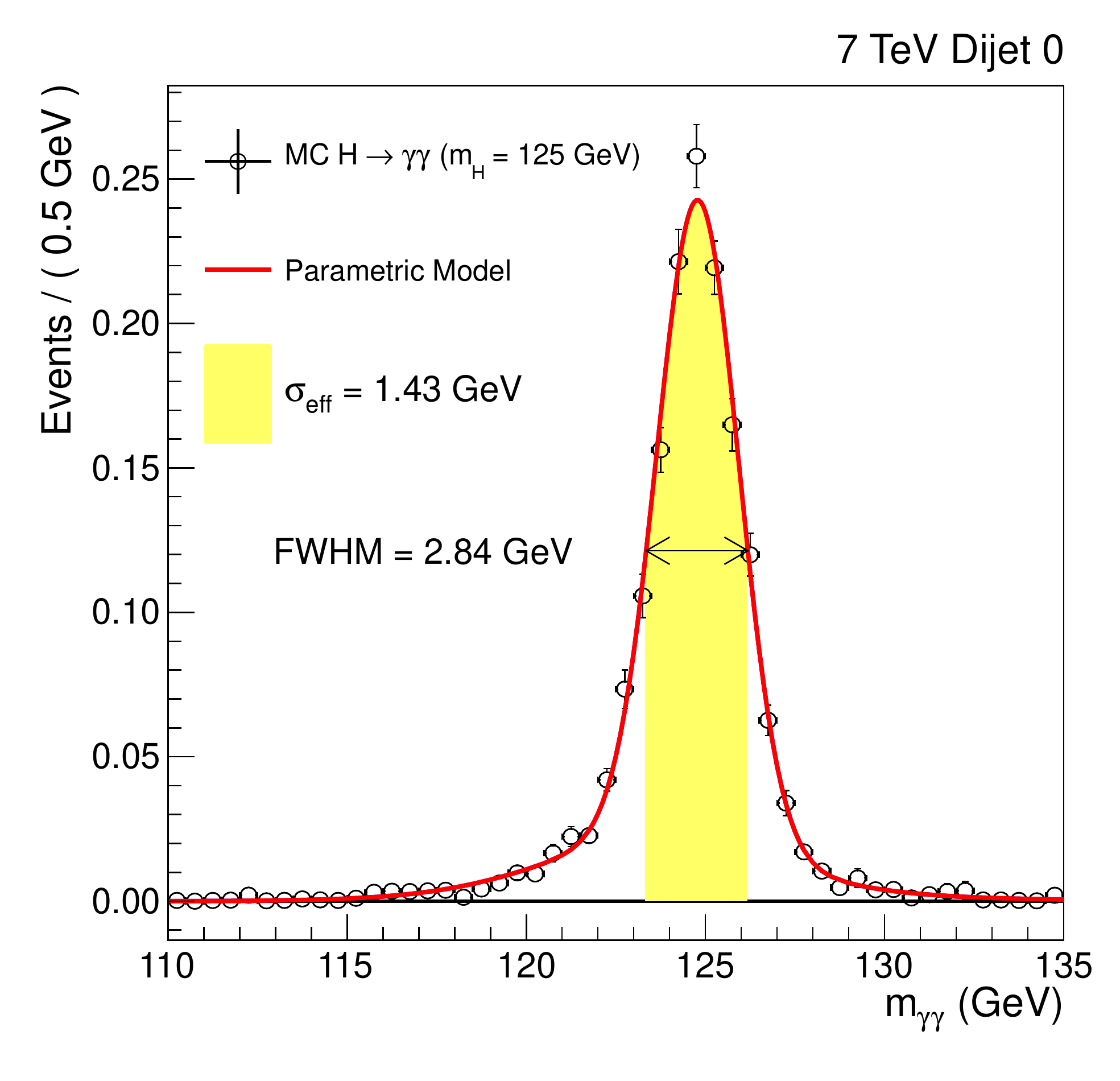}
    \includegraphics[width=0.4\textwidth]{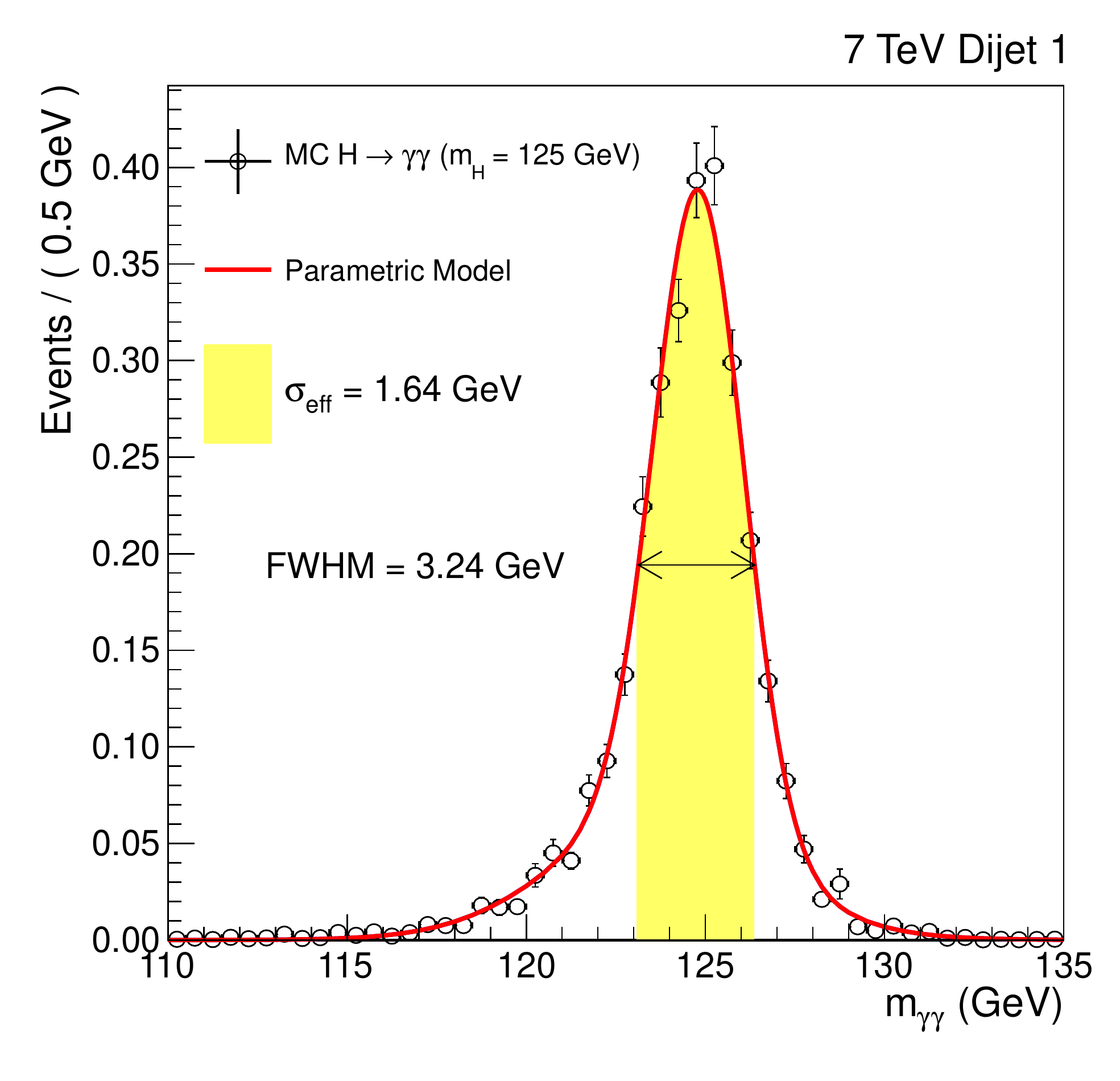}
  \end{center}
  \caption{The 7 TeV \textit{VBF} tagged classes's diphoton mass spectra (points) and the fitted distributions (red lines) of Monte Carlo $H\rightarrow \gamma\gamma$ events at a Higgs mass of 125 GeV.}
  \label{fig:sigmod vbf 7TeV}
\end{figure}
\begin{figure}[hbpt] 
  \begin{center}
    \includegraphics[width=0.4\textwidth]{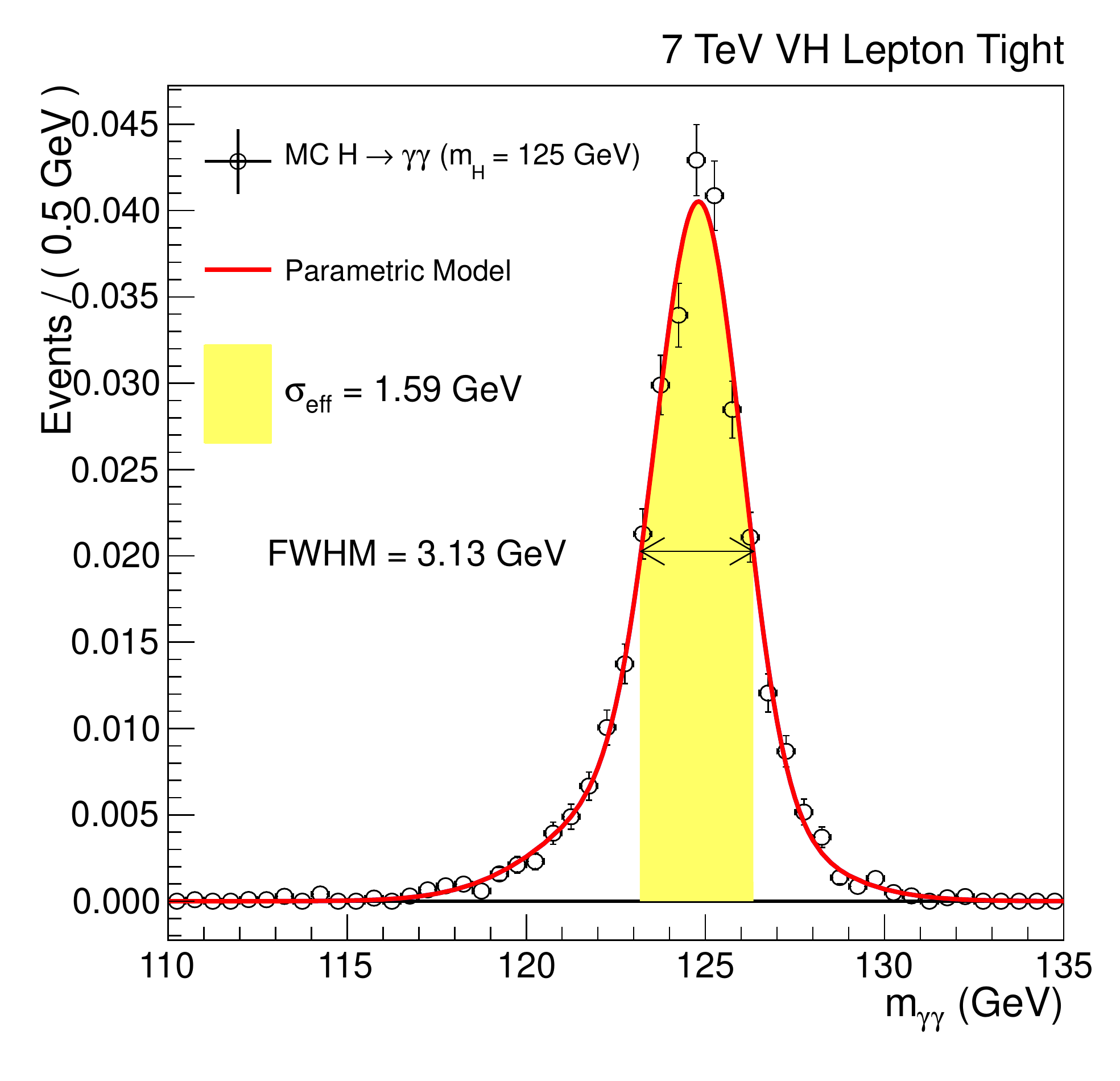}
    \includegraphics[width=0.4\textwidth]{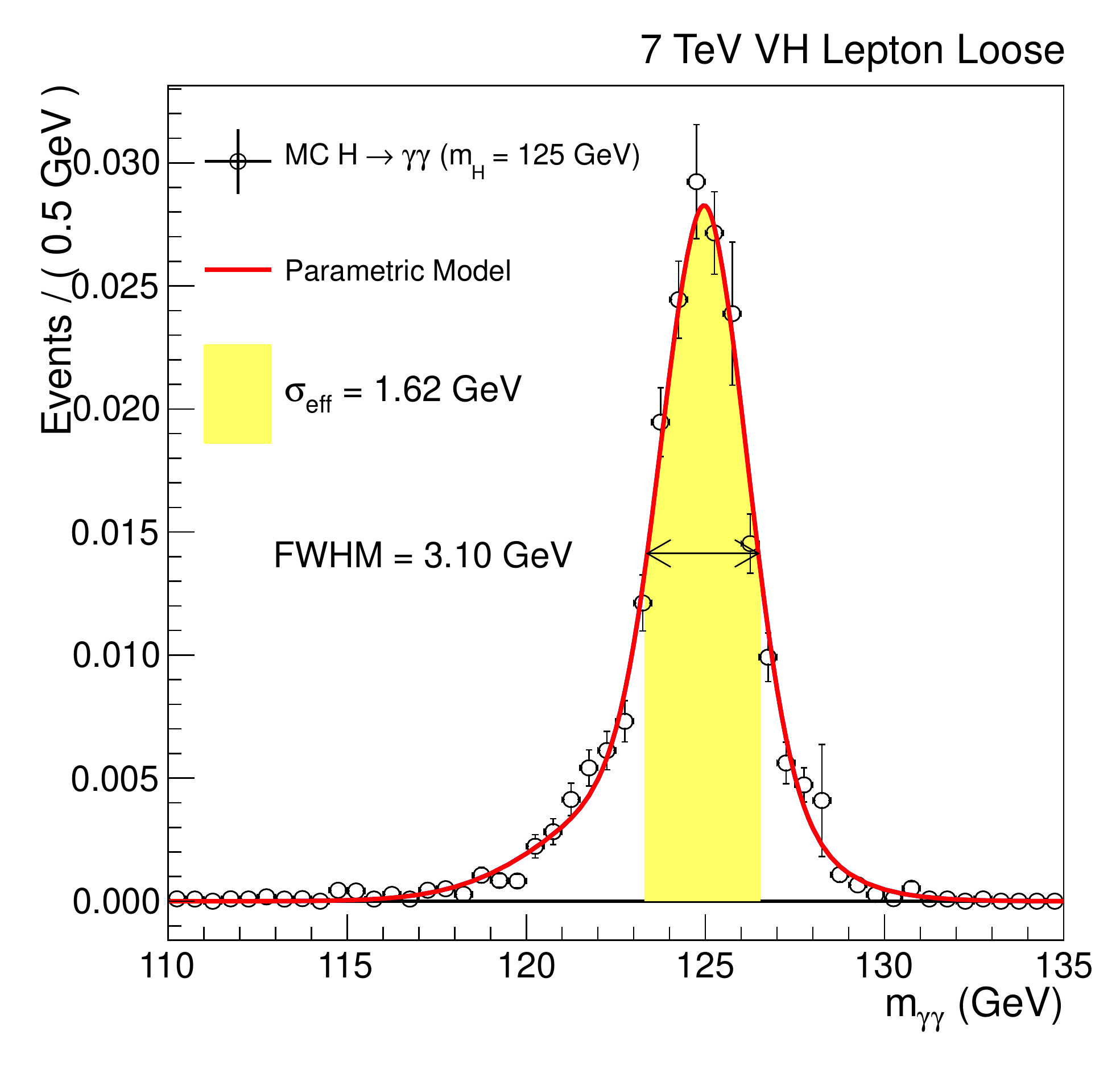}
    \includegraphics[width=0.4\textwidth]{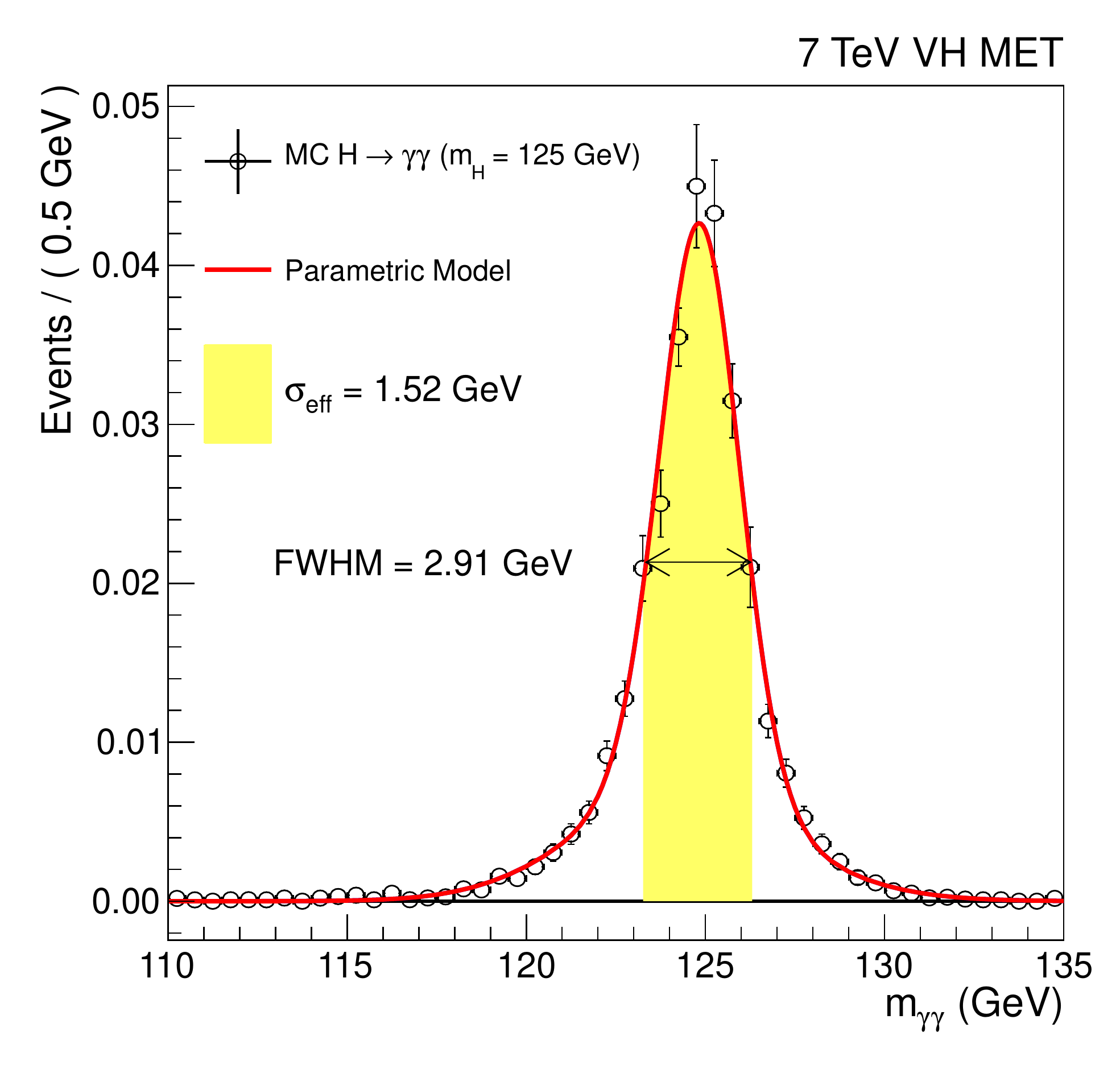}
    \includegraphics[width=0.4\textwidth]{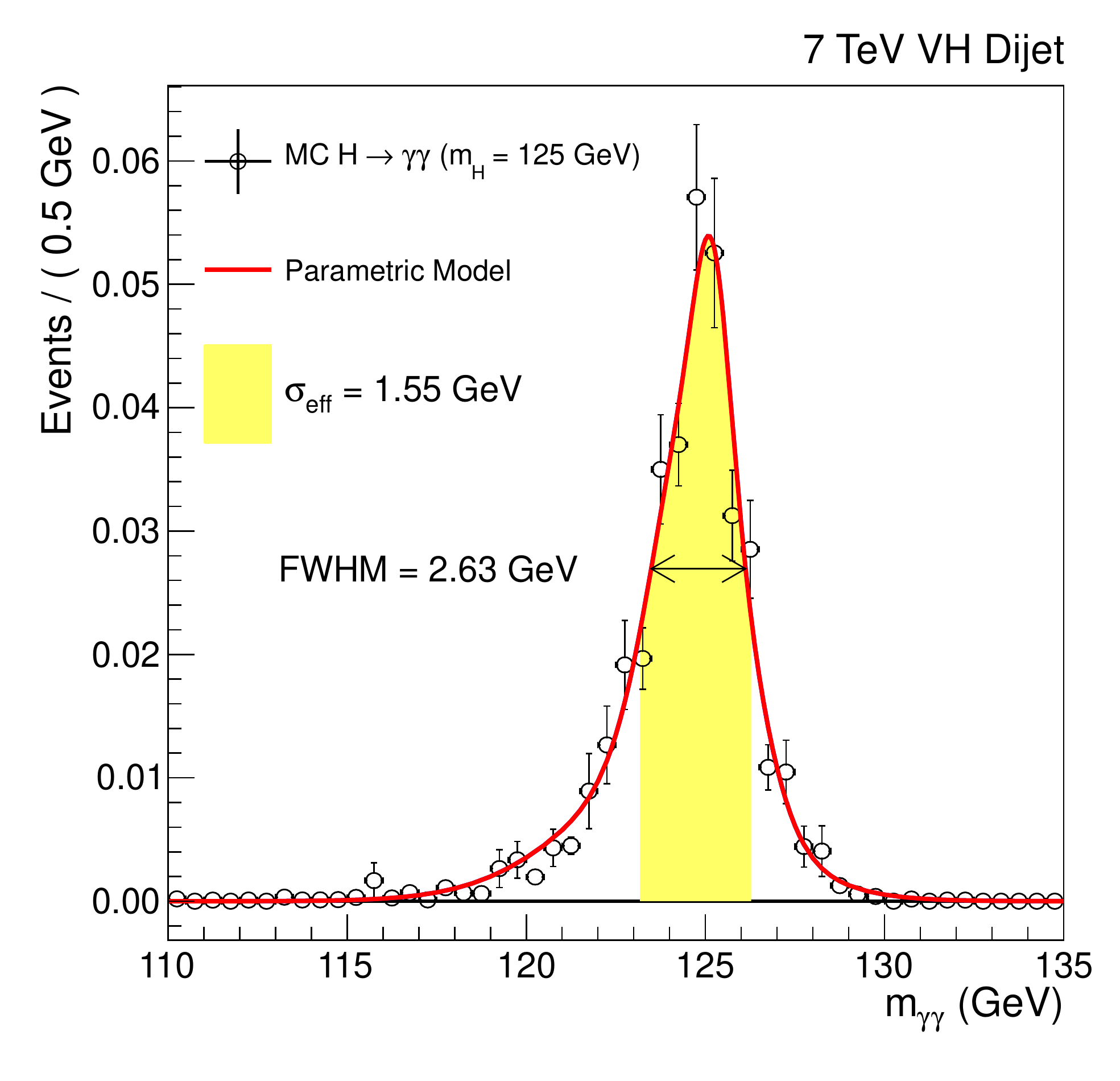}
  \end{center}
  \caption{The 7 TeV \textit{VH} tagged classes's diphoton mass spectra (points) and the fitted distributions (red lines) of Monte Carlo $H\rightarrow \gamma\gamma$ events at a Higgs mass of 125 GeV.}
  \label{fig:sigmod vh 7TeV}
\end{figure}
\begin{figure}[hbpt] 
  \begin{center}
    \includegraphics[width=0.4\textwidth]{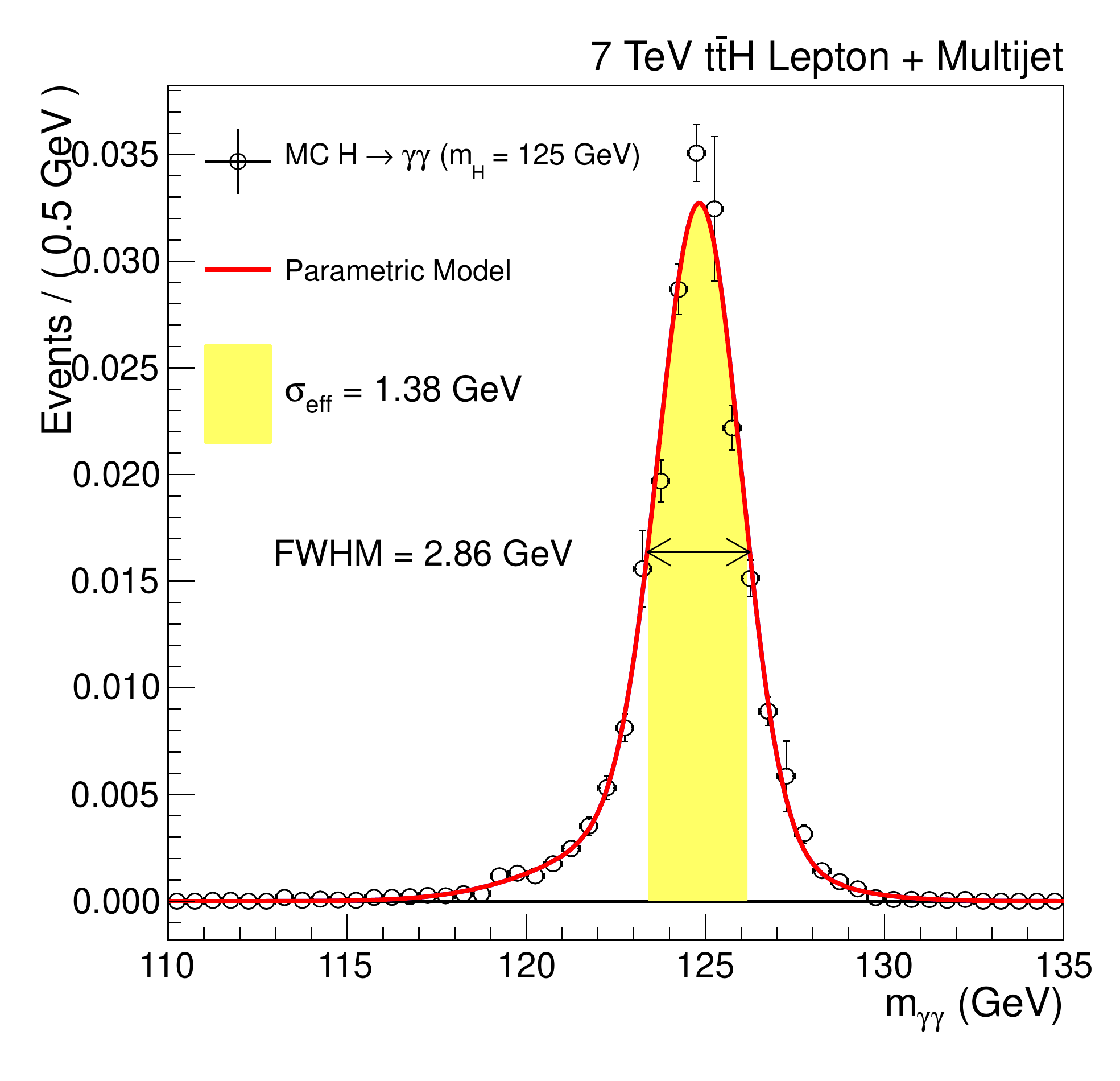}
  \end{center}
  \caption{The 7 TeV \textit{$t\overline{t}$H} tagged class's diphoton mass spectrum (points) and the fitted distribution (red line) of Monte Carlo $H\rightarrow \gamma\gamma$ events at a Higgs mass of 125 GeV.}
  \label{fig:sigmod tth 7TeV}
\end{figure}
\begin{figure}[hbpt] 
  \begin{center}
  \includegraphics[width=0.4\textwidth]{fig/sigmod/8TeV/sigmodel_125_hgg_8TeV_2013final_bdt0.pdf}
  \includegraphics[width=0.4\textwidth]{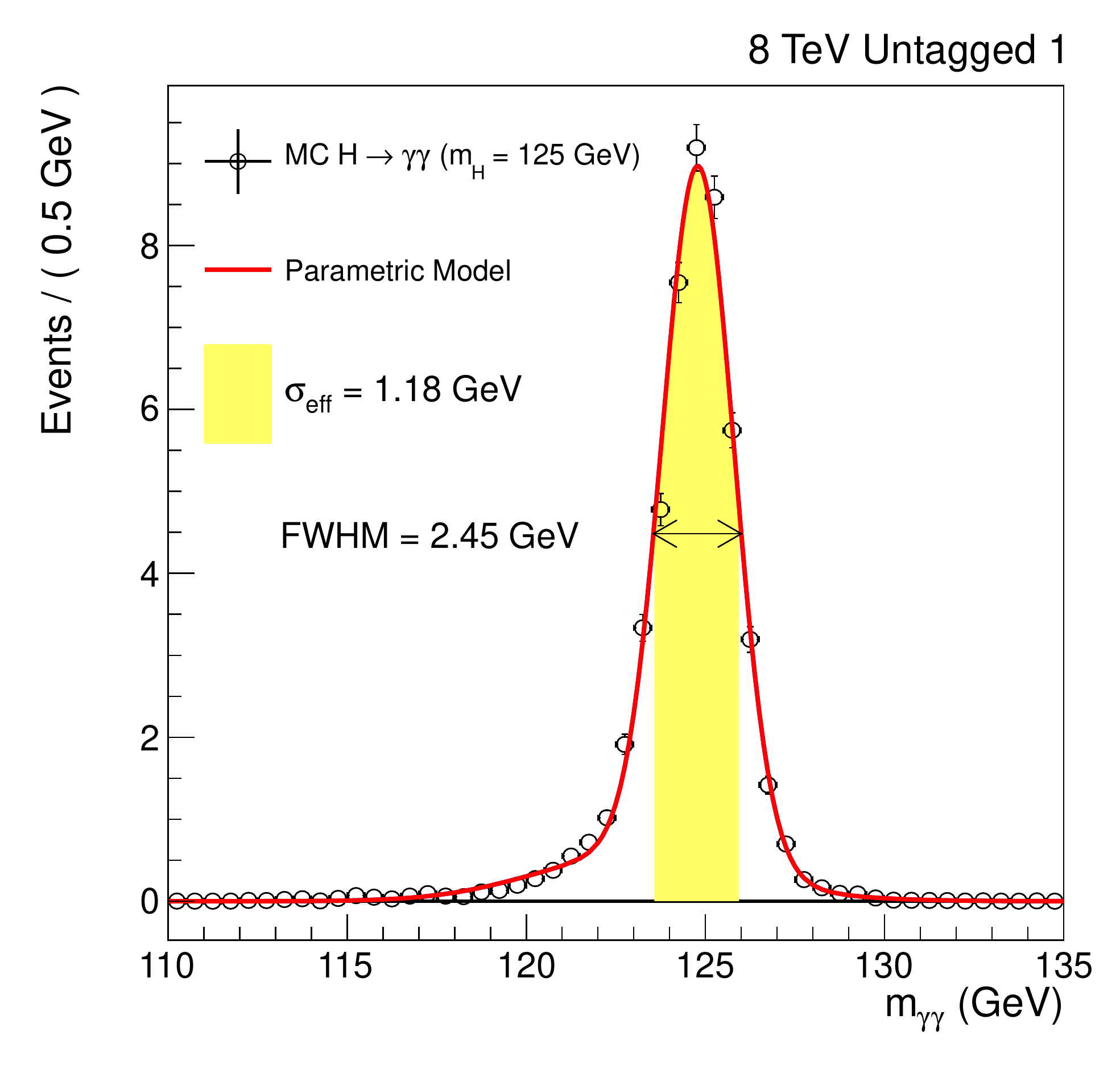}
  \includegraphics[width=0.4\textwidth]{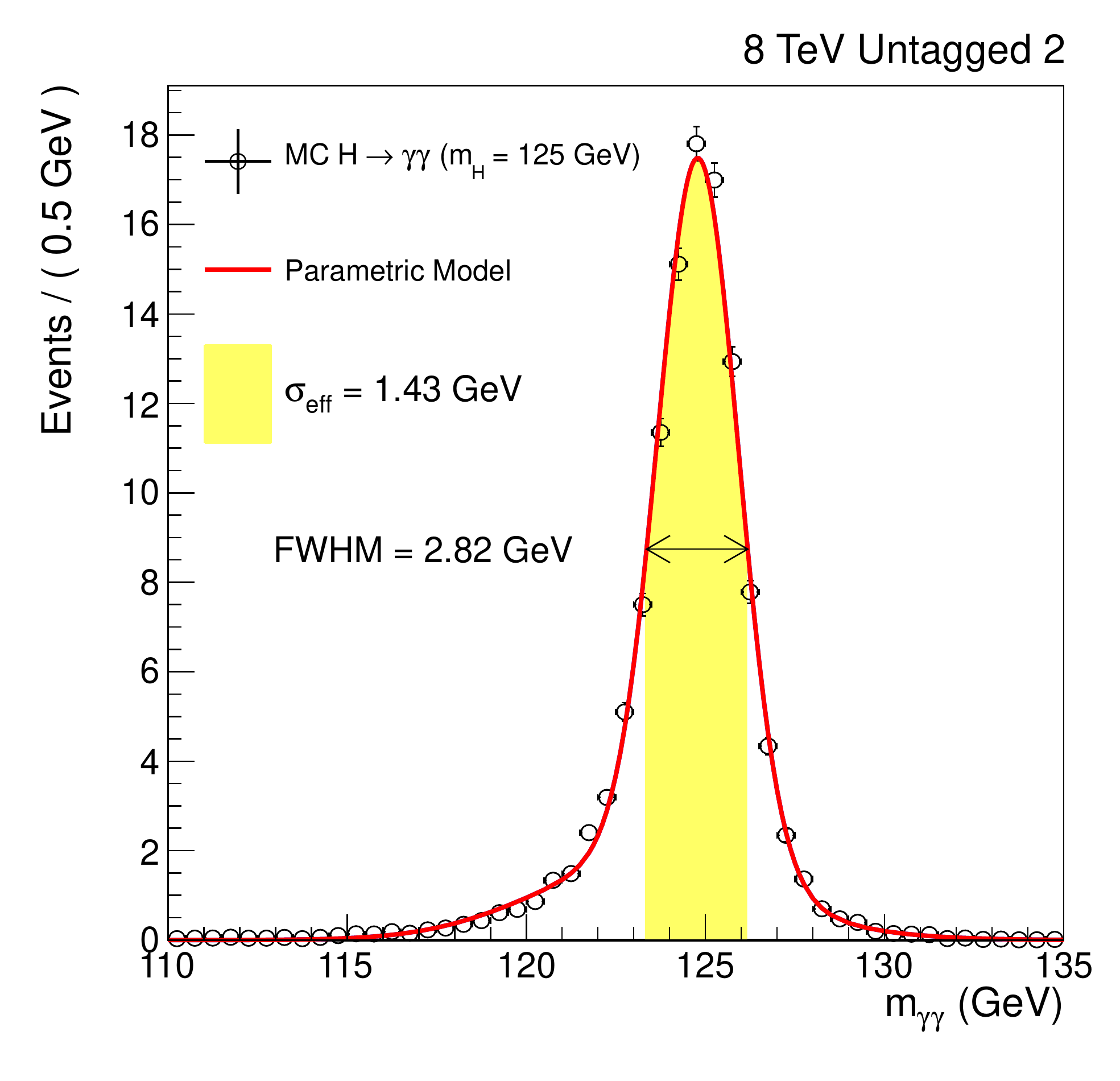}
  \includegraphics[width=0.4\textwidth]{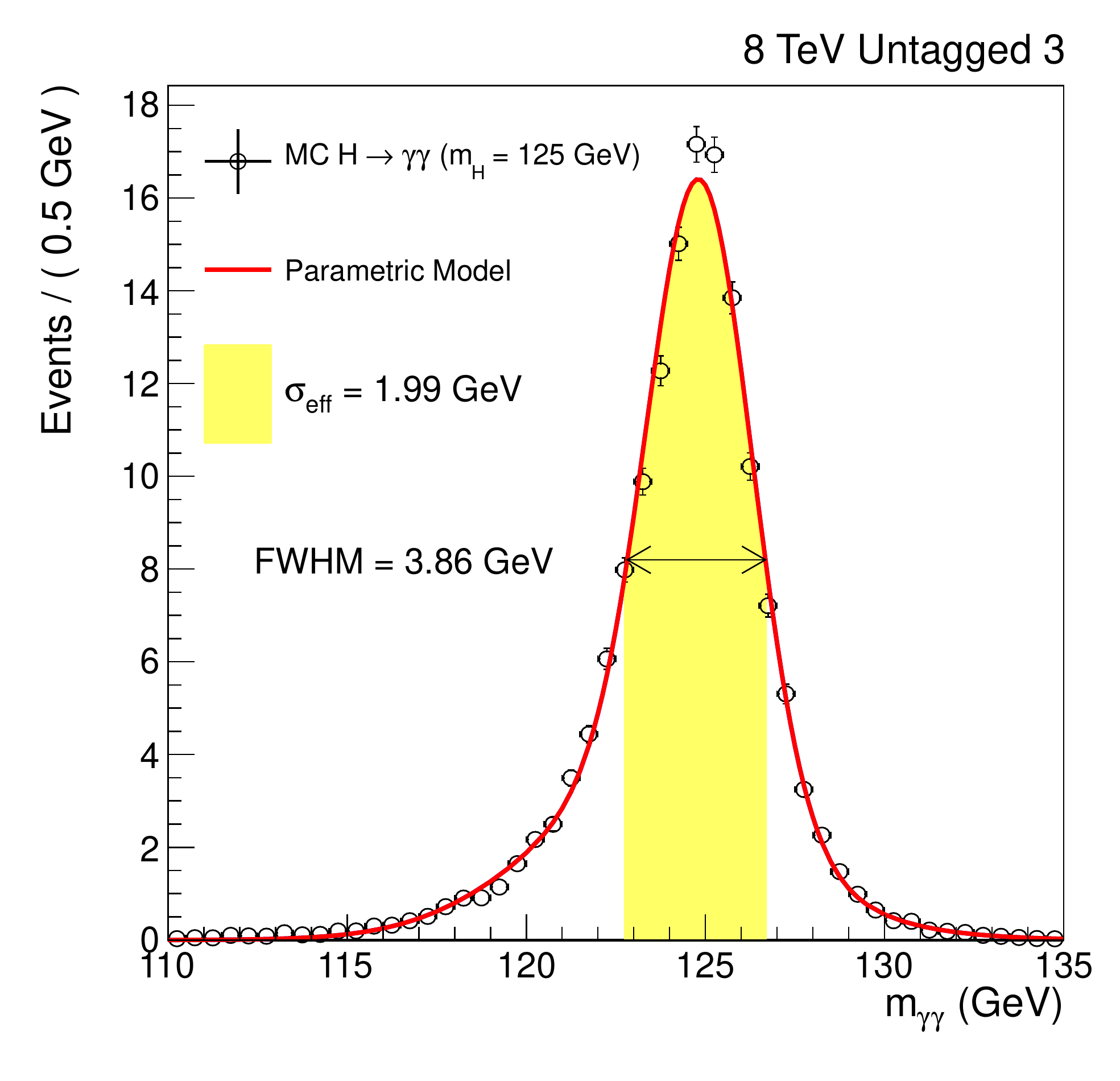}
  \includegraphics[width=0.4\textwidth]{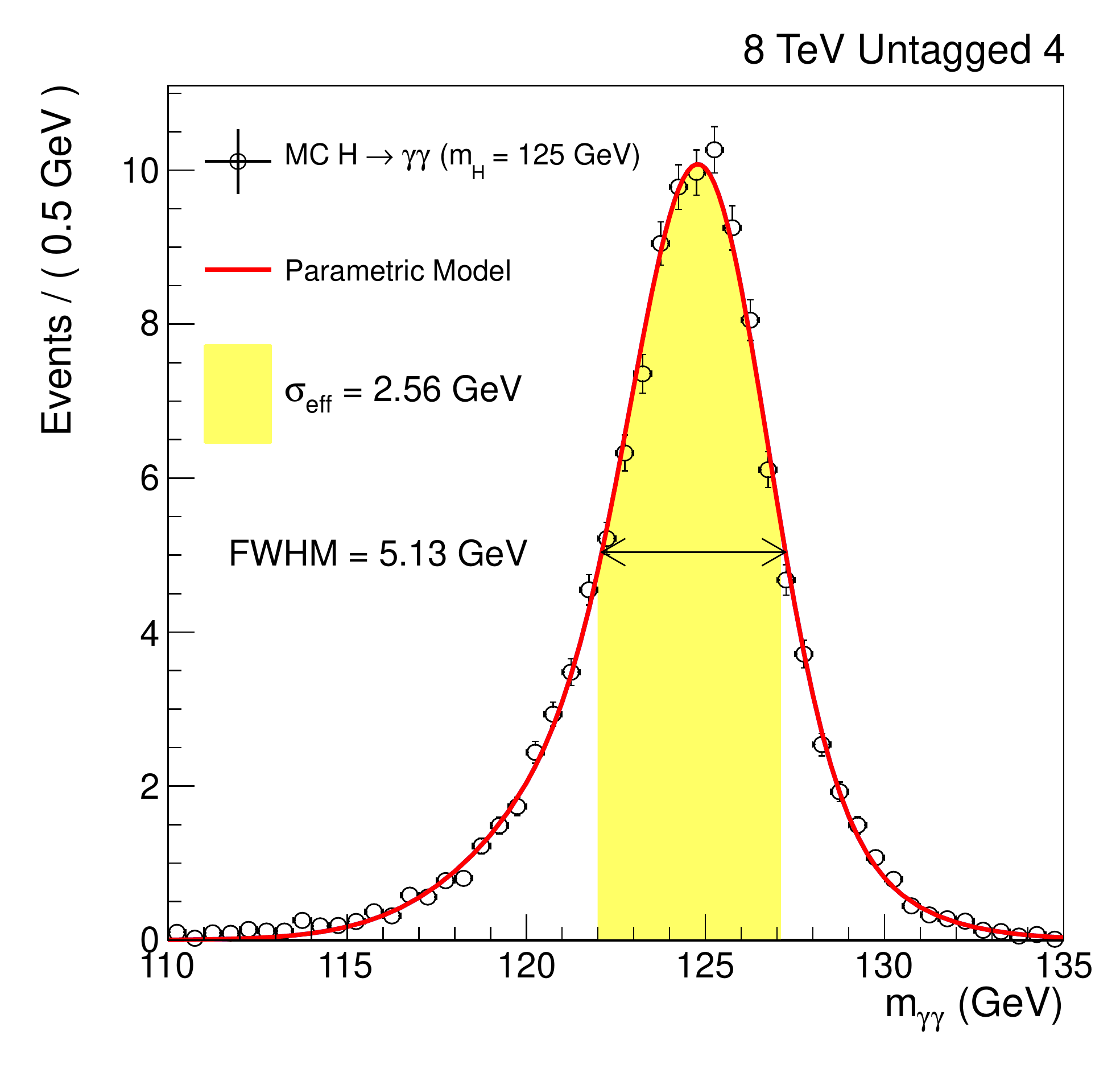}
  \end{center}
  \caption{The 8 TeV untagged classes's diphoton mass spectra (points) and the fitted distributions (red lines) of Monte Carlo $H\rightarrow \gamma\gamma$ events at a Higgs mass of 125 GeV.}
  \label{fig:sigmod inclusive 8TeV}
\end{figure}
\begin{figure}[hbpt] 
  \begin{center}
    \includegraphics[width=0.4\textwidth]{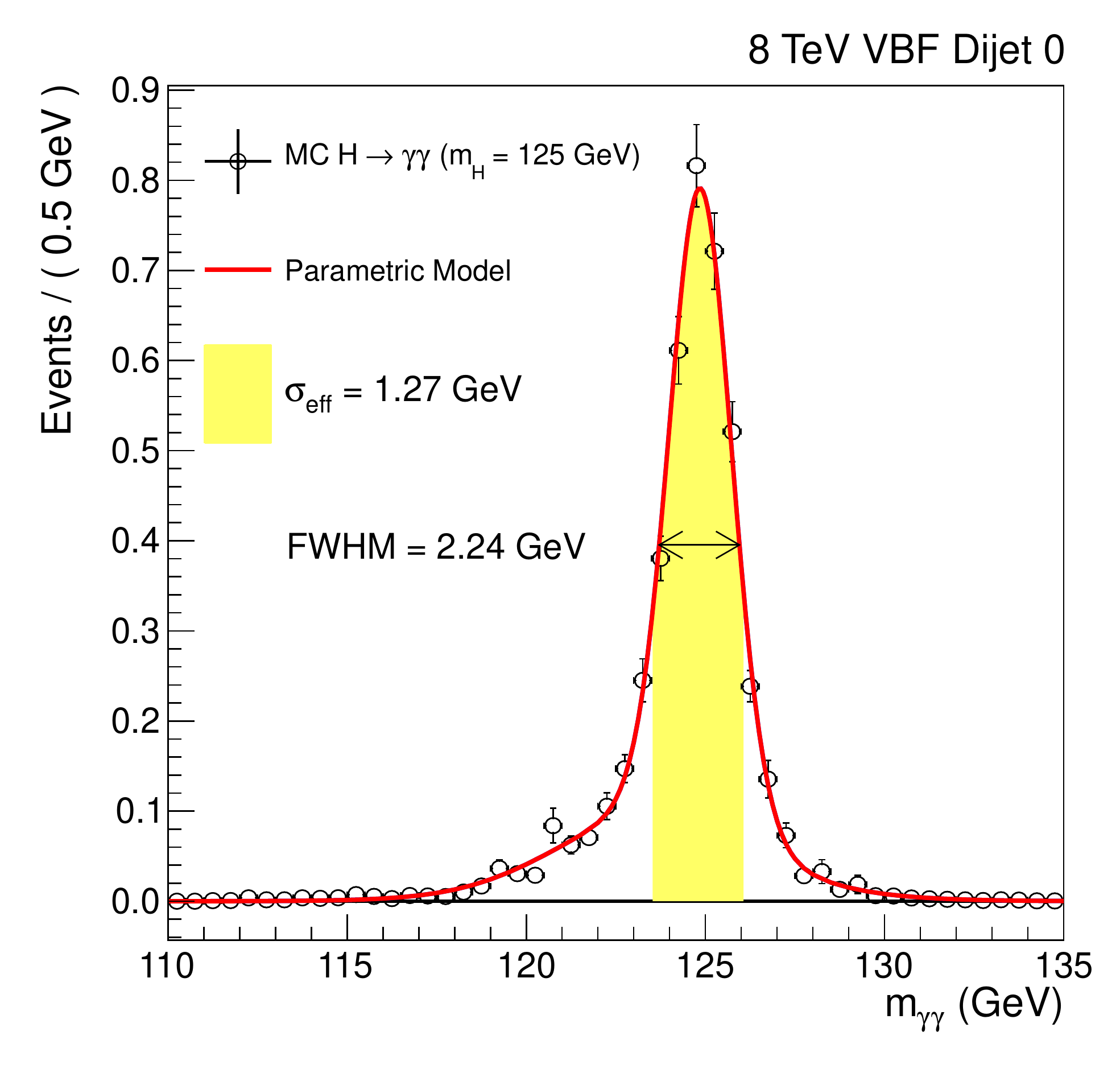}
    \includegraphics[width=0.4\textwidth]{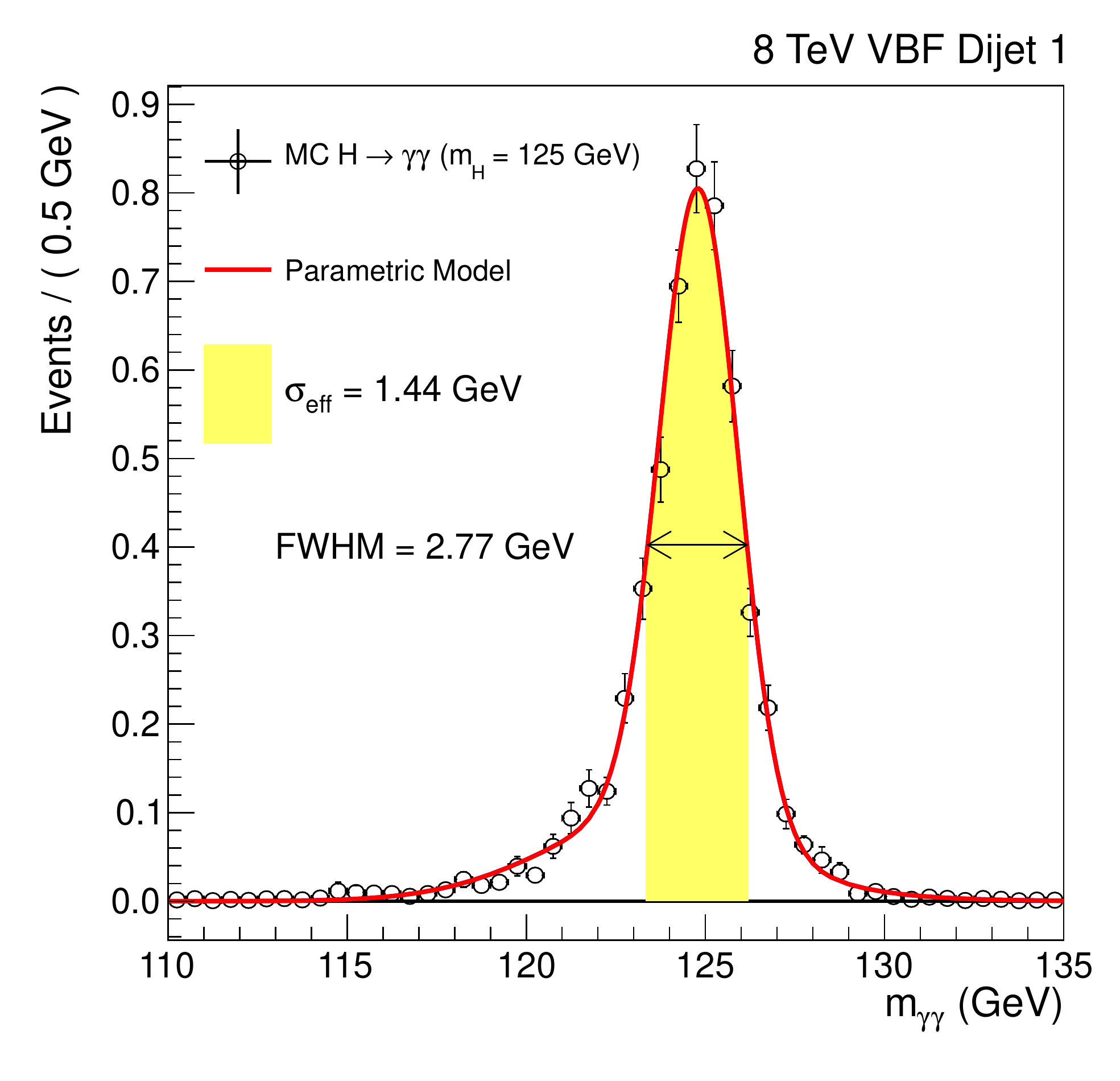}
    \includegraphics[width=0.4\textwidth]{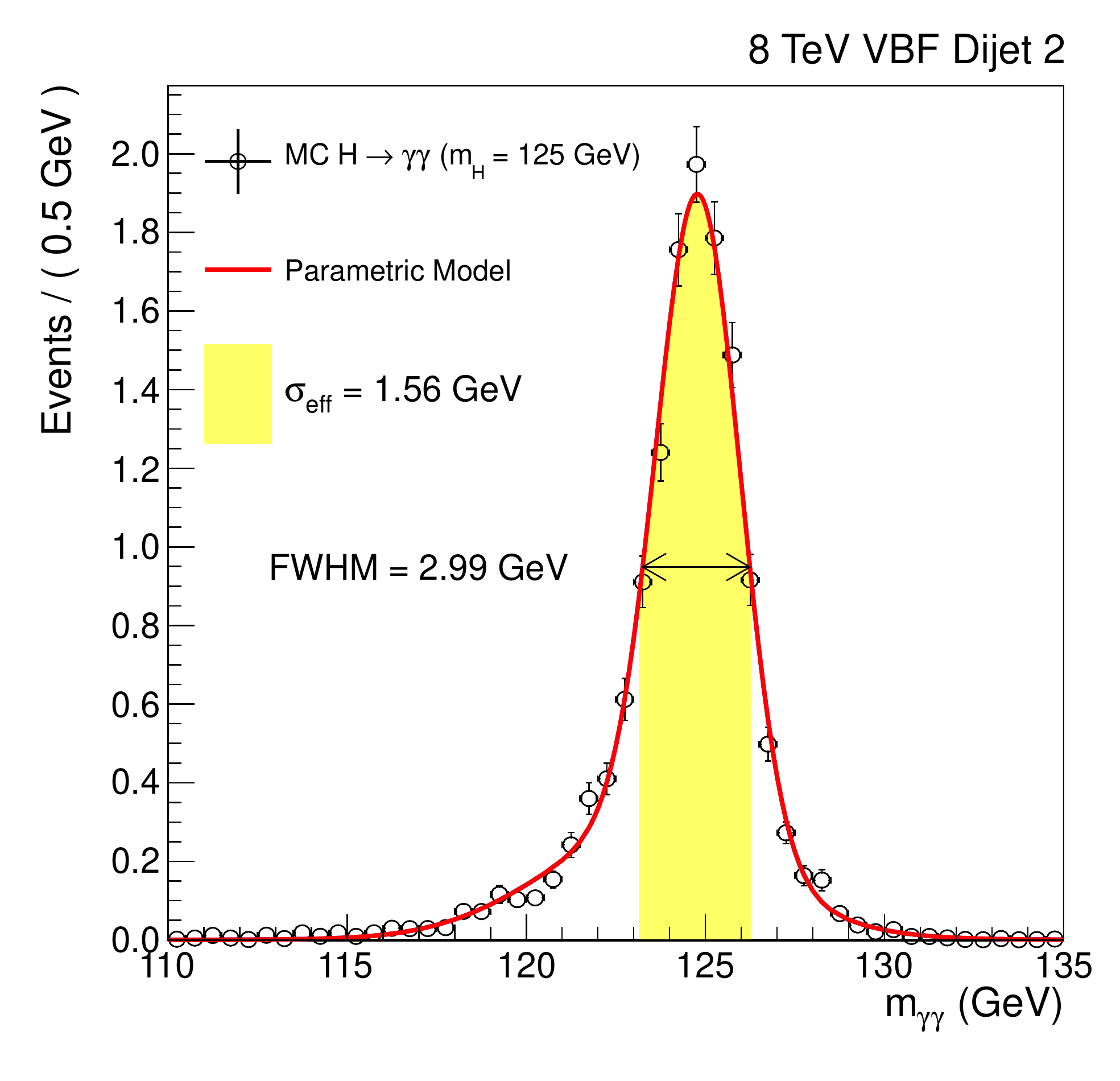}
  \end{center}
  \caption{The 8 TeV \textit{VBF} tagged classes's diphoton mass spectra (points) and the fitted distributions (red lines) of Monte Carlo $H\rightarrow \gamma\gamma$ events at a Higgs mass of 125 GeV.}
  \label{fig:sigmod vbf 8TeV}
\end{figure}
\begin{figure}[hbpt] 
  \begin{center}
    \includegraphics[width=0.4\textwidth]{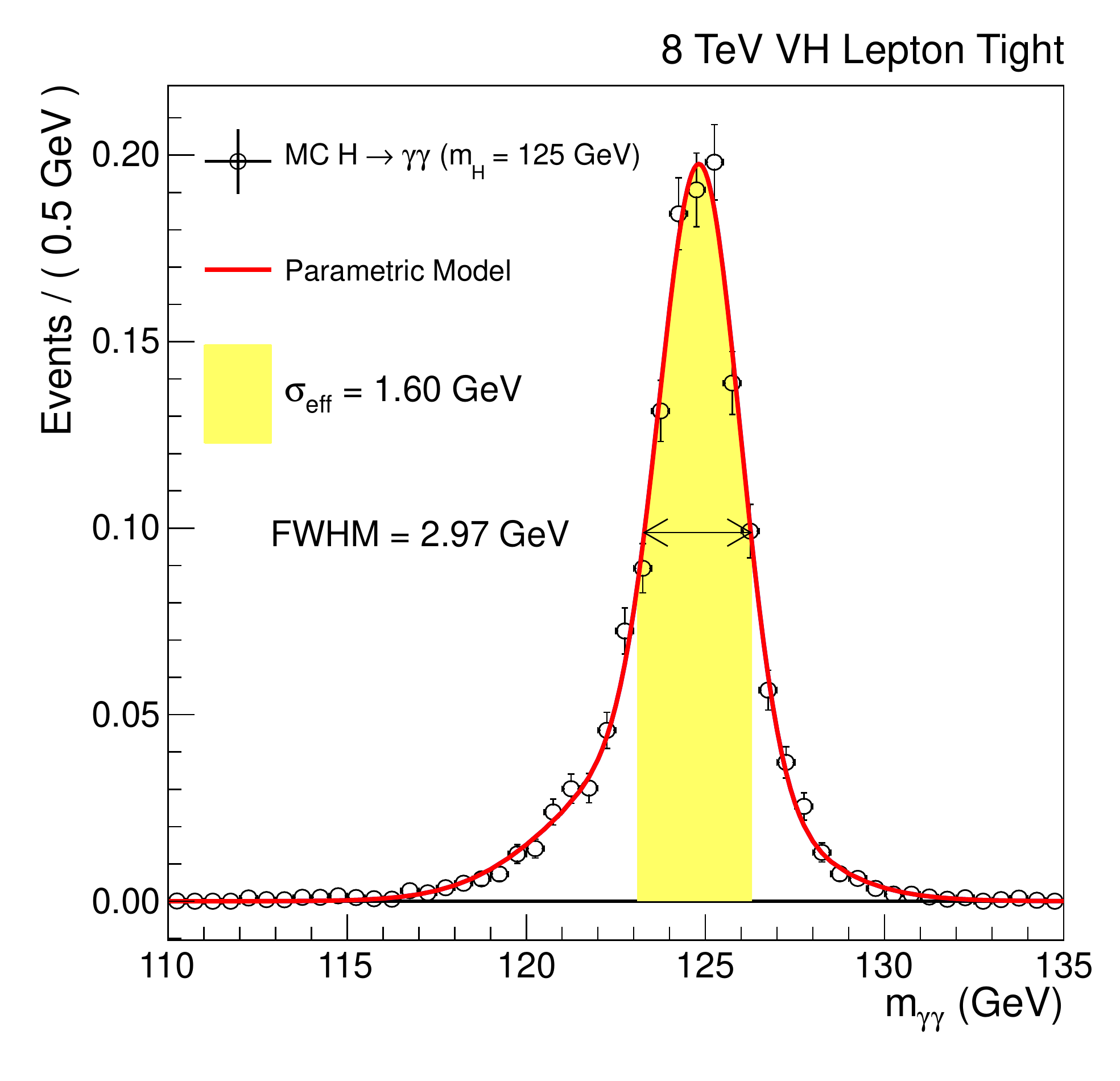}
    \includegraphics[width=0.4\textwidth]{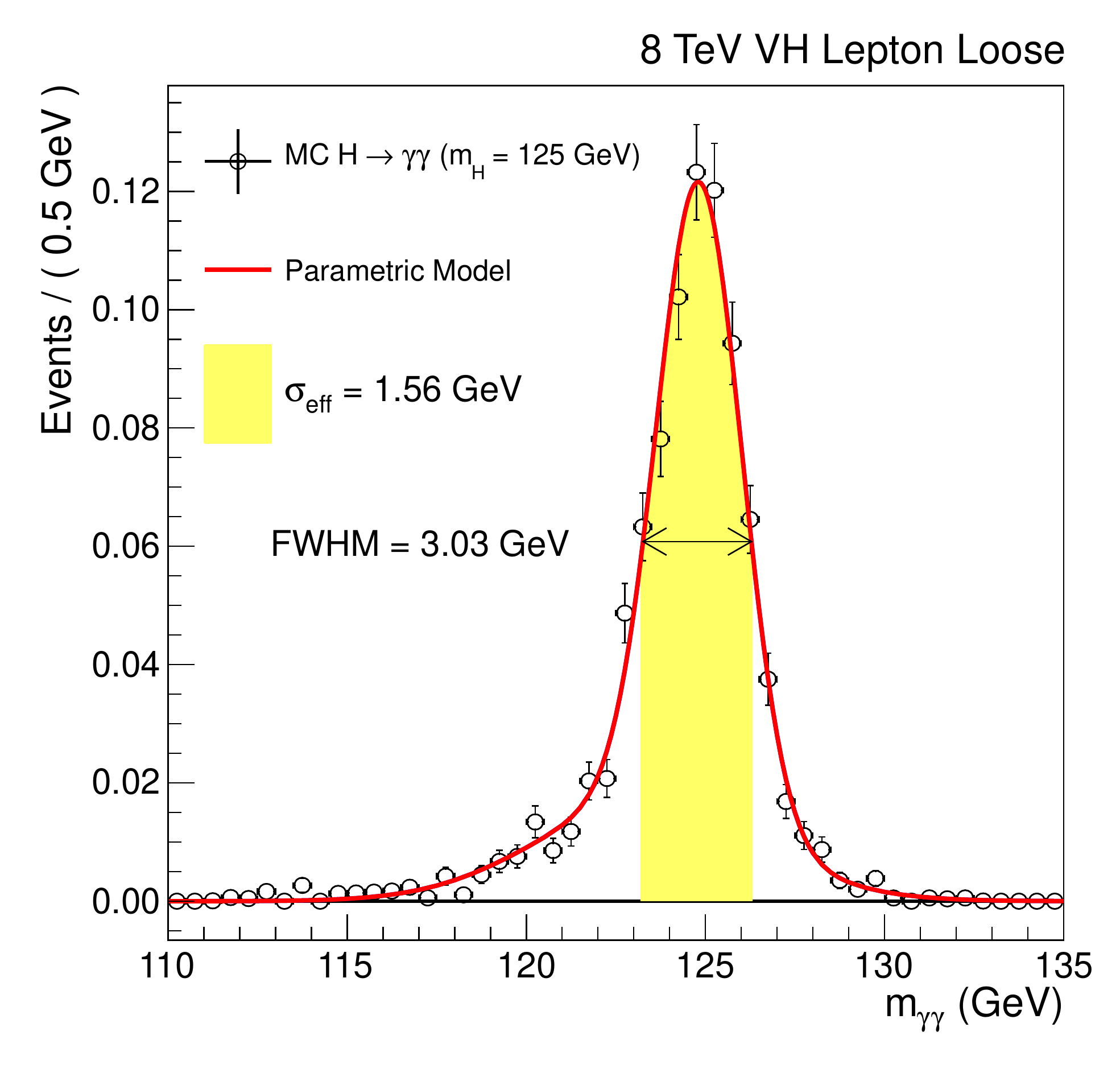}
    \includegraphics[width=0.4\textwidth]{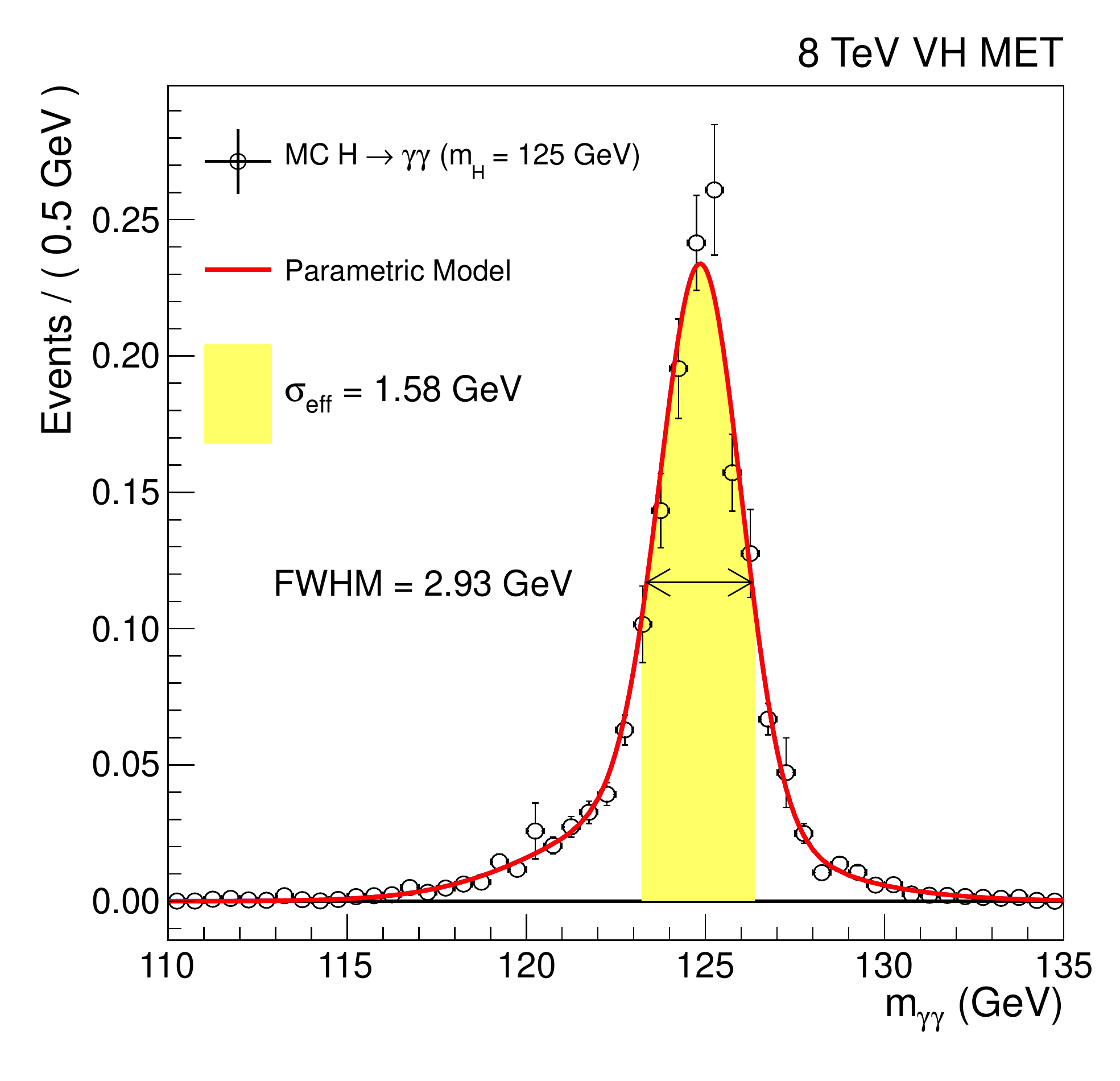}
    \includegraphics[width=0.4\textwidth]{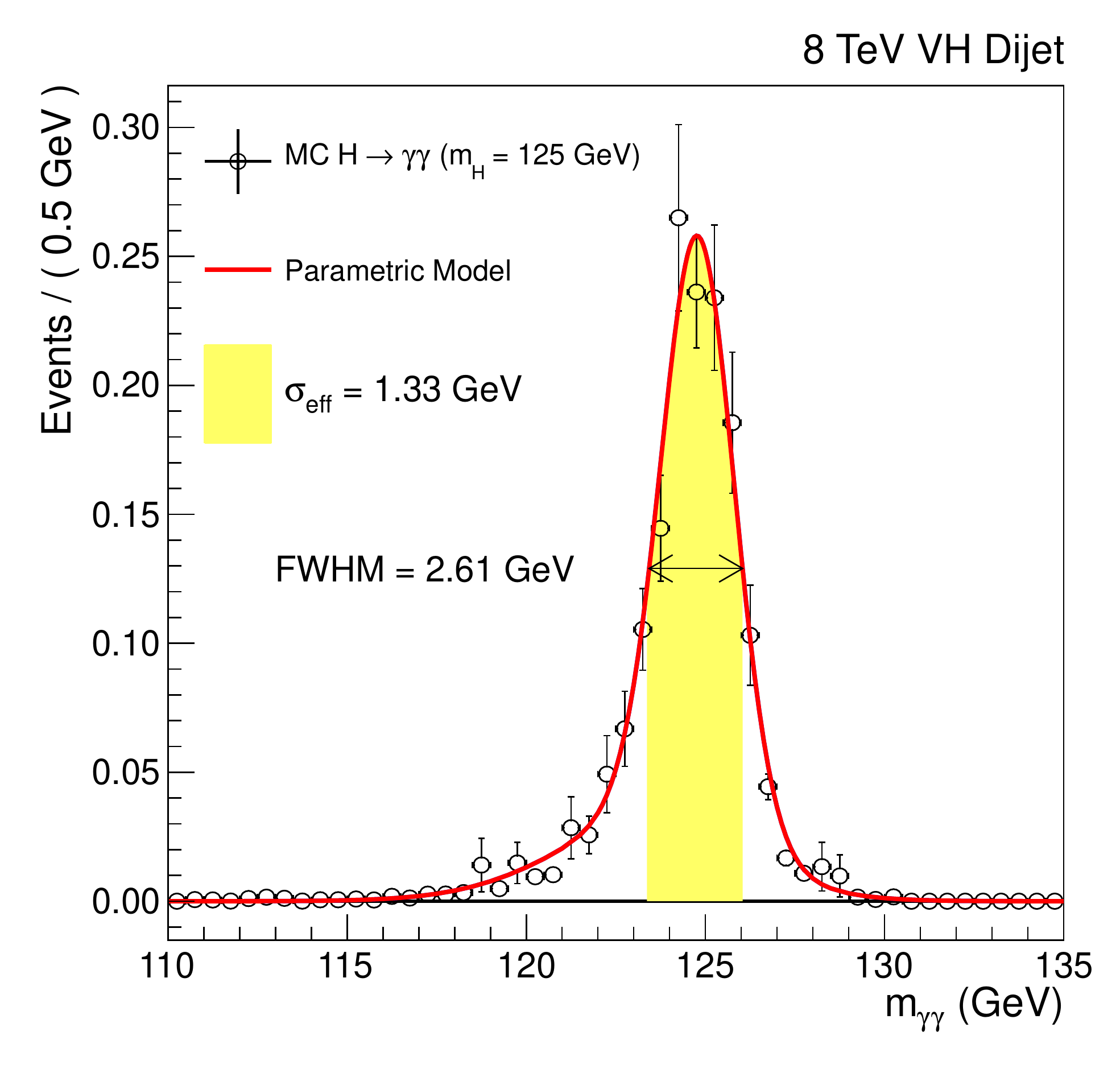}
  \end{center}
  \caption{The 8 TeV \textit{VH} tagged classes's diphoton mass spectra (points) and the fitted distributions (red lines) of Monte Carlo $H\rightarrow \gamma\gamma$ events at a Higgs mass of 125 GeV.}
  \label{fig:sigmod vh 8TeV}
\end{figure}
\begin{figure}[hbpt] 
  \begin{center}
    \includegraphics[width=0.4\textwidth]{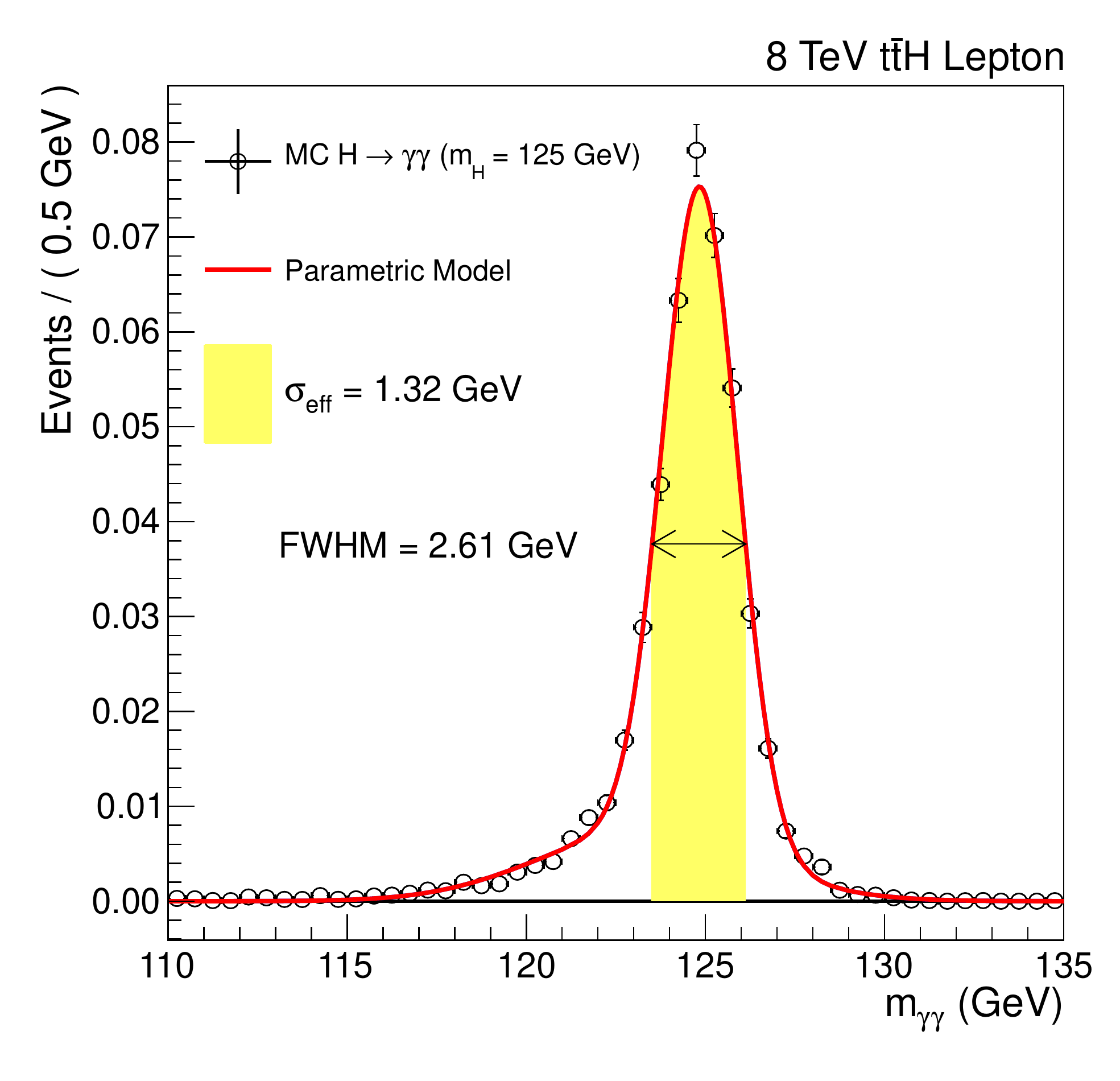}
    \includegraphics[width=0.4\textwidth]{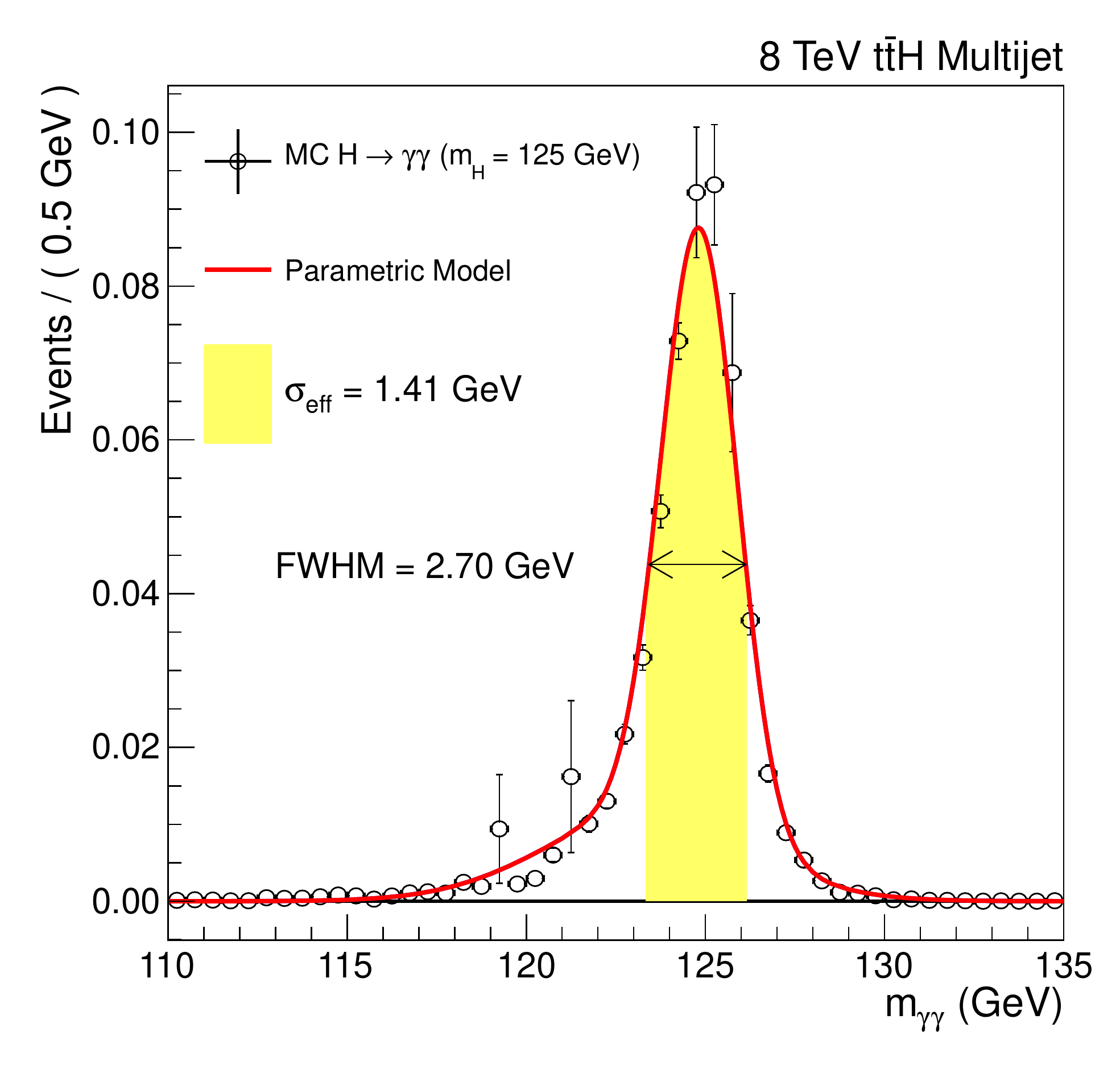}
  \end{center}
   \caption{The 8 TeV \textit{$t\overline{t}$H} tagged classes's diphoton mass spectra (points) and the fitted distributions (red lines) of Monte Carlo $H\rightarrow \gamma\gamma$ events at a Higgs mass of 125 GeV.}
  \label{fig:sigmod tth 8TeV}
\end{figure}

%% file: appb.tex
\chapter{Variables for Higgs Production Tagging}
\label{chap:Variables For Higgs Production Tagging}


\section{Variables Related to Jets}
\label{sec:Variables Related to Jets}
\begin{itemize}
\item $\sum p_{T}^{TRK PU}/\sum p_{T}^{TRK}$: the ratio between the scalar $p_{T}$ sum of tracks in the jet which match any of the pileup vertices and the scalar $p_{T}$ sum of all tracks in the jet.
\item $\sum (({p_{T}^{PF}})^{2}\cdot{\Delta{R}}^{2})/\sum ({p_{T}^{PF}})^{2}$: the average square of $\Delta{R}$ between the particle-flow candidate momentum within the jet and the jet momentum weighted by the $p_{T}^2$ of the particle-flow candidate. This measures the width of the jet. 
\item $p_{T}^{j(1,2)}$: the transverse momentum of the jet (leading, sub-leading).
\item ${\eta}^{j(1,2)}$: the pseudorapidity of the jet (leading, sub-leading).
\item $m_{jj}$: the dijet mass.
\item $N_{j}$: the number of jets.
\item $N_{B-j}$: the number of b-jets
\end{itemize}  

\section{Variables Related to Electrons}
\label{sec:Variables Related to Electrons}
\begin{itemize}
\item $d_{xy}^{e}$: the absolute impact parameter of the electron track with respect to its closest vertex in the transverse plane. 
\item $d_{z}^{e}$: the absolute impact parameter of the electron track with respect to its closest vertex in $z$.
\item $P_{ConvVtx}$: the $\chi^2$ p-value for the vertex fit of the conversion matching the electron.
\item $N_{Miss}$: the number of missing hits before the first hit of the track.
\item EleMVA: the identification score evaluated by a Multivariate Technique estimating the likelihood of being a prompt electron over the likelihood of being an electron from a jet\cite{Hzz}.
\item $\mathrm{ISO}_{RelPUCorrPFCombine03}$: the pileup corrected $p_{T}$ sum of particle-flow charged hadrons, neutral hadrons and photons within $\Delta R$ $<$ 0.3 of the electron divided by the electron $p_{T}$. The pileup contamination is estimated and subtracted by $\rho_{event}$ times an effective area.
\item  $\eta^{e}$: the pseudorapidity of electron. 
\item  $p_{T}^{e}$: the transverse momentum of electron. 
\item  $m_{ee}$: the dielectron mass.
\end{itemize}  

\section{Variables Related to Muons}
\label{sec:Variables Related to Muons}
\begin{itemize}
\item $N_{Pixel}$: the number of hits in pixel detector.
\item $N_{TRKLayer}$: the number of tracker layers with hits.
\item $N_{MuonChamber}$: the number of hits in muon chamber.
\item $N_{Matching}$: the number of muon stations with muon segments matching the tracker track.
\item $d_{xy}^{\mu}$: the absolute impact parameter of the muon track with respect to its closest vertex in the transverse plane. 
\item $d_{z}^{\mu}$: the absolute impact parameter of the muon track with respect to its closest vertex in $z$.
\item $\chi^{2}/NDF$: $\chi^{2}$ divided by number of degrees of freedom for the global muon track fit. 
\item ISO$_{RelBetaPuCorrPFCombine04}$: the pileup corrected $p_{T}$ sum of particle-flow charged hadrons, neutral hadrons and photons within $\Delta R$ $<$ 0.4 of the muon divided by the muon $p_{T}$. The pileup contamination is estimated and subtracted by 0.5 times the $p_{T}$ sum of the charged particle-flow particles within the cone associated with pileup vertices. 
\item  $\eta^{\mu}$: the pseudorapidity of muon. 
\item  $p_{T}^{\mu}$: the transverse momentum of muon. 
\item  $m_{\mu\mu}$: the dimuon mass.
\end{itemize}

\section{Variables Related to Transverse Missing Energy}
\label{sec:Variables Related to Transverse Missing Energy}
\begin{itemize}
\item $\text{MET}$: the magnitude of transverse missing energy.
\end{itemize}  

\section{Variables Related to Photons}
\label{sec:Variables Related to Photons}
\begin{itemize}
\item $|\eta_{\gamma\gamma}-\frac{\eta^{j1}+\eta^{j2}}{2}|$: the separation between the diphoton pseudorapidity and the average pseudorapidity of the dijet\cite{Rainwater:1996ud}. 
\item $\Delta\phi_{jj,\gamma\gamma}$: the separation in the azimuthal angle between dijet and diphoton. 
\item  $\Delta R_{\gamma,e}$: $\Delta R$ between photon and electron. 
\item  $\Delta R_{\gamma,etrk}$: $\Delta R$ between photon and electron track.  
\item  $m_{\gamma,e}$: the photon-electron mass. 
\item  $\Delta R_{\gamma,\mu}$: $\Delta R$ between photon and muon. 
\item  $cos(\theta^*)$: cosine of the angle between the diphoton momentum in the center of mass frame of diphoton-dijet and the total momentum of diphoton-dijet in the lab frame. 
\item $|\Delta \phi_{\gamma\gamma,j1}|$: the separation in the azimuthal angle between the diphoton and the leading jet.
\item $|\Delta \phi_{\gamma\gamma,MET}|$: the separation in the azimuthal angle between the diphoton and MET. 
\end{itemize} 

%% file: biblio.tex
\begin{singlespace}
\bibliography{main}
\bibliographystyle{ieeetr}
\end{singlespace}